%% file: crossfit-arXiv_v2.tex
\documentclass[12pt]{article}
\usepackage[margin=1in]{geometry}
\usepackage{setspace}                     
\usepackage{indentfirst}
\usepackage[round, authoryear]{natbib}
\usepackage{hyperref}

\usepackage[dvipsnames,svgnames]{xcolor}
\usepackage{color}
\usepackage{tcolorbox}

\usepackage{amsmath}
\usepackage{amssymb}
\usepackage{amsthm}
\usepackage{bbm}
\usepackage{algorithm}
\usepackage{algorithmic}

\usepackage{enumerate}
\usepackage{tikz}
\definecolor{tikzfontcolor}{HTML}{000000}
\usetikzlibrary{calc, shapes.geometric, positioning, decorations.pathreplacing, patterns, arrows.meta, backgrounds}
\usetikzlibrary{angles, calc}
\tikzset{
      dot/.style={
        circle, fill=tikzfontcolor, inner sep=1pt, outer sep=0pt
  },
  dot label/.style={
        circle, inner sep=0pt, outer sep=1pt
  },
  pics/right angle/.append style={
    /tikz/draw, /tikz/angle radius=5pt
  }
}
\newcommand{\legendBox}[1]{%
    \mbox{\tikz[baseline=-0.5ex] \node[minimum size=0.35cm, inner sep=0pt, anchor=center, #1] {};}%
}
\usepackage{standalone}

\usepackage{tabularx}
\usepackage{booktabs}
\usepackage{multirow}
\usepackage{threeparttable}
\usepackage{makecell}
\usepackage{float}

\newtheorem{theorem}{Theorem}
\newtheorem{lemma}{Lemma}
\newtheorem{corollary}{Corollary}
\newtheorem{proposition}{Proposition}

\theoremstyle{definition}
\newtheorem{definition}{Definition}
\newtheorem{condition}{Condition}
\newtheorem{example}{Example}

\theoremstyle{remark}
\newtheorem{remark}{Remark}

\definecolor{lightblue}{RGB}{0, 200, 200}
\definecolor{rosered}{RGB}{255,3,62}

\newif\ifnowrap\nowrapfalse

\newcommand{\size}[1]{\vert#1\vert}
\newcommand{\set}[1]{\{#1\}}

\newcommand{\norm}[1]{\lVert #1\rVert}
\newcommand{\bignorm}[1]{\bigl\lVert #1\bigr\rVert}
\newcommand{\Bignorm}[1]{\Bigl\lVert #1\Bigr\rVert}

\newcommand{\abs}[1]{|#1|}
\newcommand{\bigabs}[1]{\bigl| #1\bigr|}
\newcommand{\Bigabs}[1]{\Bigl| #1\Bigr|}
\newcommand{\biggabs}[1]{\biggl|#1\biggr|}

\newcommand{\id}{\boldsymbol{1}}
\newcommand{\zero}[0]{\boldsymbol{0}}

\def\hat{\widehat}
\def\f{f}
\def\hf{\hat{\f}}
\def\htau{\hat{\tau}}
\def\hlam{\hat{\lambda}}
\def\tf{\widetilde{f}}
\def\x{x}
\def\z{z}

\def\L{L}
\def\truecps{\mathcal{T}^\ast}
\def\estcps{\widehat{\mathcal{T}}}
\def\cpsset{\mathcal{T}}
\def\trunc{T}
\def\risk{\overline{L}}
\def\Lval{L}
\def\mcv{\mathsf{recv}}
\def\cv{\mathsf{cv}}
\def\cf{\mathsf{cf}}

\def\m{\mathsf{m}}

\def\snr{\mathsf{snr}}
\def\homo{\mathcal{I}^{\mathsf{nhomo}}}
\def\heter{\mathcal{I}^{\mathsf{heter}}}
\def\dacc{d_{\mathsf{acc},n}}
\def\dm{d_{\m}}
\def\leel{u}
\def\det{\mathrm{det}}

\newcommand{\Ncal}{\mathcal{N}}
\newcommand{\Kcal}{\mathcal{K}}
\newcommand{\Hcal}{\mathcal{H}}
\newcommand{\Gcal}{\mathcal{G}}
\newcommand{\Zcal}{\mathcal{Z}}
\newcommand{\Ebb}{\mathbb{E}}
\newcommand{\Pbb}{\mathbb{P}}
\newcommand{\Gbb}{\mathbb{G}}
\newcommand{\Rbb}{\mathbb{R}}
\newcommand{\qbf}{\mathbf{q}}
\newcommand{\vbf}{\mathbf{v}}
\newcommand{\Ibf}{\mathbf{I}}

\DeclareMathOperator*{\argmin}{\mathrm{arg\,min}}
\DeclareMathOperator*{\Var}{\mathrm{Var}}

\begin{document}
\title{\textbf{Changepoint Detection in Complex Models: Cross-Fitting Is Needed}}
\author{
Chengde Qian$^a$, Guanghui Wang$^b$, Zhaojun Wang$^b$, and Changliang Zou$^b$\\
{\small $^a$\it{School of Mathematical Sciences, Shanghai Jiao Tong University}}\\
{\small $^b$\it{School of Statistics and Data Science, Nankai University}}
}
\date{}

\maketitle

\begin{abstract}
Changepoint detection is commonly formulated by minimizing the sum of in-sample losses to quantify the model's overall fit. However, for flexible modeling procedures---especially those involving high-dimensional parameter spaces or hyperparameter tuning---this strategy can lead to inaccurate changepoint estimation due to over-adaptivity biases. To mitigate this issue, we propose a novel cross-fitting methodology based on out-of-sample loss evaluations, which decouples model fitting from changepoint search. We establish a general theoretical framework for consistent changepoint estimation under mild conditions, and further extend it to temporally dependent data. A key implication of the theory is that consistency depends primarily on the models' predictive accuracy over nearly homogeneous segments. Numerical experiments show that the proposed method substantially improves the reliability and adaptability of changepoint detection in complex scenarios.
\end{abstract}

\vspace{0.2cm}

\newpage
\tableofcontents
\newpage

\section{Introduction}\label{sec:intro}
In modern big data analytics, datasets often exhibit shifts in distributions over time or space.
Consider a sequence of observations $\{\z_i\}_{i=1}^n$, where each $\z_i$ takes values in $\mathcal{Z}$ and has distribution $P_i$.
We focus on the following multiple changepoint model:
\begin{align}\label{MCP}
    P_i = P_{\tau_k^\ast},\ \ i\in(\tau^\ast_{k-1},\tau^\ast_k],\ k=1,\ldots,K^\ast+1;\quad i=1,\ldots,n,
\end{align}
where $\{\tau^\ast_k\}_{k=1}^{K^\ast}$ are the $K^\ast$ changepoints that divide the data into $K^\ast+1$ contiguous segments, with the conventions $\tau^\ast_0=0$ and $\tau^\ast_{K^\ast+1}=n$.
The samples in $(\tau^\ast_{k-1},\tau^\ast_k]$ share a common distribution $P_{\tau_k^\ast}$.
We aim to detect changes in a model/parameter $\f_i^\ast$, determined by the distribution $P_i$, which may be high- or infinite-dimensional.

Changepoint detection aims to identify the true segmentation $\truecps=(\tau^\ast_1,\ldots,\tau^\ast_{K^\ast})$.
This typically involves minimizing the sum of segmentwise losses to quantify the model's overall fit across all data segments \citep{MR2796565,MR3012396,killick_optimal_2012-1,MR3210993}.
Define a potential segmentation as $\cpsset=(\tau_1,\ldots,\tau_K)$, comprising $K$ ordered integers $\{\tau_k\}_{k=1}^K$ (with $\tau_0=0$ and $\tau_{K+1}=n$), and denote its cardinality by $\size{\cpsset}$. We define the objective function by
\begin{equation}\label{equ:mcp_loss}
    \mathcal{L}(\cpsset) := \sum_{k=1}^{K+1}\Lval_{(\tau_{k-1},\tau_k]} + \gamma\size{\cpsset},
\end{equation}
where $\Lval_{I}$ represents the empirical loss over a segment $I\subset(0,n]$ and $\gamma\size{\cpsset}$ acts as a penalty to discourage over-segmentation, with $\gamma>0$ being a prespecified parameter. If the true number of changepoints is known, one can set $\gamma=0$ and $K=K^\ast$. The resulting changepoint estimator is $\estcps = (\htau_{1},\ldots,\htau_{\widehat{K}})=\argmin_{\cpsset}\mathcal{L}(\cpsset)$.

Minimizing Eq.~\eqref{equ:mcp_loss} across all potential segmentations is often solved efficiently using dynamic programming techniques \citep{MR978902,jackson2005algorithm,killick_optimal_2012-1}.
Approximate solutions can also be obtained through algorithms like binary segmentation and its variants \citep{fryzlewicz_wild_2014,kovacs_seeded_2022} or moving windows \citep{Hao+SelenaNiu+Zhang-2013,MR3706768}.
For a comprehensive review, see \cite{truong2020selective}.

Commonly, the \textit{in-sample} loss is adopted in \eqref{equ:mcp_loss}:
\begin{equation}\label{obj:inSample}
    \mathcal{L}_{\mathsf{in}}(\cpsset) := \sum_{k=1}^{K+1}\L(\z_{(\tau_{k-1},\tau_k]};\hf_{(\tau_{k-1},\tau_k]}) + \gamma\size{\cpsset},
\end{equation}
where $\hf_I$ denotes the model estimator for segment $I$ and $\L(\z_I;\f) = \sum_{i \in I} \ell(\z_i; \f)$ evaluates the fit of the model $\f$ to the data segment $\z_I=\{\z_i:i\in I\}$ with some individual loss function $\ell(\cdot; \cdot)$. Generally, $\f_i^\ast$ is defined as the population loss minimizer $\f_i^\ast = \argmin_{\f\in\mathcal{F}} \Ebb[\ell(\z_i; \f)]$, where $\mathcal{F}$ is a feasible model space.

Traditional changepoint detection approaches typically employ models characterized by finite-dimensional parameters, estimated via least squares or maximum likelihood techniques \citep{MR2743035}.
With the in-sample loss, a fundamental principle in this framework is that the estimator $\hf_I$ must uniformly approximate the target minimizer (referred to as the \textit{estimand}),
$$\f_{I}^\ast = \argmin_{\f\in\mathcal{F}} \bigl\{ \risk_I(\f) := \Ebb[\L(\z_I;\f)] \bigr\},$$
across all potential segments $I$, regardless of whether they contain changepoints \citep{MR919373,bai1998estimating,MR1782480}.
This principle of \textit{uniform consistency} is crucial for successful changepoint detection. To illustrate, consider the theoretical \textit{oracle losses} $\risk_I(\f_{I}^\ast)$.
By substituting them into \eqref{equ:mcp_loss} we get an oracle objective function
\begin{equation*}
    \overline{\mathcal{L}}_{\mathsf{oracle}}(\cpsset) := \sum_{k=1}^{K+1}\risk_{(\tau_{k-1},\tau_k]}(\f_{{(\tau_{k-1}, \tau_k]}}^\ast) + \gamma\size{\cpsset},
\end{equation*}
which is minimized precisely when the candidate changepoints match the true changepoints, i.e., $\cpsset=\truecps$ \citep{londschien2022changeforest}. Intuitively, the uniform consistency of $\hf_I$ regarding $\f_{I}^\ast$ ensures that the in-sample loss $\L(\z_I; \hf_I)$ closely tracks the oracle one $\risk_I(\f_{I}^\ast)$, thereby allowing consistent recovery of $\truecps$.

For example, consider detecting changes in the mean of a sequence of independent univariate normal observations, where $\mathcal{Z}=\mathbb{R}$ and $P_i=\Ncal(\f_i^\ast,\sigma^2)$ with time-varying means $\f_i^\ast$ and common variance $\sigma^2$.
A standard approach employs the squared loss $\L(\z_I;\f)=\sum_{i\in I}(\z_i-\f)^2$.
In this case, the estimand is the population mean over the segment: $\f_{I}^\ast = \argmin_{\f\in\mathbb{R}}\Ebb[\sum_{i\in I}(\z_i-\f)^2] = \size{I}^{-1}\sum_{i\in I}\f_i^\ast$.
The oracle objective function becomes:
\begin{align*}
    \overline{\mathcal{L}}_{\mathsf{oracle}}(\cpsset) = \sum_{k=1}^{K+1}\sum_{i\in(\tau_{k-1},\tau_k]}(\f_{i}^\ast - \f_{I}^\ast)^2 + n\sigma^2 + \gamma\size{\cpsset}.
\end{align*}
This function clearly attains its global minimum at $\cpsset=\truecps$, where $f_i^*=\f_{(\tau^\ast_{k-1},\tau^\ast_k]}^\ast$ for all $i\in(\tau^\ast_{k-1},\tau^\ast_k]$.
In this setting, the minimizer of empirical loss $\L(\z_I;\f)$ is the sample average $\hf_I = \size{I}^{-1}\sum_{i\in I}\z_i$, which uniformly converges to $f_I^{\ast}$ under mild conditions; consequently, minimizing \eqref{equ:mcp_loss} produces a consistent estimate of $\truecps$ \citep{MR919373}.

Recent years have seen a surge in incorporating flexible machine learning tools to accommodate complex data architectures in changepoint analysis.
Examples include linear regression \citep{MR3453652,leonardi_computationally_2016,kaul_efficient_2019-1,rinaldo_localizing_2021,wang_statistically_2021,xu2024change}, quantile regression \citep{MR3862349,Wang+Liu+Zhang+Liu-2024-p1}, Gaussian graphical models \citep{MR4313475,MR4353053}, vector autoregression \citep{bai2023multiple}, trace regression \citep{JMLR:v25:22-0852}, and nonparametric models \citep{MR4048973,londschien2022changeforest,li2022automatic}.
These detection approaches integrate advanced model fitting techniques such as penalized regression, random forests, kernel methods and neural networks.
{Table S1 in the Supplementary Material summarizes a range of recent complex changepoint models in which detection is based on minimizing a total in-sample loss, using either a global or a local search method.}
In many cases, the principle of uniform consistency remains relevant for consistent changepoint estimation.
For example, in high-dimensional linear models with changepoints, the lasso estimator, equipped with an appropriately prespecified regularizer, uniformly approximates its population counterpart across different segments \citep{xu2024change,qian2025reliever}.
Likewise, in other models, the estimated parameters $\hf_I$ can closely mirror their estimands $\f_{I}^\ast$, with suitable regularizers \citep{MR4353053,bai2023multiple,Wang+Liu+Zhang+Liu-2024-p1} or kernels \citep{MR3892345}.

\begin{figure}[ht]
    \centering
    \includegraphics[width=0.85\linewidth]{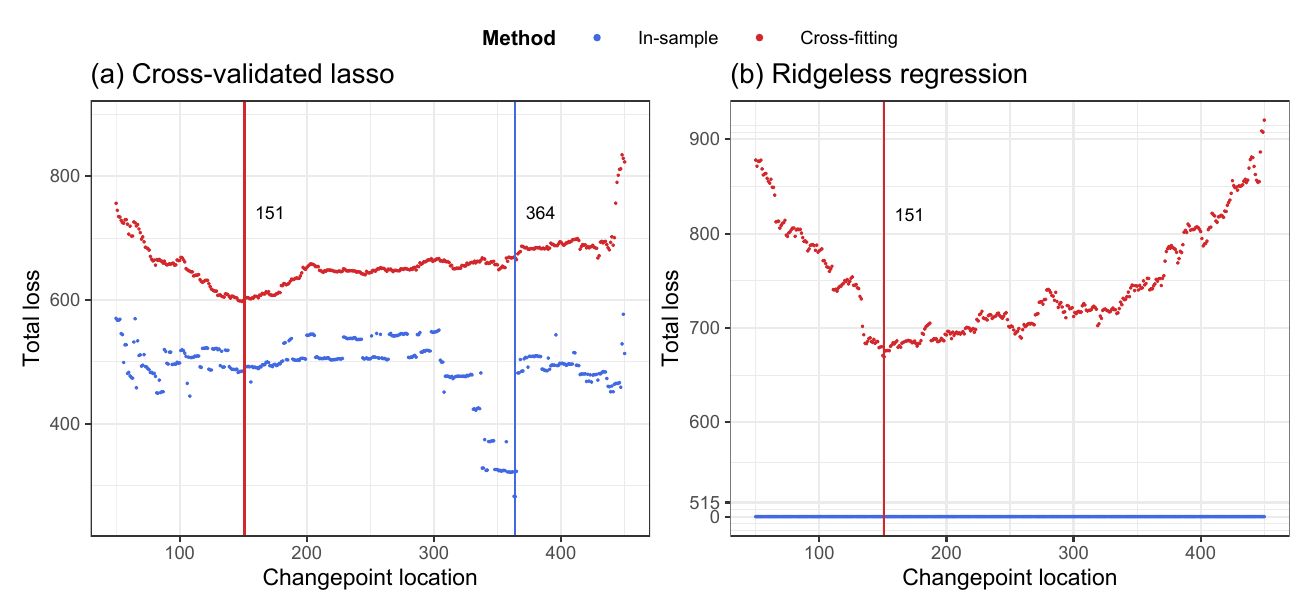}
    \caption{\small Total in-sample and out-of-sample losses as functions of the changepoint location for cross-validated lasso and ridgeless regression in high-dimensional linear models (see Section \ref{sec:linear}). See Section \ref{sec:HD-tuning} and {Section S3.1 in the Supplementary Material for the specific simulation settings.} The true changepoint is located at $150$.}
\label{fig:intro}
\end{figure}

While these methods provide enhanced modeling flexibility, they typically entail high-dimensional or even infinite-dimensional parameter spaces that pose substantial challenges for model fitting.  Consequently, data-driven procedures like cross-validation are widely adopted to enable automatic selection of tuning parameters/models in the implementation of these sophisticated learners. However, the inherent coupling of complex model fitting and changepoint searching can lead to overfitting, creating non-negligible biases that potentially undermine the consistency of the identified breaks.
An illustrative example involves overparameterized deep neural networks, which typically achieve (near) perfect fits to training data---a phenomenon known as \textit{interpolation} \citep{zhang2021understanding,MR3997901}.
Consequently, $\L(\z_I;\hf_I)\approx 0$ for any segment $I$, rendering accurate changepoint detection infeasible.
Another instance occurs with high-dimensional linear models using cross-validated lasso estimators.
Figure \ref{fig:intro}a depicts the curve of the total in-sample loss plotted against the changepoint location, which often attains a minimum far from the true changepoint.
In these examples, the uniform consistency of $\hf_I$ relies on stringent model assumptions that may not hold in practice, leading the in-sample loss $\L(\z_I; \hf_I)$ to deviate significantly from the oracle loss $\risk_I(\f_{I}^\ast)$.
This discrepancy results in unreliable changepoint estimators.
For a formal theoretical justification, refer to Section \ref{sec:framework_in_and_out}.

This raises a fundamental question: how can highly adaptive modeling approaches be used reliably for changepoint detection, and under what conditions are they effective?

\subsection{Our contributions}

In essence, popular in-sample loss minimization methods succeed in changepoint detection by closely approximating the oracle losses $\risk_I(\f_{I}^\ast) = \Ebb\{\L(\z_I;\f_{I}^\ast)\}$ for all segments $I$, thus aligning with the oracle objective function $\overline{\mathcal{L}}_{\mathsf{oracle}}(\cpsset)$.
However, they may fall short for models prone to overfitting.
Our approach introduces a simple yet effective remedy: \textit{out-of-sample} loss evaluations.
These evaluations are designed to estimate population losses using samples independent of those employed for model fitting, thereby producing unbiased estimators of expected losses conditional on model fits.
We employ a \textit{cross-fitting} strategy to facilitate efficient data reuse.
As depicted in Figure \ref{fig:intro}, the total out-of-sample loss curve attains its minimum near the actual changepoint for both the cross-validated lasso and ridgeless regression approaches.
Our contributions are as follows.
\begin{itemize}
    \item We propose a cross-fitting method adaptable to various changepoint models, where detection relies on loss minimization.
    The minimization of the cross-fitting objective function is easy to implement by incorporating diverse changepoint search algorithms, on either a global or local scale.

\item We introduce Recycled Cross-Validation (RECV), a computationally efficient implementation that seamlessly integrates cross-fitting with standard hyperparameter tuning. By repurposing models fitted during the selection process, RECV achieves the robustness of out-of-sample evaluation without incurring additional computational overhead.

    \item We delineate the high-level prerequisites under which empirical evaluations successfully approximate oracle losses, thereby ensuring consistent changepoint detection. Notably, cross-fitting alleviates some of these prerequisites, requiring only that models yield accurate predictions on nearly homogeneous data segments.

\end{itemize}

We apply the proposed methodology and theoretical framework to changepoint detection in high-dimensional linear models and multivariate nonparametric models, demonstrating the robustness and practical relevance of our findings across diverse and complex data structures. We also extend the theoretical guarantees to accommodate temporal dependence.

\subsection{Related literature}

\textbf{Cross-fitting}: Recently, cross-fitting has gained popularity as a method to enhance the efficiency and robustness of sample-splitting-based statistical inference.
This approach relaxes the requirements imposed on estimation algorithms, allowing for the utilization of machine learning techniques.
It has been widely adopted in various domains, such as double machine learning \citep{chernozhukov2018dml}, prediction-powered inference \citep{MR4430964} %\citep{MR4430964,MR4740351},
and variable importance measurement \citep{williamson2023agnostic}.

\textbf{Out-of-sample evaluations}: \cite{londschien2022changeforest} explored detecting changes in nonparametric distributions, utilizing classifier-based log-likelihood ratios.
The authors advocated for using random forests as classifiers, noting one of their attractive attributes: the ability to generate unbiased probability estimates through out-of-bag predictions.
By incorporating binary segmentation search algorithm, they demonstrated consistent changepoint detection in scenarios with a single changepoint, relying on the uniform consistency assumption for the classifier.
The out-of-bag mechanism inherent in random forests naturally provides an out-of-sample evaluation, similar to the approach we adopt here.
In our manuscript, we systematically motivate the need and advantages of out-of-sample loss evaluations, addressing a critical gap between loss evaluation and changepoint detection in a more comprehensive framework encompassing multiple changepoints with flexible learning techniques beyond random forest classifiers.
Additionally, we argue that the assumption of uniformly consistent model fits across all segments is excessively stringent.

\subsection{Structure and notation}

The remainder of this paper is structured as follows.
Section \ref{sec:framework} introduces the cross-fitting methodology and its practical implementation via Recycled Cross-Validation. Section \ref{sec:theory} establishes the general theoretical framework and provides consistency guarantees. In Section \ref{Secapp}, we apply our findings to high-dimensional and nonparametric models. Section \ref{sec:blockwise_temporal} considers the extensions to accommodate temporally dependent data.
Simulation studies and real-data analyzes are detailed in Section \ref{sec:simul}.  Section \ref{sec:conclusion} concludes the paper. All theoretical proofs are provided in the Supplementary Material.

For integer $n > 0$, let $[n]=\{1,\ldots,n\}$.
For $a,b\in\mathbb{R}$, $a\vee b=\max\{a,b\}$ and $a \wedge b = \min\{a, b\}$.
For a vector $a\in\mathbb{R}^p$ and $q\in\{1,2\}$, the $L_q$-norm of $a$ is defined by $\|a\|_q=(\sum_{j=1}^p |a_j|^q)^{1/q}$, and let $\|a\|=\|a\|_2$.
For a positive definite matrix $\Sigma\in\mathbb{R}^{p\times p}$, define the Mahalanobis norm $\|a\|_\Sigma=\sqrt{a^\top \Sigma a}$.
For a matrix $A\in\mathbb{R}^{p_1\times p_2}$, $A_{R,J}$ denotes the submatrix with rows indexed by $R$ and columns indexed by $J$.
Denote $A_{R}=A_{R,[p_2]}$ and $A_{R,-J}=A_{R,[p_2]\setminus J}$. For two sequences $a_n$ and $b_n$, $a_n\asymp b_n$ if there exist constants $c$ and $C$ such that $c\le |a_n/b_n|\le C$ for sufficiently large $n$, $a_n\gg b_n$ ($b_n \ll a_n$)  means $a_n/b_n\to\infty$, and $a_n\gtrsim b_n$ ($b_n \lesssim a_n$) means $a_n>Cb_n$ for some constant $C > 0$.
For a random variable $X$, define its sub-Weibull norm as $\norm{X}_{\Psi_\kappa} = \inf\set{t > 0: \Ebb\{\exp(|X|^{\kappa}/t)\} \le 2}$ for $\kappa > 0$. For ease of reference, Table \ref{tab:notation_loss} summarizes the main loss- and model-related notation used throughout the paper.

\begin{table}[htb]
\begin{threeparttable}
\centering
\caption{\small Summary of loss- and model-related notation. Here, $\ell$ denotes individual loss, and $L$ denotes loss over a segment. Bars indicate expectations, and superscript asterisks indicate oracle quantities.}
\label{tab:notation_loss}
\begin{tabular}{c c l}
\toprule
\textbf{Notation} & \textbf{Definition} & \textbf{Description} \\
\midrule
$\ell(\z_i; \f)$ & -- & {Individual} loss of model $\f$ at index $i$. \\
$\overline{\ell}_i(\f)$ & $\Ebb[\ell(\z_i; \f)]$ & Expected {individual} loss of \emph{fixed} model $\f$ at index $i$. \\
$L(\z_I; \f)$ & $\sum_{i \in I} \ell(\z_i; \f)$ & Empirical loss of model $\f$ over segment $I$. \\
$\overline{L}_I(\f)$ & $\Ebb[L(\z_I; \f)] = \sum_{i \in I} \overline{\ell}_i(\f)$ & Expected loss of model $\f$ over segment $I$. \\
$\f_I^\ast$ & $\argmin_{\f \in \mathcal{F}} \overline{L}_I(\f)$ & Oracle model over segment $I$. \\
$\hf_I$ & -- & Model estimator associated with segment $I$. \\
$\Lval_{I}$ & -- & Generic empirical loss over segment $I$. \\
\bottomrule
\end{tabular}
\end{threeparttable}
\end{table}

\section{General methodology}\label{sec:framework}

\subsection{Changepoint detection via cross-fitting}\label{sec:framework_method}

We now introduce the proposed cross-fitting framework.
For a fixed integer $M\ge 2$, any interval $I$ is partitioned into $M$ disjoint, approximately equal-sized folds $\set{J_{m, I}}_{m=1}^M$. Let $J_{m, I}$ and $J_{-m,I} = I \setminus J_{m, I}$ denote the validation set and the training set for the $m$-th fold, respectively.
Rather than minimizing the in-sample objective function $\mathcal{L}_{\mathsf{in}}(\cpsset)$, we propose minimizing the total out-of-sample loss via cross-fitting:
\begin{equation}\label{obj:crossFitting}
    \mathcal{L}_{\cf}(\cpsset) := \sum_{k=1}^{K+1}\sum_{m=1}^M\L(\z_{J_{m, (\tau_{k-1},\tau_k]}};\hf_{J_{-m, (\tau_{k-1},\tau_k]}}) + \gamma\size{\cpsset}.
\end{equation}
Here, $\hf_{J_{-m, I}}$ represents the estimator trained on $J_{-m, I}$.
This training process may involve feature screening, model selection, or other hyperparameter tuning methods.

\begin{remark}[On partitioning]\label{rmk:fold_partition}{
There can be many strategies for constructing these splits. One of commonly used methods is the \textit{order-preserved splitting} method \citep{wang+Samworth+2018+p57,zou_consistent_2020}. It assigns indices cyclically by setting $J_m=\{i\in[n]: (i-1) \bmod M = m - 1\}$ and defining $J_{m, I} = J_{m} \cap I$.
In practice, we recommend setting $M=5$ or $10$, consistent with standard cross-validation practices.
}
\end{remark}

Since the proposed cross-fitting approach modifies only the loss evaluation mechanism within the changepoint detection framework, the minimization of $\mathcal{L}_{\cf}(\cpsset)$ can be approached either globally or locally, employing search algorithms originally developed for optimizing the in-sample objective function $\mathcal{L}_{\mathsf{in}}(\cpsset)$. The only adjustment in those algorithms is that, for each candidate segment $(\tau_{k-1},\tau_k]$, the in-sample loss $\L(\z_{(\tau_{k-1},\tau_k]};\hf_{(\tau_{k-1},\tau_k]})$ is replaced by the corresponding cross-fitted loss $\sum_{m=1}^M\L(\z_{J_{m, (\tau_{k-1},\tau_k]}};\hf_{J_{-m, (\tau_{k-1},\tau_k]}})$.
The underlying search architecture remains unchanged, whether employing binary segmentation (BS) methods such as wild binary segmentation (WBS) \citep{fryzlewicz_wild_2014} and seeded binary segmentation (SeedBS) \citep{kovacs_seeded_2022}, or global optimization methods that precisely minimize the objective \eqref{equ:mcp_loss} via dynamic programming \citep{MR978902,jackson2005algorithm} and its pruned counterpart \citep{killick_optimal_2012-1}.
{In Section~S1 of the Supplementary Material, we illustrate how to solve the generic loss minimization problem \eqref{equ:mcp_loss} with dynamic programming and SeedBS.}
Compared to traditional in-sample evaluations that fit a single model per candidate segment, the $M$-fold cross-fitting procedure requires fitting $M$ models.
The computational complexity increases only by a constant factor.

\subsection{Cross-fitting with model selection}\label{amcv}

Consider a model fitting procedure that involves hyperparameter tuning or model selection from a pool of candidates $\{\hf_I(\lambda):\lambda\in\Lambda\}$ for a segment $I$, where $\Lambda$ is a candidate set for choosing $\lambda$. The cross-validation (CV) method has been widely used as a technique to estimate the prediction error of a model and to select the tuning parameters involved \citep{leonardi_computationally_2016,MR4767500}. Here we show that CV can be seamlessly integrated with our proposed cross-fitting-based loss evaluation framework \eqref{obj:crossFitting}.

A direct approach is to incorporate CV within the model estimation step $\{\hf_{J_{-m, I}}\}_{m=1}^M$ for a given set $I$. This typically requires further partitioning the set $J_{-m, I}$ into $B$ disjoint folds, and therefore adopting this ``naive" strategy necessitates a total of $BM\size{\Lambda}$ model fits to obtain $\{\hf_{J_{-m, I}}(\hlam_{J_{-m, I}}^{\cv})\}_{m\in[M]}$, where $\hlam_{J_{-m, I}}^\cv$ denotes the selected $\lambda$ for $J_{-m, I}$.

Notice that, in the cross-fitting procedure (\ref{obj:crossFitting}), the segment $(\tau_{k-1},\tau_k]$ has already been split into $M$ disjoint folds. This motivates a direct reuse of these folds,
without any additional partitioning, to select $\lambda$. Specifically, we define
\begin{align} \label{eq:lambda_recv}
    \hat{\lambda}_{(\tau_{k-1},\tau_k]}^\mcv =\mathop{\arg\min}_{\lambda \in \Lambda} \sum_{m=1}^M \L(\z_{J_{m, (\tau_{k-1},\tau_k]}}; \hf_{J_{-m, (\tau_{k-1},\tau_k]}}(\lambda)),
\end{align}
and accordingly the objective function becomes
\begin{align}
\mathcal{L}_{ \mcv}(\cpsset) :=& \sum_{k=1}^{K+1}\sum_{m=1}^M L(\z_{J_{m, (\tau_{k-1},\tau_k]}};\hf_{J_{-m, (\tau_{k-1},\tau_k]}}( \hat{\lambda}_{(\tau_{k-1},\tau_k]}^\mcv))+ \gamma\size{\cpsset}\label{eq:recyc_cv}\\
=&\sum_{k=1}^{K+1}\biggl\{\min_{\lambda \in \Lambda}\sum_{m=1}^M L(\z_{J_{m, (\tau_{k-1},\tau_k]}};\hf_{J_{-m, (\tau_{k-1},\tau_k]}}({\lambda}))\biggr\}+ \gamma\size{\cpsset}.\nonumber
\end{align}

 In Eq.~\eqref{eq:lambda_recv}, the CV procedure is performed directly on each segment $(\tau_{k-1},\tau_k]$: we compute the cross-fitted losses for every candidate $\lambda$ and select the minimum. In other words, for a given segment $(\tau_{k-1},\tau_k]$,
The estimators $\hf_{J_{-m, (\tau_{k-1},\tau_k]}}$ share the same $\lambda$ across all folds $m$. We term this strategy as the \textit{REcycled CV} (RECV) loss, because it recycles the losses computed during the CV process. This effectively reduces the computational burden to $M\size{\Lambda}$ model fits per segment $I$ in contrast to the naive one mentioned above which requires about $BM\size{\Lambda}$ fits. Specifically, when the in-sample loss $\mathcal{L}_{\mathsf{in}} (\cpsset)$ in (\ref{obj:inSample}) needs to employ an $M$-fold CV to obtain $\hf_{(\tau_{k-1},\tau_k]}$, the RECV strategy achieves comparable computational efficiency, providing out-of-sample robustness without additional overhead.

The generic cross-fitting/RECV procedure for changepoint detection is summarized in Algorithm \ref{alg:cf_meta}.

\begin{algorithm}[tb]
\caption{\small Cross-fitting framework for changepoint detection}
\label{alg:cf_meta}
\begin{algorithmic}[1]
\REQUIRE Data sequence $\{\z_i\}_{i=1}^n$, individual loss $\ell(\cdot; \cdot)$, penalty parameter $\gamma > 0$, number of folds $M$, and optional tuning-parameter set $\Lambda$.
\ENSURE Estimated changepoint set $\estcps$.
\STATE \textbf{Loss evaluation:} For each candidate segment $I = (s, e]$ visited by the changepoint search:
\STATE \textit{Partition} $I$ into $M$ folds $J_{m, I}$, $m \in [M]$, and let $J_{-m, I} = I \setminus J_{m, I}$.
\FOR{$m = 1$ \TO $M$}
    \STATE \textit{Estimate} $\hf_{J_{-m, I}}$ on $\z_{J_{-m, I}}$.
    \STATE \textit{Evaluate} the out-of-sample loss $L(\z_{J_{m, I}}; \hf_{J_{-m, I}}) = \sum_{i \in J_{m, I}} \ell(\z_i; \hf_{J_{-m, I}})$.
\ENDFOR
\STATE \quad \textit{Aggregate} the losses over $m$ to obtain the cross-fitting loss $\Lval_{I}$.
\STATE \quad \textit{(Optional) RECV:} If tuning parameters are involved, replace $\mathcal L_I$ by the RECV loss, that is, minimize over $\lambda \in \Lambda$ as in \eqref{eq:lambda_recv}.
\STATE \textbf{Change search:} Apply a changepoint search algorithm (e.g., dynamic programming or SeedBS; see the Supplementary Material) to solve:
\[
    \estcps = \argmin_{\cpsset}\biggl\{\sum_{k=1}^{K+1}\Lval_{(\tau_{k-1},\tau_k]} + \gamma\size{\cpsset}\biggr\}.
\]
\end{algorithmic}
\end{algorithm}

\section{A unified theory}\label{sec:theory}

\subsection{Theoretical comparison of in-sample and out-of-sample losses}\label{sec:framework_in_and_out}

Recall that $\Lval_{I}$ denotes a generic empirical loss,
including both the in-sample loss $\L(\z_I;\hf_I)$ and the cross-fitted loss $\sum_{m=1}^M\L(\z_{J_{m, I}}; \hf_{J_{-m, I}})$.
To ensure reasonable model estimates and feasible fold-splitting, as a convention \citep[e.g.,][]{bai1998estimating,baranowski_narrowest_2019, xu2024change}, we consider a minimal-segment-length threshold in the generic optimization problem \eqref{equ:mcp_loss}.
Let $\mathbb{T}(\dm)$ be the collection of all potential segmentations $\cpsset=(\tau_1,\ldots,\tau_K)$ such that the minimal segment length satisfies $\min_{k\in[K^\ast+1]}\{\tau_k-\tau_{k-1}\}\ge \dm$, where $\dm>0$ is a prespecified integer.
The optimization problem can be stated formally as:
\begin{equation}\label{equ:mcp_loss_dm}
    \estcps = (\htau_{1},\ldots,\htau_{\widehat{K}})=\argmin_{\cpsset\in\mathbb{T}(\dm)}\biggl\{\sum_{k=1}^{K+1}\Lval_{(\tau_{k-1},\tau_k]} + \gamma\size{\cpsset}\biggr\}.
\end{equation}

For clarity, we introduce several notations related to different loss metrics (see Table \ref{tab:notation_loss}) and the change signals used in our framework.
Here we consider a fixed model $\f$.
Note that in the above discussions, we keep the usage of the loss at the segment level, i.e., the empirical loss $\L(\z_I; \f) = \sum_{i \in I} \ell(\z_i; \f)$ for segment $I$.
At the individual level, we introduce several parallel notations.
The corresponding expected loss is $\overline{\ell}_i(\f)=\Ebb\{\ell(\z_i;\f)\}$, and the oracle model is $\f_{i}^\ast =\argmin_{\f\in\mathcal{F}} \overline{\ell}_i(\f)$.
Typically, the loss function $\ell$ is specified such that $\f_{i}^\ast=\f_{\tau^\ast_k}^\ast$ for $i \in (\tau^\ast_{k-1}, \tau^\ast_k]$.
Back to the segment level, for any segment $I$, the expected loss becomes
$\risk_I(\f) = \Ebb\{\L(\z_I; \f)\} = \sum_{i \in I} \overline{\ell}_i(\f)$. To quantify the signal of change at the changepoint $\tau^\ast_k$ for $k\in[K^\ast]$, we introduce
\[
\Delta_k = \{\overline{\ell}_{\tau^\ast_k}(\f_{(\tau^\ast_{k},\tau^\ast_{k+1}]}^\ast) - \overline{\ell}_{\tau^\ast_k}(\f_{(\tau^\ast_{k-1},\tau^\ast_k]}^\ast)\} \vee \{\overline{\ell}_{\tau^\ast_k+1}(\f_{(\tau^\ast_{k-1},\tau^\ast_k]}^\ast) - \overline{\ell}_{\tau^\ast_k+1}(\f_{(\tau^\ast_{k},\tau^\ast_{k+1}]}^\ast)\},
\]
with the conventions $\Delta_0 = \Delta_{K^\ast + 1} = 0$.

\begin{example}\label{ex:linear}
Take the linear models with changepoints for example \citep{bai1998estimating}. Consider pairs of response and covariates $\z_i=(y_i,\x_i)\in\mathbb{R}\times\mathbb{R}^p$, where $y_i=\x_i^\top\f_{(\tau^\ast_{k-1},\tau^\ast_k]}^\ast+\epsilon_i$, with $\epsilon_i$ representing mean-zero noise. The individual loss function is often chosen as the squared loss $\ell(\z_i;\f)=(y_i-\x_i^\top\f)^2$. Thus, we have $\overline{\ell}_i(\f) = \|\f_i^\ast-\f\|^2_\Sigma + \sigma_\epsilon^2$, minimized at $\f=\f_i^\ast=\f_{(\tau^\ast_{k-1},\tau^\ast_k]}^\ast$ for $i \in (\tau^\ast_{k-1}, \tau^\ast_k]$ over $\mathcal{F}=\mathbb{R}^p$, where $\Sigma$ is the covariance matrix of the covariates and $\sigma_\epsilon^2$ is the variance of the noise.
The signal metric simplifies to $\Delta_k=\|\f_{(\tau^\ast_{k-1},\tau^\ast_{k}]}^\ast-\f_{(\tau^\ast_{k},\tau^\ast_{k+1}]}^\ast\|_\Sigma^2$.
\end{example}

As discussed in Section~\ref{sec:intro}, traditional changepoint analysis focuses predominantly on achieving uniform parameter estimation consistency across various data segments---accurately recovering oracle models \(\f_{I}^\ast\), regardless of the presence of changepoints.
Such consistency typically ensures that the in-sample objective function $\mathcal{L}_{\mathsf{in}}(\cpsset)$ closely approximates the oracle objective function $\overline{\mathcal{L}}_{\mathsf{oracle}}(\cpsset)$.
Consequently, minimizing $\mathcal{L}_{\mathsf{in}}(\cpsset)$ tends to yield solutions that approach the actual segmentation \(\truecps\), the minimizer of \(\overline{\mathcal{L}}_{\mathsf{oracle}}(\cpsset)\) \citep{londschien2022changeforest}.
Next we offer a formal justification.

\begin{condition}[Changes]\label{cond:changes}
(a) There exists a sufficiently large constant $C_{\snr}$ and a sequence of positive constants $\dacc \gtrsim (\log n)^{1/\kappa}$ with $0 < \kappa \le 1$ such that for $k\in[K^\ast + 1]$, $\tau^\ast_k - \tau^\ast_{k-1} \ge C_{\snr} \dacc (\Delta_{k-1}^{-1} \vee 1 + \Delta_k^{-1} \vee 1)$.
(b) $\max_{k\in[K^\ast]} \Delta_k \lesssim \dacc / (\log n)^{1/\kappa}$.
(c) There exists a constant $C_{\Delta}$ such that for $k\in[K^\ast]$, $\Delta_k \le C_\Delta[\{\overline{\ell}_{\tau^\ast_k}(\f_{[\tau^\ast_k, \tau^\ast_k + 1]}^{\ast}) - \overline{\ell}_{\tau^\ast_k}(\f_{(\tau^\ast_{k-1},\tau^\ast_k])}^{\ast}) + \{\overline{\ell}_{\tau^\ast_k+1}(\f_{[\tau^\ast_k, \tau^\ast_k + 1]}^{\ast}) - \overline{\ell}_{\tau^\ast_k+1}(\f_{(\tau^\ast_{k},\tau^\ast_{k+1}]}^\ast)\}]$.
\end{condition}

Condition \ref{cond:changes}(a) imposes a minimum distance requirement between two successive changepoints.
Condition \ref{cond:changes}(b) caps the magnitude of change signals while allowing vanishing magnitudes.
Condition \ref{cond:changes}(c) is a technical requirement ensuring $\overline{\ell}_{\tau^\ast_k}(\f_{[\tau^\ast_k, \tau^\ast_k + 1]}^\ast) - \overline{\ell}_{\tau^\ast_k}(\f_{(\tau^\ast_{k-1},\tau^\ast_k]}^\ast)$
(or $\overline{\ell}_{\tau^\ast_k+1}(\f_{[\tau^\ast_k, \tau^\ast_k + 1]}^\ast) - \overline{\ell}_{\tau^\ast_k+1}(\f_{(\tau^\ast_{k},\tau^\ast_{k+1}]}^\ast)$) is comparable to $\overline{\ell}_{\tau^\ast_k}(\f_{(\tau^\ast_{k},\tau^\ast_{k+1}]}^\ast) - \overline{\ell}_{\tau^\ast_k}(\f_{(\tau^\ast_{k-1},\tau^\ast_k]}^\ast)$ (or $\overline{\ell}_{\tau^\ast_k+1}(\f_{(\tau^\ast_{k-1},\tau^\ast_k]}^\ast) - \overline{\ell}_{\tau^\ast_k+1}(\f_{(\tau^\ast_{k},\tau^\ast_{k+1}]}^\ast)$, respectively).
In the case of linear models with squared loss, $\f_{[\tau^\ast_k, \tau^\ast_k + 1]}^\ast = (\f_{(\tau^\ast_{k-1},\tau^\ast_k]}^\ast+\f_{(\tau^\ast_{k},\tau^\ast_{k+1}]}^\ast)/2$, and recalling that $\Delta_k=\|\f_{(\tau^\ast_{k-1},\tau^\ast_{k}]}^\ast-\f_{(\tau^\ast_{k},\tau^\ast_{k+1}]}^\ast\|_\Sigma^2$, Condition \ref{cond:changes}(c) is easily satisfied with $C_\Delta=2$.

The next lemma establishes a general set of sufficient conditions on the loss behavior that guarantee consistent changepoint detection. The argument relies on a careful case analysis, distinguishing between \textit{nearly homogeneous} intervals in $\homo$ and intervals in its complement, where
\begin{align*}
    \homo = \bigl\{(s,e]: \size{\truecps \cap (s,e]} \le 1; \exists k, |s-\tau^\ast_{k-1}|\le\widetilde{C}\dacc\Delta_{k-1}^{-1},|e-\tau^\ast_k|\le\widetilde{C}\dacc\Delta_k^{-1}\bigr\},
\end{align*}
with a constant $\widetilde{C} > 0$.
For segment $I$, define
\begin{align*}
    \xi_I&=\bigl\{\Lval_{I} - \sum_{i\in I}\ell(\z_i;\f_{i}^\ast)\bigr\} - \bigl\{\risk_I(\f_{I}^\ast) - \sum_{i\in I}\overline{\ell}_i(\f_{i}^\ast)\bigr\},
\end{align*}
which quantifies the approximation error of an empirical loss $\Lval_{I}$ to the oracle one $\sum_{i\in I}\ell(\z_i;\f_{i}^\ast)$, properly centered.

\begin{lemma}\label{lem:loc_err_g}
Suppose Condition \ref{cond:changes} holds. By solving the optimization problem~\eqref{equ:mcp_loss_dm} with $d_{\m} \lesssim C_{\snr} \dacc$ and $\gamma = C_{\gamma} \dacc$ for some constant $C_{\gamma}$, we have
\begin{equation*}
    \widehat{K} = K^\ast\ \text{and}\ \max_{1 \le k \le K^\ast} \min_{1 \le j \le \widehat{K}} \Delta_k \abs{\tau^\ast_k - \htau_{j}} \le \widetilde{C} \dacc,
\end{equation*}
conditional on the event $\Gbb = \Gbb^{\mathsf{nhomo}} \cap \Gbb^{-\mathsf{nhomo}}$, where
\begin{align*}
    \Gbb^{\mathsf{nhomo}} &= \bigl\{\mbox{for any }I\in \homo,\ |\xi_I| < C_{\ref*{lem:loc_err_g}.1} \dacc\bigr\}, \\
    \Gbb^{-\mathsf{nhomo}} &= \bigl\{\mbox{for any }I\not\in\mathcal{I}^{\mathsf{nhomo}}\text{ such that }\size{I} \ge \dm,\\
    &~~~~~~~~~~~~~~~~~~~~~~\xi_I > - [C_{\ref*{lem:loc_err_g}.2} \{\risk_I(\f_{I}^\ast) - \sum_{i\in I}\overline{\ell}_i(\f_{i}^\ast)\}]\vee(C_{\ref*{lem:loc_err_g}.1} \dacc)\bigr\},
                    \end{align*}
for some constants $C_{\ref*{lem:loc_err_g}.1} > 0$ and $C_{\ref*{lem:loc_err_g}.2} \in (0, 1/(1 + 4 C_{\Delta}))$. The constants $C_{\gamma}$ and $\widetilde{C}$ depend only on the constants $C_{\ref*{lem:loc_err_g}.1}$, $C_{\ref*{lem:loc_err_g}.2}$, $C_{\snr}$, $C_{\Delta}$ and the ratio $d_{\m}/(C_{\snr} \dacc)$.
\end{lemma}

Lemma \ref{lem:loc_err_g} shows that consistent changepoint detection relies on how accurately the empirical losses approximate the oracle losses across different segment types, with the approximation error  quantified by $\xi_I$.
Specifically, for nearly homogeneous segments $\homo$, the approximation error must be tightly bounded such that $\abs{\xi_I} \lesssim \dacc$.
Conversely, for all remaining segments, only a lower bound on this error is necessary.

Consequently, applying a specific changepoint model requires proving that the event $\Gbb$ holds with high probability.
For methods based on in-sample loss minimization, this is typically achieved by establishing uniform control of the parameter estimation error across all segments
\citep{MR3453652,xu2024change,MR3862349,Wang+Liu+Zhang+Liu-2024-p1,bai2023multiple,londschien2022changeforest}.
In contrast, out-of-sample loss evaluation via cross-fitting substantially simplifies the verification of $\Gbb$.
For heterogeneous segments, the event $\Gbb^{-\mathsf{nhomo}}$ holds under mild assumptions.
Since only lower bounds on the loss values are required in $\Gbb^{-\mathsf{nhomo}}$, the cross-fitting framework is robust to overfitted model estimates.
By contrast, in-sample minimization is vulnerable in this regime: overfitting can bias the in-sample loss $\L(\z_I;\hf_I)$ substantially downwards, thereby violating the lower bound condition; see Section \ref{sec:linear} for a concrete example.
For nearly homogeneous segments, the event $\Gbb^{\mathsf{nhomo}}$ holds as long as the model estimates achieve low prediction error on such segments, ensuring that the cross-fitted losses remain well behaved.

\subsection{Cross-fitting ensures consistent detection}
\label{sec:framework_theory}

Here we provide a formal justification for the preceding discussion under the assumption that the samples $\{\z_i\}_{i=1}^n$ are independent.
We begin by establishing the necessary conditions for the loss function.
Let $s_{i,\f} = \ell(\z_i; \f) - \ell(\z_i; \f_{i}^\ast) - \Ebb[\ell(\z_i; \f) - \ell(\z_i; \f_{i}^\ast)]$ denote the centered empirical excess risk at the individual level, and let $\sigma_{i, \f}^2$ denote its variance.

\begin{condition}[Smoothness of the loss]\label{cond:model_and_loss}
(a) There is an envelope parameter $m_{i, \f}$ such that the sub-Weibull tail holds: $\mathbb{P}(\abs{{s_{i, \f}}/{m_{i, \f}}} > x) \le \exp(1 - x^{\kappa_1})$ for $\kappa_1 \in [\kappa, 1]$ with $\kappa$ defined in Condition~\ref{cond:changes}.
Moreover, there is a positive sequence $1 \le c_n \le \dacc / (\log n)^{1/\kappa}$ such that $\sigma_{i, \f} \lesssim \{c_n (\overline{\ell}_i(\f) - \overline{\ell}_i(\f_{i}^\ast))\}^{{1}/{2}} \vee (\overline{\ell}_i(\f) - \overline{\ell}_i(\f_{i}^\ast)) + \dacc / n$ and $m_{i, \f} \le \sigma_{i, \f} + {\dacc}/{(\log n)^{1/\kappa}}$.
(b) For any $i, j \in [n]$, we have $\abs{\overline{\ell}_j(\f) - \overline{\ell}_j(\f_{i}^\ast)} \le C_{\ell} (\abs{\overline{\ell}_i(\f) - \overline{\ell}_i(\f_{i}^\ast)} + \abs{\overline{\ell}_j(\f_{i}^\ast) - \overline{\ell}_j(\f_{j}^\ast)})$ for some universal constant $C_{\ell} \ge 1$.
\end{condition}

Condition \ref{cond:model_and_loss} is essential for establishing non-asymptotic concentration inequalities for the empirical losses. Condition \ref{cond:model_and_loss}(a) ensures that the stochasticity of the sample $\z_i$ does not cause excessive deviation in the empirical loss. {The sub-Weibull tail condition generalizes the sub-Gaussian and sub-exponential cases and has recently been used in the changepoint literature \citep{xu2024change, li2025robust}.} Condition~\ref{cond:model_and_loss}(b) controls the sensitivity of the expected loss to shifts in the underlying distribution. For linear models with squared loss, the centered empirical risk is given by $s_{i, \f} = \{\x_i^\top(\f-\f_{i}^\ast)\}^2 - \Ebb [\{\x_i^\top(\f-\f_{i}^\ast)\}^2] - 2 \epsilon_i \{\x_i^\top(\f-\f_{i}^\ast)\}$. This formulation readily satisfies Condition \ref{cond:model_and_loss}(a) with $\sigma_{i, \f} = m_{i, \f}$ and $c_n = 1$ under the assumption that both the covariates $\x_i$ and the noise $\epsilon_i$ are sub-Gaussian \citep{vershynin_high-dimensional_2018}. Furthermore, Condition \ref{cond:model_and_loss}(b) holds with $C_{\ell} = 2$ through the ordinary decomposition of the squared loss.
Other Lipschitz-continuous loss functions, such as quantile loss or Huber loss, are also compatible with this condition.

In Condition \ref{cond:model_and_loss}, we avoid imposing a norm or distance on the  model space of $\f$.
This metric-free approach suffices to control the event $\Gbb^{-\mathsf{nhomo}}$ without bounding the model estimation error.
Meanwhile, $\Gbb^{\mathsf{nhomo}}$ is governed by Condition~\ref{cond:predict}, which only requires the cross-fitting approach to achieve low prediction error on \textit{nearly homogeneous segments}.

\begin{condition}[Model's predictive accuracy]\label{cond:predict}
With probability at least $1 - p_n$ (where $p_n\to 0$), uniformly across each $m\in[M]$ and $I \in \homo$,
$\abs{ \risk_{J_{m, I}}(\hf_{J_{-m, I}}) - \risk_{J_{m, I}}(\f_{{J_{-m, I}}}^\ast) } \lesssim \dacc$, where the expectation in $\risk_{J_{m, I}}(f)$ is taken over the samples $z_{J_{m, I}}$, with the model $f$ held fixed.
\end{condition}

Condition \ref{cond:predict} underscores that the model fitting procedure should predict accurately on nearly homogeneous segments identified within $\homo$.
These segments include intervals where each endpoint should be within a distance of $O(\dacc\Delta_{k-1}^{-1})$ or $O(\dacc\Delta_k^{-1})$ samples from a changepoint.
Notably, for linear models with squared loss, this condition manifests as $\|\hf_{J_{-m, I}}-\f_{{I}}^\ast\|^2_\Sigma\lesssim \dacc/\size{I}$.
For lasso fit with an appropriately prespecified regularizer, $\dacc=s_n\log p$, where $s_n$ denotes sparsity and $p$ is the dimension of covariates; {see, e.g., \cite{kaul_efficient_2019-1} and \cite{xu2024change}.} The rate varies depending on the model, inherently determining the precision of estimated changepoints.

\begin{theorem}\label{thm:cross-fitting}
Suppose Conditions \ref{cond:changes}, \ref{cond:model_and_loss} and \ref{cond:predict} hold with $\kappa=\kappa_1$. Assume that $d_{\m}$ and $\gamma$ are chosen as in Lemma~\ref{lem:loc_err_g}, and, in addition, that $d_{\m} \gtrsim (\log n)^{1/\kappa}$. The estimator $\widehat{\mathcal{T}}_{\cf} = \{\htau_{\cf,1}, \dots, \htau_{\cf,\widehat{K}_{\cf}}\}$, obtained by minimizing the cross-fitting objective $\mathcal{L}_{\cf}(\cpsset)$ in (\ref{obj:crossFitting}), satisfies:
\begin{equation*}
    \widehat{K}_{\cf} = K^\ast\ \text{and}\ \max_{1 \le k \le K^\ast} \min_{1 \le j \le \widehat{K}_{\cf}} \abs{\tau^\ast_k - \htau_{\cf,j}} \le \widetilde{C} \dacc\Delta_k^{-1},
\end{equation*}
with probability at least $1 - p_n - n^{-C}$ for some constant $C > 0$.
\end{theorem}

In the probability bound $1 - p_n - n^{-C}$, $p_n$ and $n^{-C}$ represent the probability bounds related to the prediction accuracy and the concentration inequalities of $\xi_I$.
Under specific models (e.g., lasso in Corollary~\ref{coro:recycled_cv_lasso}), $p_n$ takes a similar polynomial decay form of $\mathcal{O}(n^{-C})$. Although the specific constants may differ, they naturally merge into a bound of $1 - \mathcal{O}(n^{-C})$.

Theorem \ref{thm:cross-fitting} demonstrates that cross-fitting, when integrated with an appropriate algorithm that attains the minimum \citep{jackson2005algorithm,killick_optimal_2012-1}, can provide consistent estimation on both the number and locations of changepoints, given that $\dacc\Delta_k^{-1}/n\to 0$.
Unlike traditional in-sample methods that necessitate uniform parameter estimation consistency across all data segments, our approach requires accurate prediction only on \textit{nearly homogeneous} segments (Condition \ref{cond:predict}).
Further theoretical comparisons are explored in Section \ref{sec:linear}.
This methodological shift, embracing the principles of cross-fitting, moves away from the constraints typically imposed on parameter estimation in in-sample frameworks.
Our approach offers great possibilities for employing flexible model fitting procedures focused on achieving high predictive accuracy, enhancing changepoint detection capabilities.
Similar themes have been discussed in \cite{chernozhukov2018dml}, \cite{MR4430964} and \cite{williamson2023agnostic}, though in different contexts.

\subsection{Theoretical guarantees for recycled cross-validation}
\label{sec:recycled_theory}

We now establish the theoretical foundation for the RECV procedure defined in \eqref{eq:recyc_cv}. A significant theoretical advantage of the RECV is the relaxation of the predictive accuracy requirement. Unlike the cross-fitting approach discussed in Section \ref{sec:framework_theory}, which requires a specific pre-selected estimator to perform well, the RECV framework only requires the \textit{existence} of a sufficiently accurate candidate within the hyperparameter pool $\Lambda$.

\begin{condition}[Model's predictive accuracy, existence]\label{cond:predict_exist}
Consider the set of candidate model estimates $\{\hf_{J_{-m, I}}(\lambda)\}_{\lambda \in \Lambda, m \in [M]}$ with $\size{\Lambda} \le n^c$ for some constant $c > 0$. With probability at least $1 - p_n$ (where $p_n\to 0$), uniformly for each nearly homogeneous segment $I \in \homo$, there is an oracle parameter $\lambda_{I} \in \Lambda$ satisfying
$    \abs{ \risk_{J_{m, I}}(\hf_{J_{-m, I}}(\lambda_I)) - \risk_{J_{m, I}}(\f_{{J_{-m, I}}}^\ast) } \lesssim \dacc$,
for all $m \in [M]$.
\end{condition}

By substituting Condition \ref{cond:predict} with the relaxed Condition \ref{cond:predict_exist}, we have the following consistency result for changepoint detection utilizing the RECV.

\begin{theorem}\label{thm:cvloss}
Suppose Conditions \ref{cond:changes}, \ref{cond:model_and_loss} and \ref{cond:predict_exist} hold with $\kappa = \kappa_1$. Assume that $d_{\m}$ and $\gamma$ are chosen as in Lemma~\ref{lem:loc_err_g}, and, in addition, that $d_{\m} \gtrsim (\log n)^{1/\kappa}$. Then, the estimator $\widehat{\mathcal{T}}_{\mcv} = \{\htau_{\mcv,1}, \dots, \htau_{\mcv,\widehat{K}_{\mcv}}\}$, obtained by minimizing the RECV objective $\mathcal{L}_{ \mcv}(\cpsset)$ in (\ref{eq:recyc_cv}), satisfies:
\begin{equation*}
    \widehat{K}_{\mcv} = K^\ast\ \text{and}\ \max_{1 \le k \le K^\ast} \min_{1 \le j \le \widehat{K}_{\mcv}} \abs{\tau^\ast_k - \htau_{\mcv,j}} \le \widetilde{C} \dacc\Delta_k^{-1},
\end{equation*}
with probability at least $1 - p_n - n^{-C}$ for some constant $C > 0$.
\end{theorem}
The core mechanism of Theorem \ref{thm:cvloss} is that the RECV minimization ($\min_{\lambda \in \Lambda}$) automatically makes the loss concentrates on the desired loss value with the ``oracle'' parameter $\lambda_I$ described in Condition \ref{cond:predict_exist}.
Consequently, the CV loss asymptotically matches that of the best-performing candidate. Provided that a suitable candidate exists in the search grid, RECV adaptively exploits it for detection.
In contrast, traditional methods often decouple model selection and detection, requiring uniformly pre-specified hyperparameters across all segments.

\section{Applications}\label{Secapp}

Theorem \ref{thm:cvloss} outlines high-level prerequisites for the changepoint model, loss function, and model fitting procedure to ensure accurate changepoint detection via the proposed cross-fitting-based RECV method.
To elucidate the practical implications of these conditions and conclusions, this section applies the general framework to two specific example: high-dimensional linear models and multivariate nonparametric models.

\subsection{Changepoint detection in high-dimensional linear models}\label{sec:linear}

Consider the linear model with changepoints described in Example \ref{ex:linear}. The objective is to detect changes in the regression coefficients $\f_{(\tau^\ast_{k-1},\tau^\ast_k]}^\ast$ under scenarios where {$p\gtrsim n$}.
This changepoint model has received significant attention in recent years; see, for example, \cite{MR3453652}, \cite{kaul_efficient_2019-1}, \cite{wang_statistically_2021} and \cite{xu2024change}.
Condition \ref{cond:model_and_loss_lasso} is similar to the common assumptions in the literature, ensuring that Condition \ref{cond:model_and_loss} is satisfied.

\begin{condition}[Linear model and loss]\label{cond:model_and_loss_lasso}
(a) The covariates $\x_i$'s are i.i.d. sub-Gaussian with zero mean and covariance matrix $\Sigma$, and $\norm{\Sigma^{-{1}/{2}}\x_i}_{\Psi_2} \le C_x$ for some constant $C_x>0$;
(b) The noises $\epsilon_i$'s are i.i.d. sub-Gaussian with zero mean, variance $\sigma_\epsilon^2$, and $\norm{\epsilon_i}_{\Psi_2} \le C_{\epsilon}$ for some constant $C_{\epsilon}>0$;
{(c) There exists a sequence of integers $s_n$ such that $\size{\set{j\in[p]: \f_{k,j}^\ast \neq 0}} \le s_n < p$.
In addition, $\abs{\f_{k,j}^\ast} \le C_\f$ for some constant $C_\f>0$;}
(d) The loss function $\ell(\z_i;\f)$ is taken as $(y_i-\x_i^\top\f)^2$.
\end{condition}

To examine the impact of model overfitting on changepoint detection, particularly in high-dimensional settings, we explore three distinct model fitting procedures that utilize in-sample loss minimization.
Let $y=(y_1,\ldots,y_n)^\top$ and $X=(\x_1^\top,\ldots,\x_n^\top)^\top$.
\begin{itemize}
    \item \textbf{Least squares on selected variables:} This procedure involves selecting a set of potentially relevant variables, denoted as $X_{I,\widehat{S}_I}$, with an index subset $\widehat{S}_I\subset[p]$, and performing least squares regression on these selected variables.
    It yields the parameter estimator $\hf_I^{\rm ls}$, with $\hf_{I,\widehat{S}_I}^{\rm ls} = (X_{I,\widehat{S}_I}^\top X_{I,\widehat{S}_I})^{-1} X_{I,\widehat{S}_I}^\top y_I$ and $\hf_{I,-\widehat{S}_I}^{\rm ls}=0$.
    \item \textbf{Cross-validated lasso:} A lasso estimator with a prespecified regularizer $\lambda$ is defined by:
    \begin{equation}\label{eq:lasso}
        \hf_I^\mathsf{lasso}(\lambda)\in\argmin_{\f\in\mathbb{R}^p}\big\{\L(\z_I;\f)+\lambda\sqrt{\size{I}}\|\f\|_1\big\},
    \end{equation}
    where $\L(\z_I;f)=\|y_I-X_I \f\|^2$.
    The optimal regularizer, $\hlam_I^{\cv}$, is selected from a candidate set $\Lambda$ through standard CV.     The cross-validated lasso estimator is defined as $\hf_I^\mathsf{lasso}(\hlam_I^{\cv})$.
    \item \textbf{Ridgeless regression:} The estimator is the minimum $L_2$-norm least squares solution $\hf_I^{\rm rl}=(X_I^\top X_I)^+X_I^\top y_I$ \citep{bartlett2020benign, hastie2022surprises}, where $A^+$ denotes the Moore-Penrose pseudoinverse of matrix $A$.
\end{itemize}

Ridgeless regression models interpolate the data such that $\L(\z_I;\hf_I^{\rm rl})=0$.
Consequently, $\mathcal{L}_{\mathsf{in}}(\cpsset)=0$ for all candidate segmentations $\cpsset$, rendering them indistinguishable.
To analyze the impact of model selection and hyperparameter tuning on the other two procedures, we introduce a formula that characterizes the approximation error $\xi_I$, as in Lemma \ref{lem:loc_err_g}.

\begin{proposition}\label{prop:linear}
Under Condition \ref{cond:model_and_loss_lasso}, for $\hf_I\in\{\hf_I^{\rm ls},\hf_I^\mathsf{lasso}(\lambda),\hf_I^\mathsf{lasso}(\hlam_I^{\cv})\}$, the approximation error $\xi_I$ when the in-sample loss is employed is given by
\begin{align*}
    \xi_{I,\mathsf{in}} = \underbrace{\mathcal{O}_p\big(\sqrt{\sum_{i\in I}\|\f_{i}^\ast-\f_{I}^\ast\|^2_\Sigma}\cdot\sqrt{s_n \log n}\big)}_\mathsf{Concentration\ error} \underbrace{-2\underbrace{u_I^\top X_I(\hf_I-\f_{I}^\ast)}_\mathsf{Cross} + \underbrace{\|X_I(\hf_I-\f_{I}^\ast)\|^2}_\mathsf{Squared}}_\mathsf{Bias},
\end{align*}
where $u_I=y_I-X_I\f_{I}^\ast$.
\end{proposition}

The concentration error quantifies the rate at which the empirical loss $\sum_{i\in I} \{(y_i-\x_i^\top\f_{I}^\ast)^2-\epsilon_i^2\}$ concentrates around its expectation $\sum_{i\in I}\|\f_{i}^\ast-\f_{I}^\ast\|^2_\Sigma$.
The bias measures the impact of evaluating the in-sample loss at $\hf_I$ rather than at $\f_{I}^\ast$.
According to Lemma \ref{lem:loc_err_g}, it is generally required that $|\xi_I|\lesssim \dacc$ for nearly homogeneous segments and $\xi_I>-(C_{\ref*{lem:loc_err_g}.2} \sum_{i\in I}\|\f_{i}^\ast-\f_{I}^\ast\|^2_\Sigma)\vee(C_{\ref*{lem:loc_err_g}.1} \dacc)$ for the remaining segments, to ensure that changepoints can be detected within the precision $\Delta_k^{-1}\dacc$.

If the bias is $\mathcal{O}_p(s_n\log p)$---for instance, when employing $\hf_I^\mathsf{lasso}(\lambda^\ast)$ with $\lambda^\ast\asymp\sqrt{\log p}$ \citep{MR3453652,leonardi_computationally_2016}---this detection precision is guaranteed with $\dacc=s_n\log p$, aligning with nearly optimal detection rates \citep{rinaldo_localizing_2021}. However, there are instances where the bias could become the dominating term, exceeding $s_n\log p$. This is due to the dependence between $u_I^\top X_I$ and $\hf_I$ in the cross term, which is of order $\mathcal{O}_p(\sqrt{(\size{\widehat{S}_I} + s_n) \size{I} \log p}~\norm{\hf_I - \f_I^\ast})$, where $\widehat{S}_I$ is the active set associated with $\hf_I$.
For example, consider $\hf_I=\hf_I^{\rm ls}$, assuming the sure screening property holds \citep{MR2530322} and $\size{\widehat{S}_I} \ge s_n$.
The total bias term becomes ``$-{u}_I^\top X_{I,\widehat{S}_I} (X_{I,\widehat{S}_I}^\top X_{I,\widehat{S}_I})^{-1} X_{I,\widehat{S}_I}^\top{u}_I$''.
For a \textit{given} homogeneous segment $I$, this bias is $\mathcal{O}_p(\size{\widehat{S}_I}\log p)$, as demonstrated by \cite{fan2012variance} in the context of error variance estimation.
Similarly, for a homogeneous segment $I$ and $\hf_I=\hf_I^\mathsf{lasso}(\hlam_I^{\cv})$ with an active set $\widehat{S}_I$, the bias term is of order $\mathcal{O}_p(\sqrt{(\size{\widehat{S}_I}+s_n)s_n}\log^{3/2}p + s_n\log^2 p)$, as noted in \cite{chetverikov2021cross}.
Consequently, when the number of selected variables $\size{\widehat{S}_I}$ substantially exceeds $s_n$---perhaps due to a lenient screening rule or a small regularizer---the bias term may dominate.
This dominance can compromise the near-optimal detection rate, and a large change signal might be necessary to enable consistent changepoint detection, albeit at a reduced level of precision.

In contrast, examine a parallel version of Proposition \ref{prop:linear}
in the cross-fitting evaluation:
\begin{align*}
\xi_{I,\mathsf{cf}}
    =& \mathcal{O}_p\big(\sqrt{\sum_{i\in I}\|\f_{i}^\ast-\f_{I}^\ast\|^2_\Sigma}\cdot\sqrt{s_n \log n}\big)\\
    &\underbrace{
    -\sum_{m=1}^M\{2\underbrace{u_{J_{m, I}}^\top X_{J_{m, I}}(\hf_{J_{-m, I}}-\f_{I}^\ast)}_\mathsf{Cross}\} + \sum_{m=1}^M\underbrace{\|X_{J_{m, I}}(\hf_{J_{-m, I}}-\f_{I}^\ast)\|^2}_\mathsf{Squared}
    }_\mathsf{Bias},
\end{align*}
where $\hf_{J_{-m, I}}$ can be specified as $\hf_{J_{-m, I}}^{\rm ls}$, $\hf_{J_{-m, I}}^\mathsf{lasso}(\lambda)$ or $\hf_{J_{-m, I}}^\mathsf{lasso}(\hlam_{J_{-m, I}}^{\cv})$.
By independence between $\z_{J_{m, I}}$ and $\hf_{J_{-m, I}}$, each cross term has a mean zero and is of order $\mathcal{O}_p(\sqrt{\size{I}}~\norm{\hf_{J_{-m, I}}-\f_{I}^\ast})$, where the selection-dependence factor $\sqrt{\size{\widehat{S}_I} + s_n}$ that appears in the in-sample case is absent.
For the cross-validated lasso estimator $\hf_{J_{-m, I}}^\mathsf{lasso}(\hlam_{J_{-m, I}}^{\cv})$, the total bias term becomes $\mathcal{O}_p(s_n\log^2 p)$ \citep{chetverikov2021cross}.
{Table S2 in the Supplementary Material} compares the contributors to bias in both in-sample and out-of-sample loss evaluations, demonstrating how out-of-sample loss evaluations could mitigate the effects of bias on detection precision through these cross terms.

Next, for a more precise justification, we present a result that establishes the consistency of changepoint detection by employing RECV strategy in this setting.
We need the following condition regarding the signal strength tailored for the high-dimensional linear models.

\begin{condition}[Changes]\label{cond:changes_lasso}
There exists a sufficiently large constant $C_{\snr}$ such that for $k\in[K^\ast + 1]$, $\tau^\ast_k - \tau^\ast_{k-1} \ge C_{\snr} s_n(\log^2 p) (\Delta_{k-1}^{-1} \vee 1 + \Delta_k^{-1} \vee 1)$.
\end{condition}

\begin{corollary}\label{coro:recycled_cv_lasso}
Suppose that Conditions \ref{cond:model_and_loss_lasso} and \ref{cond:changes_lasso} hold, and that the candidate set $\Lambda$ for the lasso estimator \eqref{eq:lasso} satisfies $\size{\Lambda} \le n^{c}$ for some constant $c > 0$ and contains some $\lambda^\ast\asymp \sqrt{\log p}$.
Assume that $d_{\m}$ and $\gamma$ are chosen as $d_{\m} = C_{\m} \log n$ and $\gamma = C_{\gamma} s_n\log p$ for some constants $C_{\m}, C_{\gamma} > 0$.
Then, the estimator $\widehat{\mathcal{T}}_{\mcv} = \{\htau_{\mcv,1}, \dots, \htau_{\mcv,\widehat{K}_{\mcv}}\}$, obtained by minimizing the RECV objective $\mathcal{L}_{ \mcv}(\cpsset)$ in (\ref{eq:recyc_cv}) based on cross-validated lasso fits, satisfies:
\begin{equation*}
    \widehat{K}_{\mcv} = K^\ast\ \text{and}\ \max_{1 \le k \le K^\ast} \min_{1 \le j \le \widehat{K}_{\mcv}} \Delta_k\abs{\tau^\ast_k - \htau_{\mcv,j}} \lesssim s_n\log p,
\end{equation*}
with probability at least $1 - n^{-C}$ for some constant $C > 0$.
\end{corollary}
This result is a corollary to Theorem \ref{thm:cvloss}.
Condition~\ref{cond:predict_exist} posits the existence of a tuning parameter $\lambda^\ast \in \Lambda$ that yields low prediction error.
This is a standard property of lasso that holds under Condition~\ref{cond:model_and_loss_lasso} \citep[see, e.g.,][]{qian2025reliever}.
Unlike previous methods requiring carefully prespecified regularizers \citep{MR3453652,wang_statistically_2021,xu2024change}, Corollary \ref{coro:recycled_cv_lasso} demonstrates that RECV achieves the minimax-optimal detection rate up to a logarithmic factor \citep{rinaldo_localizing_2021}.
This tuning flexibility allows RECV to adapt to heterogeneity across segments.
Consequently, RECV provides a compelling strategy when the model fitting procedure inherently involves model selection.

\begin{remark}[On ridgeless fits]
For a given segment $I$ without changepoints, \cite{bartlett2020benign} shows that the ridgeless estimator can achieve consistent prediction accuracy; specifically, $\|\hf_I^{\rm rl}-\f_{I}^\ast\|_\Sigma$ can be controlled, provided that  $\Sigma$ is \textit{benign} (cf. Definition 4 in \cite{bartlett2020benign}).
It may be possible to extend the techniques in \cite{bartlett2020benign} to establish the predictive accuracy required in Condition \ref{cond:predict} on nearly homogeneous segments.
Extending the present theory to ridgeless fits is a promising direction for future research.
\end{remark}

\subsection{Changepoint detection in multivariate nonparametric models}\label{sec:kde_application}

Let $\z_i \in \mathbb R^p$ be independent observations, and let $f_i^\ast$ denote the density of $P_i$. Under \eqref{MCP}, $f_i^\ast = f_{(\tau_{k-1}^\ast,\tau_k^\ast]}^\ast$ for $i \in (\tau_{k-1}^\ast,\tau_k^\ast]$ and $k=1,\dots,K^\ast+1$. Our goal is to detect the changepoints in these segment-wise densities.
Multivariate nonparametric changepoint detection via kernel methods has been studied in, for example, \citet{MR4048973}, \citet{padilla2021optimal} and \citet{padilla2023change}.
Here we use a negative log-likelihood loss, i.e., $\ell(\z_i;f) = -\log f(\z_i)$, which fits directly into the general framework of Section~\ref{sec:framework_theory}.
For an interval $I\subset (0, n]$ and bandwidth $h>0$, consider the kernel density estimator (KDE)
$$
\hf_{I, h}(\cdot) = \max\big\{(|I| h^p)^{-1} \sum_{i \in I} \mathcal{K}(h^{-1}\{(\cdot) - \z_i\}), t_n\big\},\quad t_n = n^{-\beta},
$$
for a sufficiently large constant $\beta > 2$. The lower truncation avoids the singularity of the logarithmic loss when the estimated density is close to zero; similar technical devices are standard in the literature of nonparametric estimation \citep{hardle2012nonparametric}.

Then, by definition, we have, for a given segment $I$,
\[
\L(\z_I;\f) = \sum_{i \in I} \ell(\z_i; \f)= -\sum_{i \in I}\log \hf_{I, h}(\z_i)
\]
which evaluates the fit of the model $\hf_{I, h}(\cdot)$ to the data  $\z_I$. Consequently, $\mathcal{L}_{\mathsf{in}}(\cpsset)$ and $\mathcal{L}_{\mathsf{cf}} (\cpsset)$ in (\ref{obj:inSample}) and (\ref{obj:crossFitting}) can be defined receptively. Specially, the RECV objective in (\ref{eq:recyc_cv}) is given by
\begin{align}\label{eq:recv_kde}
\mathcal{L}_{ \mcv}(\cpsset)=\sum_{k=1}^{K+1}\left\{\min_{h \in \Lambda}\sum_{m=1}^M L(\z_{J_{m, (\tau_{k-1},\tau_k]}};\hf_{J_{-m, (\tau_{k-1},\tau_k]},h})\right\}+ \gamma\size{\cpsset},
\end{align}
where $\Lambda$ is a candidate bandwidth set.

\begin{remark}
This is closely related to \citet{londschien2022changeforest}, where a classifier is used to learn a likelihood ratio for changepoint detection.
In contrast, we directly estimate the segment density and evaluate it by out-of-sample log-likelihood, so that bandwidth selection is naturally incorporated into the RECV objective.
\end{remark}

Next, we verify the conditions of the general theory in Section 3 for the KDE-based RECV estimator.
Condition~\ref{cond:model_and_loss_kernel} collects standard regularity conditions for multivariate kernel density estimation, tailored to our changepoint setting; see also \citet{MR1876841} and \citet{padilla2021optimal}.

\begin{condition}[Density and kernel]\label{cond:model_and_loss_kernel}
    (a) For every $i \in [n]$, the density $\f_{i}^\ast$ satisfies $\sup_{\z\in\Zcal} \f_{i}^\ast(\z) \le C_f$ and $\int_{\Zcal} \{\f_{i}^\ast(\z)\}^{-1} \mathrm{d} \z \le C_{\mathsf{inv}}$, where all $\f_{i}^\ast$ share a common support set $\Zcal \subset \Rbb^p$ and $C_f,C_{\mathsf{inv}}$ are some positive constants. The support set $\Zcal$ has bounded Lebesgue measure and diameter.
    (b) For every $i \in [n]$, $\f_{i}^\ast$ belongs to the Holder class $\Hcal(r, L, \Rbb^p)$, i.e., $\f_{i}^\ast$ is $\lfloor r \rfloor$-times continuously differentiable and its Taylor polynomial of degree $\lfloor r \rfloor$ at $\z_0$, $\f_{i, \z_0}^{\ast, (\lfloor r \rfloor)}(\z)$, satisfies that $\abs{\f_{i, \z_0}^{\ast (\lfloor r \rfloor)}(\z) - \f_i^\ast(\z)} \le L \norm{\z - \z_0}^{r}$ for all $\z, \z_0 \in \Rbb^p$.
    (c) The kernel $\Kcal$ is Lipschitz continuous and bounded, with $\sup_{\z \in \Rbb^p} \abs{\Kcal(\z)} \le C_{K}$, and satisfies $\Kcal(\z) \le C \exp(- \norm{\z}_2)$ for all $\z \in \Rbb^p$ and some constant $C > 0$.
    (d) The kernel $\Kcal(\cdot)$ is adaptive to $\Hcal(r, L, \Rbb^p)$, i.e., for every $f \in \Hcal(r, L, \Rbb^p)$,
    \begin{equation*}
        \sup_{\z \in \Rbb^p} \Bigabs{\int_{\Rbb^p} h^{-p} \Kcal\Bigl(\frac{\z - \x}{h}\Bigr) f(\x) \mathrm{d} x - \f(\z)} \le C_r h^{r},
    \end{equation*}
    for some constant $C_r > 0$ and all $h > 0$.
    (e) For each interval $I$, the candidate bandwidth set $\Lambda$ satisfies $\log \size{\Lambda} \lesssim \log n$, and all its elements lie in $[2n^{-{1}/{p} - {1}/({2r + p})} (\log n)^{{1}/{p}}, C_h]$ for some constant $C_h > 1$. Moreover, $\Lambda$ contains some $h_I^\ast$ such that $h_I^\ast \asymp \size{I}^{- {1}/({2r + p})}$.
\end{condition}

\begin{lemma}[Kernel density estimation]\label{lem:kde_err}
    Suppose Condition~\ref{cond:model_and_loss_kernel} holds. Uniformly for every nearly homogeneous segment $I \in \homo$, there exists $h_I^\ast \in \Lambda$ with $h_I^\ast \asymp \size{I}^{-{1}/({2r+p})}$ such that for every $m\in[M]$,
    \begin{equation*}
        \sup_{\z \in \Zcal} |\hf_{J_{-m, I}, h_I^\ast}(\z) - \f_{{J_{-m, I}}}^\ast(\z)| \lesssim C \size{J_{-m, I}}^{-\frac{r}{p + 2r}} (\log n)^{{1}/{2}},
    \end{equation*}
    with probability at least $1 - n^{-C}$ for some constants $C > 0$.
\end{lemma}

Lemma~\ref{lem:kde_err} extends the classical convergence theory of KDE to nearly homogeneous segments in our setting; in particular, the KDE attains the usual nonparametric minimax rate \citep[e.g.,][]{{padilla2023change}}, up to logarithmic factors, on such segments. As a consequence, Condition \ref{cond:predict_exist} holds with $\dacc=n^{p/(2r+p)} \log n$. Moreover, Condition~\ref{cond:changes}(c) and Condition~\ref{cond:model_and_loss} hold under Condition~\ref{cond:model_and_loss_kernel}. See Section~S2.7 in the Supplementary Material for details. In particular, the signal quantity $\Delta_k$ in Section~\ref{sec:framework_theory} specializes to
\[
    \Delta_k = \mathrm{KL}(\f_{(\tau_{k-1}^\ast,\tau_k^\ast]}^\ast\|\f_{(\tau_k^\ast,\tau_{k+1}^\ast]}^\ast)
    \vee
    \mathrm{KL}(\f_{(\tau_k^\ast,\tau_{k+1}^\ast]}^\ast\|\f_{(\tau_{k-1}^\ast,\tau_k^\ast]}^\ast),
\]
where $\mathrm{KL}(f||g) = \Ebb_{\z \sim f}[\log\{f(\z) / g(\z)\}]$ denotes the Kullback-Leibler divergence. It remains to impose Conditions~\ref{cond:changes}(a) and \ref{cond:changes}(b), which are summarized in the following condition.

\begin{condition}[Nonparametric changes]\label{cond:changes_kde}
There exists a sufficiently large constant $C_{\snr}>0$ such that for $k\in[K^*+1]$, $\tau^\ast_k - \tau^\ast_{k-1} \ge C_{\snr} {n^{{p}/({2r+p})}\log^2 n} (\Delta_{k-1}^{-1} \vee 1 + \Delta_k^{-1} \vee 1)$, and $\max_{k \in [K^\ast]} \Delta_k \lesssim  n^{{p}/({2r+p})}$.
\end{condition}

We now state the consistency result of the RECV procedure.
\begin{corollary}\label{coro:cvloss_kde}
Suppose that Conditions \ref{cond:model_and_loss_kernel} and \ref{cond:changes_kde} hold. Let $\dm = C_{\m} \log n$ and $\gamma = C_{\gamma} {n^{{p}/({2r+p})} \log^2 n}$ for some constants $C_{\m}, C_{\gamma} > 0$. The estimator $\widehat{\mathcal{T}}_{\mcv} = \{\htau_{\mcv,1}, \dots, \htau_{\mcv,\widehat{K}_{\mcv}}\}$, obtained by minimizing $\mathcal{L}_{ \mcv}(\cpsset)$ in \eqref{eq:recv_kde}, satisfies:
\begin{equation*}
    \widehat{K}_{\mcv} = K^\ast\ \text{and}\ \max_{1 \le k \le K^\ast} \min_{1 \le j \le \widehat{K}_{\mcv}} \Delta_k \abs{\tau^\ast_k - \htau_{\mcv,j}} \lesssim n^{{p}/({2r+p})} \log^2 n,
\end{equation*}
with probability at least $1 - n^{-C}$ for some constant $C > 0$.
\end{corollary}

Corollary~\ref{coro:cvloss_kde} is a direct consequence of Theorem~\ref{thm:cvloss}. It shows that the proposed cross-fitting framework with RECV implementation attains a localization rate {$\dacc=n^{{p}/({2r+p})}\log^2 n$}. Compared with \citet{padilla2023change}, where the bandwidth is pre-specified in the detection step, the present result allows bandwidth selection from a data-driven candidate set through RECV while retaining the same minimax rate order (up to logarithmic factors).

\section{Extension to temporally dependent data}\label{sec:blockwise_temporal}

In many applications, changepoint detection is performed on time-series data where observations exhibit temporal dependence. In such settings,  the choice of splitting scheme for cross-fitting becomes critical.
The order-preserved splitting method, discussed in Remark \ref{rmk:fold_partition} assigns observations to folds in an interleaved manner. It is less appropriate for temporally dependent sequences, because the training and validation observations remain temporally adjacent.

To this end, we consider a simple variant. For any segment $I$ with $\size{I}\geq M$, we partition it into $M$ contiguous folds $\{J_{1,I}, \dots, J_{M,I}\}$ of approximately equal length, so that $\size{J_{m,I}} \approx \size{I}/M$ for $m\in[M]$. Since observations near the boundaries of the $m$-th validation fold $J_{m,I} = (a, b]$ remain highly correlated with adjacent observations in the training set $J_{-m, I} = I \setminus J_{m, I} = I \setminus (a, b]$,
we further incorporate a \textit{buffer} mechanism inspired by \citet{RACINE200039}.
Specifically, when evaluating the loss on $J_{m,I}$, we use the modified training set $J_{-m,I}^{v} = I \setminus (a - v, b + v]$, which is obtained by excluding from the original training set $J_{-m,I}$ a buffer of size $v\ge 0$ on both sides of the validation fold $J_{m,I}=(a,b]$.
This creates a minimum temporal separation between the validation set $J_{m, I}$ and the training set $J_{-m, I}^{v}$, thereby mitigating short-range dependence and improving the reliability of the out-of-sample loss evaluation. We next provide theoretical guarantees for this strategy.

In the independent setting, the model estimator $\hf_{J_{-m, I}}$ is strictly independent of the validation set $J_{m, I}$.
This allows us to apply Bernstein's inequality to the centered empirical excess risk $\sum_{i \in J_{m, I}} s_{i, \hf_{J_{-m, I}}}$, conditional on $\hf_{J_{-m, I}}$, in order to derive the bounds on $\xi_I$ required in Lemma~\ref{lem:loc_err_g}. In the temporally-dependent setting, the model estimator is replaced by $\hf_{J_{-m, I}^{v}}$.
Intuitively, when $v$ is sufficiently large, the temporal dependence between $\hf_{J_{-m, I}^{v}}$ and $\{\z_i\}_{i \in J_{m, I}}$ becomes negligible.

To rigorously establish the consistency of the cross-fitting estimator, we adopt the functional dependence measure (FDM) framework \citep{wu2005nonlinear}.
A sequence $\{\z_i\}_{i=-\infty}^{\infty}$ is called a functional-dependent sequence if it admits the representation $\z_i = g_i(\{\zeta_{j}\}_{j \le i})$, where $\{\zeta_{i}\}_{i=-\infty}^{\infty}$ are i.i.d. innovations and $\{g_i(\cdot)\}_{i=-\infty}^{\infty}$ are measurable functions describing the underlying dependence mechanism.
The FDM relies on a coupling argument: let $\widetilde{\z}_{i, \{i-v\}}$ be a coupled version of $\z_i$ in which the lag-$v$ innovation $\zeta_{i-v}$ is replaced by an independent copy $\widetilde{\zeta}_{i-v}$. Then, $\widetilde{\z}_{i, \{i-v\}}$ is identically distributed to $\z_i$ but independent of $\zeta_{i-v}$, and the effect of $\zeta_{i-v}$ on $\z_i$ is measured by $\z_{i} - \widetilde{\z}_{i, \{i - v\}}$.
Consequently, the $q$-th moment FDM and its cumulative version are defined by $\theta_{\z, q}(v) = \sup_{i \in \mathbb{Z}} \Ebb \norm{\z_i - \widetilde{\z}_{i, \{i - v\}}}_q$ and $\Theta_{\z, q}(v) = \sum_{j = v}^{\infty} \theta_{\z, q}(j)$, respectively.

For any index set $J \subset \mathbb{Z}$, define similarly the coupled observation $\widetilde{\z}_{i, J}$, in which the innovations $\{\zeta_{j}\}_{j \in J}$ are replaced by independent copies.
Let $\widetilde{s}_{i, \f, J}$ be the coupled version of the centered empirical excess risk $s_{i, \f}$, with $\z_i$ replaced by $\widetilde{\z}_{i, J}$.
To control the approximation error term $\xi_I$ under temporal dependence, we impose the following conditions on the  dependence structure.

\begin{condition}\label{cond:fdm_stability}
(a) (Functional dependence decay) Assume that the cumulative FDM satisfies $\Theta_{\z, q}(v) \le \exp(-c v^{\kappa_2})$ for some constants $c, \kappa_2 > 0$.

(b) (Feature-space Lipschitz envelope)
There exists a feature map $\phi$ from $\mathcal{Z}$ to a normed space $\mathcal{B}$ with norm $\norm{\cdot}_{\mathcal{B}}$ such that for any perturbation index set $J \subset \mathbb{Z}$ and $i \in [n]$, it holds that $\sigma_{i, \f}^{-1} \abs{s_{i, \f} - \widetilde{s}_{i, \f, J}} \le \bignorm{\phi(\z_i) - \phi(\widetilde{\z}_{i, J})}_{\mathcal{B}}$.
Moreover, the effect of perturbing a single innovation decays fast enough in the sense that $[\mathbb{E}\bignorm{\phi(\z_i) - \phi(\widetilde{\z}_{i, \{i-v\}})}_{\mathcal{B}}^2]^{1/2} \le \theta_{\z, q}(v)$.

(c) (Mean-square relative algorithmic stability)
For any segment $I$ and fold $m$, let the validation set be $J_{m,I} = (a, b]$ and the model estimator $\hf := \hf_{J_{-m,I}^v}$.
Let $\tf$ denote the decoupled model estimator trained on $\{\widetilde{\z}_{i, (a-v, b]}\}_{i \in J_{-m, I}^{v}}$.
For $i \in J_{m, I}$, these two model estimators satisfy the following mean-square algorithmic stability:
\begin{equation*}
    \max \Biggl\{ \mathbb{E} \Bigabs{ \frac{s_{i, \hf} - s_{i, \tf}}{\sigma_{i, \hf}} }^2, \; \mathbb{E} \Bigabs{ \frac{\sigma_{i, \hf} - \sigma_{i, \tf}}{\sigma_{i, \hf}} }^2, \; \mathbb{E} \Bigabs{ \frac{m_{i, \hf} - m_{i, \tf}}{m_{i, \hf}} }^2 \Biggr\} \le C_{\mathsf{stab}}^2 \Theta_{\z, q}^2(v),
\end{equation*}
for some universal constant $C_{\mathsf{stab}} > 0$.
\end{condition}

\begin{remark}[Feature map for squared loss]
For the squared loss $\ell(\z; \f) = (y - \x^\top \f)^2$ with $\z = (\x, y)$ and $y = \x^\top \f_{i}^\ast + \epsilon$, the variance proxy satisfies $\sigma_{i, \f} \asymp \norm{\f - \f_{i}^\ast}_2^2 \vee \norm{\f - \f_{i}^\ast}_2$.
The triangle inequality yields:
\begin{equation*}
    \bigabs{s_{i, \f} - \widetilde{s}_{i, \f, J}} \lesssim \norm{\f - \f_{i}^\ast}_2^2 \bignorm{\x_i \x_i^\top - \widetilde{\x}_{i, J} \widetilde{\x}_{i, J}^\top}_{\mathrm{op}} + \norm{\f - \f_{i}^\ast}_2 \bignorm{\epsilon_i \x_i - \widetilde{\epsilon}_{i, J} \widetilde{\x}_{i, J}}_2.
\end{equation*}
This identifies the feature map $\phi(\z) = (\x \x^\top, \epsilon \x)$ and the induced norm:
\begin{equation*}
    \bignorm{\phi(\z_i) - \phi(\widetilde{\z}_i)}_{\mathcal{B}} \asymp \bignorm{\x_i \x_i^\top - \widetilde{\x}_{i, J} \widetilde{\x}_{i, J}^\top}_{\mathrm{op}} + \bignorm{\epsilon_i \x_i - \widetilde{\epsilon}_{i, J} \widetilde{\x}_{i, J}}_2.
\end{equation*}
Verifying Condition~\ref{cond:fdm_stability}(b) therefore requires controlling $\mathbb{E}\bignorm{\x_i \x_i^\top - \widetilde{\x}_{i, J} \widetilde{\x}_{i, J}^\top}_{\mathrm{op}}^2$, which involves fourth-degree cross-terms of the covariates.
Hence one typically needs finite fourth moments, that is, $q = 4$. In particular, Condition~\ref{cond:fdm_stability}(b) follows if both $\{\x_i\}$ and $\{\epsilon_i\}$ are functional-dependent sequences with exponentially-decay fourth-moment FDM.
Condition~\ref{cond:fdm_stability}(c) can be verified under similar assumptions.
\end{remark}

Condition \ref{cond:fdm_stability} formalizes the role of temporal dependence under blockwise splitting with a buffer. Together, these conditions ensure that the dependence between the fitted model and the validation data is weak enough for the cross-fitting argument to remain valid.

\begin{theorem}\label{thm:cross_fitting_temporal}
Suppose Conditions \ref{cond:changes}, \ref{cond:model_and_loss}, \ref{cond:predict}, and \ref{cond:fdm_stability} hold with $\kappa = (\kappa_1^{-1} + \kappa_{2}^{-1})^{-1} < 1$.
Assume that $d_{\m}$ and $\gamma$ are chosen as in Lemma~\ref{lem:loc_err_g},  that $d_{\m}$ further satisfies $d_{\m} \gtrsim (\log n)^{1/\kappa}$, and that the buffer size satisfies $v \gtrsim (\log n)^{1/\kappa_2}$.
Then, the estimator $\widehat{\mathcal{T}}_{\cf} = \{\htau_{\cf,1}, \dots, \htau_{\cf,\widehat{K}_{\cf}}\}$, obtained by minimizing the cross-fitting objective $\mathcal{L}_{\cf}(\cpsset)$, satisfies:
\begin{equation*}
    \widehat{K}_{\cf} = K^\ast\ \text{and}\ \max_{1 \le k \le K^\ast} \min_{1 \le j \le \widehat{K}_{\cf}} \abs{\tau^\ast_k - \htau_{\cf,j}} \le \widetilde{C} \dacc\Delta_k^{-1},
\end{equation*}
with probability at least $1 - p_n - n^{-C}$ for some constant $C > 0$.

Similarly, if Condition~\ref{cond:predict} is replaced by Condition~\ref{cond:predict_exist}, the estimator $\widehat{\mathcal{T}}_{\mcv}$, obtained by minimizing the RECV objective $\mathcal{L}_{ \mcv}(\cpsset)$, satisfies:
\begin{equation*}
    \widehat{K}_{\mcv} = K^\ast\ \text{and}\ \max_{1 \le k \le K^\ast} \min_{1 \le j \le \widehat{K}_{\mcv}} \abs{\tau^\ast_k - \htau_{\mcv,j}} \le \widetilde{C} \dacc\Delta_k^{-1},
\end{equation*}
with probability at least $1 - p_n - n^{-C}$ for some constant $C > 0$.
\end{theorem}

Theorem~\ref{thm:cross_fitting_temporal} extends the consistency guarantees to temporally dependent regimes.
By taking the buffer size $v \gtrsim (\log n)^{1/\kappa_2}$, the exponential decay of $\Theta_{\z, q}(v)$ is sufficiently fast, ensuring that the effect of temporal dependence can be rigorously controlled.

\section{Numerical studies}\label{sec:simul}

\subsection{Common experimental setup}\label{sec:simul_setup}

We begin by describing several aspects of the experimental setup that are common across the simulations and real-data analyses.

To assess the accuracy of changepoint estimation, we use the Hausdorff distance between the estimated changepoint set and the true changepoint set:
\[
\max\biggl\{\max_{1 \le j \le \hat K} \min_{1 \le k \le K^\ast} |\tau_k^\ast - \htau_j|, \max_{1 \le k \le K^\ast} \min_{1 \le j \le\hat K} |\tau_k^\ast - \htau_j|\biggr\},
\]
averaged over $500$ replications. In some settings, however, the true number of changepoints is known in advance. In such cases, we choose the penalty parameter $\gamma$ so that the resulting estimator satisfies $\hat K=K^\ast$ when solving the objective function \eqref{equ:mcp_loss}. The Hausdorff distance remains the evaluation metric in this setting; for example, in the single-changepoint case, it reduces to $|\tau_1^\ast-\widehat\tau_1|$.

We next summarize the competing methods. For model-fitting procedures without tuning parameters, we consider the in-sample and cross-fitting approaches obtained by minimizing the objective functions in \eqref{obj:inSample} and \eqref{obj:crossFitting}, respectively. We refer to these methods as \textbf{in} and \textbf{cf}.

For model-fitting procedures involving a tuning parameter $\lambda\in\Lambda$, we consider two strategies for selecting $\lambda$, for a given value of $\gamma$.
\begin{itemize}
    \item The first uses a \textit{common} tuning parameter across all segments. For the in-sample approach, we adopt an order-preserved hold-out split of the data into training and validation sets \citep{leonardi_computationally_2016,rinaldo_localizing_2021,xu2024change}. For each $\lambda\in\Lambda$, we estimate changepoints by minimizing the in-sample objective on the training set, evaluate the resulting segmentation through its hold-out prediction error, and select the value of $\lambda$ with the smallest hold-out error. We then refit the estimator on the full data using the selected $\lambda$. We denote this method by \textbf{in-ho}. For the cross-fitting approach, the splitting scheme already provides natural hold-out sets. We therefore minimize the cross-fitting objective \eqref{obj:crossFitting} on the full data for each $\lambda \in \Lambda$, and select the value of $\lambda$ with the smallest cross-fitting loss together with the corresponding changepoint estimate. We denote this method by \textbf{cf-ho}.
    \item The second strategy allows the tuning parameter to \textit{vary} across segments. For both the in-sample and cross-fitting approaches, we use segmentwise cross-validation to select the tuning parameter, as described in Section \ref{amcv}. The two versions differ only in the loss used for evaluation: the in-sample version uses \(L(\z_I;\hat f_I(\hat\lambda_I^{\rm cv}))\), whereas the cross-fitting version uses \(\sum_{m=1}^M L(\z_{J_{m,I}};\hat f_{J_{-m,I}}(\hat\lambda_{J_{-m,I}}^{\rm cv}))\). We refer to them as \textbf{in-cv} and \textbf{cf-cv}, respectively. Finally, the \textbf{recv} method is obtained by minimizing the objective function in \eqref{eq:recyc_cv}.
\end{itemize}

When the penalty parameter $\gamma$ is not fixed in advance, we select it using a hold-out strategy for all methods \citep{zou_consistent_2020}. For in-sample methods, we compute the hold-out prediction error over a candidate set of $\gamma$ values, select the value with the smallest error, and then refit the estimator on the full data. For cross-fitting methods, we optimize the corresponding cross-fitting or RECV objective for each candidate $\gamma$, select the value with the smallest prediction error, and retain the associated changepoint estimate.

\subsection{High-dimensional linear models}\label{sec:simul_linear}

In this subsection, we consider high-dimensional linear models with changepoints as in Section \ref{sec:linear}. Throughout, we use the squared loss and lasso-based model fitting, with the different methods distinguished by their loss evaluation and tuning strategy.

\subsubsection{Impact of common tuning under a single-changepoint model}\label{sec:HD-tuning}

We first examine how a common tuning parameter $\lambda$ affects changepoint localization under a single-changepoint model, with $(n,p,K^\ast,\tau^\ast_1) = (500,1000,1,150)$.
The covariates $\x_i$ are i.i.d. from $\Ncal(0, \Sigma)$ with $\Sigma_{ij} = 0.2^{\abs{i-j}}$, and the noises $\epsilon_i$ are i.i.d. from $\Ncal(0,1)$.
The regression coefficients change from $\f_{(\tau^\ast_{0},\tau^\ast_{1}]}^\ast = \id_{5}$ to $\f_{(\tau^\ast_{1},\tau^\ast_{2}]}^\ast = (1+\sqrt{b/5}) \id_5$, with $b\in\set{0.5, 1, 5}$ to reflect changes ranging from mild to substantial.
Here, $\id_{s}$ denotes a vector in $\Rbb^p$ with the first $s$ entries being one and the rest being zero.

\begin{figure}[tb]
    \centering
    \includegraphics[width=0.85\linewidth]{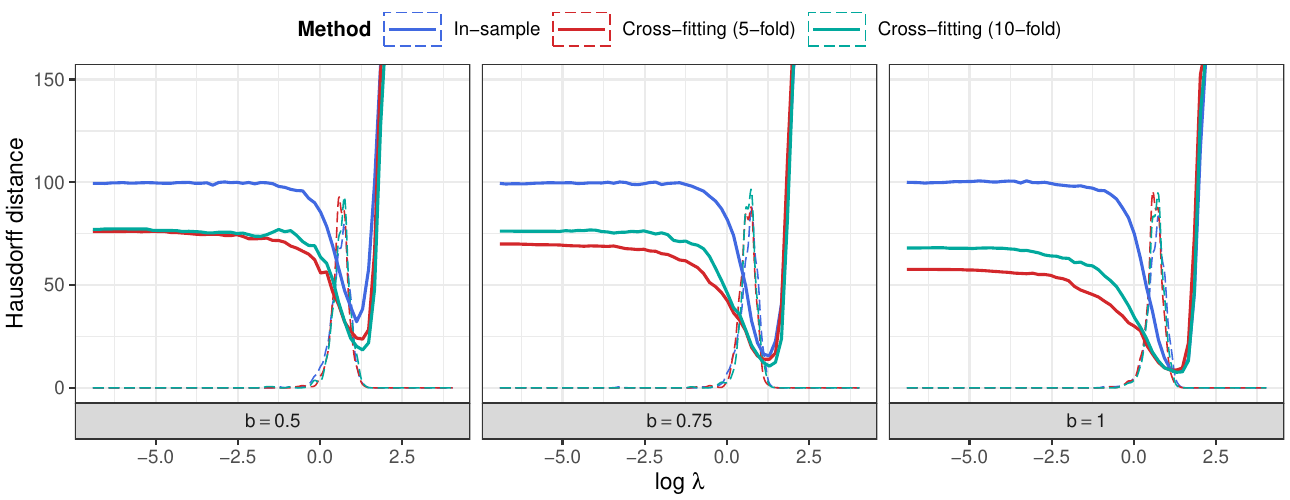}
    \caption{\small Influence of a common tuning parameter on changepoint localization in a high-dimensional linear model with a single changepoint.}
\label{fig:sensitivity}
\end{figure}

Figure \ref{fig:sensitivity} plots the average Hausdorff distances against $\log\lambda$, where smaller values of $\lambda$ correspond to more overfitting lasso fits.
We compare the in-sample and cross-fitting losses, with $5$ or $10$ folds used in the latter.
Across all settings, the cross-fitting approach matches or improves the in-sample approach, with the gain being most visible in strongly overfitting regimes.
This advantage is more pronounced when the change is larger, where overfitted lasso estimators may still achieve reasonable prediction performance while producing many false positives.
This observation aligns with the insights from Proposition \ref{prop:linear}.
The dotted curves in Figure \ref{fig:sensitivity} show the empirical distributions of the tuning parameter selected by \textbf{in-ho} and \textbf{cf-ho}. These selections tend to concentrate in regions where cross-fitting provides smaller localization error.
This suggests that, under a common data-driven tuning parameter, out-of-sample evaluation can yield more accurate localization than its in-sample counterpart.

\begin{figure}[tb]
    \centering
    \includegraphics[width=0.85\linewidth, height=4.2cm]{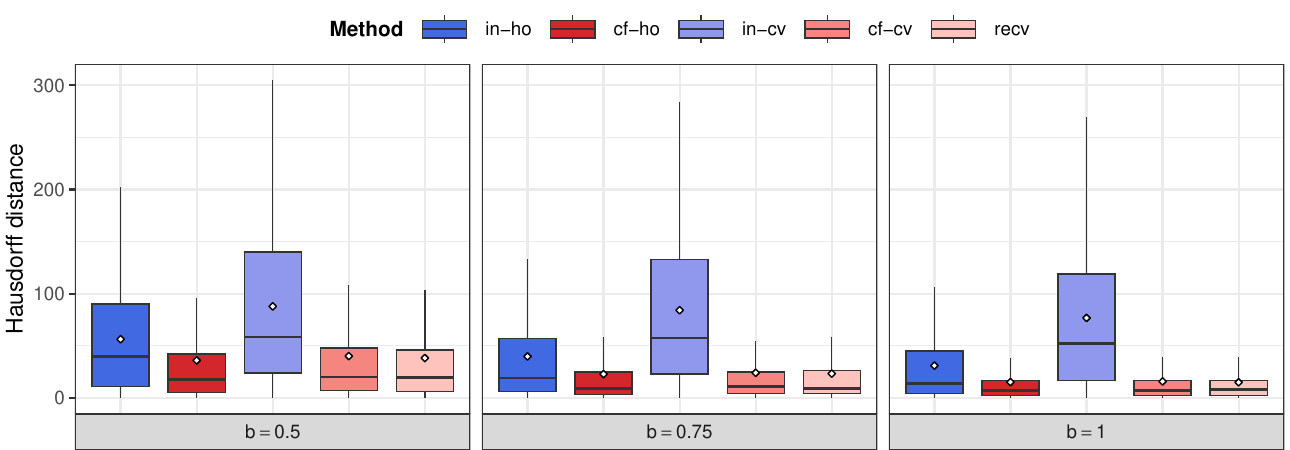}
    \caption{\small Boxplots of Hausdorff distances for data-driven methods in a high-dimensional linear model with a single changepoint.}
\label{fig:loc_err_scp_homo}
\end{figure}

Figure \ref{fig:loc_err_scp_homo} reports boxplots of the Hausdorff distances for methods with data-driven tuning parameter selection.
The in-sample methods \textbf{in-ho} and \textbf{in-cv} generally exhibit  larger localization errors than their cross-fitting counterparts \textbf{cf-ho} and \textbf{cf-cv}. The \textbf{recv} method achieves comparable performance to \textbf{cf-cv}, while avoiding the additional computational cost of a separate inner cross-validation step.

\subsubsection{Segment-specific versus global tuning under multiple-changepoint models}\label{sec:HD-common-vs-vary}

We next compare segment-specific tuning with global tuning under multiple-changepoint models, in addition to the comparison between in-sample and cross-fitting losses. In the notation of Section \ref{sec:simul_setup}, \textbf{cf-cv} and \textbf{recv} select tuning parameters separately across segments, whereas \textbf{cf-ho} uses a common tuning parameter.
As in the single-changepoint setting, we assess localization accuracy with the true number of changepoints treated as known, in order to avoid confounding the comparison with errors in estimating the number of changepoints (see Section \ref{sec:HD-ncp}).
We consider two data generating processes (DGPs), with the second introducing additional heterogeneity through segment-dependent noise levels.
In both setups, we take $(n,p,K^\ast)=(1000,1000,3)$, with changepoints at $\set{350, 500, 880}$. The covariates $\x_i$ are generated i.i.d. from $\Ncal(0, \Sigma)$, where $\Sigma_{ij} = 0.2^{\abs{i-j}}$.
\begin{itemize}
    \item \textbf{DGP 1:} The regression coefficients are given by $\f_{(\tau^\ast_{0},\tau^\ast_{1}]}^\ast = \f_{(\tau^\ast_{2},\tau^\ast_{3}]}^\ast = \id_{5}$, $\f_{(\tau^\ast_{1},\tau^\ast_{2}]}^\ast = (1+\sqrt{b/5}) \id_5$, and $\f_{(\tau^\ast_{3},\tau^\ast_{4}]}^\ast = (1-\sqrt{b/5}) \id_5$, where $b\in\{1.0,1.25,1.5\}$ represents different levels of signal strength. The noises $\epsilon_i$ are i.i.d. from $\Ncal(0,1)$.
    \item \textbf{DGP 2:} The regression coefficients are given by $\f_{(\tau^\ast_{0},\tau^\ast_{1}]}^\ast = \f_{(\tau^\ast_{2},\tau^\ast_{3}]}^\ast = \id_5$, $\f_{(\tau^\ast_{1},\tau^\ast_{2}]}^\ast = \id_5 + \qbf / \norm{\qbf}_2$, and $\f_{(\tau^\ast_{3},\tau^\ast_{4}]}^\ast = \id_5 - 0.9 \qbf / \norm{\qbf}_2$, where $\qbf=(1, -1.3, 1, -1.3, 1, 0, \ldots, 0)^\top$.
    The noises $\epsilon_i$ are Gaussian with mean zero, and their variances are chosen so that the response variance remains constant across segments.
        The minimum noise standard deviation, $\mathrm{se} = \min_{i}\{\Var(\epsilon_i)\}^{\frac{1}{2}}$, varies over $\set{0.8, 0.5, 0.2}$ to represent different levels of signal strength.
    \end{itemize}

\begin{figure}[tb]
    \centering
    \includegraphics[width=0.85\linewidth]{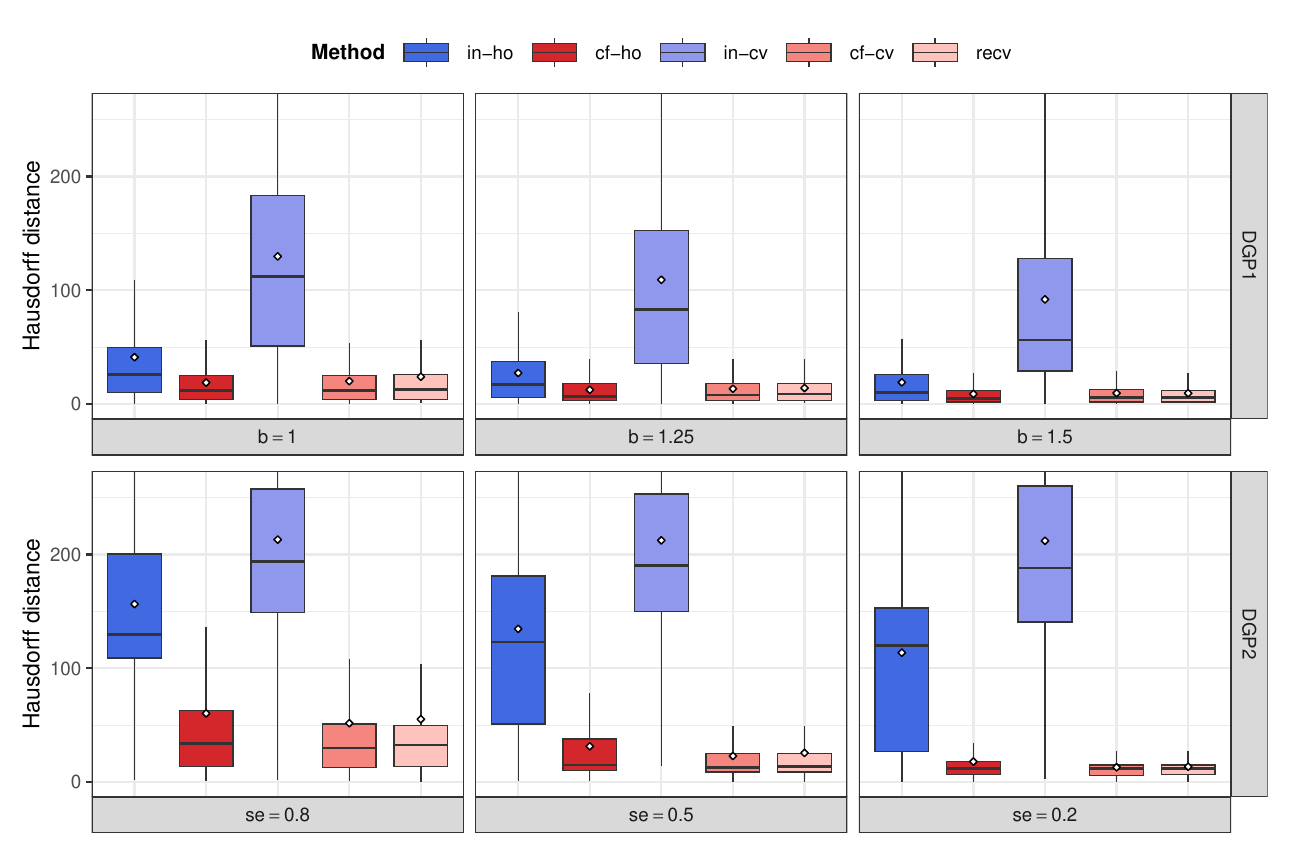}
\caption{\small Boxplots of Hausdorff distances for global and segment-specific tuning methods in high-dimensional linear models with multiple changepoints.}
\label{fig:loc_err_lasso_homo_and_heter}
\end{figure}

Figure~\ref{fig:loc_err_lasso_homo_and_heter} reports boxplots of the Hausdorff distance for different tuning strategies under both DGPs.
Cross-fitting-based methods consistently improve upon their in-sample counterparts: \textbf{cf-ho} outperforms \textbf{in-ho}, and \textbf{cf-cv} outperforms \textbf{in-cv}.
The \textbf{recv} method again achieves performance comparable to \textbf{cf-cv}.
The comparison between global and segment-specific tuning depends on the degree of heterogeneity in the data. Under DGP 1, the segment-specific tuning approaches \textbf{cf-cv} and \textbf{recv} perform similarly, on average, to the global tuning method \textbf{cf-ho}. Under DGP 2, however, they yield consistently smaller localization error than \textbf{cf-ho}, indicating that segment-specific tuning is particularly beneficial when the segments differ more substantially.

\subsubsection{Data-driven estimation of the number of changepoints}\label{sec:HD-ncp}

We now examine the performance of the fully data-driven procedures in estimating the number of changepoints. Unlike Sections \ref{sec:HD-tuning} and \ref{sec:HD-common-vs-vary}, where localization accuracy is assessed with the true number of changepoints treated as known, here the penalty parameter is selected in a fully data-driven manner, as described in \ref{sec:simul_setup}.

\begin{figure}[tb]
    \centering
    \includegraphics[width=0.85\linewidth]{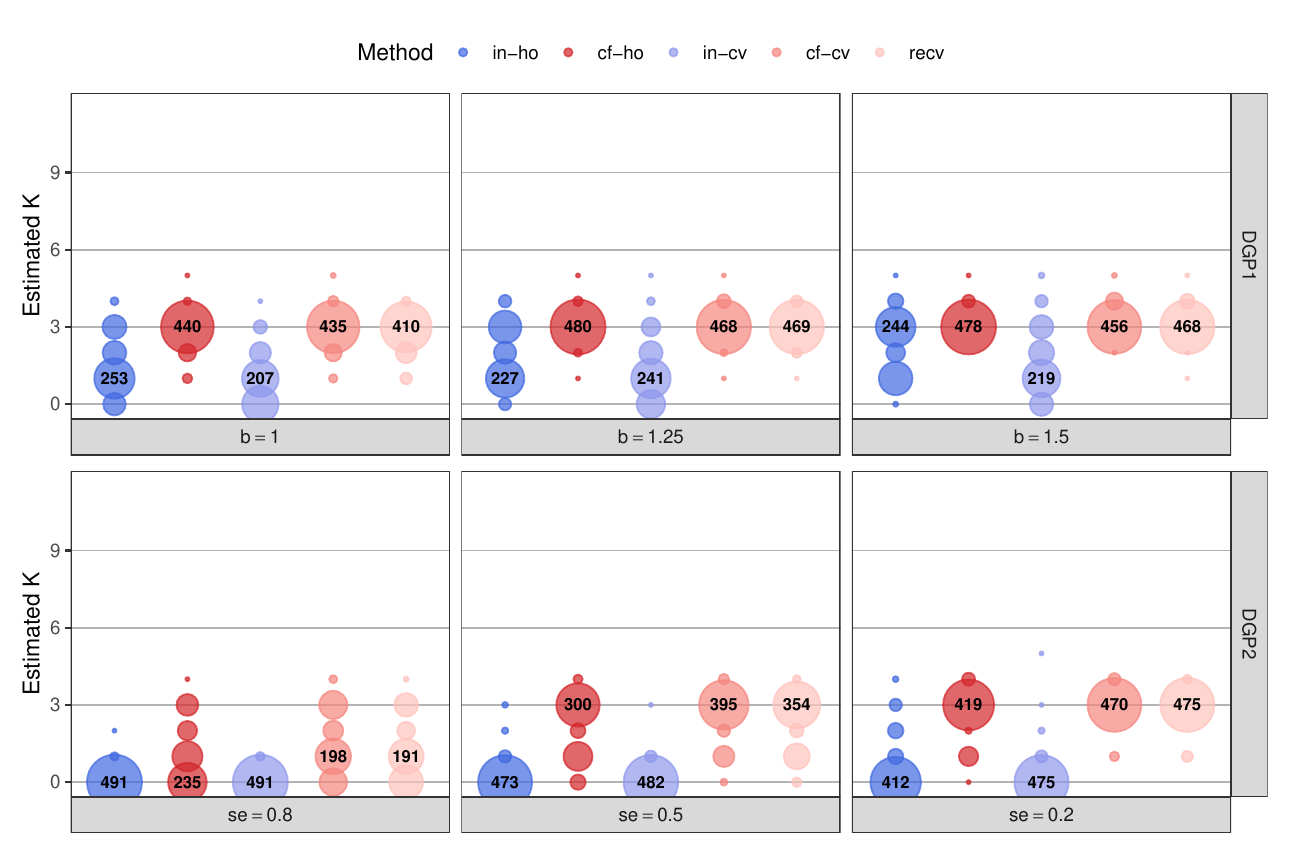}
\caption{\small Bubble plot of $\hat K$ for data-driven methods in high-dimensional linear models with multiple changepoints, where $K^\ast=3$.}
\label{fig:Kest_lasso_homo_and_heter_bubble}
\end{figure}

Figure~\ref{fig:Kest_lasso_homo_and_heter_bubble} summarizes the empirical distribution of  $\hat K$ over $500$ replications. Overall, the cross-fitting-based methods provide more accurate estimation of $K^\ast=3$ than their in-sample counterparts. In particular, when the change signal is sufficiently strong, the cross-fitting methods combined with the fully data-driven selection procedure recover $K^\ast$ with high frequency. This provides further evidence that the proposed framework improves not only localization accuracy, but also the practical estimation of the number of changepoints. Developing a theory for joint data-driven selection of both the number and locations of changepoints under the proposed framework remains an important direction for future research.

\subsubsection{Temporal dependence}

We also examine a temporally dependent setting based on DGP 2 in Section \ref{sec:HD-common-vs-vary}. Specifically, the baseline observations $\{(\x_i, \epsilon_i)\}$ are first generated as in DGP 2 and then transformed through an AR(2) recursion to induce temporal dependence: $\x_{i}' = \sqrt{0.6} \x_{i} + \sqrt{0.2} (\x_{i-1}' + \x_{i-2}')$ and $\epsilon_{i}' = \sqrt{0.6} \epsilon_{i} + \sqrt{0.2} (\epsilon_{i-1}' + \epsilon_{i-2}')$.

\begin{figure}[tb]
    \centering
    \includegraphics[width=0.85\linewidth]{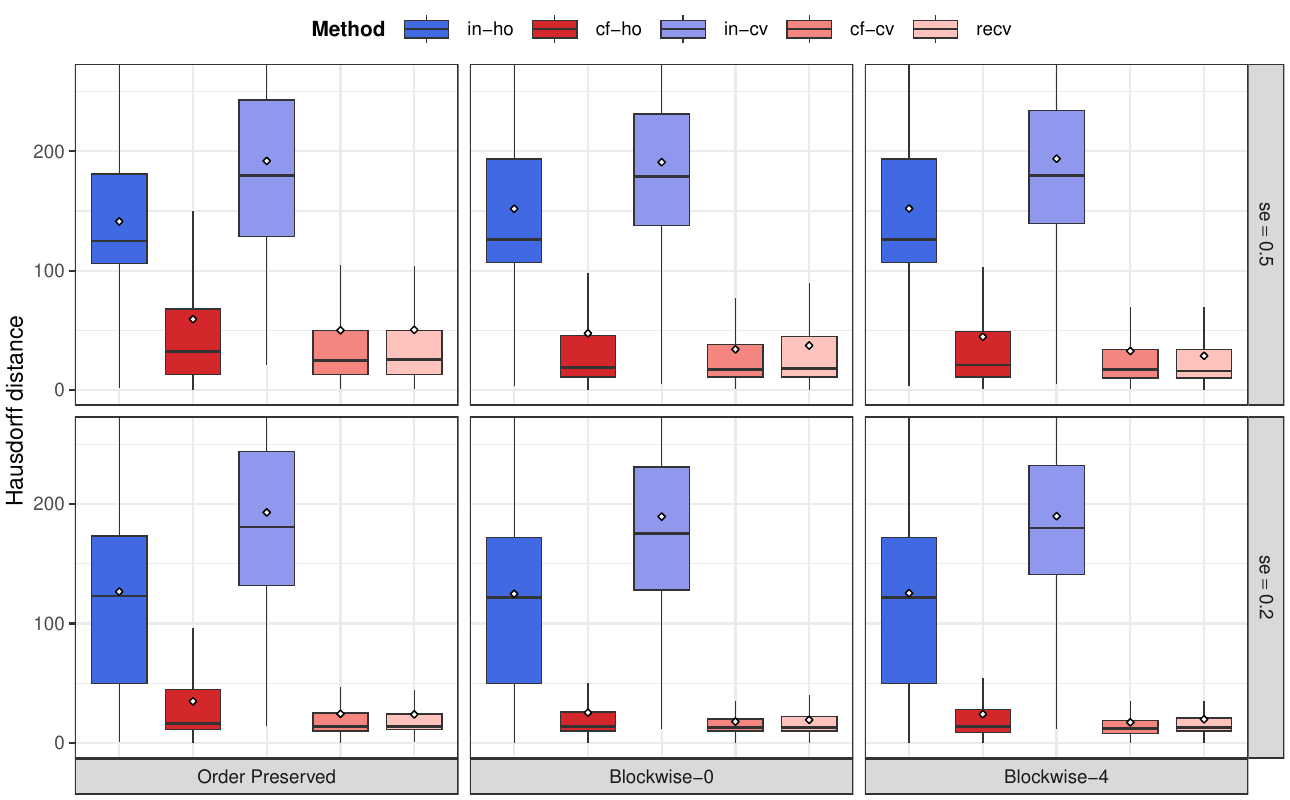}
\caption{\small Boxplots of Hausdorff distances for in-sample and cross-fitting methods in high-dimensional linear models with multiple changepoints under AR(2) dependence.}
\label{fig:loc_err_lasso_heter_temporal}
\end{figure}

As in Section \ref{sec:HD-common-vs-vary}, we treat the true number of changepoints as known, so that the comparison focuses on localization accuracy. Figure~\ref{fig:loc_err_lasso_heter_temporal} reports boxplots of the Hausdorff distance for $\mathrm{se} \in \{0.5, 0.2\}$ under several data-splitting strategies, including the original cross-fitting method, as well as the blockwise splitting methods proposed in Section \ref{sec:blockwise_temporal} with buffer sizes $v=0$ and $v=4$.
Temporal dependence reduces the accuracy of the original cross-fitting method, but the resulting localization error remains moderate. By contrast, the blockwise splitting methods mitigate the effect of temporal dependence and further improve performance.
Across all settings considered here, the cross-fitting approaches remain more accurate than their in-sample counterparts.
In addition, we similarly study the fully data-driven estimates of $K^\ast$ as in Section \ref{sec:HD-common-vs-vary}. The results are consistent with those in the independent case, and the blockwise strategy also offers certain improvements; see Figure S6 in the Supplementary Material.

\subsection{A real-data application: Honey bee waggle dance}

We illustrate the proposed method using the Bee-Dance dataset \citep{oh2008learning}, which is available at \url{https://sites.cc.gatech.edu/~borg/ijcv_psslds/}.
The data record a bee's pixel locations in the $x$- and $y$-directions over time, together with the body angle $\theta$, to study the waggle dance of bees.
From a behavioral perspective, the waggle dance alternates among three distinct states, and the goal is to detect the changepoints corresponding to transitions between these states.
These transitions are of interest because the dance is used to communicate the location of food or water to other bees.

To reduce the effect of the bee's changing absolute position in the image, we first difference the trajectory and then construct three motion features from the differenced coordinates and the body angle.
Let $\Delta x_t = x_{t+1} - x_t$ and $\Delta y_t = y_{t+1} - y_t$. We define the forward speed $v_{f,t} = \Delta x_t \cos(\theta_t) + \Delta y_t \sin(\theta_t)$, the sideways speed $v_{s,t} = -\Delta x_t \sin(\theta_t) + \Delta y_t \cos(\theta_t)$, and the turning speed $\omega_t = \theta_{t+1} - \theta_{t}$. We then analyze the normalized trivariate series $z_t=(v_{f,t},v_{s,t},\omega_t)$. These features describe local movement more directly than the original pixel coordinates.

We use the nonparametric loss in Section~\ref{sec:kde_application}, $\ell(z_i;f)=-\log f(z_i)$, with a kernel density estimator for the segmentwise density. For the proposed method, bandwidth selection is data-driven and implemented through the RECV procedure for temporally dependent data introduced in Section~\ref{sec:blockwise_temporal}. As a benchmark, we also consider the corresponding in-sample objective based on the same loss, with bandwidth selected in a similarly data-driven way.

For this dataset, prior information from video analysis suggests using $K=15$ changepoints \citep{oh2008learning}. We therefore fix $K=15$ for both the in-sample and RECV procedures, and use these $15$ annotated changepoints as reference changepoints for comparison.

\begin{figure}[htbp]
\centering
\includegraphics[width=0.9\textwidth]{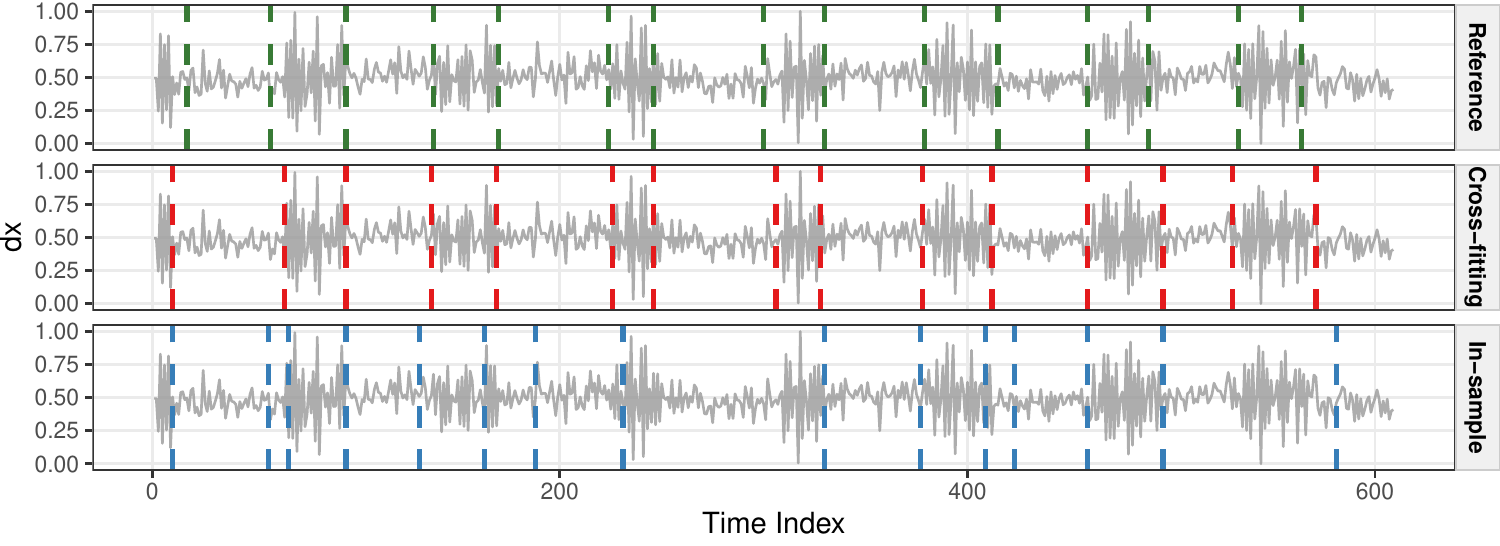}
\caption{\small Comparison of changepoint detection methods on the Bee-Dance dataset. The panels display the reference (top), the estimates from the cross-fit method (middle), and the estimates from the in-sample method (bottom).}
\label{fig:beedance}
\end{figure}

The results show that the RECV-based estimates are closer to the reference changepoints than those from the in-sample approach. The in-sample method tends to react to small local fluctuations and misses two changepoints in the middle part of the sequence, whereas RECV gives a more stable segmentation overall. This example supports the use of out-of-sample loss evaluation in this application.

\section{Concluding remarks}\label{sec:conclusion}

We conclude with several remarks. First, in practice, changepoint search algorithms typically require specifying either the number of changepoints or the penalty value in the objective function. Developing principled data-driven procedures for selecting the number of changepoints \citep{zou_consistent_2020,MR4815989} remains an important direction for future research. By treating the penalty parameter $\gamma$ as an additional tuning parameter, the RECV strategy could potentially be extended to achieve simultaneous consistency in estimating both the number and locations of changepoints.

Second, it would be of interest to extend the present loss-minimization framework to discrepancy-based changepoint detection \citep{wang+Samworth+2018+p57,MR4399097,MR4767500}. Such methods typically rely on local comparisons of estimators and are often combined with localized search schemes such as BS or WBS. Since the cross-fitting principle is naturally compatible with such localized procedures, developing a unified theory and efficient implementation that bridges out-of-sample evaluation with localized search is another promising direction.

Finally, it is important to highlight that the proposed cross-fitting framework is inherently loss-agnostic. In applications with heavy-tailed data, the squared loss may be replaced by more robust alternatives, such as Huber or rank-based
losses \citep{fearnhead2019changepoint, cui2024robust}. Such substitutions can preserve favorable tail behavior of the centered excess risk, which is the core component in our analysis, even when the raw data satisfy only polynomial moment conditions. This modularity allows the proposed methodology to be robustly extended to a wide range of non-vanilla distributional settings.

\newpage
\appendix

\def\thesection{S\arabic{section}}
\setcounter{section}{0}
\def\theequation{S\arabic{equation}}
\setcounter{equation}{0}
\def\thelemma{S\arabic{lemma}}
\setcounter{lemma}{0}
\def\thetheorem{S\arabic{theorem}}
\setcounter{theorem}{0}
\def\thecondition{S\arabic{condition}}
\setcounter{condition}{0}
\def\theproposition{S\arabic{proposition}}
\setcounter{proposition}{0}
\def\thetable{S\arabic{table}}
\setcounter{table}{0}
\def\thefigure{S\arabic{figure}}
\setcounter{figure}{0}
\def\theassumption{S\arabic{assumption}}
\def\thedefinition{S\arabic{definition}}
\def\thealgorithm{S\arabic{algorithm}}
\setcounter{definition}{0}
\setcounter{algorithm}{0}

\begin{center}
\bf\Large Supplementary Material for\\
``Changepoint Detection in Complex Models: Cross-Fitting Is Needed''
\end{center}

This supplementary material contains algorithmic details, theoretical proofs, and additional numerical results.

\section{Computational algorithms}
\label{sec:appendix_optimization}

\subsection{Generic loss minimization}

To detect the true segmentation $\truecps=(\tau^\ast_1,\ldots,\tau^\ast_{K^\ast})$, our methodology requires solving the discrete optimization problem \eqref{equ:mcp_loss_dm} over the space of potential segmentations:
\begin{equation}\label{equ:appendix_mcp_loss}
    \estcps = \argmin_{\cpsset\in\mathbb{T}(d_{\mathsf{m}})}\biggl\{\sum_{k=1}^{K+1}\Lval_{(\tau_{k-1},\tau_k]} + \gamma\size{\cpsset}\biggr\},
\end{equation}
where $\mathbb{T}(d_{\mathsf{m}})$ denotes the collection of all valid segmentations with a minimum segment length $d_{\mathsf{m}}$, $\Lval_{I}$ represents the evaluated loss (e.g., the in-sample loss, the cross-fitted loss, or the RECV loss) on segment $I$, and $\gamma > 0$ acts as the penalty factor.
In Table~\ref{tab:models}, we summarize several changepoint detection tasks for complex models via such loss minimization.

Existing search algorithms originally developed for traditional in-sample losses, such as exact dynamic programming \citep{jackson2005algorithm}, or efficient binary-segmentation-type methods \citep{fryzlewicz_wild_2014,kovacs_seeded_2022}, and other computationally accelerated variants \citep{killick_optimal_2012-1,li2023divide,qian2025reliever}, can be seamlessly adopted.
One simply needs to replace the standard in-sample loss evaluation with our proposed cross-fitting or RECV loss in these algorithms to obtain the cross-fitting-based changepoint estimates.

Below, we illustrate this algorithmic adaptability by detailing two representative algorithms: the optimal partitioning algorithm for exact dynamic programming, and the seeded binary segmentation algorithm for highly efficient approximation.

\begin{table}[H]
\setlength\tabcolsep{0pt}
\begin{threeparttable}
\caption{Changepoint detection for complex models via loss minimization.}
\label{tab:models}
\centering
\begin{tabular*}{\textwidth}{c@{\extracolsep{\fill}}*{4}{c}}
\toprule
$\z_i\in\mathcal{Z}$ & $\f_i^\ast$ & $\L(\z_I;\f)$ & $\hf_I$ \\
\midrule
\multicolumn{4}{l}{\underline{High-dimensional models}}\\
$z_i=(y_i,\x_i)\in\mathbb{R}\times\mathbb{R}^p$ & $\Ebb(y_i\mid\x_i)=\x_i^\top\f_i^\ast$ & $\sum_{i\in I}(y_i-\x_i^\top\f)^2$ & $\hf_I^\mathsf{lasso}$\tnote{$\mathparagraph$} \\
\\
& $F_{y_i\mid\x_i}^{-1}(\alpha)=\x_i^\top\f_i^\ast$\tnote{$\mathsection$} & $\sum_{i\in I}\rho_\alpha(y_i-\x_i^\top\f)$\tnote{$\mathsection\mathsection$} & $\hf_I^\mathsf{lasso}$ \\
\\
$\z_i\in\mathbb{R}^p$ & \makecell{$\f_i^\ast=(\mu_i,\Omega_i)$,\\ $P_i=\Ncal(\mu_i,\Omega_i^{-1})$} & \makecell{$\size{I}\left\{{\rm tr}(\Omega^\top S_I)-\log\det(\Omega)\right\}$,\\ $S_I=\sum_{i\in I}(\z_i-\mu)(\z_i-\mu)^\top$,\\ $\f=(\mu,\Omega)$} & $(\size{I}^{-1}\sum_{i\in I}\z_i,\widehat{\Omega}_I^\mathsf{glasso})$ \\
\\
$\z_i\in\mathbb{R}^{p_1\times p_2}$ & \makecell{$\f_i^\ast=(L_i,S_i)$,\\ $\z_i\sim\Ncal((L_i+S_i)\z_{i-1},\sigma^2I)$\\ (Low-rank $L_i$, sparse $S_i$)} & \makecell{$\sum_{i\in I}\|\z_i-(L+S)\z_{i-1}\|^2$,\\ $\f=(L,S)$} & $(\widehat{L}_I^\mathsf{nuc},\widehat{S}_I^\mathsf{lasso})$ \\
\\
\multicolumn{4}{l}{\underline{Nonparametric models}}\\
General objects & $\f_i^\ast=dP_i/d(n^{-1}\sum_{j=1}^nP_j)$ & $-\sum_{i\in I}\log \f(z_i)$ & $\hf_I(\cdot)$\tnote{$\dagger$} \\
\\
& $\int k_\gamma(z_i,\cdot)dP_i$\tnote{$\ddagger$} & $\sum_{i\in I}\|k_\gamma(\z_i,\cdot)-f\|_{\mathcal{H}_\gamma}^2$ & $\size{I}^{-1}\sum_{j\in I}k_\gamma(\z_j,\cdot)$ \\
\bottomrule
\end{tabular*}
\begin{tablenotes}[flushleft]\footnotesize
    \item[$\mathparagraph$] The superscripts lasso/glasso (nuc) generally refer to $L_1$- (nuclear-)norm regularized estimators, varying from context to context.
    \item[$\mathsection$] For $\alpha\in(0,1)$, $F_{y\mid\x}^{-1}(\alpha)$ denotes the $\alpha$-quantile of the conditional distribution $F_{y\mid\x}$ of $y$ given $\x$.
    \item[$\mathsection\mathsection$] $\rho_\alpha(u)=u(\alpha-1_{\{u<0\}})$ is the quantile loss function.
    \item[$\dagger$] $\hf_I(\z_i)$ denotes the predicted class-$1$ probability for $\z_i$, scaled by $n\size{I}^{-1}$, with the classifier trained based on inputs $\{\z_i\}_{i=1}^n$ and artificially generated labels $1$ for $i\in I$ and $0$ otherwise; refer to Section \ref{sec:simul_nonpara}.
    \item[$\ddagger$] $k_\gamma(\cdot,\cdot)$ denotes the kernel in a reproducing kernel Hilbert space, with a tuning hyperparameter $\gamma$.
\end{tablenotes}
\end{threeparttable}
\end{table}

\subsection{Exact global search via optimal partitioning}

Optimal partitioning \citep[OP,][]{jackson2005algorithm} utilizes dynamic programming to recursively find the exact global minimum of the objective function. Let $F(t)$ denote the minimum optimal cost for segmenting the data subsequence up to index $t$. The algorithm relies on the Bellman equation to recursively evaluate all valid preceding changepoint locations $s$:
\begin{equation}
    F(t) = \min_{0 \le s \le t - d_{\mathsf{m}}} \left\{ F(s) + \Lval_{(s, t]} + \gamma \right\},
\end{equation}
with the boundary condition $F(0) = -\gamma$ to cancel out the penalty for the first segment. By recording the minimizer $s$ at each step $t$, the estimated changepoint set $\estcps$ can be exactly reconstructed through a backward pass starting from $n$. The process is detailed in Algorithm \ref{alg:opart}.

\begin{algorithm}[H]
\caption{Optimal Partitioning for Changepoint Detection}
\label{alg:opart}
\begin{algorithmic}[1]
\REQUIRE Data length $n$, minimum segment length $d_{\mathsf{m}}$, penalty parameter $\gamma > 0$, loss evaluation function $\Lval_{(\cdot, \cdot]}$.
\ENSURE Estimated changepoint set $\estcps = (\htau_{1},\ldots,\htau_{\widehat{K}})$.
\STATE Initialization: $F(0) \gets -\gamma$;\; $\mathcal{P} \gets \zero_{n}$ (Array to store the optimal predecessor index)
\FOR{$t = 1$ \TO $d_{\mathsf{m}} - 1$}
    \STATE $F(t) \gets \infty$ (Invalidate segments shorter than $d_{\mathsf{m}}$)
\ENDFOR
\FOR{$t = d_{\mathsf{m}}$ \TO $n$}
    \STATE $F(t) \gets \min_{0 \le s \le t - d_{\mathsf{m}}} \left\{ F(s) + \Lval_{(s, t]} + \gamma \right\}$;\; $\mathcal{P}(t) \gets \argmin_{0 \le s \le t - d_{\mathsf{m}}} \left\{ F(s) + \Lval_{(s, t]} + \gamma \right\}$
\ENDFOR
\STATE $\estcps \gets \emptyset$;\; $\tau \gets \mathcal{P}(n)$ (Backtracking to reconstruct the segmentation)
\WHILE{$\tau > 0$}
    \STATE $\estcps \gets \estcps \cup \{\tau\}$;\; $\tau \gets \mathcal{P}(\tau)$
\ENDWHILE
\RETURN $\estcps$
\end{algorithmic}
\end{algorithm}

\subsection{Efficient approximation via seeded binary segmentation}\label{sec:seedbs}

When the loss evaluation $\Lval_{(s, t]}$ is computationally expensive, evaluating $\mathcal{O}(n^2)$ possible segments in OP becomes intractable. To this end, seeded binary segmentation \citep[SeedBS]{kovacs_seeded_2022} serves as a computationally accelerated version, motivated by wild binary segmentation \citep[WBS]{fryzlewicz_wild_2014}.

At a high level, rather than exhaustively evaluating all possible segments, SeedBS operates on a deterministic multi-scale grid of background intervals, denoted as the seeded interval collection $\mathcal{I}_{\mathrm{seed}}$ (Please refer to \citet{kovacs_seeded_2022} for the construction details). Begin with the full data sequence $(0, n]$, the algorithm evaluates each background interval $(s, e] \in \mathcal{I}_{\mathrm{seed}}$ by finding a single optimal split point $\tau$ that maximizes the loss reduction $\Lval_{(s, e]} - \Lval_{(s, \tau]} - \Lval_{(\tau, e]}$. It then identifies the most significant split point $\htau$ that yields the maximum overall loss reduction $\Delta_{\max}$ across all intervals in $\mathcal{I}_{\mathrm{seed}}$. If this maximum reduction satisfies $\Delta_{\max} \le \gamma$, the algorithm terminates. Otherwise, $\htau$ is accepted as a valid changepoint, the sequence is partitioned into two sub-intervals $(0, \htau]$ and $(\htau, n]$, and the SeedBS procedure is recursively applied to both the two sub-intervals.

Algorithm \ref{alg:seedbs_search} details this recursive changepoint search utilizing the maximum loss reduction across the candidate intervals in $\mathcal{I}_{\mathrm{seed}}$.

\begin{figure}[H]
    \centering
    \includegraphics[width=0.8\linewidth]{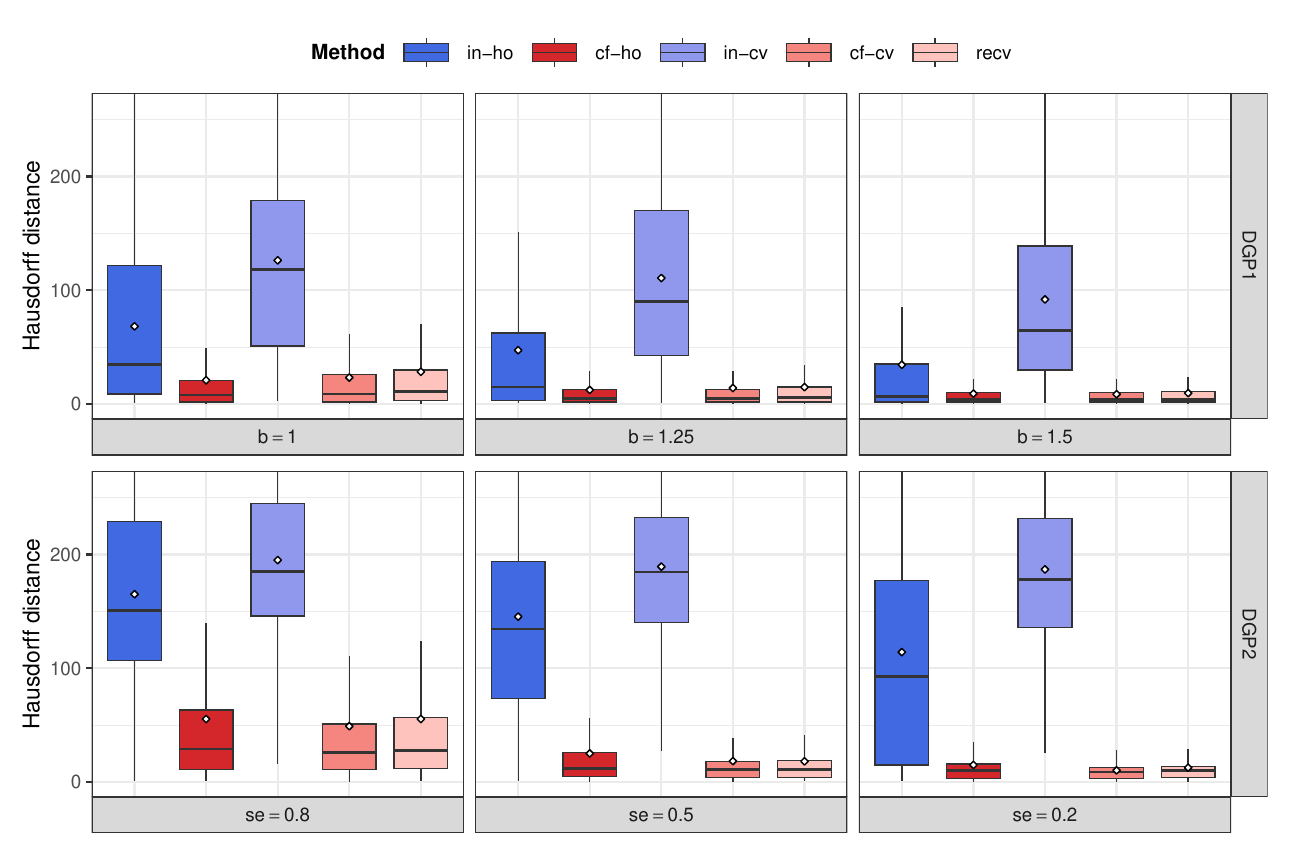}
\caption{\small Boxplot of empirical Hausdorff distances in high-dimensional linear models with multiple changepoints using the SeedBS algorithm.}
\label{fig:loc_err_lasso_homo_and_heter_sbs}
\end{figure}

\begin{figure}[H]
    \centering
    \includegraphics[width=0.8\linewidth]{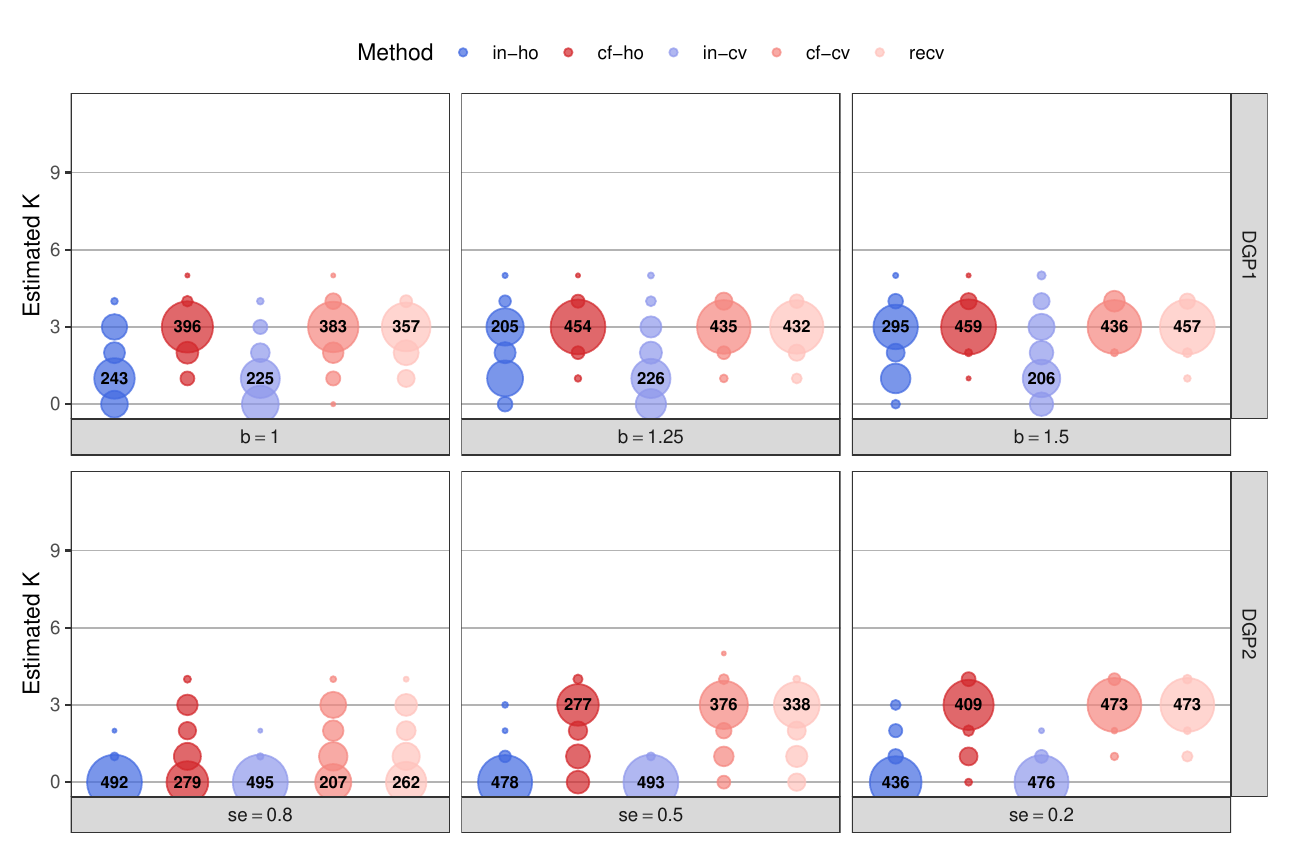}
\caption{\small Bubble plot of $\widehat{K}$ in high-dimensional linear models with multiple changepoints using the SeedBS algorithm.}
\label{fig:Kest_lasso_homo_and_heter_bubble_sbs}
\end{figure}

For experiments in the main text, we employ OP to solve the optimization problem. Here, we repeat the experiments in Section~\ref{sec:HD-common-vs-vary} using SeedBS as the optimizer. Figures~\ref{fig:loc_err_lasso_homo_and_heter_sbs}--\ref{fig:Kest_lasso_homo_and_heter_bubble_sbs} show that the Hausdorff distance and $\widehat{K}$ closely match the exact solutions obtained via OP.

\begin{algorithm}[htb]
\caption{Seeded Binary Segmentation for Changepoint Detection}
\label{alg:seedbs_search}
\begin{algorithmic}[1]
\REQUIRE Data length $n$, minimum segment length $d_{\mathsf{m}}$, penalty parameter $\gamma > 0$, loss evaluation function $\Lval_{(\cdot, \cdot]}$, current search domain $(a, b] \subseteq (0, n]$ (initialized with $a=0, b=n$), seeded interval set $\mathcal{I}_{\mathrm{seed}}$.
\ENSURE The set of detected changepoints $\text{SeedBS}((a, b], d_{\mathsf{m}}, \mathcal{I}_{\mathrm{seed}}, \gamma)$ within $(a, b]$ and the final estimated changepoint set $\estcps = (\htau_{1},\ldots,\htau_{\widehat{K}}) = \text{SeedBS}((0, n], d_{\mathsf{m}}, \mathcal{I}_{\mathrm{seed}}, \gamma)$.
\IF{$b - a < 2 d_{\mathsf{m}}$}
    \RETURN $\emptyset$
\ENDIF
\STATE Extract valid intervals: $\mathcal{I}_{a,b} \gets \{ (s, e] \in \mathcal{I}_{\mathrm{seed}} \mid a \le s \text{ and } e \le b \} \cup \{(a, b]\}$
\STATE Identify the maximum loss reduction $\Delta_{\max}$ across all candidate intervals and valid split points:

\quad $\Delta_{\max} = \max_{(s,e] \in \mathcal{I}_{a,b}} \max_{s + d_{\mathsf{m}} \le \tau \le e - d_{\mathsf{m}}} \{ \Lval_{(s, e]} - ( \Lval_{(s, \tau]} + \Lval_{(\tau, e]} ) \}$

\STATE Let $\htau$ be the optimal split point that achieves $\Delta_{\max}$
\IF{$\Delta_{\max} > \gamma$}
    \STATE $\cpsset_{\mathsf{L}} \gets \text{SeedBS}((a, \htau], d_{\mathsf{m}}, \mathcal{I}_{\mathrm{seed}}, \gamma)$ \COMMENT{Recursive search on the left}
    \STATE $\cpsset_{\mathsf{R}} \gets \text{SeedBS}((\htau, b], d_{\mathsf{m}}, \mathcal{I}_{\mathrm{seed}}, \gamma)$ \COMMENT{Recursive search on the right}
    \RETURN $\cpsset_{\mathsf{L}} \cup \{\htau\} \cup \cpsset_{\mathsf{R}}$
\ELSE
    \RETURN $\emptyset$
\ENDIF
\end{algorithmic}
\end{algorithm}

\subsection{Relieving the computation burden of multiple model fits in cross-fitting}

To ensure scalability to large datasets, we integrate the cross-fitting framework with the \textit{Reliever} technique \citep{qian2025reliever}.
The Reliever method reduces the computational burden by constructing a small set of proxy segments, denoted as the relief set $\mathcal{R}$.
For each candidate interval $I$, there exists a relief interval $R \in \mathcal{R}$ such that $R \subset I$ and $\size{R} / \size{I} \ge r$. Here $r \in [0, 1]$ represents the coverage ratio of the constructed set $\mathcal{R}$.
The Reliever strategy only needs to fit models in the set $\mathcal{R}$, rather than all the sub-intervals in $(0, n]$, which substantially reduces the computation time in model-fitting.

In the original in-sample setting, Reliever replaces the in-sample loss $L(\z_I; \hf_I)$ by $L(\z_I; \hf_R)$.
In both theory and practice,
$L(\z_I; \hf_R)$ can well approximate the original loss $L(\z_I; \hf_I)$ and provide desired changepoint estimation.
We refer to \citet{qian2025reliever} for technical details including the construction of $\mathcal{R}$ and the theoretical results.

For the cross-fitting alternative, we still split the interval $I$ into $M$ folds $\{J_{m, I}\}$ and denote $J_{-m, I} = I \setminus J_{m, I}$.
With specific choice of the splitting method, we can ensure that $J_{m,R} = J_{m, I} \cap R$ and $J_{-m, R} = J_{-m, I} \cap R$. Examples include the order-preserved splitting, and the blockwise splitting as illustrated in Figure~\ref{fig:illus_reliever}.
The reliever-version cross-fitting loss is then defined as
$$L_I = \sum_{m \in [M]} L(\z_{J_{m, I}}; \hf_{J_{-m, R}}).$$

Once the loss function is well-defined, search algorithms such as OP and SeedBS can be naturally incorporated. By combining the fast Reliever loss-evaluation strategy with an efficient search algorithm to solve the loss minimization problem, the computational burden can be effectively resolved.

\begin{figure}[t]
   \centering
    \resizebox{\linewidth}{!}{%
        \input{figures/fold_illus_reliever.tex}
    }
         \label{fig:illus_reliever}
\end{figure}

\section{Theoretical proofs}
\subsection{Proof of Lemma \ref{lem:loc_err_g}}\label{sec:proof_loc_err_g}

\begin{proof}[Proof of Lemma \ref{lem:loc_err_g}]

    We first introduce some notations.
    For a given changepoint estimation $\tau \in [n]$ and a set of changepoints $\cpsset = \set{0  < \tau_1 < \dots < \tau_K < \tau_{K+1} < n}$, denote $\Kcal_{+}(\tau,\cpsset) \triangleq \min_{k}\set{k : \tau_k > \tau}$ and $\Kcal_{-}(\tau,\cpsset) \triangleq \max_{k} \set{k: \tau_k < \tau}$.
    For simplicity, further denote $k_{\tau,+}^\ast = \Kcal_{+}(\tau, \truecps)$, $\hat{k}_{\tau,+} = \Kcal_{+}(\tau, \estcps)$, $k_{\tau,-}^{\ast} = \Kcal_{-}(\tau, \truecps)$ and $\hat{k}_{\tau,-} = \Kcal_{-}(\tau, \estcps)$.
    Let $\estcps = \set{\htau_1,\dots,\htau_{\widehat{K}}}$ be the minimizer of $\mathcal{L}_{\mathsf{in}}(\cpsset)$.
    Denote $d_k = \widetilde{C} \dacc \Delta_k^{-1}$ and the set of heterogeneous intervals $\heter = \set{(s, e]: \exists h \in [K^\ast], \min(\tau_h^\ast - s, e - \tau_h^\ast) > d_h}$.

    Assume that $\Hcal = \set{(\htau_a, \htau_{a+1}]: (\htau_a, \htau_{a+1}] \in \heter} \neq \emptyset$, i.e., $\exists h \in [K^\ast]$ such that $\estcps \cap [\tau_h^\ast - d_h, \tau_h^\ast + d_h] = \emptyset$.
    For such $h$ and $a$, without loss of generality assume that $\tau_h^\ast - \htau_a > d_h$, it can be observed that $(\tau_h^\ast - d_h, \tau_h^\ast + d_h] \subset (\htau_a, \htau_{a+1}]$ and $(\htau_{a}, \htau_{a+1}] \notin \homo$.

    Recall that the change magnitude $\Delta_k$ is defined by the loss differences on the samples $\{\z_{\tau_{k}}, \z_{\tau_{k+1}}\}$ and models $\f_{(\tau^\ast_{k-1},\tau^\ast_{k}]}^\ast$ and $\f_{(\tau^\ast_{k},\tau^\ast_{k+1}]}^\ast$.
    Here for the simplicity of notations, we generalize the definition of change magnitude to general sample sets.
    For an index set $I \subseteq (0, n]$, denote the oracle loss $\Lval_{I}^\ast = \sum_{i \in I} \ell(\z_i, \f_{i}^\ast)$ and its expectation $\overline{\Lval}_{I}^{\ast} = \Ebb[\Lval_{I}^\ast]$.
    The change magnitude of the set $I$ is denoted by $\Delta_{I} = \size{I}^{-1} \{\risk_I(\f_{I}^\ast) - \overline{\Lval}_{I}^{\ast}\}$.

    By Condition~\ref{cond:changes}(c) and the fact that $(\tau_h^\ast - d_h, \tau_h^\ast + d_h] \subset (\htau_a, \htau_{a+1}]$, We have $\Delta_{(\htau_a, \htau_{a+1}]} (\htau_{a+1} - \htau_a) \ge 2 d_h \Delta_{(\tau_h^\ast - d_h, \tau_h^\ast + d_h]} \ge C_{\Delta}^{-1} d_h \Delta_{h} = C_{\Delta}^{-1} \widetilde{C} \dacc$.

    We need the following definitions to divide $\Hcal$ into four groups to move further.
    \begin{definition}[Separability of a point]
        For a changepoint estimation $\tau$ and the true changepoint set $\truecps$, let $u = k_{\tau,-}^\ast$ and $v = k_{\tau,+}^\ast$.
        We say that $\tau$ is separable from the left if $\tau - \tau_u^\ast > d_u \vee d_{\mathsf{m}}$ and separable from the right if $\tau_v^\ast - \tau > d_v \vee d_{\mathsf{m}}$.
        Otherwise, $\tau$ is inseparable from the left (right).
    \end{definition}
    \begin{definition}[Separability of an interval]
        For the intervals $(\tau_l,\tau_r] \in \Hcal$, we make the following definitions,
        \begin{enumerate}[$\Hcal_1:$]
            \item $(\tau_l,\tau_r] \in (0, n]$ is separable if $\tau_l$ is separable from the right and $\tau_r$ is separable from the left.
            \item $(\tau_l,\tau_r] \in (0, n]$ is left-separable if $\tau_l$ is separable from the right and $\tau_r$ is inseparable from the left.
            \item $(\tau_l,\tau_r] \in (0, n]$ is right-separable if $\tau_l$ is inseparable from the right and $\tau_r$ is separable from the left.
            \item $(\tau_l,\tau_r] \in (0, n]$ is inseparable if $\tau_l$ is inseparable from the right and $\tau_r$ is inseparable from the left.
        \end{enumerate}
    \end{definition}
    Now the sub-intervals in $\Hcal$ have been classified into four groups $\Hcal = \Hcal_1 \cup \Hcal_2 \cup \Hcal_3 \cup \Hcal_4$.
    By emptying these groups, we will show that $\Hcal = \emptyset$.

    \subsubsection*{Case 1: $\Hcal_1 = \emptyset$}

    \begin{figure}[H]
        \centering
        \resizebox{\textwidth}{!}{
        \begin{tikzpicture}
            \path coordinate (A) at (0, 0)
                  coordinate (E) at (15, 0);
            \path coordinate (B) at ($ (A)!.25!(E) $)
                  coordinate (C) at ($ (A)!.5!(E) $)
                  coordinate (D) at ($ (A)!.75!(E) $);

            \draw[line width=.5pt]
               (A) -- (B) -- (C) -- (D) -- (E);
            \node[dot, label = {$\htau_{a}$}] at (A) {};
            \node[dot, label = {$\tau_{h}^\ast$}] at (B) {};
            \node[dot, label = {$\cdots$}] at (C) {};
            \node[dot, label = {$\tau_{h+t}^\ast$}] at (D) {};
            \node[dot, label = {$\htau_{a+1}$}] at (E) {};
        \end{tikzpicture}}
    \end{figure}

    For $(\htau_a, \htau_{a+1}] \in \Hcal_1$, let $h = k_{\htau_a,+}^\ast$, $\cpsset_a = \set{\tau_{h}^\ast,\dots,\tau_{h+t}^\ast}=\truecps \cap (\htau_a, \htau_{a+1})$ and $\widetilde{\cpsset} = \estcps \cup \cpsset_a$.
    Since $\gamma = C_{\gamma} \dacc$,
    \begin{align*}
        &\mathcal{L}(\estcps) - \mathcal{L}(\widetilde{\cpsset}) = \Lval_{(\htau_a, \htau_{a+1}]} - \Bigl\{\Lval_{(\htau_a, \tau_h^\ast]} + \Lval_{(\tau_{h+t}^\ast, \htau_{a+1}]} + \sum_{j=h}^{h+t-1} \Lval_{(\tau_j^\ast, \tau_{j+1}^\ast]} + (t+1)\gamma\Bigr\}\\
        > & (1 - C_{\ref*{lem:loc_err_g}.2}) \Delta_{(\htau_a, \htau_{a+1}]}^2 (\htau_{a+1} - \htau_a) - C_{\ref*{lem:loc_err_g}.1} \dacc - (t+2)C_{\ref*{lem:loc_err_g}.1} \dacc - (t+1) \gamma\\
        \ge & \Bigl\{(1 - C_{\ref*{lem:loc_err_g}.2}) (t + 1) C_{\Delta}^{-1} \widetilde{C} - (t+3) C_{\ref*{lem:loc_err_g}.1} - (t+1) C_{\gamma}\Bigr\} \dacc > 0,
    \end{align*}
    provided that $\widetilde{C} \ge (1 - C_{\ref*{lem:loc_err_g}.2})^{-1} C_{\Delta} (3 C_{\ref*{lem:loc_err_g}.1} + C_{\gamma})$.
    The first inequality is from the fact that $(\htau_a, \htau_{a+1}] \in \heter$ and other intervals are belong to $\homo$.
    Therefore $\Hcal_1 = \emptyset$.

    \subsubsection*{Case 2: $\Hcal_2 = \Hcal_3 = \emptyset$}

    \begin{figure}[H]
        \centering
        \resizebox{\textwidth}{!}{
        \begin{tikzpicture}
            \path coordinate (A) at (0, 0)
                  coordinate (F) at (15, 0);
            \path coordinate (G) at ($ (A)!.02!(F) $)
                  coordinate (B) at ($ (A)!.25!(F) $)
                  coordinate (C) at ($ (A)!.27!(F) $)
                  coordinate (D) at ($ (A)!.5!(F) $)
                  coordinate (E) at ($ (A)!.75!(F) $);
            \draw[line width=.5pt]
               (A) -- (G) -- (B) -- (C) -- (D) -- (E) -- (F);
            \node[dot, label = {-90:$\htau_{a-1}$}] at (A) {};
            \node[dot, label = {10:$\tau_{h-1}^\ast$}] at (G) {};
            \node[dot, label = {-90:$\htau_{a}$}] at (B) {};
            \node[dot, label = {$\tau_{h}^\ast$}] at (C) {};
            \node[dot, label = {$\cdots$}] at (D) {};
            \node[dot, label = {$\tau_{h+t}^\ast$}] at (E) {};
            \node[dot, label = {$\htau_{a+1}$}] at (F) {};
        \end{tikzpicture}}
    \end{figure}

    Without loss of generality, by the symmetry of $\Hcal_2$ and $\Hcal_3$, we only show that $\Hcal_3 = \emptyset$.
    If the claim does not hold, one can choose $(\htau_a, \htau_{a+1}] \in \Hcal_3$ to be the leftmost one.
    Hence $\htau_a$ must be separable from the left by Condition \ref{cond:changes}.
    Since $\Hcal_1 = \emptyset$ and $(\htau_a, \htau_{a+1}]$ is the leftmost interval in $\Hcal_3$, one obtains $(\htau_{a-1}, \htau_a] \not\in \Hcal$.
    Denote $h = k_{\htau_a,+}^\ast$ and $\cpsset_a = \truecps \cap (\htau_a + d_{\mathsf{m}}, \htau_{a+1}-d_{\mathsf{m}}) = \set{\tau_{h+1}^\ast,\dots,\tau_{h+t}^\ast}$ ($t=0$ if $\cpsset_a = \emptyset$).
    Let $\widetilde{\cpsset} = (\estcps \setminus {\htau_a}) \cup {\tau_h^\ast} \cup \cpsset_a = (\estcps \setminus {\htau_a}) \cup \set{\tau_j^\ast}_{j=h}^{h+t}$.
    \begin{align}\label{equ:case2_low_diff}
        \mathcal{L}(\estcps) - \mathcal{L}(\widetilde{\cpsset}) = & \Lval_{(\htau_a, \htau_{a+1}]} + \bigl(\Lval_{(\htau_{a-1}, \htau_a]} - \Lval_{(\htau_{a-1}, \tau_h^\ast]}\bigr) - \Bigl\{\sum_{j=h}^{h+t-1} \Lval_{(\tau_j^\ast, \tau_{j+1}^\ast]} + \Lval_{(\tau_{h+t}^\ast, \htau_{a+1}]} + t \gamma \Bigr\}\nonumber\\
        > & (1 - C_{\ref*{lem:loc_err_g}.2})\Delta_{(\htau_a, \htau_{a+1}]}^2 (\htau_{a+1} - \htau_a) - [(t+1) C_{\ref*{lem:loc_err_g}.1} + t C_{\gamma}] \dacc \nonumber\\
        + & \Bigl(\Lval_{(\htau_a, \tau_h^\ast]}^\ast + \Lval_{(\htau_{a-1}, \htau_a]} - \Lval_{(\htau_{a-1}, \tau_h^\ast]}\Bigr).
    \end{align}
    Since $(\htau_{a-1}, \htau_a] \not\in \Hcal$ and $0 < \tau_h^\ast - \htau_a < d_{\mathsf{m}}$, one must obtain that either $(\htau_{a-1}, \htau_a) \cap \cpsset^\ast = \emptyset$ or $0 < \tau_{h-1}^\ast - \htau_{a-1} < d_{h-1} = \widetilde{C} \Delta_{h-1}^{-1} \dacc$.

    For the first scenario, under $\Gbb^{\mathsf{nhomo}}$, $\Lval_{(\htau_{a-1}, \tau_h^\ast]} - \Lval_{(\htau_{a-1}, \tau_h^\ast]}^{\ast} \le C_{\ref*{lem:loc_err_g}.1} \dacc$.
    Under $\Gbb^{-\mathsf{nhomo}}$, $\Lval_{(\htau_{a-1}, \htau_{a}]} - \Lval_{(\htau_{a-1}, \htau_{a}]}^\ast > -C_{\ref*{lem:loc_err_g}.1} \dacc$ because $\risk_{(\htau_{a-1}, \htau_{a}]}(\f_{(\htau_{a-1}, \htau_{a}]}) - \risk_{(\htau_{a-1}, \htau_{a}]}^\ast = 0$.

    In summary, under $\Gbb^{\mathsf{nhomo}} \cap \Gbb^{-\mathsf{nhomo}}$,
    \begin{equation}\label{equ:case2_1}
        \Lval_{(\htau_a, \tau_h^\ast]}^\ast + \Lval_{(\htau_{a-1}, \htau_a]} - \Lval_{(\htau_{a-1}, \tau_h^\ast]} > - 2 C_{\ref*{lem:loc_err_g}.1} \dacc.
    \end{equation}
    Therefore,
    \begin{align*}
        \mathcal{L}(\estcps) - \mathcal{L}(\widetilde{\cpsset}) & > (1 - C_{\ref*{lem:loc_err_g}.2})\Delta_{(\htau_{a}, \htau_{a+1}]} (\htau_{a+1} - \htau_{a}) - [(t+3) C_{\ref*{lem:loc_err_g}.1} + t C_{\gamma}] \dacc \\
        & \ge \bigl\{(1 - C_{\ref*{lem:loc_err_g}.2})(t \vee 1)C_{\Delta}^{-1}\widetilde{C} - (t + 4) C_{\ref*{lem:loc_err_g}.1} - t C_{\gamma} \bigr\} \dacc > 0,
    \end{align*}
    provided that $\widetilde{C} \ge (1 - C_{\ref*{lem:loc_err_g}.2})^{-1} C_{\Delta} (5 C_{\ref*{lem:loc_err_g}.1} + C_{\gamma})$.

    For the second scenario, let $I_1 = (\htau_{a-1}, \htau_a]$ and $I_2 = (\htau_{a-1}, \tau_h^\ast]$.
    Firstly, we will show that the difference of variations $\Delta_{I_2} \size{I_2} - \Delta_{I_1} \size{I_1}$ is small.
    Since $I_1 \subset I_2$, we have $\Delta_{I_2} \size{I_2} - \Delta_{I_1} \size{I_1} \ge 0$.

    Since $0 < \tau_{h-1}^\ast - \htau_{a-1} < d_{h-1} = \widetilde{C} \Delta_{h-1}^{-1} \dacc$, we have
    \begin{align*}
        \Delta_{I_1} \size{I_1} &\le \Delta_{I_2} \size{I_2} \le \risk_{I_2}(\f_{(\tau^\ast_{h-1},\tau^\ast_{h}]}^\ast) - \overline{\Lval}_{I_2}^{\ast} \\
        &= \risk_{(\htau_{a-1}, \tau_{h-1}^\ast]}(\f_{(\tau^\ast_{h-1},\tau^\ast_{h}]}^\ast) - \risk_{(\htau_{a-1}, \tau_{h-1}^\ast]}(\f_{(\tau^\ast_{h-2},\tau^\ast_{h-1}]}^\ast) \\
        &\le   d_{h-1} \Delta_{h-1} = \widetilde{C} \dacc.
    \end{align*}
    Therefore under $\Gbb^{-\mathsf{nhomo}}$,
    $$\Lval_{I_1} - \overline{\Lval}_{I_1}^\ast - \Delta_{I_1} \size{I_1} > - \{(C_{\ref*{lem:loc_err_g}.2} \widetilde{C}) \vee C_{\ref*{lem:loc_err_g}.1} \} \dacc \ge - (C_{\ref*{lem:loc_err_g}.2} \widetilde{C} + C_{\ref*{lem:loc_err_g}.1} ) \dacc.$$
    Since $I_2 \in \homo$, we have $\Lval_{I_2} - \Lval_{I_2}^\ast - \Delta_{I_2} \size{I_2} \le C_{\ref*{lem:loc_err_g}.1} \dacc$ under $\Gbb^{\mathsf{nhomo}}$.

    Denote $b_1 = \tau_{h-1}^\ast - \htau_{a-1}$, $b_2 = \htau_a - \tau_{h-1}^\ast$ and $b_3 = \tau_h^\ast - \htau_a < d_{\mathsf{m}}$.
    By the definition of $\Delta_{(\cdot)}$, we have
    $$\Delta_{I_1} \size{I_1} \ge \frac{b_2}{b_2 + b_3} \Delta_{I_2} \size{I_2}.$$
    Hence
    \begin{align*}
        0 &\le \Delta_{I_2} \size{I_2} - \Delta_{I_1} \size{I_1} \le \Delta_{I_2} \size{I_2} \{1 - \frac{b_2}{b_2 + b_3}\} \\
        &= \Delta_{I_2} \size{I_2} \frac{b_3}{b_2 + b_3} \le \widetilde{C} \dacc \frac{b_3}{b_2 + b_3} \le \frac{d_{\m}}{C_{\snr} \dacc} \widetilde{C} \dacc.
    \end{align*}
    Denote $C_{\m, 1} = \frac{d_{\m}}{C_{\snr} \dacc}$ for short.
    Since $C_{\snr}$ is sufficiently large and $\frac{d_{\m}}{C_{\snr} \dacc}$ is sufficiently small, it holds that $0 < C_{\m, 1} \ll 1$ and we can set
    \begin{equation}\label{equ:Cm1_bound}
        0 < C_{\m, 1} < (4 C_{\Delta})^{-1} \{1 - (1 + 4 C_{\Delta}) C_{\ref*{lem:loc_err_g}.2}\}.
    \end{equation}

    In summary, under $\Gbb^{\mathsf{nhomo}} \cap \Gbb^{-\mathsf{nhomo}}$,
    \begin{equation}\label{equ:case2_2}
        \Lval_{(\htau_a, \tau_h^\ast]}^\ast + \Lval_{(\htau_{a-1}, \htau_a]} - \Lval_{(\htau_{a-1}, \tau_h^\ast]} > - \{2 C_{\ref*{lem:loc_err_g}.1} + (C_{\ref*{lem:loc_err_g}.2} + C_{\m, 1}) \widetilde{C} \}\dacc.
    \end{equation}

    Combining Eq.~\eqref{equ:case2_2} with Eq.~\eqref{equ:case2_low_diff}, under $\Gbb^{\mathsf{nhomo}} \cap \Gbb^{-\mathsf{nhomo}}$,
    \begin{align*}
        \mathcal{L}(\estcps) - \mathcal{L}(\widetilde{\cpsset}) & > (1 - C_{\ref*{lem:loc_err_g}.2} )\Delta_{(\htau_{a}, \htau_{a+1}]}^2 (\htau_{a+1} - \htau_{a}) - [(t + 3) C_{\ref*{lem:loc_err_g}.1} + t C_{\gamma} + (C_{\ref*{lem:loc_err_g}.2} + C_{\m, 1}) \widetilde{C}] \dacc \\
        & \ge \bigl\{(1 - C_{\ref*{lem:loc_err_g}.2})(t \vee 1) C_{\Delta}^{-1} \widetilde{C} - (C_{\ref*{lem:loc_err_g}.2} + C_{\m, 1}) \widetilde{C} - (t + 4) C_{\ref*{lem:loc_err_g}.1} - t C_{\gamma}\bigr\} \dacc > 0,
    \end{align*}
    provided that $\widetilde{C} \ge \{1 - (1+C_{\Delta}) C_{\ref*{lem:loc_err_g}.2} - C_{\Delta} C_{\m, 1}\}^{-1} C_{\Delta} (5 C_{\ref*{lem:loc_err_g}.1} + C_{\gamma})$.
    Hence $\Hcal_2 \cup \Hcal_3 = \emptyset$.

    \subsubsection*{Case 3: $\Hcal_4 = \emptyset$}

    Similar to Case 2, let $(\htau_a, \htau_{a+1}] \in \Hcal_4$, then $\htau_a$ is separable from the left and $\htau_{a+1}$ is separable from the right.

    \begin{figure}[H]
        \centering
        \resizebox{\textwidth}{!}{
        \begin{tikzpicture}
            \path coordinate (A) at (0, 0)
                  coordinate (F) at (15, 0);
            \path coordinate (H) at ($ (A)!.02!(F) $)
                  coordinate (B) at ($ (A)!.25!(F) $)
                  coordinate (C) at ($ (A)!.27!(F) $)
                  coordinate (D) at ($ (A)!.5!(F) $)
                  coordinate (E) at ($ (A)!.75!(F) $)
                  coordinate (G) at ($ (A)!.77!(F) $)
                  coordinate (I) at ($ (A)!.98!(F) $);
            \draw[line width=.5pt]
               (A) -- (H) -- (B) -- (C) -- (D) -- (E) -- (G) -- (I) -- (F);
            \node[dot, label = {-90:$\htau_{a-1}$}] at (A) {};
            \node[dot, label = {10:$\tau_{h-1}^\ast$}] at (H) {};
            \node[dot, label = {-90:$\htau_{a}$}] at (B) {};
            \node[dot, label = {$\tau_{h}^\ast$}] at (C) {};
            \node[dot, label = {$\cdots$}] at (D) {};
            \node[dot, label = {90:$\tau_{h+t}^\ast$}] at (E) {};
            \node[dot, label = {-87:$\htau_{a+1}$}] at (G) {};
            \node[dot, label = {150:$\tau_{h+t+1}^\ast$}] at (I) {};
            \node[dot, label = {-80:$\htau_{a+2}$}] at (F) {};
        \end{tikzpicture}}
    \end{figure}

    By the fact that $\Hcal_1 \cup \Hcal_2 \cup \Hcal_3 = \emptyset$, we also obtain $(\htau_{a-1}, \htau_a] \not\in \Hcal$ and $(\htau_{a+1}, \htau_{a+2}] \not\in \Hcal$.
    Let $h = k_{\htau_a,+}^\ast$ and $h + t = k_{\htau_{a+1},-}^\ast$.
    Denote $\cpsset_a = \set{\tau_h^\ast,\dots,\tau_{h+t}^\ast}$ and $\widetilde{\cpsset} = (\estcps \setminus \set{\htau_a, \htau_{a+1}} \cup \cpsset_a$.
    We have
    \begin{align*}
        \mathcal{L}(\estcps) - \mathcal{L}(\widetilde{\cpsset}) =& \Lval_{(\htau_a, \htau_{a+1}]} +[\Lval_{(\htau_{a-1}, \htau_a]} + \Lval_{(\htau_{a+1}, \htau_{a+2}]} - \Lval_{(\htau_{a-1}, \tau_{h}^\ast]} - \Lval_{(\tau_{h+t}^\ast, \htau_{a+2}]}]\\
        &- \sum_{j=h}^{h+t-1} \Lval_{(\tau_j^\ast, \tau_{j+1}^\ast]} - (t-1)\gamma\\
        > & (1 - C_{\ref*{lem:loc_err_g}.2})\Delta_{(\htau_a, \htau_{a+1}]} (\htau_{a+1} - \htau_a) - \{ (t + 1) C_{\ref*{lem:loc_err_g}.1} + (t-1) C_{\gamma}\} \dacc \\
        &+ [\Lval_{(\htau_a, \tau_h^\ast] \cup (\tau_{h+1}^\ast, \htau_{a+1}]}^\ast + \Lval_{(\htau_{a-1}, \htau_a]} + \Lval_{(\htau_{a+1}, \htau_{a+2}]} - \Lval_{(\htau_{a-1}, \tau_{h}^\ast]} - \Lval_{(\tau_{h+t}^\ast, \htau_{a+2}]}].
    \end{align*}

    Follow the same discussion in Case 2, see Eq.~\eqref{equ:case2_2}, we have
    \begin{align*}
        & \Lval_{(\htau_a, \tau_h^\ast] \cup (\tau_{h+1}^\ast, \htau_{a+1}]}^\ast+ \Lval_{(\htau_{a-1}, \htau_a]} + \Lval_{(\htau_{a+1}, \htau_{a+2}]} - \Lval_{(\htau_{a-1}, \tau_{h}^\ast]} - \Lval_{(\tau_{h+t}^\ast, \htau_{a+2}]} \\
        > & - 2 \{2 C_{\ref*{lem:loc_err_g}.1} + (C_{\ref*{lem:loc_err_g}.2} + C_{\m, 1}) \widetilde{C} \}\dacc.
    \end{align*}
    Hence,
    \begin{align*}
        & \mathcal{L}(\estcps) - \mathcal{L}(\widetilde{\cpsset})\\
        > & (1 - C_{\ref*{lem:loc_err_g}.2})\Delta_{(\htau_a, \htau_{a+1}]}^2 (\htau_{a+1} - \htau_a) - [(t+5) C_{\ref*{lem:loc_err_g}.1} + (t-1) C_{\gamma} + 2 (C_{\ref*{lem:loc_err_g}.2} + C_{\m, 1}) \widetilde{C}] \dacc \\
        \ge & \Bigl\{(1 - C_{\ref*{lem:loc_err_g}.2}) C_{\Delta}^{-1} [(t-1) \vee 1] \widetilde{C} - 2 (C_{\ref*{lem:loc_err_g}.2} + C_{\m, 1}) \widetilde{C} - (t + 5) C_{\ref*{lem:loc_err_g}.1} - (t-1) C_{\gamma}\Bigr\} \dacc \ge 0
    \end{align*}
    provided that $\widetilde{C} \ge \{1 - (1 + 2 C_{\Delta}) C_{\ref*{lem:loc_err_g}.2} - 2 C_{\Delta} C_{\m, 1}\}^{-1} C_{\Delta} (7 C_{\ref*{lem:loc_err_g}.1} + C_{\gamma} + 2 C_{\m, 1})$.

    In summary, we obtain $\Hcal = \emptyset$ provided that $\widetilde{C} \ge \{1 - (1 + C_{\Delta}) C_{\ref*{lem:loc_err_g}.2} - 2 C_{\Delta} C_{\m, 1}\}^{-1} C_{\Delta} (7 C_{\ref*{lem:loc_err_g}.1} + C_{\gamma} + 2 C_{\m, 1})$.
    Hence $\max_{1 \le j \le K^\ast} \min_{1 \le k \le \widehat{K}} \Delta_{j}^2 |\tau_j^\ast - \htau_k| \le \widetilde{C} \dacc$.
    It also implies that $\widehat{K} \ge K^\ast$.

    It remains to show that $\widehat{K} \le K^\ast$.
    Otherwise, assume that $\widehat{K} > K^\ast$.
    Then there must be $j \in [0, K^\ast]$ and $k \in [1, \widehat{K}]$ such that $\htau_{k - 1} \in [\tau_j^\ast - d_j, \tau_j^\ast + d_j]$ and $\htau_{k+t} \in [\tau_{j+1}^\ast - d_{j+1}, \tau_{j+1}^\ast + d_{j+1}]$ for some $t \ge 1$.
    Without loss of generality, assume that $t = 1$.
    Similar to the decomposition of $\Hcal$, we can divide the event into four groups.
    \begin{enumerate}[$\Gcal_1:$]
        \item $\tau_j^\ast \le \htau_{k-1} < \htau_k < \htau_{k + 1} \le \tau_{j+1}^\ast$.
        \item $\tau_j^\ast - d_j \le \htau_{k-1} < \tau_j^\ast$ and $\tau_j^\ast \le \htau_k < \htau_{k+1} \le \tau_{j+1}^\ast$.
        \item $\tau_j^\ast \le \htau_{k-1} < \htau_k \le \tau_{j+1}^\ast$ and $\tau_{j+1}^\ast < \htau_{k+1} \le \tau_{j+1}^\ast + d_{j+1}$.
        \item $\tau_j^\ast - d_j \le \htau_{k-1} < \tau_j^\ast \le \htau_k \le \tau_{j+1}^\ast < \htau_{k+1} \le \tau_{j+1}^\ast + d_{j+1}$
    \end{enumerate}

    \subsubsection*{Case 1: $\Gcal_1 = \emptyset$}

    \begin{figure}[H]
        \centering
        \resizebox{\textwidth}{!}{
        \begin{tikzpicture}
            \path coordinate (A) at (0, 0)
                  coordinate (F) at (15, 0);
            \path coordinate (B) at ($ (A)!.05!(F) $)
                  coordinate (C) at ($ (A)!.5!(F) $)
                  coordinate (D) at ($ (A)!.95!(F) $);

            \draw[line width=.5pt]
               (A) -- (B) -- (C) -- (D) -- (F);
            \node[dot, label = {$\tau_j^\ast$}] at (A) {};
            \node[dot, label = {-90:$\htau_{k-1}$}] at (B) {};
            \node[dot, label = {-90:$\htau_{k}$}] at (C) {};
            \node[dot, label = {-90:$\htau_{k+1}$}] at (D) {};
            \node[dot, label = {$\tau_{j+1}^\ast$}] at (F) {};
        \end{tikzpicture}}
    \end{figure}

    Let $\widetilde{\cpsset} = \estcps \setminus \set{\htau_k}$.
    We have
    \begin{align*}
        \mathcal{L}(\widetilde{\cpsset}) - \mathcal{L}(\estcps) &= \Lval_{(\htau_{k-1}, \htau_{k+1}]} - \Lval_{(\htau_{k-1}, \htau_k]} -  \Lval_{(\htau_k, \htau_{k+1}]} - \gamma \\
        &< (3 C_{\ref*{lem:loc_err_g}.1} - C_{\gamma}) \dacc \le 0,
    \end{align*}
    provided that $C_{\gamma} \ge 3 C_{\ref*{lem:loc_err_g}.1}$.

    \subsubsection*{Case 2: $\Gcal_2 \cup \Gcal_3 = \emptyset$}
    We will show that $\Gcal_2 = \emptyset$ because the proof for $\Gcal_3 = \emptyset$ is the same by symmetry.
    Assume that $\{j, k\}$ is the leftmost case that satisfies $\Gcal_2$.
    It implies that $\htau_{k-2} \in [\tau_{j-1}^\ast - d_{j-1}, \tau_{j-1}^\ast + d_{j-1}]$.
    Otherwise assume $\htau_{k-2} > \tau_{j-1}^\ast + d_{j-1}$.
    Since $\max_{1 \le j \le K^\ast} \min_{1 \le k \le \widehat{K}} \Delta_{j}^2 \abs{\tau_j^\ast - \htau_k} \le \widetilde{C} \dacc / 2$, there must be $\htau_{k - h} \in [\tau_{j-1}^\ast - d_{j-1}, \tau_{j-1}^\ast + d_{j-1}]$ for some $h > 2$.
    It contradicts with $\Gcal_1 = \emptyset$ and the choice of $\{j, k\}$.

    \begin{figure}[H]
        \centering
        \resizebox{\textwidth}{!}{
        \begin{tikzpicture}
            \path coordinate (A) at (0, 0)
                  coordinate (F) at (15, 0);
            \path coordinate (B) at ($ (A)!.02!(F) $)
                  coordinate (C) at ($ (A)!.25!(F) $)
                  coordinate (D) at ($ (A)!.27!(F) $)
                  coordinate (E) at ($ (A)!.5!(F) $)
                  coordinate (G) at ($ (A)!.97!(F) $);
            \draw[line width=.5pt]
               (A) -- (B) -- (C) -- (D) -- (E) -- (G) -- (F);
            \node[dot, label = {-90:$\htau_{k-2}$}] at (A) {};
            \node[dot, label = {$\tau_{j-1}^\ast$}] at (B) {};
            \node[dot, label = {-90:$\htau_{k-1}$}] at (C) {};
            \node[dot, label = {$\tau_{j}^\ast$}] at (D) {};
            \node[dot, label = {-80:$\htau_{k}$}] at (E) {};
            \node[dot, label = {-90:$\htau_{k+1}$}] at (G) {};
            \node[dot, label = {$\tau_{j+1}^\ast$}] at (F) {};
        \end{tikzpicture}}
    \end{figure}

    Let $\widetilde{\cpsset} = \set{\tau_j^\ast} \cup \estcps \setminus \set{\htau_{k-1}, \htau_k}$.
    \begin{align*}
        \mathcal{L}(\widetilde{\cpsset}) - \mathcal{L}(\estcps) &= \Lval_{(\htau_{k-2}, \tau_j^\ast]} + \Lval_{(\tau_j^\ast, \htau_{k+1}]} - \Bigl\{\sum_{t = k-2}^{k} \Lval_{(\htau_t, \htau_{t+1}]} + \gamma \Bigr\} \\
        & = [\Lval_{(\htau_{k-2}, \tau_j^\ast]} - \Lval_{(\htau_{k-2}, \htau_{k-1}]}] + \Lval_{(\tau_j^\ast, \htau_{k+1}]} - \Bigl\{\sum_{t = k-1}^{k} \Lval_{(\htau_t, \htau_{t+1}]} + \gamma \Bigr\}\\
        & < \Bigl\{ \Lval_{(\htau_{k-2}, \tau_j^\ast]} - \Lval_{(\htau_{k-2}, \htau_{k-1}]} - \Lval_{(\htau_{k-1}, \tau_j^\ast]}^\ast \Bigr\} + ( 3 C_{\ref*{lem:loc_err_g}.1} - C_{\gamma}) \dacc \\
        &\le \{5 C_{\ref*{lem:loc_err_g}.1} + (C_{\ref*{lem:loc_err_g}.2} + C_{\m, 1}) \widetilde{C} - C_{\gamma}\} \dacc \le 0,
    \end{align*}
    provided that $C_{\gamma} \ge 5 C_{\ref*{lem:loc_err_g}.1} + (C_{\ref*{lem:loc_err_g}.2} + C_{\m, 1}) \widetilde{C}$.
    The first inequality is from the facts that $(\tau_j^\ast, \htau_{k+1}] \in \homo$, $(\htau_k, \htau_{k+1}]$ contains no changepoints and for $I = (\htau_{k-1}, \htau_{k}] \notin \homo$, $\Lval_{I} - \Lval_{I}^\ast > \Delta_{I} \size{I} - (C_{\ref*{lem:loc_err_g}.2} \Delta_{I} \size{I}) \vee (C_{\ref*{lem:loc_err_g}.1} \dacc) > - C_{\ref*{lem:loc_err_g}.1} \dacc$ because $0 < C_{\ref*{lem:loc_err_g}.2} < 1$.
    The second inequality is from Eq.~\eqref{equ:case2_2}.

    \subsubsection*{Case 3: $\Gcal_4 = \emptyset$}
    \begin{figure}[H]
        \centering
        \resizebox{\textwidth}{!}{
        \begin{tikzpicture}
            \path coordinate (A) at (0, 0)
                  coordinate (F) at (15, 0);
            \path coordinate (B) at ($ (A)!.02!(F) $)
                  coordinate (C) at ($ (A)!.25!(F) $)
                  coordinate (D) at ($ (A)!.27!(F) $)
                  coordinate (E) at ($ (A)!.5!(F) $)
                  coordinate (G) at ($ (A)!.75!(F) $)
                  coordinate (H) at ($ (A)!.78!(F) $)
                  coordinate (I) at ($ (A)!.98!(F) $);
            \draw[line width=.5pt]
               (A) -- (B) -- (C) -- (D) -- (E) -- (G) -- (H) -- (I) -- (F);
            \node[dot, label = {-90:$\htau_{k-2}$}] at (A) {};
            \node[dot, label = {$\tau_{j-1}^\ast$}] at (B) {};
            \node[dot, label = {-90:$\htau_{k-1}$}] at (C) {};
            \node[dot, label = {$\tau_{j}^\ast$}] at (D) {};
            \node[dot, label = {-80:$\htau_{k}$}] at (E) {};
            \node[dot, label = {$\tau_{j+1}^\ast$}] at (G) {};
            \node[dot, label = {-80:$\htau_{k+1}$}] at (H) {};
            \node[dot, label = {$\tau_{j+2}^\ast$}] at (I) {};
            \node[dot, label = {-80:$\htau_{k+2}$}] at (F) {};
        \end{tikzpicture}
        }
    \end{figure}
    Now $\Gcal_1 \cup \Gcal_2 \cup \Gcal_3 = \emptyset$.
    Assume that $\set{\htau_{k-1}, \htau_k, \htau_{k+1}}$ satisfies $\Gcal_4$.
    Similar to the analysis of $\Gcal_2 = \emptyset$, we have $\htau_{k-2} \in [\tau_{j-1}^\ast - d_{j-1}, \tau_{j-1}^\ast + d_{j-1}]$ and $\htau_{k+2} \in [\tau_{j+1}^\ast + d_{j+1}, \tau_{j+1}^\ast + d_{j+1}]$.
    Follow the same arguments in the proof for $\Gcal_2 \cup \Gcal_3 = \emptyset$, we can set $\widetilde{\cpsset} = \set{\tau_j^\ast, \tau_{j+1}^\ast} \cup \estcps \setminus \set{\htau_{k-1}, \htau_k, \htau_{k+1}}$ and obtain
    \begin{align*}
        \mathcal{L}(\widetilde{\cpsset}) - \mathcal{L}(\estcps) &= \Lval_{(\htau_{k-2}, \tau_j^\ast]} + \Lval_{(\tau_j^\ast, \tau_{j+1}^\ast]} + \Lval_{(\tau_{j+1}^\ast, \htau_{k+2}]} - \Bigl\{\sum_{t = k-2}^{k+1} \Lval_{(\htau_t, \htau_{t+1}]} + \gamma \Bigr\} \\
        & = \{\Lval_{(\htau_{k-2}, \tau_j^\ast]} - \Lval_{(\htau_{k-2}, \htau_{k-1}]} + \Lval_{(\tau_{j+1}^\ast, \htau_{k+2}]} - \Lval_{(\htau_{k+1}, \htau_{k+2}]}\} \\
        & ~~~~+ \Lval_{(\tau_j^\ast, \tau_{j+1}^\ast]} - \Bigl\{\sum_{t = k-1}^{k} \Lval_{(\htau_t, \htau_{t+1}]} + \gamma \Bigr\}\\
        & < \Lval_{(\htau_{k-2}, \tau_j^\ast]} - \Lval_{(\htau_{k-2}, \htau_{k-1}]} + \Lval_{(\tau_{j+1}^\ast, \htau_{k+2}]} - \Lval_{(\htau_{k+1}, \htau_{k+2}]} -\Lval_{(\htau_{k-1}, \tau_j^\ast] \cup (\tau_{j+1}^\ast, \htau_{k+1}]}^\ast \\
        & ~~~~+ (3 C_{\ref*{lem:loc_err_g}.1} - C_{\gamma}) \dacc \le \{7 C_{\ref*{lem:loc_err_g}.1} + 2 (C_{\ref*{lem:loc_err_g}.2} + C_{\m, 1}) \widetilde{C} - C_{\gamma}\} \dacc \le 0,
    \end{align*}
    provided $C_{\gamma} \ge 7 C_{\ref*{lem:loc_err_g}.1} + 2 (C_{\ref*{lem:loc_err_g}.2} + C_{\m, 1}) \widetilde{C}$.

    Combining the proof in the $\Hcal$ and $\Gcal$ parts, we can determine the two constants by solving the following inequalities:
    \begin{equation}
        \left\{
        \begin{aligned}
            &C_{\gamma} \ge 7 C_{\ref*{lem:loc_err_g}.1} + 2 (C_{\ref*{lem:loc_err_g}.2} + C_{\m, 1}) \widetilde{C}, \\
            &\widetilde{C} \ge \{1 - (1 + 2 C_{\Delta}) C_{\ref*{lem:loc_err_g}.2} - 2 C_{\Delta} C_{\m, 1}\}^{-1} C_{\Delta} (7 C_{\ref*{lem:loc_err_g}.1} + C_{\gamma} + 2 C_{\m, 1}).
        \end{aligned}
        \right.
    \end{equation}

    Note that by setting $C_{\snr}$ sufficiently large, we have $C_{\m, 1}$ is sufficiently small and Eq.~\eqref{equ:Cm1_bound} holds.
    Let $C_{\gamma} = 7 C_{\ref*{lem:loc_err_g}.1} + 2 (C_{\ref*{lem:loc_err_g}.2} + C_{\m, 1}) \widetilde{C}$, one obtains the following inequality w.r.t. $\widetilde{C}$,
    \begin{equation*}
        \{1 - (1 + 2 C_{\Delta}) C_{\ref*{lem:loc_err_g}.2} - 2 C_{\Delta} C_{\m, 1}\} \widetilde{C} \ge C_{\Delta} \{ 14 C_{\ref*{lem:loc_err_g}.1} + 2 C_{\m, 1} + 2 (C_{\ref*{lem:loc_err_g}.2} + C_{\m, 1}) \widetilde{C}\}.
    \end{equation*}
        We can find the feasible region
    $$\widetilde{C} \ge \{1 - (1 + 4 C_{\Delta}) C_{\ref*{lem:loc_err_g}.2} - 4 C_{\Delta} C_{\m, 1}\}^{-1} C_{\Delta} (14 C_{\ref*{lem:loc_err_g}.1} + 2 C_{\m, 1}).$$
                With the solution of $\widetilde{C}$ and $C_{\gamma}$, we have $\estcps = \widetilde{\cpsset}$ and the final result, i.e.,
    \begin{equation}
        \widehat{K} = K^\ast;\; \max_{1 \le k \le K^\ast} \min_{1 \le j \le \widehat{K}} \Delta_k |\tau^\ast_k - \htau_j| \le \widetilde{C} \dacc.
    \end{equation}
\end{proof}

\subsection{Proof of Theorem \ref{thm:cross-fitting}}

We first present the following lemmas concerning expected excess risks, which are instrumental in proving Theorem~\ref{thm:cross-fitting} and verifying the conditions of it.
Their proofs are deferred to the end of this section.

\begin{lemma}\label{lem:mix_optimal}
Assume that for $t_1=\tau_k^\ast$ and $t_2 = \tau_{k}^\ast + 1$, Condition~\ref{cond:model_and_loss}(b) holds. Denote $\f_{{\alpha}}^\ast = \argmin_{\f} (1-\alpha) \overline{\ell}_{t_1}(\f) + \alpha \overline{\ell}_{t_2}(\f)$ with $\alpha \in (0, 1/2]$. We have
\begin{equation}\label{equ:mix_optimal_1}
\overline{\ell}_{t_1}(\f_{{\alpha}}^\ast) - \overline{\ell}_{t_1}(\f_{{t_1}}^\ast) \lesssim \alpha \Delta_k,
\end{equation}
and for the expected excess risk on $t_2$,
\begin{equation}\label{equ:mix_optimal_2}
\overline{\ell}_{t_2}(\f_{{\alpha}}^\ast) - \overline{\ell}_{t_2}(\f_{{t_2}}^\ast) \lesssim \Delta_k.
\end{equation}
\end{lemma}

\begin{lemma}\label{lem:risk_bound_equiv}
    Suppose that Condition~\ref{cond:model_and_loss}(b) holds.
For $I \in \homo$, we have $\abs{\risk_{J_{-m, I}}(\f) - \risk_{J_{-m, I}}(\f_{J_{-m, I}}^\ast)} \lesssim \dacc$ is equivalent to $\abs{\risk_{J_{m, I}}(\f) - \risk_{J_{m, I}}(\f_{J_{-m, I}}^\ast)} \lesssim \dacc$.
\end{lemma}

Now we are ready to prove Theorem~\ref{thm:cross-fitting}.

\begin{proof}[Proof of Theorem \ref{thm:cross-fitting}]
    We revisit some essential notations here.
    For interval $I$, denote $\set{J_{m, I}}_{m=1}^M$ be the $M$-fold splits and $\set{J_{-m,I} = I \setminus J_{m, I}}$ be the complements.
    For example, in the order-preserved splitting method,
    we set $J_m=\{i\in[n]: (i-1) \bmod M = m - 1\}$ and the $M$-fold splits of interval $I$ as $J_{m, I} := J_{m} \cap I$ for $m\in[M]$.
    For an index set $I$ (not necessary a interval), $\Lval_{I}^\ast = \sum_{i \in I} \ell(\z_i, \f_{i}^\ast)$ and $\overline{\Lval}_{I}^\ast = \sum_{i \in I} \overline{\ell}_i(\f_{i}^\ast)$ are defined in Section~\ref{sec:proof_loc_err_g}.
    We further denote $\Delta_{I, \f} = \size{I}^{-1} \{\risk_I(\f) - \overline{\Lval}_{I}^{\ast}\}$ which extends the notation $\Delta_{I}$ by $\Delta_{I, \f_{I}^\ast} = \Delta_{I}$.

    It is sufficient to check the events $\Gbb^{\mathsf{nhomo}}$ and $\Gbb^{-\mathsf{nhomo}}$ in Lemma~\ref{lem:loc_err_g} hold with high probability.
    The proof is mainly about deriving concentration bounds for the empirical losses based on the following Bernstein's inequality whose proof can be found in \citet{boucheron2013concentration}.

    \begin{lemma}[Bernstein's inequality]\label{lem:bernstein}
        For independent random variables $\set{X_i}_{i=1}$ that satisfy the sub-Weibull tail $\mathbb{P}(\abs{X_i / m_i} \ge x) \le \exp(1 - x^{\kappa})$ with variance $\sigma_{i}^2$ and envelope parameter $m_{i}$ and $0 < \kappa \le 1$,
        \[
            \Pbb\Bigl(\Bigabs{\sum_{i} X_i} \ge t\Bigr) \le 2 \exp\Bigl\{-c \min\Bigl(\frac{t^2}{\sum_{i} \sigma_{i}^2}, \frac{t^{\kappa}}{\max_{i} m_{i}^{\kappa}}\Bigr)\Bigr\}.
        \]
    \end{lemma}

We consider $\Gbb^{\mathsf{nhomo}}$ at first.

For $I = (s, e] \in \homo$, we have $\Delta_{J_{m, I}} \size{J_{m, I}} \le \Delta_{J_{m, I}, \f_{I}^\ast} \size{J_{m, I}} \le \Delta_{I} \size{I} \le 2 \widetilde{C} \dacc$.
Our immediate goal is to bound the total excess risk evaluated on the validation fold, defined as $\Delta_{J_{m, I}, \hf_{J_{-m, I}}} \size{J_{m, I}} = \risk_{J_{m, I}}(\hf_{J_{-m, I}}) - \overline{\Lval}_{J_{m, I}}^{\ast}$.
This term can be decomposed into:
\begin{equation}\label{equ:risk_err_joint}
\Delta_{J_{m, I}, \hf_{J_{-m, I}}} \size{J_{m, I}} = \{ \risk_{J_{m, I}}(\hf_{J_{-m, I}}) - \risk_{J_{m, I}}(\f_{J_{-m, I}}^\ast) \} + \Delta_{J_{m, I}, \f_{J_{-m, I}}^\ast} \size{J_{m, I}}.
\end{equation}
The first term is bounded by $\dacc$ directly from Condition~\ref{cond:predict}. For the second term, we discuss two cases based on the presence of a changepoint.

If $\truecps \cap (s, e) = \emptyset$, the distributions are homogeneous, yielding $\f_{J_{-m, I}}^\ast = \f_I^\ast$. Thus, $\Delta_{J_{m, I}, \f_{J_{-m, I}}^\ast} \size{J_{m, I}} = \Delta_{J_{m, I}, \f_{I}^\ast} \size{J_{m, I}} \le \Delta_I \size{I} \le 2 \widetilde{C} \dacc$.

Otherwise, without loss of generality, there is exact one changepoint $k \in [K^\ast]$ such that $\truecps \cap (s, e) = \set{\tau_k^\ast}$ with $\abs{e - \tau_k^\ast} \le \widetilde{C} \dacc \Delta_k^{-1}$.

We divide $I$ into $I_1 = (s, \tau_k^\ast]$ and $I_2 = (\tau_k^\ast, e]$. Let $t_1=\tau_k^\ast$ and $t_2 = \tau_{k}^\ast + 1$. Denote the mixture weight $\alpha = \size{I_2 \cap J_{-m, I}} / \size{J_{-m, I}}$.
Note that $\size{I_2 \cap J_{-m, I}} \le \size{I_2} \le \widetilde{C} \dacc \Delta_k^{-1}$. By Condition~\ref{cond:changes}, $\size{J_{-m, I}} \ge \frac{M-1}{M} \size{I} \ge \frac{M-1}{M} C_{\snr} \dacc \Delta_k^{-1}$. Thus, $\alpha \le \frac{\widetilde{C} M}{(M-1) C_{\snr}} \le 1/2$ since $C_{\snr}$ is sufficiently large.
Applying Lemma~\ref{lem:mix_optimal} to the training fold $J_{-m, I}$, we obtain:
\begin{align*}
\Delta_{J_{m, I}, \f_{J_{-m, I}}^\ast} \size{J_{m, I}}
&\lesssim \size{J_{m, I} \cap I_1} \alpha \Delta_k + \size{J_{m, I} \cap I_2} \Delta_k \\
&= \size{J_{m, I} \cap I_1} \frac{\size{J_{-m, I} \cap I_2}}{\size{J_{-m, I}}} \Delta_k + \size{J_{m, I} \cap I_2} \Delta_k \\
&\le \frac{\size{J_{m, I}}}{\size{J_{-m, I}}} \size{I_2} \Delta_k + \size{I_2} \Delta_k = \frac{M}{M-1} \size{I_2} \Delta_k \lesssim \widetilde{C} \dacc.
\end{align*}
Therefore,
\begin{equation}\label{equ:risk_err_oracle_1}
\Delta_{J_{m, I}, \hf_{J_{-m, I}}} \size{J_{m, I}} = \risk_{J_{m, I}}(\hf_{J_{-m, I}}) - \overline{\Lval}_{J_{m, I}}^{\ast} \lesssim \widetilde{C} \dacc
\end{equation}
holds for all $I \in \homo$.
By similarly evaluating the oracle $\f_I^\ast$ on $I_1$ and $I_2$, we also establish the following auxiliary bounds on sub-segments:
\begin{equation*}
\risk_{J_{m, I}}(\f_{I}^\ast) - \overline{\Lval}_{J_{m, I}}^{\ast} \lesssim \widetilde{C} \dacc, \quad \risk_{J_{m, I_1}}(\f_{I}^\ast) - \overline{\Lval}_{J_{m, I_1}}^{\ast} \lesssim \widetilde{C} \dacc.
\end{equation*}

With these results, we next determine the parameters $\{\sigma_i, m_i\}$ to apply Lemma~\ref{lem:bernstein}.
For $i \in J_{m, I}$, denote $s_i = \ell(\z_i; \hf_{J_{-m, I}}) - \ell(\z_i; \f_{i}^\ast) - \overline{\ell}_i(\hf_{J_{-m, I}}) + \overline{\ell}_{i}(\f_{i}^\ast)$. Let $\sigma_i^2 = \sigma_{i, \hf_{J_{-m, I}}}^2$ and $m_i = m_{i, \hf_{J_{-m, I}}}$.
We will first show that $\max_{i \in J_{m, I}} \{\overline{\ell}_i(\hf_{J_{-m, I}}) - \overline{\ell}_i(\f_{i}^\ast) \}\lesssim \dacc / (\log n)^{1/\kappa}$.

If $\truecps \cap (s, e) = \emptyset$, we have:
$$\max_{i \in J_{m, I}} \{\overline{\ell}_i(\hf_{J_{-m, I}}) - \overline{\ell}_i(\f_{i}^\ast) \} = \Delta_{J_{m, I}, \hf_{J_{-m, I}}}.$$
Since $\Delta_{J_{m, I}, \hf_{J_{-m, I}}} \size{J_{m, I}} \lesssim \widetilde{C} \dacc$ and $\size{J_{m, I}} \ge 1 + M^{-1}\size{I} \gtrsim (\log n)^{1/\kappa}$, dividing the total sum yields:
$$\max_{i \in J_{m, I}} \{\overline{\ell}_i(\hf_{J_{-m, I}}) - \overline{\ell}_i(\f_{i}^\ast) \} \lesssim \dacc / (\log n)^{1/\kappa}.$$

Otherwise, it is the case that $\truecps \cap (s, e) = \set{\tau_k^\ast}$. By Eq.~\eqref{equ:risk_err_oracle_1}, we have:
$$\risk_{J_{m, I_1}}(\hf_{J_{-m, I}}) - \overline{\Lval}_{J_{m, I_1}}^{\ast} \le \risk_{J_{m, I}}(\hf_{J_{-m, I}}) - \overline{\Lval}_{J_{m, I}}^{\ast} \lesssim \widetilde{C} \dacc.$$
By Condition~\ref{cond:changes}(a), $\size{J_{m, I}} \ge \size{J_{m, I_1}} \gtrsim C_{\snr} (\log n)^{1/\kappa}$.
Thus, for $i \in J_{m, I_1}$, the point-wise error is bounded by the average: $\overline{\ell}_i(\hf_{J_{-m, I}}) - \overline{\ell}_i(\f_{i}^\ast) = \size{J_{m, I_1}}^{-1} \{\risk_{J_{m, I_1}}(\hf_{J_{-m, I}}) - \overline{\Lval}_{J_{m, I_1}}^{\ast}\} \lesssim \dacc / (\log n)^{1/\kappa}$.
For the minority segment $i \in J_{m, I_2}$, we apply the decomposition:
$$\overline{\ell}_i(\hf_{J_{-m, I}}) - \overline{\ell}_i(\f_{i}^\ast) = \{\overline{\ell}_i(\hf_{J_{-m, I}}) - \overline{\ell}_i(\f_{{I_1}}^\ast)\} + \{\overline{\ell}_i(\f_{{I_1}}^\ast) - \overline{\ell}_i(\f_{{I_2}}^\ast)\}.$$
For the second summand, $\overline{\ell}_i(\f_{{I_1}}^\ast) - \overline{\ell}_i(\f_{{I_2}}^\ast) \le \Delta_k \lesssim \dacc / (\log n)^{1/\kappa}$.
For the first summand, using Condition~\ref{cond:model_and_loss}(b) with any $j \in I_1$, we have $\overline{\ell}_i(\hf_{J_{-m, I}}) - \overline{\ell}_i(\f_{{I_1}}^\ast) \lesssim \Delta_k + \overline{\ell}_j(\hf_{J_{-m, I}}) - \overline{\ell}_j(\f_{{I_1}}^\ast) \lesssim \dacc / (\log n)^{1/\kappa}$.
These upper bounds collectively guarantee the desired result.
In summary, we have that for any $I \in \homo$ and $m \in [M]$,
$$\max_{i \in J_{m, I}} \{\overline{\ell}_i(\hf_{J_{-m, I}}) - \overline{\ell}_i(\f_{i}^\ast) \} \lesssim \frac{\dacc}{(\log n)^{1/\kappa}}.$$

By Condition~\ref{cond:model_and_loss}(a), we have
    \begin{align*}
        \max_{i \in J_{m, I}} \sigma_i
        &\lesssim \max_{i \in J_{m, I}} \{\overline{\ell}_i(\hf_{J_{-m, I}}) - \overline{\ell}_i(\f_{i}^\ast) \} \vee \Bigl[c_n \max_{i \in J_{m, I}} \{\overline{\ell}_i(\hf_{J_{-m, I}}) - \overline{\ell}_i(\f_{i}^\ast) \}\Bigr]^{\frac{1}{2}} \\
        &\lesssim \frac{\dacc}{(\log n)^{1/\kappa}}
    \end{align*}
and
\begin{align*}
    \sum_{i \in J_{m, I}} \sigma_i^2
        &\lesssim c_n \Delta_{J_{m, I}, \hf_{J_{-m, I}}} \size{J_{m, I}} + \max_{i \in J_{m, I}} \{\overline{\ell}_i(\hf_{J_{-m, I}}) - \overline{\ell}_i(\f_{i}^\ast) \} \Delta_{J_{m, I}, \hf_{J_{-m, I}}} \size{J_{m, I}} \\
    &\lesssim c_n \widetilde{C}\dacc + \widetilde{C} \frac{\dacc^2}{(\log n)^{1/\kappa}} \lesssim  \widetilde{C} \frac{\dacc^2}{(\log n)^{1/\kappa}},
\end{align*}
where the last step is due to the fact that $c_n \lesssim \dacc / (\log n)^{1/\kappa}$ and $0 < \kappa \le 1$.

    It also holds that
    $$\max_{i \in J_{m, I}} m_{i} \le \max_{i \in J_{m, I}} \sigma_i + \frac{\dacc}{(\log n)^{1/\kappa}} \lesssim \frac{\dacc}{(\log n)^{1/\kappa}}.$$
    By Lemma~\ref{lem:bernstein}, there exists a universal constant $C>0$, such that with probability at least $1 - \exp(- C \log n)$,
    $$\sum_{i \in J_{m, I}} s_i \lesssim \widetilde{C}^{\frac{1}{2}}\dacc.$$

    By the assumption that $\abs{\risk_{J_{m, I}}(\hf_{J_{-m, I}}) - \risk_{J_{m, I}}(\f_{{I}}^\ast)} \lesssim \dacc$, and taking the summation over $m \in [M]$ and the union bound over $I \in \homo$, there exists a universal constant $C>0$ such that  with probability at least $1 - \exp(- C \log n)$, uniformly for all $I \in \homo$,
    $$\xi_I = \sum_{m \in [M]} \{\L(\z_{J_{m, I}}; \hf_{J_{-m, I}})\} - \Lval_{I}^\ast - \Delta_I \size{I} \lesssim \widetilde{C}^{\frac{1}{2}}\dacc.$$
    Similarly, the lower bound holds, that is,
    $$\xi_I = \sum_{m \in [M]} \{\L(\z_{J_{m, I}}; \hf_{J_{-m, I}})\} - \Lval_{I}^\ast - \Delta_I \size{I} \gtrsim -\widetilde{C}^{\frac{1}{2}}\dacc.$$

    Next we study $\Gbb^{-\mathsf{nhomo}}$. Actually, we can provide the lower bound result for any $I = (s, e] \subset (0, n]$ with $\size{I} \ge C_{\m} (\log n)^{1/\kappa}$.
    For notation simplicity, we set $\Delta = \Delta_{J_{m, I}, \hf_{J_{-m, I}}} = \size{J_{m, I}}^{-1}\sum_{i \in J_{m, I}} \{\overline{\ell}_i(\hf_{J_{-m, I}}) - \overline{\ell}_i(\f_{i}^\ast)\}$ in this part.

    We first consider the upper bound of $\max_{i \in J_{m, I}} \{\overline{\ell}_i(\hf_{J_{-m, I}}) - \overline{\ell}_i(\f_{i}^\ast)\}$ in the following three situations.

    (a) If $\size{I \cap \truecps} = 0$, it holds that $\max_{i \in J_{m, I}} \{\overline{\ell}_i(\hf_{J_{-m, I}}) - \overline{\ell}_i(\f_{i}^\ast)\} = \Delta$.

    (b) If $I \cap \truecps = \{\tau_k^\ast\}$, there must be $i \in J_{m, I}$ such that $\{\overline{\ell}_i(\hf_{J_{-m, I}}) - \overline{\ell}_i(\f_{i}^\ast)\} \le \Delta$. Without loss of generality, we assume that this index $i \in (s, \tau_k^\ast]$. For any $j \in (\tau_k^\ast, e]$, by Condition~\ref{cond:model_and_loss}(b), $\{\overline{\ell}_j(\hf_{J_{-m, I}}) - \overline{\ell}_j(\f_{j}^\ast)\} \lesssim \Delta_k + \Delta$.

    (c) If $\size{I \cap \truecps} \ge 2$ and $I \cap \truecps = \{\tau_k^\ast,\cdots, \tau_{k+t}^\ast\}$, for every $i \in (\tau_{k+u}^\ast, \tau_{k+u+1}^\ast]$ with $u \in \set{0, \cdots, t-1}$, we have $\{\overline{\ell}_i(\hf_{J_{-m, I}}) - \overline{\ell}_i(\f_{i}^\ast)\} \le \Delta \size{J_{m, I}} / (C_{\snr} \dacc)$.
    For the rest $i \in (s, \tau_k^\ast] \cup (\tau_{k+t}^\ast, e]$, similarly to case~(b), we have $\{\overline{\ell}_i(\hf_{J_{-m, I}}) - \overline{\ell}_i(\f_{i}^\ast)\} \lesssim \Delta \size{J_{m, I}} / (C_{\snr} \dacc) + \Delta_k \vee \Delta_{k+t} \le \Delta \size{J_{m, I}} / (C_{\snr} \dacc) + \dacc/(\log n)^{1/\kappa}$.

    In summary of cases (a)---(c), we have
    \begin{equation}\label{equ:max_excess_loss}
        \max_{i \in J_{m, I}} \{\overline{\ell}_i(\hf_{J_{-m, I}}) - \overline{\ell}_i(\f_{i}^\ast)\} \lesssim \frac{\Delta \size{J_{m, I}}}{C_{\snr} \dacc} + \Delta + \frac{\dacc}{(\log n)^{1/\kappa}}.
    \end{equation}

    By Eq.~\eqref{equ:max_excess_loss} and Condition~\ref{cond:model_and_loss}(a) with $\kappa = \kappa_1$, we have for any $I \notin \homo$ with $\size{I} \ge C_{\m} (\log n)^{1/\kappa}$ and any $m \in [M]$,
        $$\max_{i \in J_{m, I}}\sigma_i \lesssim \frac{\Delta \size{J_{m, I}}}{C_{\snr} \dacc} + \Delta + \frac{\dacc}{(\log n)^{1/\kappa}},$$
    and
    \begin{align*}
        \sum_{i \in J_{m, I}} \sigma_i^2 & \lesssim c_n \sum_{i \in J_{m, I}} \{\overline{\ell}_i(\hf_{J_{-m, I}}) - \overline{\ell}_i(\f_{i}^\ast)\} \\
        &~~~~+ \max_{i \in J_{m, I}} \{\overline{\ell}_i(\hf_{J_{-m, I}}) - \overline{\ell}_i(\f_{i}^\ast)\} \sum_{i \in J_{m, I}} \{\overline{\ell}_i(\hf_{J_{-m, I}}) - \overline{\ell}_i(\f_{i}^\ast)\} \\
        & = c_n \Delta \size{J_{m, I}} + \max_{i \in J_{m, I}} \{\overline{\ell}_i(\hf_{J_{-m, I}}) - \overline{\ell}_i(\f_{i}^\ast)\} \cdot \Delta \size{J_{m, I}} \\
        & \lesssim \frac{\Delta^2 \size{J_{m, I}}^2}{C_{\snr} \dacc} + \Delta^2 \size{J_{m, I}} + \frac{\Delta \size{J_{m, I}} \dacc}{(\log n)^{1/\kappa}}.
    \end{align*}

    It also holds that
    $$\max_{i \in J_{m, I}}m_i \lesssim \frac{\Delta \size{J_{m, I}}}{C_{\snr} \dacc} + \Delta + \frac{\dacc}{(\log n)^{1/\kappa}}.$$

    The above two inequalities make it valid to apply the Bernstein's inequality (Lemma~\ref{lem:bernstein}). Uniformly for all $I \notin \homo$ with $\size{I} \ge C_{\m} (\log n)^{1/\kappa}$ and any $m \in [M]$, with probability at least $1 - \exp(-C \log n)$,
    \begin{equation}\label{equ:onefold_lower}
        \sum_{i \in J_{m, I}} s_i \ge -C_2 \Bigl\{(C_{\m}^{-\frac{1}{2}} + \widetilde{C}^{-\frac{1}{2}})  \Delta_{J_{m, I}, \hf_{J_{-m, I}}} \size{J_{m, I}} + \widetilde{C}^{\frac{1}{2}} \dacc\Bigr\},
    \end{equation}
    for some universal constants $C, C_2 > 0$.

    Now we consider the lower bound of $\sum_{m \in [M]} \Delta_{J_{m, I}, \hf_{J_{-m, I}}} \size{J_{m, I}}$.

    When $\{J_{m, I}\}_{m=1}^M$ are constructed by the order-preserved splitting method, by Condition~\ref{cond:changes}(b) and Condition~\ref{cond:model_and_loss}(b), we have that for general $\f$, $|\Delta_{J_{m, I}, \f} - \Delta_{I, \f}| \le C \size{I}^{-1} \max_{k} \Delta_k \le C \size{I}^{-1} \dacc / (\log n)^{1/\kappa}$ with some universal constant $C > 0$.
    Combining the fact that $\Delta_{I, \f} \ge \Delta_{I}$, we have:
    \begin{equation}\label{equ:lower_bound_oracle}
        \sum_{m \in [M]} \Delta_{J_{m, I}, \hf_{J_{-m, I}}} \size{J_{m, I}} \ge \Delta_{I} \size{I} - C \dacc.
    \end{equation}

    Here we also discuss other splitting strategies. For example, when $\{J_{m, I}\}_{m=1}^M$ are constructed by the blockwise splitting method. By the definition of $\Delta_{I, \f}$ and $\f_{I}^\ast$, we have that $\Delta_{I, \f_{{J_{-m, I}}}^\ast} \size{I} \ge \Delta_{I} \size{I} = \Delta_{I, \f_{I}^\ast} \size{I}$ and $\Delta_{J_{-m, I}, \f_{{J_{-m, I}}}^\ast} \size{J_{-m, I}} \le \Delta_{J_{-m, I}, \f_{I}^\ast} \size{J_{-m, I}}$. Therefore
    \begin{align*}
        & \Delta_{J_{m, I}, \f_{{J_{-m, I}}}^\ast} \size{J_{m, I}} \\
        = & \Delta_{I, \f_{{J_{-m, I}}}^\ast} \size{I} - \Delta_{J_{-m, I}, \f_{{J_{-m, I}}}^\ast} \size{J_{-m, I}} \\
        \ge & \Delta_{I, \f_{I}^\ast} \size{I} - \Delta_{J_{-m, I}, \f_{I}^\ast} \size{J_{-m, I}} \\
        = & \Delta_{J_{m, I}, \f_{{I}}^\ast} \size{J_{m, I}}.
    \end{align*}
    By summation over $m \in [M]$,
    \begin{equation*}
        \sum_{m \in [M]} \Delta_{J_{m, I}, \f_{{J_{-m, I}}}^\ast} \size{J_{m, I}} \ge \sum_{m \in [M]} \Delta_{J_{m, I}, \f_{{I}}^\ast} \size{J_{m, I}} = \Delta_{I} \size{I}.
    \end{equation*}
Note that the above inequality holds for any fixed-design splits $\set{J_{m, I}}$.
Furthermore, since $\hf_{J_{-m, I}}$ is the empirical counterpart targeting the oracle $\f_{J_{-m, I}}^\ast$, its expected out-of-sample risk should not be fundamentally lower than that of the oracle model. Therefore, provided that $\risk_{J_{m, I}}(\hf_{J_{-m, I}}) \ge \risk_{J_{m, I}}(\f_{J_{-m, I}}^\ast) - \dacc$, we obtain \eqref{equ:lower_bound_oracle} in this setting.

    Subsequently, by taking the summation of Eq.~\eqref{equ:onefold_lower} over $m \in [M]$, we have
    \begin{equation}\label{equ:lower_bound_general}
        \sum_{m \in [M]} \L(\z_{J_{m, I}}; \hf_{J_{-m, I}}) - \Lval_{I}^\ast - \Delta_{I} \size{I} \ge - C_2 (C_{\m}^{-\frac{1}{2}} + \widetilde{C}^{-\frac{1}{2}}) \Delta_{I} \size{I} - M C_2 \widetilde{C}^{\frac{1}{2}} \dacc.
    \end{equation}
    Overall, we have $\Pbb(\Gbb^{\mathsf{nhomo}} \cap \Gbb^{-\mathsf{nhomo}}) \ge 1 - \exp(-C\log n)$ with $C_{\ref*{lem:loc_err_g}.1} = C' \widetilde{C}^{\frac{1}{2}}$ and $C_{\ref*{lem:loc_err_g}.2} = C' (C_{\m}^{-\frac{1}{2}} + \widetilde{C}^{-\frac{1}{2}})$ for some universal constants $C,C' > 0$, where the events $\Gbb^{\mathsf{nhomo}}$ and $\Gbb^{-\mathsf{nhomo}}$ are defined in Lemma~\ref{lem:loc_err_g}.

    Based on the choices of $C_{\ref*{lem:loc_err_g}.1}$ and $C_{\ref*{lem:loc_err_g}.2}$, the feasible region in the proof of Lemma~\ref{lem:loc_err_g}, i.e.,
    $$\widetilde{C} \ge \{1 - (1 + 4 C_{\Delta}) C_{\ref*{lem:loc_err_g}.2} - 4 C_{\Delta} C_{\m, 1}\}^{-1} C_{\Delta} (14 C_{\ref*{lem:loc_err_g}.1} + 2 C_{\m, 1})$$
    is non-empty.
    It means that there is a constant $\widetilde{C}$ that only depends on $C_{\m, 1} = \frac{d_{\m}}{C_{\snr} \dacc}$, $C_{\m}$, $C_{\snr}$, $C_{\Delta}$ so that Theorem~\ref{thm:cross-fitting} holds.

                        \end{proof}

\begin{proof}[Proof of Lemma~\ref{lem:mix_optimal}]
    By the definition of $\f_{\alpha}^\ast$,
    \begin{equation*}
        (1 -\alpha) \overline{\ell}_{t_1}(\f_{\alpha}^\ast) + \alpha \overline{\ell}_{t_2}(\f_{\alpha}^\ast) \le (1 - \alpha) \overline{\ell}_{t_1}(\f_{t_1}^\ast) + \alpha \overline{\ell}_{t_2}(\f_{{t_1}}^\ast).
    \end{equation*}
    Rearranging the terms to isolate the excess risk on $t_1$, we have:
    \begin{equation*}
        (1 - \alpha) \{ \overline{\ell}_{t_1}(\f_{\alpha}^\ast) - \overline{\ell}_{t_1}(\f_{{t_1}}^\ast) \} \le \alpha \{ \overline{\ell}_{t_2}(\f_{{t_1}}^\ast) - \overline{\ell}_{t_2}(\f_{\alpha}^\ast) \}.
    \end{equation*}
    Since the left hand side is non-negative, applying Condition~\ref{cond:model_and_loss}(b) to the right hand side yields:
    \begin{align*}
        (1 - \alpha) \{ \overline{\ell}_{t_1}(\f_{\alpha}^\ast) - \overline{\ell}_{t_1}(\f_{{t_1}}^\ast) \} &\le \alpha \abs{ \overline{\ell}_{t_2}(\f_{{t_1}}^\ast) - \overline{\ell}_{t_2}(\f_{\alpha}^\ast) } \\
        &\le \alpha C_{\ell} \{ \overline{\ell}_{t_1}(\f_{\alpha}^\ast) - \overline{\ell}_{t_1}(\f_{{t_1}}^\ast) \} + \alpha C_{\ell} \{ \overline{\ell}_{t_2}(\f_{{t_1}}^\ast) - \overline{\ell}_{t_2}(\f_{{t_2}}^\ast) \} \\
        &\le \alpha C_{\ell} \{ \overline{\ell}_{t_1}(\f_{\alpha}^\ast) - \overline{\ell}_{t_1}(\f_{{t_1}}^\ast) \} + \alpha C_{\ell} \Delta_k.
    \end{align*}
    Because $\alpha \le \frac{\widetilde{C}}{M C_{\snr}}$ and $C_{\snr}$ is sufficiently large, we can guarantee that $\alpha(1 + C_{\ell}) \le 1/2$. Moving the risk term to the left side:
    \begin{equation*}
        \frac{1}{2} \{ \overline{\ell}_{t_1}(\f_{\alpha}^\ast) - \overline{\ell}_{t_1}(\f_{{t_1}}^\ast) \} \le (1 - \alpha - \alpha C_{\ell}) \{ \overline{\ell}_{t_1}(\f_{\alpha}^\ast) - \overline{\ell}_{t_1}(\f_{{t_1}}^\ast) \} \le \alpha C_{\ell} \Delta_k.
    \end{equation*}
    Hence,
    \begin{equation*}
        \overline{\ell}_{t_1}(\f_{\alpha}^\ast) - \overline{\ell}_{t_1}(\f_{{t_1}}^\ast) \le 2 C_{\ell} \alpha \Delta_k \lesssim \alpha \Delta_k.
    \end{equation*}
    For the second inequality, applying Condition~\ref{cond:model_and_loss}(b) directly yields:
    \begin{equation*}
        \overline{\ell}_{t_2}(\f_{\alpha}^\ast) - \overline{\ell}_{t_2}(\f_{{t_2}}^\ast) \le C_{\ell} \{ \overline{\ell}_{t_1}(\f_{\alpha}^\ast) - \overline{\ell}_{t_1}(\f_{{t_1}}^\ast) \} + C_{\ell} \Delta_k \lesssim \alpha \Delta_k + \Delta_k \lesssim \Delta_k.
    \end{equation*}
\end{proof}

\begin{proof}[Proof of Lemma~\ref{lem:risk_bound_equiv}]
    We show that $\abs{\risk_{J_{-m, I}}(\f) - \risk_{J_{-m, I}}(\f_{J_{-m, I}}^\ast)} \lesssim \dacc$ implies $\abs{\risk_{J_{m, I}}(\f) - \risk_{J_{m, I}}(\f_{J_{-m, I}}^\ast)} \lesssim \dacc$.
    For brevity, let $\hf = \f$ and $\f^\ast = \f_{J_{-m, I}}^\ast$.

    If $I$ contains no changepoint, the data distributions are identical ($\f_I^\ast = \f_{J_{m, I}}^\ast = \f^\ast$).
    It trivially holds that $\abs{\risk_{J_{m, I}}(\hf) - \risk_{J_{m, I}}(\f^\ast)} = \frac{\size{J_{m, I}}}{\size{J_{-m, I}}} \abs{\risk_{J_{-m, I}}(\hf) - \risk_{J_{-m, I}}(\f^\ast)} \lesssim \dacc$.

    Otherwise, without loss of generality, there is exact one changepoint $k \in [K^\ast]$ such that $\truecps \cap (s, e) = \set{\tau_k^\ast}$ with $\abs{e - \tau_k^\ast} \le \widetilde{C} \dacc \Delta_k^{-1}$.
    We divide $I$ into $I_1 = (s, \tau_k^\ast]$ and $I_2 = (\tau_k^\ast, e]$. Let $t_1=\tau_k^\ast$ and $t_2 = \tau_{k}^\ast + 1$. For the training fold $J_{-m, I}$, we define its sub-segments as $J_1 = J_{-m, I} \cap I_1$ and $J_2 = J_{-m, I} \cap I_2$.

    Applying Condition~\ref{cond:model_and_loss}(b), we have:
    \begin{align*}
        \abs{\risk_{J_{-m, I}}(\hf) - \risk_{J_{-m, I}}(\f^\ast)}
        &= \abs{ \size{J_1} \{\overline{\ell}_{t_1}(\hf) - \overline{\ell}_{t_1}(\f^\ast)\} + \size{J_2} \{\overline{\ell}_{t_2}(\hf) - \overline{\ell}_{t_2}(\f^\ast)\} } \\
        &\ge \size{J_1} \abs{\overline{\ell}_{t_1}(\hf) - \overline{\ell}_{t_1}(\f^\ast)} - \size{J_2} \abs{\overline{\ell}_{t_2}(\hf) - \overline{\ell}_{t_2}(\f^\ast)} \\
        &\ge (\size{J_1} - C_{\ell} \size{J_2}) \abs{\overline{\ell}_{t_1}(\hf) - \overline{\ell}_{t_1}(\f^\ast)} - C_{\ell} \size{J_2} \Delta_k.
    \end{align*}
    By the definition of $\homo$, we have $\size{J_1}$ dominates $J_2$ ($\size{J_1} \asymp \size{I} \gg \size{J_2}$), we can control the point-wise error on $t_1$:
    \begin{equation*}
        \abs{\overline{\ell}_{t_1}(\hf) - \overline{\ell}_{t_1}(\f^\ast)} \lesssim \frac{\dacc + C_{\ell} \size{J_2} \Delta_k}{\size{J_1} - C_{\ell} \size{J_2}} \lesssim \frac{\dacc}{\size{J_1}}.
    \end{equation*}
    By Condition~\ref{cond:model_and_loss}(b), the point-wise error on $t_2$ is automatically bounded: $\abs{\overline{\ell}_{t_2}(\hf) - \overline{\ell}_{t_2}(\f^\ast)} \lesssim \abs{\overline{\ell}_{t_1}(\hf) - \overline{\ell}_{t_1}(\f^\ast)} + \Delta_k \lesssim \Delta_k$.
    Finally, summing these point-wise bounds over the validation set $J_{m, I}$ yields the desired result:
    \begin{align*}
        \abs{\risk_{J_{m, I}}(\hf) - \risk_{J_{m, I}}(\f^\ast)}
        &\le \size{J_{m, I} \cap I_1} \abs{\overline{\ell}_{t_1}(\hf) - \overline{\ell}_{t_1}(\f^\ast)} + \size{J_{m, I} \cap I_2} \abs{\overline{\ell}_{t_2}(\hf) - \overline{\ell}_{t_2}(\f^\ast)} \\
        &\lesssim \size{J_{m, I} \cap I_1} \frac{\dacc}{\size{J_1}} + \size{I_2} \Delta_k \lesssim \dacc.
    \end{align*}
    The reverse direction holds analogously.
\end{proof}

\subsection{Proof of Theorem \ref{thm:cvloss}}
\begin{proof}[Proof of Theorem \ref{thm:cvloss}]
It is sufficient to show that $\Gbb^{\mathsf{nhomo}} \cap \Gbb^{-\mathsf{nhomo}}$ holds with high probability, which is equivalent to developing the upper and lower bounds of the approximation error when the RECV loss is adopted in Lemma~\ref{lem:loc_err_g}.

Let $\Lval_{I, \lambda} = \sum_{m \in [M]} \{\L(\z_{J_{m, I}}; \hf_{J_{-m, I}}(\lambda))\}$ and $\xi_{I, \lambda} = \Lval_{I, \lambda} - \Lval_{I}^\ast - \Delta_I \size{I}$, for any fixed hyperparameter $\lambda \in \Lambda$. For the RECV loss, we define:
$$\xi_{I, \mcv} = \min_{\lambda \in \Lambda} \xi_{I, \lambda} \quad\mbox{and}\quad \Lval_{I, \mcv} = \min_{\lambda \in \Lambda} \Lval_{I, \lambda}.$$

\noindent\textit{Step 1: Upper Bound}

We verify the upper bound for $\Gbb^{\mathsf{nhomo}}$. For any nearly homogeneous segment $I \in \homo$, we must show:
\[
\xi_{I,\mcv} \le C_{\ref*{lem:loc_err_g}.1} \dacc
\]
holds with probability at least $1 - \exp(-C \log n)$.

Recall that the RECV loss is $\Lval_{I,\mcv} = \min_{\lambda \in \Lambda} \Lval_{I, \lambda}$.
By the definition of the minimum, we have:
\[
\Lval_{I,\mcv} \le \Lval_{I, \lambda} \text{ and } \xi_{I, \mcv} \le \xi_{I, \lambda} \quad \text{for every } \lambda \in \Lambda.
\]
Following the same arguments in the proof of Theorem~\ref{thm:cross-fitting}, by Condition \ref{cond:predict_exist}, it is guaranteed that there exists a ``good'' hyperparameter $\lambda_I \in \Lambda$ such that for every $m \in [M]$:
\[
\abs{ \risk_{J_{m, I}}(\hf_{J_{-m, I}}(\lambda_I)) - \risk_{J_{m, I}}(\f_{{I}}^\ast) } \lesssim \dacc.
\]
This "good" model $\hf_{J_{-m, I}}(\lambda_I)$ satisfies the requirements of Condition \ref{cond:predict} that used in the proof of Theorem~\ref{thm:cross-fitting}. It means that with probability at least $1 - \exp(-C\log n)$, uniformly for every $I \in \homo$,
$$\xi_{I, \mcv} \le \xi_{I, \lambda_I} \lesssim \widetilde{C}^{\frac{1}{2}}\dacc.$$
where the term, $\xi_{I, \lambda_I}$, is precisely the approximation error for the cross-fitting loss with a fixed and given hyperparameter $\lambda_I$, which is the exact quantity that was well-controlled in the proof of Theorem~\ref{thm:cross-fitting}.
It establishes the required upper bound for $\Gbb^{\mathsf{nhomo}}$.

\noindent\textit{Step 2: Lower Bound}

We verify the lower bound for $\Gbb^{\mathsf{nhomo}}$ and $\Gbb^{-\mathsf{nhomo}}$. For every nearly homogeneous segment $I \in \homo$, we must show:
\[
\xi_{I,\mcv} \ge -C_{\ref*{lem:loc_err_g}.1} \dacc
\]
holds with probability at least $1 - \exp(-C \log n)$.

From the proof of Theorem~\ref{thm:cross-fitting}, specifically the derivation of the lower bound in Eq.~\eqref{equ:lower_bound_general}, we have a general bound on the concentration term for every \textit{fixed} $\lambda$.
Recall that in Condition~\ref{cond:predict_exist}, we assume the set of candidate hyperparameters $\Lambda$ is finite with $\log |\Lambda| \lesssim \log n$. By applying a union bound over all $\lambda \in \Lambda$, \textit{uniformly} for all $\lambda \in \Lambda$ and $I$ with $\size{I} \ge C_{\m} \dacc$, with probability at least $1 - \exp(-C \log n)$:
\begin{equation}\label{equ:lower_bound_general_fixLam}
        \sum_{m \in [M]} \L(\z_{J_{m, I}}; \hf_{J_{-m, I}}(\lambda)) - \Lval_{I}^\ast - \Delta_{I} \size{I} \ge -C_A \Delta_{I} \size{I} - \widetilde{C}^{\frac{1}{2}} \dacc.
\end{equation}
where $C_A = \{C_3 + (1 - C_3) (C_{\m}^{-\frac{1}{2}} + \widetilde{C}^{-\frac{1}{2}})\} \in [0, 1)$.

For $I \in \homo$, since $\Delta_I \size{I} \le 2 \widetilde{C} \dacc$ and $C_{\m}$ is sufficiently large, we have
\begin{equation*}
        \sum_{m \in [M]} \L(\z_{J_{m, I}}; \hf_{J_{-m, I}}(\lambda)) - \Lval_{I}^\ast - \Delta_{I} \size{I} \gtrsim - \widetilde{C}^{\frac{1}{2}} \dacc,
\end{equation*}
uniformly holds over $I \in \homo, \lambda \in \Lambda$. It implies that,
\begin{equation*}
    \xi_{I, \mcv} \gtrsim - \widetilde{C}^{\frac{1}{2}} \dacc,
\end{equation*}
for all $I \in \homo$.
This establishes the required lower bound for $\Gbb^{\mathsf{nhomo}}$. Combining the result in Step 1, we have $P(\Gbb^{\mathsf{nhomo}}) \ge 1 - \exp(-C\log n)$ with $C_{\ref*{lem:loc_err_g}.1} = C' \widetilde{C}^{\frac{1}{2}}$ for some constant $C' > 0$.

Similarly, for $I \notin \homo$ with $\size{I} \ge d_{\m} = C_{\m} \dacc$, Eq.~\eqref{equ:lower_bound_general_fixLam} implies that
\begin{equation*}
    \xi_{I, \mcv} \gtrsim - (C_{\m}^{-\frac{1}{2}} + \widetilde{C}^{-\frac{1}{2}}) \Delta_{I} \size{I} - \widetilde{C}^{\frac{1}{2}} \dacc.
\end{equation*}
Therefore we have $\Pbb(\Gbb^{-\mathsf{nhomo}}) \ge 1 - \exp(-C\log n)$ with $C_{\ref*{lem:loc_err_g}.2} = C' (C_{\m}^{-\frac{1}{2}} + \widetilde{C}^{-\frac{1}{2}})$ for some constant $C' > 0$.

In summary, $\Pbb(\Gbb^{\mathsf{nhomo}} \cap \Gbb^{-\mathsf{nhomo}}) \ge 1 - \exp(-\Omega(\log n))$. Based on the choices of $C_{\ref*{lem:loc_err_g}.1}$ and $C_{\ref*{lem:loc_err_g}.2}$, the feasible region in the proof of Lemma~\ref{lem:loc_err_g}, i.e.,
    $$\widetilde{C} \ge \{1 - (1 + 4 C_{\Delta}) C_{\ref*{lem:loc_err_g}.2} - 4 C_{\Delta} C_{\m, 1}\}^{-1} C_{\Delta} (14 C_{\ref*{lem:loc_err_g}.1} + 2 C_{\m, 1})$$
is non-empty.
It means that there is a constant $\widetilde{C}$ that only depends on $C_{\m, 1}, C_{\m}, C_{\snr}, C_{\Delta}$ so that Theorem~\ref{thm:cvloss} holds.

\end{proof}

\subsection{Proof of Theorem \ref{thm:cross_fitting_temporal}}

In this subsection, we present the proof of Theorem~\ref{thm:cross_fitting_temporal}, which is based on a new variance-adaptive Bernstein-type inequality for functional dependence sequence (formally introduced in Section~\ref{sec:bernMix}).

\begin{proof}[Proof of Theorem \ref{thm:cross_fitting_temporal}]

    For any segment $I$ and candidate model $\f$, let $S_I(\f) = \sum_{i \in I} s_{i, \f}$ denote the block sum of the centered empirical excess risk defined in Condition~\ref{cond:model_and_loss}(a). Define its variance proxy as $\mathcal{V}_I(\f) = \sum_{i \in I} \sigma_{i, \f}^2$ and the self-normalized empirical process as $Z_I(\f) = S_I(\f) / \mathcal{V}_I(\f)^{1/2}$.
Let $J = J_{m, I} = (a, b]$ denote the validation block of size $|J|$.
Recall that $\tf$ denotes the decoupled model estimator trained on $\{\widetilde{\z}_{i, (a-v, b]}\}_{i \in J_{-m, I}^{v}}$. We further denote a decoupled validation set $\widetilde{D}_{J} = \{\widetilde{\z}_{i, (-\infty, a-v]}\}_{i \in J} := \{\widetilde{\z}_{i}\}_{i \in J}$. Correspondingly, let $\tilde{s}_{i, \f}$ be the centered excess risk evaluated on the decoupled observation $\widetilde{\z}_i$ and set $\widetilde{S}_{J}(\tf) = \sum_{i \in J} \tilde{s}_{i, \tf}$ and $\widetilde{Z}_{J}(\tf) = \widetilde{S}_{J}(\tf) / \mathcal{V}_J(\tf)^{1/2}$.

To derive the concentration bounds for the approximation error $\xi_I$ with the cross-fitting loss and the RECV loss, it suffices to uniformly bounding the large deviation of the self-normalized cross-validation empirical process $Z_{J}(\hf)$.
Once this concentration bound is rigorously established, the remainder of the proof seamlessly identically follows that of the independent case.
We now detail the derivation of this concentration inequality utilizing the decoupling strategy and the mean-square relative stability.

The major tasks lie in controlling the coupling difference term $\abs{Z_{J}(\hf) - \widetilde{Z}_{J}(\tf)}$ and deriving a concentration bound for the decoupled self-normalized sum $\widetilde{Z}_{J}(\tf)$.
This second task is achieved using a novel variance-adaptive and self-normalized Bernstein-type inequality for mixing sequences, which is developed in Section~\ref{sec:bernMix}.

To control the first part, we will rely on the adaptive truncation operation.
We define the truncation parameter as
\begin{equation}\label{equ:defTrunc}
    \trunc := 2 \bigl\{c_{\trunc} (\log(e |J|) + x^{\kappa_1}) \bigr\}^{1/\kappa_1},
\end{equation}
where $c_{\trunc} \ge 2$ is a sufficiently large absolute structural constant.
For any fixed hypothesis $\f \in \mathcal{F}$, we define the truncated centered variable $s_{\trunc, i, \f} = \varphi_{\trunc m_{i, \f}}(s_{i, \f}) - \mathbb{E}\bigl[\varphi_{\trunc m_{i, \f}}(s_{i, \f})\bigr]$, where $\varphi_{\trunc}(z) = (z \wedge \trunc) \vee (-\trunc)$, and its block sum $S_{\trunc, J}(\f) = \sum_{i \in J} s_{\trunc, i, \f}$.
Denote the truncation bias by $b_{\trunc, i, \f} = \mathbb{E}[\varphi_{\trunc m_{i, \f}}(s_{i, \f})]$.
When the truncation does not happen, we deterministically have $s_{i, \f} = s_{\trunc, i, \f} + b_{\trunc, i, \f}$.
Let $E_{\trunc} := \set{\max_{i \in J} \abs{s_{i, \hf}} / m_{i, \hf} \le \trunc}$ denote the event where the learned model avoids truncation on the original validation data.
Conditional on the event $E_{\trunc}$, we have $S_J(\hf) = S_{\trunc, J}(\hf) - \sum_{i \in J} b_{\trunc, i, \hf}$.

This uniform truncation bias is controlled by the extreme tail behavior of the un-truncated loss.
Using the sub-Weibull tail assumption $\mathbb{P}(\abs{s_{i, \f}}/m_{i, \f} > z) \le \exp(1 - z^{\kappa_1})$, the expected bias is bounded by $\abs{b_{\trunc, i, \f}} \le C_{b} m_{i, \f} \trunc^{1-\kappa_1} \exp(-\trunc^{\kappa_1})$.
Denote $\bar{m}_{s, \sigma} = \sup_{i, \f} m_{i, \f} / \sigma_{i, \f}$.
We have:
\begin{equation*}
    \sup_{\f \in \mathcal{F}} \frac{\sum_{i \in J} \bigabs{b_{\trunc, i, \f}}}{\mathcal{V}_J(\f)^{1/2}} \le C_{b} \bar{m}_{s, \sigma} \sqrt{|J|} \trunc^{1 - \kappa_1} \exp(- \trunc^{\kappa_1}).
\end{equation*}
By the definition of $\trunc$ in Eq.~\eqref{equ:defTrunc}, for $x$ bounded away from zero, the right hand side of the above inequality decays at an overwhelmingly faster rate of $\mathcal{O}\bigl(|J|^{1 - c_{\trunc}^{\kappa_1}} \exp(-c_{\trunc}^{\kappa_1} x^{\kappa_1})\bigr)$, which is trivially absorbed by $x/4$ for sufficiently large $c_{\trunc}$.
Consequently, applying the union bound yields:
\begin{equation}\label{eq:prob_split_initial}
    \mathbb{P}\bigl(S_J(\hf) \ge x \mathcal{V}_J(\hf)^{1/2}\bigr) \le \mathbb{P}\bigl( Z_{\trunc, J}(\hf) \ge \frac{3x}{4} \bigr) + \mathbb{P}(E_{\trunc}^{c}),
\end{equation}
where $Z_{\trunc, J}(\hf) = S_{\trunc, J}(\hf) / \mathcal{V}_J(\hf)^{1/2}$ is the standardized empirical process.

Note that the trained model $\hf = \hf_{J_{-m,I}^v}$ is separated from $J$ by two-sided exclusion buffers of both size $v$.
We bound the extreme event $\mathbb{P}(E_{\trunc}^c)$ using the decoupled variables to decouple the data-dependent selection bias.
For $i \in J$, let $\widetilde{s}_{i, \f}$ denote the centered empirical excess risk evaluated on the decoupled observation $\widetilde{\z}_i$.
The triangle inequality yields:
\begin{equation*}
    \mathbb{P}(E_{\trunc}^c) \le \mathbb{P}\Bigl(\max_{i \in J} \frac{\abs{\widetilde{s}_{i,\tf}}}{m_{i, \tf}} > \frac{\trunc}{2}\Bigr) + \mathbb{P}\Bigl(\max_{i \in J}\Bigabs{\frac{s_{i,\hf}}{m_{i, \hf}} - \frac{\widetilde{s}_{i,\tf}}{m_{i, \tf}}} > \frac{\trunc}{2}\Bigr).
\end{equation*}
Because the decoupled model $\tf$ and the decoupled validation block $\widetilde{D}_J$ are independent, conditioning on $\tf$ reduces it to a deterministic hypothesis, validating the union bound over the sub-Weibull tails:
\begin{equation*}
    \mathbb{P}\Bigl(\max_{i \in J} \frac{\abs{\widetilde{s}_{i,\tf}}}{m_{i, \tf}} > \frac{\trunc}{2}\Bigr) \le |J| \exp\bigl(1 - (\trunc/2)^{\kappa_1}\bigr) \le \exp(-x^{\kappa_1}).
\end{equation*}
For the coupling difference, we deploy the mean-square relative bounds established by Condition~\ref{cond:fdm_stability}(c) and the Cauchy-Schwarz inequality. The algorithmic shift of the normalized loss bypasses the denominator dependency.
Letting $c_2 = \sup_{\f} \mathbb{E}[ (s_{i, \f} / m_{i, \f})^2 ] = \mathcal{O}(1)$ be the bounded second moment under the sub-Weibull tail assumption, we have:
\begin{align*}
    \mathbb{E} \Bigabs{\frac{s_{i,\hf}}{m_{i, \hf}} - \frac{s_{i,\tf}}{m_{i, \tf}}}
    &\le \mathbb{E} \Bigabs{\frac{s_{i,\hf} - s_{i,\tf}}{m_{i, \hf}}} + \mathbb{E} \Bigabs{\frac{s_{i,\tf}}{m_{i, \tf}} \cdot \frac{m_{i,\tf} - m_{i,\hf}}{m_{i, \hf}}} \\
    &\le \biggl( \mathbb{E} \Bigabs{\frac{s_{i,\hf} - s_{i,\tf}}{m_{i, \hf}}}^2 \biggr)^{1/2} + \sqrt{c_2} \biggl( \mathbb{E} \Bigabs{\frac{m_{i,\hf} - m_{i,\tf}}{m_{i, \hf}}}^2 \biggr)^{1/2} \\
    &\le (1 + \sqrt{c_2}) C_{\mathrm{stab}} \Theta_{\z, q}(v).
\end{align*}
Combined with the data coupling bound from Condition~\ref{cond:fdm_stability}(b), applying Markov's inequality bounds the deviation:
\begin{align*}
    \mathbb{P}\Bigl(\max_{i \in J}\Bigabs{\frac{s_{i,\hf}}{m_{i, \hf}} - \frac{\widetilde{s}_{i,\tf}}{m_{i, \tf}}} > \frac{\trunc}{2}\Bigr) &\le \frac{2}{\trunc} \mathbb{E} \Biggl[ \sum_{i \in J} \Bigabs{\frac{s_{i, \hf}}{m_{i, \hf}} - \frac{s_{i, \tf}}{m_{i, \tf}}} + \sum_{i \in J} \Bigabs{\frac{s_{i, \tf}}{m_{i, \tf}} - \frac{\widetilde{s}_{i, \tf}}{m_{i, \tf}}} \Biggr] \\
    &\le \frac{2 \bigl( (1 + \sqrt{c_2})C_{\mathrm{stab}} + 1 \bigr) |J| \Theta_{\z, q}(v)}{\trunc}.
\end{align*}

To bound the core process $\mathbb{P}\bigl( Z_{\trunc, J}(\hf) \ge 3x/4 \bigr)$, we allocate $x/4$ to the algorithmic decoupling error and $x/2$ to the decoupled empirical process.
Let $\widetilde{S}_{\trunc, J}(\f)$ denote the block sum of the dynamically truncated variables evaluated on the decoupled observations, and define the decoupled self-normalized process as $\widetilde{Z}_{\trunc, J}(\f) = \widetilde{S}_{\trunc, J}(\f) / \mathcal{V}_J(\f)^{1/2}$.
This probability splitting yields:
\begin{equation*}
    \mathbb{P}\bigl( Z_{\trunc, J}(\hf) \ge \frac{3x}{4} \bigr) \le \mathbb{P}\bigl( \widetilde{Z}_{\trunc, J}(\tf) \ge \frac{x}{2} \bigr) + \mathbb{P}\Bigl( \bigabs{Z_{\trunc, J}(\hf) - \widetilde{Z}_{\trunc, J}(\tf)} \ge \frac{x}{4} \Bigr).
\end{equation*}

For the algorithmic perturbation, we bound its expectation utilizing the pointwise relative stability constraints. By leveraging the joint $1$-Lipschitz property of the truncation operator $\varphi_c(x)$, yielding $\abs{\varphi_{c_1}(x_1) - \varphi_{c_2}(x_2)} \le \abs{x_1 - x_2} + \abs{c_1 - c_2}$, the point-wise shift of the dynamically truncated variables is bounded by:
\begin{equation*}
    \bigabs{s_{\trunc, i, \hf} - s_{\trunc, i, \tf}} \le 2\bigabs{s_{i, \hf} - s_{i, \tf}} + 2 \trunc \bigabs{m_{i, \hf} - m_{i, \tf}}.
\end{equation*}
We decouple the truncated normalized process into:
\begin{equation*}
    \bigabs{Z_{\trunc, J}(\hf) - Z_{\trunc, J}(\tf)} \le \frac{\bigabs{S_{\trunc,J}(\hf) - S_{\trunc,J}(\tf)}}{\mathcal{V}_J(\hf)^{1/2}} + \bigabs{Z_{\trunc, J}(\tf)} \cdot \Bigabs{ \frac{\mathcal{V}_J(\tf)^{1/2}}{\mathcal{V}_J(\hf)^{1/2}} - 1 }.
\end{equation*}
Taking expectations and applying the Cauchy-Schwarz decomposition similarly on the variance ratio yields:
\begin{align*}
    \mathbb{E}\bigabs{Z_{\trunc, J}(\hf) - Z_{\trunc, J}(\tf)}
    &\le \mathbb{E} \Biggl[ \sum_{i \in J} \biggl( 2 \bar{m}_{s, V} \frac{\abs{s_{i, \hf} - s_{i, \tf}}}{m_{i, \hf}} + 2 \trunc \bar{m}_{s, V} \frac{\abs{m_{i, \hf} - m_{i, \tf}}}{m_{i, \hf}} \biggr) \Biggr] \\
    &\quad + \bigl( \mathbb{E}[Z_{\trunc, J}(\tf)^2] \bigr)^{1/2} \biggl( \mathbb{E} \Bigabs{ \frac{\mathcal{V}_J(\tf)^{1/2} - \mathcal{V}_J(\hf)^{1/2}}{\mathcal{V}_J(\hf)^{1/2}} }^2 \biggr)^{1/2},
\end{align*}
where $\bar{m}_{s, V} = \sup_{i \in J, \f} m_{i, \f} / \mathcal{V}_J(\f)^{1/2}$.
Here we apply the property $1/\mathcal{V}_J(\hf)^{1/2} \le \bar{m}_{s, V} / m_{i, \hf}$.
Because the truncation boundary $\trunc$ grows at a logarithmic rate $\trunc \asymp (\log |J|)^{1/\kappa_1}$, it is overwhelmingly absorbed by the exponential decay of $\Theta_{\z, q}(v)$. Applying the data decoupling bound from Condition~\ref{cond:fdm_stability}(b) and absorbing the scale-free constants $c_2$ and $\bar{m}_{s, V}$ into an enlarged stability constant $C_{\mathrm{stab}}^\prime > 0$, we control the shift:
\begin{align*}
    \mathbb{E}\bigabs{Z_{\trunc, J}(\hf) - \widetilde{Z}_{\trunc, J}(\tf)} &\le \mathbb{E}\bigabs{Z_{\trunc, J}(\hf) - Z_{\trunc, J}(\tf)} + \mathbb{E} \Biggl[ \sum_{i \in J} \frac{\sigma_{i, \tf}}{\mathcal{V}_J(\tf)^{1/2}} \norm{\phi(\z_i) - \phi(\widetilde{\z}_i)} \Biggr] \\
    &\le C_{\mathrm{stab}}^\prime |J| \Theta_{\z, q}(v).
\end{align*}
Applying Markov's inequality directly secures a tight bound on the perturbation cost:
\begin{equation*}
    \mathbb{P}\Bigl( \bigabs{Z_{\trunc, J}(\hf) - \widetilde{Z}_{\trunc, J}(\tf)} \ge \frac{x}{4} \Bigr) \le \frac{4 C_{\mathrm{stab}}^\prime |J| \Theta_{\z, q}(v)}{x}.
\end{equation*}

The remaining term evaluates the tail probability of the decoupled process $\widetilde{Z}_{\trunc, J}(\tf)$.
Because $\tf$ is deterministically decoupled from the validation dataset $\widetilde{D}_J$, by conditioning on $\tf$ and applying Theorem~\ref{thm:bern_fdm} that introduced in the next subsection,
we obtain:
\begin{align*}
    \mathbb{P}\Bigl( \widetilde{Z}_{\trunc, J}(\tf) \ge \frac{x}{2} \Bigr)
    &\le |J| \exp\biggl(-\frac{x^\kappa}{C_1 \bar{m}_{s, V}^\kappa}\biggr) + \exp\biggl(-\frac{x^2}{C_2}\biggr) \nonumber \\
    &\quad + \exp\Biggl(-\frac{x^2}{C_3 |J| \bar{m}_{s, V}^2} \exp\biggl( \frac{x^{\kappa(1-\kappa)}}{C_4 \bar{m}_{s, V}^{\kappa(1-\kappa)} \log^\kappa(x/\bar{m}_{s, V})} \biggr) \Biggr),
\end{align*}
for some constants $C_1, \dots, C_4 > 0$.

Synthesizing the unified tail probability by aggregating the probability splits, and leveraging that $\exp(-x^{\kappa_1})$ is asymptotically dominated by the tail $\exp(-x^\kappa / C_1 \bar{m}_{s, V}^\kappa)$, the probability is bounded by:
\begin{align*}
    \mathbb{P}\bigl(S_J(\hf) \ge x \mathcal{V}_J(\hf)^{1/2}\bigr)
    &\le |J| \exp\biggl(-\frac{x^\kappa}{C_1 \bar{m}_{s, V}^\kappa}\biggr) + \exp\biggl(-\frac{x^2}{C_2}\biggr) \\
    &\quad + \exp\Biggl(-\frac{x^2}{C_3 |J| \bar{m}_{s, V}^2} \exp\biggl( \frac{x^{\kappa(1-\kappa)}}{C_4 \bar{m}_{s, V}^{\kappa(1-\kappa)} \log^\kappa(x/\bar{m}_{s, V})} \biggr) \Biggr) \\
    &\quad + \frac{4C_{\mathrm{stab}}^{\prime} |J| \Theta_{\z, q}(v)}{x}.
\end{align*}
This completes the derivation of the self-normalized concentration bound. In the changepoint localization proof, the required large deviation threshold scales as $x \gtrsim (\log n)^{1/\kappa}$.
By scaling the separation buffer $v \gtrsim (\log n)^{1/\kappa_2}$, the exponential decay of the functional dependence measure $\Theta_{\z, q}(v)$ ensures that the algorithmic perturbation term $\mathcal{O}(|J| \Theta_{\z, q}(v) / x)$ decays overwhelmingly faster than any polynomial $n^{-C}$, thereby acting as a negligible residual.
The leading term in the above concentration bound will be $|J| \exp(-\frac{x^\kappa}{C_1 \bar{m}_{s, V}^\kappa}) + \exp(-\frac{x^2}{C_2})$.

The rest of the proof will similarly follow the same procedure of the proofs of Theorem~\ref{thm:cross-fitting} and Theorem~\ref{thm:cvloss}, which we omit here.

\end{proof}

\subsubsection{A variance-adaptive Bernstein's inequality for functionally-dependent sequences}\label{sec:bernMix}

To establish the bounds for $\widetilde{Z}_{\trunc, J}(\tf)$ in the above subsection under non-stationary and heteroskedastic temporal data,
we adopt the functional dependence framework \citep{wu2005nonlinear, xu2024change} and develop a novel variance-adaptive Bernstein-type inequality.

For a better readability, we introduce several notations under the functional dependence framework.
Let $\{X_i\}_{i \in \mathbb{Z}}$ be a sequence of centered real-valued random variables. Under the functional dependence framework, each observation is generated by a measurable function $g_i$ taking an independent innovation sequence as input:
\begin{equation*}
    X_i = g_i(\mathcal{F}_i^X), \quad \text{where} \quad \mathcal{F}_i^X = \{X_j^{\mathrm{ind}}\}_{j \le i},
\end{equation*}
and $\{X_j^{\mathrm{ind}}\}_{j \in \mathbb{Z}}$ are independent random elements.
To measure the temporal dependence, we quantify the impact of replacing a past innovation. Let $\mathcal{F}_{i, \{i-v\}}^X$ be the same filtration as $\mathcal{F}_i^X$ except that the innovation $X_{i-v}^{\mathrm{ind}}$ is replaced by an independent copy $\widetilde{X}_{i-v}^{\mathrm{ind}}$. The coupled variable is defined as $\widetilde{X}_{i, \{i-v\}} = g_i(\mathcal{F}_{i, \{i-v\}}^X)$.

Let $\sigma_i^2 = \Var(X_i)$ be the variance and denote the scaled variables $Z_i^{(\sigma)} = X_i / \sigma_{i}$.
For $q > 0$, the FDM for the scaled variables and its cumulative version are defined as:
\begin{equation*}
    \theta_{Z^{(\sigma)},q}(v) = \sup_{i \in \mathbb{Z}} \bignorm{(X_i - \widetilde{X}_{i, \{i-v\}})/\sigma_{i}}_q \quad \mbox{and} \quad \Theta_{Z^{(\sigma)},q}(v) = \sum_{j = v}^{\infty} \theta_{Z^{(\sigma)},q}(j).
\end{equation*}

We impose the following regularity conditions on the tail distributions and the decay rate of the temporal dependence.

\begin{condition}\label{cond:subweibull_fdm}
There exist sequences of positive parameters $\{\sigma_i\}_{i \in \mathbb{Z}}$ and $\{m_i\}_{i \in \mathbb{Z}}$ such that $\sigma_i^2 = \Var(X_i)$ and $\sigma_i \le m_i \le \bar{m} < \infty$. We assume there exist positive shape parameters $\kappa_1, \kappa_2 > 0$ such that:
\begin{itemize}
    \item[(a)] (Sub-Weibull Tails). The standardized sequence $Z_i = X_i / m_i$ satisfies $\sup_{i>0}\mathbb{P}(|Z_{i}|>t)\le \exp (1-t^{\kappa_1})$.
    \item[(b)] (Exponential Dependence Decay). The cumulative FDM of the variance-standardized sequence $Z_i^{(\sigma)} = X_i / \sigma_i$ satisfies $\Theta_{Z^{(\sigma)}, q}(v) \le \exp(-c v^{\kappa_2})$ for some absolute constant $c > 0$, $q \ge 2$ and $v \ge 1$.
\end{itemize}
\end{condition}

For any set $\mathcal{K} \subset \mathbb{Z}$, denote the effective variance of the block sum as
\begin{equation}\label{def:effect_var}
    \mathcal{V}_{\mathcal{K}} = \Var\Bigl(\sum_{i \in \mathcal{K}} X_i\Bigr).
\end{equation}
To explicitly bound this variance under temporal dependence, we utilize the martingale difference projection operator $\mathcal{P}_k (\cdot) = \mathbb{E}[\cdot \mid \mathcal{F}_k^X] - \mathbb{E}[\cdot \mid \mathcal{F}_{k-1}^X]$.
By the orthogonal decomposition $X_i = \sum_{k \le i} \mathcal{P}_k X_i$, the covariance between any two observations is bounded by the Cauchy-Schwarz inequality:
\begin{equation*}
    \bigabs{\mathrm{Cov}(X_i, X_j)} \le \sum_{k \le i \wedge j} \bigabs{\mathbb{E}[(\mathcal{P}_k X_i)(\mathcal{P}_k X_j)]} \le \sum_{k \le i \wedge j} \norm{\mathcal{P}_k X_i}_2 \norm{\mathcal{P}_k X_j}_2.
\end{equation*}
Applying Theorem 1 in \citet{wu2005nonlinear}, the $L_2$ norm of the projection is bounded by the coupling distance: $\norm{\mathcal{P}_k X_i}_2 \le \norm{X_i - \widetilde{X}_{i, \{k\}}}_2 = \sigma_i \theta_{Z^{(\sigma)},2}(i-k)$.
Consequently, utilizing the inequality $2\sigma_i\sigma_j \le \sigma_i^2 + \sigma_j^2$, the total effective variance is bounded by aggregating the functional dependence measure:
\begin{align*}
    \mathcal{V}_{\mathcal{K}} &= \sum_{i \in \mathcal{K}} \sigma_i^2 + \sum_{i, j \in \mathcal{K}, i \neq j} \mathrm{Cov}(X_i, X_j) \\
    &\le \sum_{i \in \mathcal{K}} \sigma_i^2 + \sum_{i, j \in \mathcal{K}, i \neq j} \frac{\sigma_i^2 + \sigma_j^2}{2} \sum_{k \le i \wedge j} \theta_{Z^{(\sigma)}, 2}(i-k) \theta_{Z^{(\sigma)}, 2}(j-k) \\
    &\le \sum_{i \in \mathcal{K}} \sigma_i^2 + \sum_{i \in \mathcal{K}} \sigma_i^2 \Bigl(\sum_{m \ge 0} \theta_{Z^{(\sigma)}, 2}(m) \Bigr)^2 \\
    &= \bigl( 1 + \Theta_{Z^{(\sigma)},2}^2(0) \bigr) \sum_{i \in \mathcal{K}} \sigma_i^2.
\end{align*}
By Condition~\ref{cond:subweibull_fdm}(b), the cumulative FDM $\Theta_{Z^{(\sigma)},2}(0)$ is bounded by an absolute constant.
Thus, we rigorously verify that $\mathcal{V}_{\mathcal{K}} \le C_V \sum_{i \in \mathcal{K}} \sigma_i^2$, where the structural constant is given by $C_V = 1 + \Theta_{Z^{(\sigma)},2}^2(0)$.

Let $\kappa = (\kappa_1^{-1}+\kappa_2^{-1})^{-1}$ represent the unified tail parameter characterizing the interaction between the heavy-tailed marginals and the strong temporal correlations. We have the following variance-adaptive Bernstein-type inequality.

\begin{theorem}
\label{thm:bern_fdm}
Let $(X_{i})_{i \in \mathbb{Z}}$ be a sequence of centered real-valued random variables. Assume that Condition~\ref{cond:subweibull_fdm} is satisfied and $n \ge 4$. For $\kappa < 1$, there exist absolute positive constants $C_{1}$, $C_{2}$, $C_{3}$ and $C_{4}$ depending only on $c$, $\kappa$ and $\kappa_2$ such that, for any threshold $x > 0$:
\begin{align*}
    \mathbb{P}\Bigl(\sup_{j \le n}\Big|\sum_{i=1}^j X_i\Big|\ge x\Bigr)
    &\le n\exp \Bigl(-\frac{x^{\kappa}}{C_{1} \bar{m}^{\kappa}}\Bigr) + \exp \Bigl(-\frac{x^{2}}{C_{2} \sum_{i=1}^n \sigma_i^2}\Bigr) \\
    &\quad + \exp \Biggl(-\frac{x^{2}}{C_{3}n \bar{m}^{2}}\exp \biggl(\frac{x^{\kappa (1-\kappa)} (\log(\frac{x}{\bar{m}}))^{-\kappa}}{C_{4} \bar{m}^{\kappa (1-\kappa)}}\biggr)\Biggr).
\end{align*}
\end{theorem}

To match the heterogeneous nature of changepoint problem,
the inequality in Theorem~\ref{thm:bern_fdm} depends on the variance sum $\sum_{i=1}^n \sigma_i^2$ rather than the worst case bound $n \max_{i=1}^n \sigma_i^2$ derived in
\citet{merlevede2011weakly} and \citet{xu2024change}.
To achieve this, we adopt the multi-level Cantor-tree construction technique in \citet{merlevede2011weakly} with a new adaptive truncation operation and a novel random shifting modification of the Cantor trees.
Please refer to Propositions \ref{propinter2}--\ref{propinter_block_unified} and the proof of Theorem \ref{thm:bern_fdm} below for more details.

\subsubsection{Proof of Theorem \ref{thm:bern_fdm}}

For any positive $\trunc$ and any positive integer $i$, we define the truncated and centered variable as
\begin{equation*}
X_{\trunc, i} = m_i \varphi_{\trunc}(Z_i) - \mathbb{E}[m_i \varphi_{\trunc}(Z_i)] \,,
\end{equation*}
where $Z_i = X_i / m_i$ and $\varphi_{\trunc}(x) = (x \wedge \trunc) \vee (-\trunc)$.

Before proving Theorem~\ref{thm:bern_fdm}, we introduce two supporting propositions that control the log-moment generating function of temporal-mixing sums.

The first proposition is a modification of Proposition 1 from \citet{merlevede2011weakly}.
Instead of bounding the variance by the global maximum $\max_{i} \sigma_i^2$, we retain the local variance structure of the Cantor set blocks.

\begin{proposition}
\label{propinter2} Let $(X_{i})_{i \ge 1}$ be a sequence of centered and real valued random variables satisfying Condition~\ref{cond:subweibull_fdm} and $0 < \kappa < 1$.
Let $n_{0}$ and $\bar{\leel}$ be two positive integers such that $n_{0}2^{-\bar{\leel}}\ge (1\vee 2c_{0}^{-1})$ with
\begin{equation}
c_0 = \frac{1}{4(2^{1/(\kappa)} - 1)} \Bigl( \sum_{\leel=1}^{\bar{\leel}} 2^{\leel(1 - \frac{1}{\kappa})} \Bigr)^{-1}.
\end{equation}
Let $\trunc=H^{-1}(\tau_{Z^{(\sigma)}}(c^{-\frac{1}{\kappa_2}}n_{0}))$ and for any $i$, set $X_{\trunc, i}=m_i \varphi_{\trunc}(Z_{i})-\mathbb{E}[m_i \varphi_{\trunc}(Z_{i})]$.

\noindent Then, there exists a subset $\mathcal{K}_{n_{0}}^{(\bar{\leel})}$ of the set $[n_{0}]$ with $\size{\mathcal{K}_{n_{0}}^{(\bar{\leel})}}\ge 3 n_{0}/4$, such that for any positive $t\le \bar{m}^{-1} (2^{-(1+\frac{2\kappa_{1}}{\kappa})}c_{1}^{\kappa_{1}}) (n_{0}^{\kappa-1}\wedge (2^{\bar{\leel}}/n_{0}))^{\frac{\kappa_{1}}{\kappa}}$,
\begin{equation}\label{resultpropinter2}
\log \mathbb{E} \exp \biggl(t \sum_{i \in \mathcal{K}_{n_{0}}^{( \bar{\leel})}} X_{\trunc, i} \biggr) \le t^2 \mathcal{V}_{\mathcal{K}_{n_{0}}^{(\bar{\leel})}} + t^2 \bar{m}^2 \bigl(\bar{\leel} (2{n_{0}})^{1 + \frac{\kappa_2}{\kappa}} + 2^{2+\frac{2\kappa_2}{\kappa}}{n_{0}}^{\kappa+\frac{2\kappa_2}{\kappa}} \bigr) \exp \Bigl(-\frac{1}{2}\Bigl(\frac{c_{1}{n_{0}}}{2^{\bar{\leel}}}\Bigr)^{\kappa_{1}}\Bigr),
\end{equation}
where $c_{1}=\min(c^{\frac{1}{\kappa_{1}}}c_{0}/4,2^{-\frac{1}{\kappa}})$.

\end{proposition}

\begin{proof}[Proof of Proposition \ref{propinter2}]

\noindent\textit{Step 1. The construction of $\mathcal{K}_{n_{0}}^{(\bar{\leel})}$}.
To manage the trade-off between the mixing effect and the sample efficiency, we construct the set $\mathcal{K}_{n_{0}}^{(\bar{\leel})}$ using a recursive, multi-level ``Cantor-type'' procedure.
The construction begins at level $\leel = 0$ with a single initial block of size $n_0$.
At each subsequent level $\leel \ge 1$, every continuous block of size $n_{\leel-1}$ retained from the previous level is symmetrically divided into two smaller children blocks of size $n_\leel$.
This division is achieved by removing a gap of size $d_{\leel-1}$ from the exact center of the parent block, subject to the constraint $n_{\leel-1} = 2 n_{\leel} + d_{\leel-1}$.
Specifically, the gap sequence is chosen as
\begin{equation}\label{equ:def_d\leel}
    d_{\leel-1} \in \bigl\{\lfloor c_{0}{n_{0}} 2^{-(\bar{\leel} \wedge \frac{\leel - 1}{\kappa})} \rfloor, \lfloor c_{0}{n_{0}} 2^{-(\bar{\leel} \wedge \frac{\leel - 1}{\kappa})} \rfloor - 1 \bigr\}.
\end{equation}
Notice that $d_{\leel-1}$ and $n_{\leel}$ decrease as $\leel$ increases, with $n_{\leel} < n_{\leel-1} / 2$.
Completing level $\leel$ yields $2^\leel$ disjoint continuous blocks, interspaced by the gaps introduced across all current and prior levels.
Iterating this procedure up to the targeted depth $\bar{\leel}$ produces a finite union of $2^{\bar{\leel}}$ elementary blocks, which form the highly decoupled base sets used in our variance bounding.
This fractal-like structure ensures that the fraction of indices removed, $\sum_{\leel=0}^{\bar{\leel}-1} 2^\leel d_\leel$, remains bounded by ${n_{0}}/4$, under the choice of $c_0$.
Consequently, it preserves the statistical properties of the original sum while achieving near-independence for the elementary blocks $I_{\bar{\leel}, j}$.
Figure~\ref{fig:cantor_set} illustrates the case $\bar{\leel} = 2$.
We summarize the construction details here:
\begin{enumerate}
    \item \textbf{Initialization (Level 0):} We start with the full set of indices $I_{0,1} = [{n_{0}}]$.
    \item \textbf{First Split (Level 1):} We remove a central gap of size $d_0$ to split $I_{0,1}$ into two sub-blocks $I_{1,1}$ and $I_{1,2}$ of equal size $n_1$.
    It holds that $2 n_1 + d_0 = n_0$.
    \item \textbf{Recursive Step (Level $\leel$):} At step $\leel$, each block $I_{\leel-1, j}$ from the previous level is split into two smaller blocks $I_{\leel, 2j-1}$ and $I_{\leel, 2j}$ of size $n_\leel$, separated by a gap $d_{\leel-1}$.
\end{enumerate}

\begin{figure}[H]
\centering
\input{figures/cantor_set.tex}
\caption{Visualization of the Cantor-type block construction for $\bar{\leel}=2$.
The shaded regions are the blocks kept in the set $\mathcal{K}_{n_{0}}^{(\bar{\leel})}$, while hatched regions are the gaps removed to ensure independence.}
\label{fig:cantor_set}
\end{figure}

The final set $\mathcal{K}_{n_{0}}^{(\bar{\leel})}$ is the union of the $2^{\bar{\leel}}$ disjoint blocks at the bottom level:
\[
\mathcal{K}_{n_{0}}^{(\bar{\leel})} = \bigcup_{j=1}^{2^{\bar{\leel}}} I_{\bar{\leel}, j}.
\]

To proceed with the decoupling, we introduce notation for the intermediate sets generated during the recursive construction.
For any level $\leel \in \{0, \dots, \bar{\leel}\}$ and any branch index $j \in \{1, \dots, 2^\leel\}$, let $\mathcal{K}_{{n_{0}}, \leel, j}^{(\bar{\leel})} = \mathcal{K}_{I_{\leel, j}}^{(\bar{\leel} - \leel)}$ denote the subset of the final Cantor set $\mathcal{K}_{n_{0}}^{(\bar{\leel})}$ that descends from the $j$-th block at level $\leel$.
Specifically, $\mathcal{K}_{{n_{0}}, \leel, j}^{(\bar{\leel})}$ is the union of the elementary blocks $I_{\bar{\leel}, i}$ that fall within the scope of the $j$-th branch at level $\leel$:
\begin{equation}  \label{eq:defKset}
\mathcal{K}_{{n_{0}}, \leel, j}^{(\bar{\leel})} = \bigcup_{i=(j-1)2^{\bar{\leel} - \leel} +1}^{j2^{\bar{\leel} - \leel}}I_{\bar{\leel} , i} \, .
\end{equation}
Observe that:
\begin{itemize}
    \item At $\leel=0$ (top level), $\mathcal{K}_{{n_{0}}, 0, 1}^{(\bar{\leel})} = \mathcal{K}_{n_{0}}^{(\bar{\leel})}$ is the entire set.
    \item At $\leel=\bar{\leel}$ (bottom level), $\mathcal{K}_{{n_{0}}, \bar{\leel}, j}^{(\bar{\leel})} = I_{\bar{\leel}, j}$ are the fundamental disjoint blocks.
    \item For any level $\leel < \bar{\leel}$, the set $\mathcal{K}_{{n_{0}}, \leel, j}^{(\bar{\leel})}$ is the disjoint union of its two children at level $\leel+1$:
    \[
    \mathcal{K}_{{n_{0}}, \leel, j}^{(\bar{\leel})} = \mathcal{K}_{{n_{0}}, \leel+1, 2j-1}^{(\bar{\leel})} \cup \mathcal{K}_{{n_{0}}, \leel+1, 2j}^{(\bar{\leel})}.
    \]
\end{itemize}
These two children are separated by a gap of size $d_\leel$.
We use this structure to recursively decouple the variables.
Using Lemma \ref{lemma:mix_ineq_mgf} (the decoupling lemma), we can approximate the exponential moment of the sum over the parent set by the product of the exponential moments of the sums over the children sets.
The error introduced at level $\leel$ is controlled by the mixing coefficient $\tau_{Z^{(\sigma)}}(d_{\leel})$.

\noindent\textit{Step 2. Proof of Inequality (\ref{resultpropinter2})}.
Without loss of generality, assume that $\bar{m} = 1$ in the proof.
To establish the bound on the log-moment generating function of the sum over the Cantor-type set $\mathcal{K}_{n_{0}}^{(\bar{\leel})}$, we recursively decouple the blocks.
Recall that the block size $n_\leel$ at level $\leel$ satisfies $n_{\leel} = (n_{\leel-1} - d_{\leel-1})/2$.
For any integer $\leel \in [0, \bar{\leel}]$, we define the decreased truncation level sequence $\{\trunc_{\leel}\}$ as:
\begin{equation}\label{defTruncj}
\trunc_{\leel}=H^{-1}\bigl(\tau_{Z^{(\sigma)}}(c^{-\frac{1}{\kappa_2}}{n_{0}}2^{-(\bar{\leel} \wedge \frac{\leel}{\kappa})})\bigr)\,.
\end{equation}
Notice that the parameter $\trunc$ in the statement of Proposition \ref{propinter2} corresponds exactly to $\trunc_0$.
For any positive $\trunc$ and an index set $\mathcal{K}$, we define the sum of the truncated and centered random variables over $\mathcal{K}$ as:
\begin{equation*}
S_{\trunc}(\mathcal{K})=\sum_{i\in \mathcal{K}}X_{\trunc, i} = \sum_{i\in \mathcal{K}} \bigl(m_i \varphi_{\trunc}(Z_{i})-\mathbb{E}[m_i \varphi_{\trunc}(Z_{i})]\bigr) \,.
\end{equation*}

Further denote $S_{\sigma}(\mathcal{K}) = \sum_{i \in \mathcal{K}} \sigma_i$. By the truncation, $|X_{\trunc_{0}, i}|\le 2 m_i \trunc_{0} \le 2 \trunc_{0}$.
Consequently, by Lemma \ref{lemma:mix_ineq_mgf}, we derive that for any positive $t$,
\begin{align*}
& \Bigabs{\mathbb{E}\exp \big (tS_{\trunc_{0}}(\mathcal{K}_{n_{0}}^{(\bar{\leel})})\big)-\prod_{j=1}^{2}\mathbb{E}\exp \big (tS_{\trunc_{0}}(\mathcal{K}_{{n_{0}},1,j}^{(\bar{\leel})})\big)} \\
\le & t S_{\sigma}(\mathcal{K}_{{n_{0}},1,2}^{(\bar{\leel})}) \tau_{Z^{(\sigma)}}(d_{0}) \exp (2t{n_{0}} \trunc_{0})\,.
\end{align*}
Since the random variables $S_{\trunc_{0}}(\mathcal{K}_{n_{0}}^{(\bar{\leel})})$ and $S_{\trunc_{0}}(\mathcal{K}_{{n_{0}},1,i}^{(\bar{\leel})})$ are centered, their moment generating functions are not smaller than one.
Hence applying the elementary inequality
\begin{equation}
\abs{\log x-\log y} \le \abs{x-y}\text{ for }x\ge 1\text{ and }y\ge 1,
\label{inelog}
\end{equation}
we get that, for any positive $t$,
\begin{align*}
    & \Bigabs{\log \mathbb{E}\exp \big (tS_{\trunc_{0}}(\mathcal{K}_{n_{0}}^{(\bar{\leel})})\big)-\sum_{i=1}^{2}\log \mathbb{E}\exp \big (tS_{\trunc_{0}}(\mathcal{K}_{{n_{0}},1,i}^{(\bar{\leel})})\big)} \\
    \le & t S_{\sigma}(\mathcal{K}_{{n_{0}},1,2}^{(\bar{\leel})}) \tau_{Z^{(\sigma)}}(d_{0})\exp (2t{n_{0}} \trunc_{0}) \le \frac{1}{2} t n_{0} \tau_{Z^{(\sigma)}}(d_{0})\exp (2t{n_{0}} \trunc_{0})\,.
\end{align*}

The next step is to compare $\mathbb{E}\exp \big (tS_{\trunc_{0}}(\mathcal{K}_{{n_{0}},1,i}^{(\bar{\leel})})\big)$ with $\mathbb{E}\exp \big (tS_{\trunc_{1}}(\mathcal{K}_{{n_{0}},1,i}^{(\bar{\leel})})\big)$ for $i=1,2$.
The random variables $S_{\trunc_{0}}(\mathcal{K}_{{n_{0}},1,i}^{(\bar{\leel})})$ and $S_{\trunc_{1}}(\mathcal{K}_{{n_{0}},1,i}^{(\bar{\leel})})$ have absolute values bounded by $n_0 \trunc_0$, hence applying the inequality
\begin{equation*}
|e^{tx}-e^{ty}|\le |t||x-y|(e^{|tx|}\vee e^{|ty|})\,,
\end{equation*}
we obtain that, for any positive $t$,
\begin{align*}
    & \big |\mathbb{E}\exp \big (tS_{\trunc_{0}}(\mathcal{K}_{{n_{0}},1,i}^{(\bar{\leel})})\big)-\mathbb{E}\exp \big (tS_{\trunc_{1}}(\mathcal{K}_{{n_{0}},1,i}^{(\bar{\leel})})\big)\big | \\
    \le & t\exp(2t S_{\sigma}(\mathcal{K}_{{n_{0}},1,i}^{(\bar{\leel})}) \trunc_0)\mathbb{E}\big |S_{\trunc_{0}}(\mathcal{K}_{{n_{0}},1,i}^{(\bar{\leel})})-S_{\trunc_{1}}(\mathcal{K}_{{n_{0}},1,i}^{(\bar{\leel})})\big |\,. \\
    \le & t\exp(t n_0 \trunc_0)\mathbb{E}\big |S_{\trunc_{0}}(\mathcal{K}_{{n_{0}},1,i}^{(\bar{\leel})})-S_{\trunc_{1}}(\mathcal{K}_{{n_{0}},1,i}^{(\bar{\leel})})\big |\,. \\
\end{align*}
Notice that
\begin{equation*}
\mathbb{E}\big |S_{\trunc_{0}}(\mathcal{K}_{{n_{0}},1,i}^{(\bar{\leel})})-S_{\trunc_{1}}(\mathcal{K}_{{n_{0}},1,i}^{(\bar{\leel})})\big |\le 2\sum_{j\in \mathcal{K}_{{n_{0}},1,i}^{(\bar{\leel})}}m_j \mathbb{E}|(\varphi_{\trunc_{0}}-\varphi_{\trunc_{1}})(Z_{j})|\,.
\end{equation*}
Since $|(\varphi_{\trunc_{0}}-\varphi_{\trunc_{1}})(x)|\le \trunc_{0} 1_{\set{|x|>\trunc_{1}}}$ for all $x\in {\mathbb{R}}$, by Eq.~\eqref{defTruncj}, we get that
\begin{equation*}
\mathbb{E}|(\varphi_{\trunc_{0}}-\varphi_{\trunc_{1}})(Z_{j})|\le \trunc_{0}\mathbb{P}(|Z_{j}|>\trunc_{1})\le \trunc_{0} \tau_{Z^{(\sigma)}}(c^{-\frac{1}{\kappa_{1}}}{n_{0}}2^{-(\bar{\leel} \wedge \frac{1}{\kappa})})\,.
\end{equation*}
For any $i=1,2$ and any positive $t$,
\begin{align*}
    & \big |\mathbb{E}\exp \big (tS_{\trunc_{0}}(\mathcal{K}_{{n_{0}},1,i}^{(\bar{\leel})})\big)-\mathbb{E}\exp \big (tS_{\trunc_{1}}(\mathcal{K}_{{n_{0}},1,i}^{(\bar{\leel})})\big)\big |\\
    \le & t n_{0} \trunc_{0} \exp(t{n_{0}} \trunc_{0})\tau_{Z^{(\sigma)}}(c^{-\frac{1}{\kappa_{1}}}{n_{0}}2^{-(\bar{\leel} \wedge \frac{1}{\kappa})})\\
    \le & \exp(2t{n_{0}} \trunc_{0})\tau_{Z^{(\sigma)}}(c^{-\frac{1}{\kappa_{1}}}{n_{0}}2^{-(\bar{\leel} \wedge \frac{1}{\kappa})})
\end{align*}
Using again the fact that the variables are centered, and taking Eq.(\ref{inelog}) into account, we derive that for any $i=1,2$ and any positive $t$,
\begin{align*}
    & \big |\log \mathbb{E}\exp \big (tS_{\trunc_{0}}(\mathcal{K}_{{n_{0}},1,i}^{(\bar{\leel})})\big)-\log \mathbb{E}\exp \big (tS_{\trunc_{1}}(\mathcal{K}_{{n_{0}},1,i}^{(\bar{\leel})})\big)\big |\\
    \le & \exp(2 t{n_{0}} \trunc_{0})\tau_{Z^{(\sigma)}}(c^{-\frac{1}{\kappa_{1}}}{n_{0}}2^{-(\bar{\leel} \wedge \frac{1}{\kappa})})\,.
\end{align*}

Now for any $\leel=1,\dots ,\bar{\leel}$ and any $i=1,\dots ,2^{\leel}$, $\size{\mathcal{K}_{{n_{0}},\leel,i}^{(\bar{\leel})}}\le 2^{-\leel}{n_{0}}$.
By iterating the above procedure and summing across all $2^{\leel-1}$ branches, we get for any $\leel=1,\dots ,\bar{\leel}$, and any positive $t$,
\begin{align*}
& \biggabs{\sum_{i=1}^{2^{\leel-1}}\log \mathbb{E}\exp \big (tS_{\trunc_{\leel-1}}(\mathcal{K}_{{n_{0}},\leel-1,i}^{(\bar{\leel})})\big) - \sum_{i=1}^{2^{\leel}}\log \mathbb{E}\exp \big (tS_{\trunc_{\leel-1}}(\mathcal{K}_{{n_{0}},\leel,i}^{(\bar{\leel})})\big)} \\
\le & t \sum_{i=1}^{2^{\leel-1}} S_{\sigma}(\mathcal{K}_{{n_{0}},\leel,2i}^{(\bar{\leel})}) \tau_{Z^{(\sigma)}}(d_{\leel-1}) \exp\Bigl( \frac{t{n_{0}} \trunc_{\leel-1}}{2^{\leel-2}} \Bigr) \\
\le & t \sum_{i=1}^{2^{\leel-1}} \frac{n_{0}}{2^\leel} \tau_{Z^{(\sigma)}}(d_{\leel-1})\exp\Bigl( \frac{t{n_{0}} \trunc_{\leel-1}}{2^{\leel-2}} \Bigr) \le \frac{1}{2} t n_{0} \tau_{Z^{(\sigma)}}(d_{\leel-1})
\exp\Bigl( \frac{t{n_{0}} \trunc_{\leel-1}}{2^{\leel-2}} \Bigr)\,,
\end{align*}
and for the corresponding truncation components, summing over all $2^{\leel}$ branches,
\begin{align*}
& \sum_{i=1}^{2^\leel} \bigabs{\log \mathbb{E}\exp \big (tS_{\trunc_{\leel-1}}(\mathcal{K}_{{n_{0}},\leel,i}^{(\bar{\leel})})\big)-\log \mathbb{E}\exp \big (tS_{\trunc_{\leel}}(\mathcal{K}_{{n_{0}},\leel,i}^{(\bar{\leel})})\big)} \\
\le & \sum_{i=1}^{2^\leel} \frac{t n_{0} \trunc_{\leel-1}}{2^{\leel-1}}\tau_{Z^{(\sigma)}}(c^{-\frac{1}{\kappa_{1}}}{n_{0}}2^{-(\bar{\leel} \wedge \frac{\leel}{\kappa})}) \exp\Bigl( \frac{t{n_{0}} \trunc_{\leel-1}}{2^{\leel-1}} \Bigr) \\
\le & 2^\leel \tau_{Z^{(\sigma)}}(c^{-\frac{1}{\kappa_{1}}}{n_{0}}2^{-(\bar{\leel} \wedge \frac{\leel}{\kappa})}) \exp\Bigl( \frac{t{n_{0}} \trunc_{\leel-1}}{2^{\leel-2}} \Bigr)\,.
\end{align*}
Hence finally, we get that for any $\leel=1,\dots ,\bar{\leel}$, and any positive $t$,
\begin{align*}
&\biggabs{\sum_{i=1}^{2^{\leel-1}}\log \mathbb{E}\exp \bigl(tS_{\trunc_{\leel-1}}(\mathcal{K}_{{n_{0}},\leel-1,i}^{(\bar{\leel})}) \bigr) - \sum_{i=1}^{2^{\leel}}\log \mathbb{E}\exp \bigl(tS_{\trunc_{\leel}}(\mathcal{K}_{{n_{0}},\leel,i}^{(\bar{\leel})})\bigr)} \\
\le & \Bigl\{\frac{1}{2} t n_{0} \tau_{Z^{(\sigma)}}(d_{\leel-1}) + 2^{\leel} \tau_{Z^{(\sigma)}}(c^{-\frac{1}{\kappa_{1}}}{n_{0}}2^{-(\bar{\leel} \wedge \frac{\leel}{\kappa})}) \Bigr\} \exp\Bigl(\frac{t{n_{0}} \trunc_{\leel-1}}{2^{\leel-2}}\Bigr),
\end{align*}
Set $\tilde{\leel}=\sup \{\leel\in \mathbb{N}\,,\,\leel/(\kappa) <\bar{\leel} \}\,$. Notice that $0\le \tilde{\leel}\le \bar{\leel} -1$.
Since $\mathcal{K}_{n_{0}}^{(\bar{\leel})}=\mathcal{K}_{{n_{0}},0,1}^{(\bar{\leel})}$, aggregating sequentially yields:
\begin{align}
&\Bigabs{\log \mathbb{E}\exp \bigl(tS_{\trunc_{0}}(\mathcal{K}_{n_{0}}^{(\bar{\leel})})\bigr)-\sum_{i=1}^{2^{\tilde{\leel}+1}}\log \mathbb{E}\exp \bigl(tS_{\trunc_{\tilde{\leel}+1}}(\mathcal{K}_{{n_{0}},\tilde{\leel}+1,i}^{(\bar{\leel})})\bigr)}
\notag  \label{1step} \\
\le & \frac{t n_{0}}{2} \sum_{\leel=0}^{\tilde{\leel}}\tau_{Z^{(\sigma)}}(d_{\leel})\exp \Bigl( \frac{t{n_{0}} \trunc_{\leel}}{2^{\leel-1}} \Bigr) + \sum_{\leel=0}^{\tilde{\leel}-1} 2^{\leel+1} \tau_{Z^{(\sigma)}}(c^{-\frac{1}{\kappa_2}}{n_{0}}2^{-\frac{\leel+1}{\kappa}})\exp \Bigl(\frac{t{n_{0}} \trunc_{\leel}}{2^{\leel-1}}\Bigr)  \notag \\
& \quad + 2^{\tilde{\leel} + 1} \tau_{Z^{(\sigma)}}(c^{-\frac{1}{\kappa_2}}{n_{0}}2^{-\bar{\leel}})\exp\Bigl(\frac{t{n_{0}} \trunc_{\tilde{\leel}}}{2^{\tilde{\leel}-1}}\Bigr)\,.
\end{align}

Notice now that for any $i=1,\dots ,2^{\tilde{\leel}+1}$, $S_{\trunc_{\tilde{\leel}+1}}(\mathcal{K}_{{n_{0}},\tilde{\leel}+1,i}^{(\bar{\leel})})$ is a sum of $2^{\bar{\leel} -\tilde{\leel}-1}$ blocks $\{I_{\bar{\leel}, j}\}_{j=1}^{2^{\bar{\leel} - \tilde{\leel} - 1}}$, each of size $n_{\bar{\leel}}$ and bounded by $2 \trunc_{\tilde{\leel}+1}n_{\bar{\leel}}$.
In addition the blocks are equidistant and there is a gap of size $d_{\tilde{\leel}+1}$ between two blocks.
Consequently, by using Lemma \ref{lemma:mix_ineq_mgf} along with Inequality (\ref{inelog}) and the fact that the variables are centered, we get that
\begin{align*}
&\Bigabs{\log \mathbb{E}\exp \big (tS_{\trunc_{\tilde{\leel}+1}}(\mathcal{K}_{{n_{0}},\tilde{\leel}+1,i}^{(\bar{\leel})})\big) - \sum_{j=(i-1)2^{\bar{\leel} -\tilde{\leel}-1}+1}^{i2^{\bar{\leel} -\tilde{\leel}-1}}\log \mathbb{E}\exp \bigl(tS_{\trunc_{\tilde{\leel}+1}}(I_{\bar{\leel} ,j})\bigr)} \\
\le & t S_{\sigma}(\mathcal{K}_{{n_{0}},\tilde{\leel}+1,i}^{(\bar{\leel})}) \tau_{Z^{(\sigma)}}(d_{\tilde{\leel}+1}) \exp (t \trunc_{\tilde{\leel}+1}n_{\bar{\leel}}2^{\bar{\leel} -\tilde{\leel}})\\
\le & 2^{-\tilde{\leel} - 1} t n_{0} \tau_{Z^{(\sigma)}}(d_{\tilde{\leel}+1})\exp (t \trunc_{\tilde{\leel}+1}n_{\bar{\leel}}2^{\bar{\leel} -\tilde{\leel}})\,.
\end{align*}
Summing this over $i=1, \dots, 2^{\tilde{\leel}+1}$, the linear coefficients naturally aggregate to reconstruct the global sum $S_{\sigma}$:
\begin{equation}\label{2step}
    \begin{aligned}
        & \sum_{i=1}^{2^{\tilde{\leel}+1}} \Bigabs{\log \mathbb{E}\exp \big (tS_{\trunc_{\tilde{\leel}+1}}(\mathcal{K}_{{n_{0}},\tilde{\leel}+1,i}^{(\bar{\leel})})\big)-\sum_{j=(i-1)2^{\bar{\leel} -\tilde{\leel}-1}+1}^{i2^{\bar{\leel} -\tilde{\leel}-1}}\log \mathbb{E}\exp \bigl(tS_{\trunc_{\tilde{\leel}+1}}(I_{\bar{\leel} ,j})\bigr)} \\
        \le & \sum_{i=1}^{2^{\tilde{\leel}+1}} 2^{-\tilde{\leel} - 1} t n_{0} \tau_{Z^{(\sigma)}}(d_{\tilde{\leel}+1})\exp (t \trunc_{\tilde{\leel}+1}n_{\bar{\leel}}2^{\bar{\leel} -\tilde{\leel}})\\
        \le & t n_{0} \tau_{Z^{(\sigma)}}(d_{\tilde{\leel}+1})\exp (t \trunc_{\tilde{\leel}+1}n_{\bar{\leel}}2^{\bar{\leel} -\tilde{\leel}}).
    \end{aligned}
\end{equation}

Starting from (\ref{1step}) and using (\ref{2step}) together with the fact that $n_{\bar{\leel}}\le {n_{0}}2^{-\bar{\leel}}$, we obtain:
\begin{equation}\label{3step}
    \begin{aligned}
&\Bigabs{\log \mathbb{E}\exp \bigl(tS_{\trunc_{0}}(\mathcal{K}_{n_{0}}^{(\bar{\leel})})\bigr)-\sum_{j=1}^{2^{\bar{\leel}}}\log \mathbb{E}\exp \bigl(tS_{\trunc_{\tilde{\leel}+1}}(I_{\bar{\leel} ,j})\bigr)} \\
\le & \frac{t n_{0}}{2} \sum_{\leel=0}^{\tilde{\leel}}\tau_{Z^{(\sigma)}}(d_{\leel})\exp\Bigl( \frac{t{n_{0}} \trunc_{\leel}}{2^{\leel-1}} \Bigr)+ \sum_{\leel=0}^{\tilde{\leel}-1} 2^{\leel+1} \tau_{Z^{(\sigma)}}(c^{-\frac{1}{\kappa_2}}{n_{0}}2^{-\frac{\leel+1}{\kappa}})
\exp\Bigl( \frac{t{n_{0}} \trunc_{\leel}}{2^{\leel-1}} \Bigr) \\
& + 2^{\tilde{\leel} + 1} \tau_{Z^{(\sigma)}}(c^{-\frac{1}{\kappa_2}}{n_{0}}2^{-\bar{\leel}})\exp\Bigl(\frac{t{n_{0}} \trunc_{\tilde{\leel}}}{2^{\tilde{\leel}-1}}\Bigr) + t n_{0} \tau_{Z^{(\sigma)}}(d_{\tilde{\leel}+1})\exp (t \trunc_{\tilde{\leel}+1}n_{\bar{\leel}}2^{\bar{\leel} -\tilde{\leel}})\,.
\end{aligned}
\end{equation}

Notice that by Eq.\eqref{equ:def_d\leel}, for any $\leel=0,\dots ,\bar{\leel} -1$, we have $d_{\leel}+1\ge \lfloor c_{0}{n_{0}}2^{-(\bar{\leel} \wedge \frac{\leel}{\kappa})} \rfloor$ and $c_{0}{n_{0}}2^{-(\bar{\leel} \wedge \frac{\leel}{\kappa})}\ge 2$.
Whence $d_{\leel}\ge c_{0}{n_{0}}2^{-(\bar{\leel} \wedge \frac{\leel}{\kappa})-2}$.
Consequently setting $c_{1}=\min (\frac{1}{4}c^{\frac{1}{\kappa_{1}}}c_{0},2^{-\frac{1}{\kappa}}) < 1$ and using the sub-Weibull tail condition, we derive that for any positive $t$,
\begin{align*}
&\Bigabs{\log \mathbb{E}\exp \bigl(tS_{\trunc_{0}}(\mathcal{K}_{n_{0}}^{(\bar{\leel})})\bigr)-\sum_{j=1}^{2^{\bar{\leel}}}\log \mathbb{E}\exp \bigl(tS_{\trunc_{\tilde{\leel}+1}}(I_{\bar{\leel} ,j})\bigr)} \\
&\le \frac{t n_{0}}{2} \sum_{\leel=0}^{\tilde{\leel}}\exp \Big(-\Bigl(\frac{c_{1}{n_{0}}}{2^{\frac{\leel}{\kappa}}}\Bigr)^{\kappa_{1}}+\frac{t{n_{0}} \trunc_{\leel}}{2^{\leel-1}}\Big)+2\sum_{\leel=0}^{\tilde{\leel}-1}2^{\leel}\exp \Big(-\Bigl(\frac{c_{1}{n_{0}}}{2^{\frac{\leel}{\kappa}}}\Bigr)^{\kappa_{1}}+\frac{t{n_{0}} \trunc_{\leel}}{2^{\leel-1}}\Big) \\
&\quad \quad +2^{\tilde{\leel}+1}\exp \Big(-\Bigl(\frac{{n_{0}}}{2^{\bar{\leel}}}\Bigr)^{\kappa_{1}} + \frac{t{n_{0}} \trunc_{\tilde{\leel}}}{2^{\tilde{\leel}-1}}\Big)+t n_{0} \exp \Big(-\Bigl(\frac{c_{1}{n_{0}}}{2^{\bar{\leel}}}\Bigr)^{\kappa_{1}}+\frac{t{n_{0}} \trunc_{\tilde{\leel}+1}}{2^{\tilde{\leel}}}\Big)\,.
\end{align*}

By Eq.\eqref{defTruncj}, $\trunc_{\leel}\le \big (2{n_{0}}2^{-(\bar{\leel} \wedge \frac{\leel}{\kappa})}\big)^{\frac{\kappa_{1}}{\kappa_{2}}}$. By the definition of $\tilde{\leel}$, we have $\frac{\leel}{\kappa} < \bar{\leel}$ for $0\le \leel \le \tilde{\leel}$.
Notice the exponent identity $1 + \frac{\kappa_2}{\kappa_1} = \frac{\kappa_2}{\kappa}$. Since $\frac{1}{\kappa} \ge 1$, we analytically have $1 + \frac{\kappa_2}{(\kappa)\kappa_1} \ge \frac{\kappa_2}{\kappa}$.
It rigorously follows that for $0\le \leel \le \tilde{\leel}$,
\begin{equation*}
{n_{0}} \trunc_{\leel}2^{-\leel+1} \le 2^{1+\frac{\kappa_2}{\kappa_1}} {n_{0}}^{\frac{\kappa_2}{\kappa}} 2^{-\leel \bigl(1 + \frac{\kappa_2}{(\kappa)\kappa_1}\bigr)} \le 2^{\frac{\kappa_{1}}{\kappa}}(2^{-\leel}{n_{0}})^{\frac{\kappa_{1}}{\kappa}}\,.
\end{equation*}
In addition, since $\tilde{\leel}+1\ge (\kappa) \bar{\leel}$, we get that
\begin{equation*}
\trunc_{\tilde{\leel}+1}\le (2{n_{0}}2^{-\bar{\leel}})^{\frac{\kappa_{1}}{\kappa_{2}}}\le (2{n_{0}}2^{-(\kappa) \bar{\leel}})^{\frac{\kappa_{1}}{\kappa_{2}}}\,.
\end{equation*}
Whence, using the relation $-\tilde{\leel}-1 \le -(\kappa)\bar{\leel}$,
\begin{equation*}
\trunc_{\tilde{\leel}+1}{n_{0}}2^{-\tilde{\leel}}=2 \trunc_{\tilde{\leel}+1}{n_{0}}2^{-(\tilde{\leel}+1)}\le 2^{\frac{\kappa_{1}}{\kappa}}{n_{0}}^{\frac{\kappa_{1}}{\kappa}}2^{-\frac{\kappa_{1}}{\kappa}\bar{\leel}(\kappa)}\,.
\end{equation*}
In addition, applying the bound for $\leel \le \tilde{\leel}$ established above,
\begin{equation*}
{n_{0}} \trunc_{\tilde{\leel}}2^{-\tilde{\leel}+1}\le 2^{\frac{\kappa_{1}}{\kappa}} ({n_0}2^{-\tilde{\leel}})^{\frac{\kappa_2}{\kappa}} = 2^{\frac{2\kappa_{1}}{\kappa}}({n_{0}}2^{-\tilde{\leel}-1})^{\frac{\kappa_{1}}{\kappa}}\le 2^{\frac{2\kappa_{1}}{\kappa}}{n_{0}}^{\frac{\kappa_{1}}{\kappa}}2^{-\frac{\kappa_{1}}{\kappa}\bar{\leel}(\kappa)}\,.
\end{equation*}
Hence, if $t\le c_{2}{n_{0}}^{\kappa_{1}(\kappa -1)/\kappa} 2^{-\bar{\leel} \kappa_2 (\frac{1}{\kappa} - \frac{1}{\kappa})}$ with $c_{2} = 2^{-(1+\frac{2\kappa_{1}}{\kappa})}c_{1}^{\kappa_{1}}$, we derive that
\begin{align*}
&\Bigabs{\log \mathbb{E}\exp \big (tS_{\trunc_{0}}(\mathcal{K}_{n_{0}}^{(\bar{\leel})})\big)-\sum_{j=1}^{2^{\bar{\leel}}}\log \mathbb{E}\exp \big (tS_{\trunc_{\tilde{\leel}+1}}(I_{\bar{\leel} ,j})\big)} \\
&\le \frac{t n_{0}}{2} \sum_{\leel=0}^{\tilde{\leel}}\exp \Big(-\frac{1}{2}\Bigl(\frac{c_{1}{n_{0}}}{2^{\frac{\leel}{\kappa}}}\Bigr)^{\kappa_{1}}\Big)+2\sum_{\leel=0}^{\tilde{\leel}-1}2^{\leel}\exp \Big(-\frac{1}{2}\Bigl(\frac{c_{1}{n_{0}}}{2^{\frac{\leel}{\kappa}}}\Bigr)^{\kappa_{1}}\Big) \\
&\quad \quad +(2^{\tilde{\leel}+1}+t n_{0})\exp \Bigl(-\frac{1}{2}\Bigl(\frac{c_{1}{n_{0}}}{2^{\bar{\leel}}}\Bigr)^{\kappa_{1}}\Bigr)\,.
\end{align*}
Since $2^{\tilde{\leel}}\le 2^{\bar{\leel} (\kappa)}\le {n_{0}}^{\kappa} \le n_0^\kappa$, it follows that for $0 < t\le c_{2}{n_{0}}^{\kappa_{1}(\kappa -1)/\kappa} 2^{-\bar{\leel} \kappa_2 (\frac{1}{\kappa} - \frac{1}{\kappa})}$,
\begin{equation}\label{equ:4step}
\begin{aligned}
    & \Bigabs{\log \mathbb{E}\exp \bigl(tS_{\trunc_{0}}(\mathcal{K}_{n_{0}}^{(\bar{\leel})})\bigr)-\sum_{j=1}^{2^{\bar{\leel}}}\log \mathbb{E}\exp \bigl(tS_{\trunc_{\tilde{\leel}+1}}(I_{\bar{\leel} ,j})\bigr)}\\
    \le & (2\bar{\leel} t n_{0} + 4{n_{0}}^{\kappa})\exp \Bigl(-\frac{1}{2}\Bigl(\frac{c_{1}{n_{0}}}{2^{\bar{\leel}}}\Bigr)^{\kappa_{1}}\Bigr)\,.
\end{aligned}
\end{equation}

Finally, we bound the log moment generating function of each $S_{\trunc_{\tilde{\leel}+1}} (I_{\bar{\leel} , j})$. Notice that the function $x\mapsto g (x)=x^{-2}(e^x - x - 1)$ is increasing on ${\mathbb{R}}$.
Hence, for any centered random variable $U$ such that $\|U\|_{\infty} \le \trunc$, and any positive $t$,
\begin{equation}\label{psi}
\mathbb{E} \exp ( t U) \le 1 + t^2 g(t\trunc) \mathbb{E} (U^2) \, .
\end{equation}
We distinguish two cases for the range of $t$.

\noindent\textbf{Case 1: $t \le (2 {n_{0}})^{-\kappa_2/\kappa}$.}
In this case, we apply inequality (\ref{psi}) to the whole sum over $\mathcal{K}_{n_0}^{(\bar{\leel})}$.
Notice that since $\bar{m} = 1$, we have $\| S_{\trunc_0} (\mathcal{K}_{n_{0}}^{(\bar{\leel})}) \|_{\infty} \le 2 \trunc_0 {n_{0}} \le 2^{\kappa_2/\kappa}{n_{0}}^{\kappa_2/\kappa}$.
Hence $t \| S_{\trunc_0} (\mathcal{K}_{n_{0}}^{(\bar{\leel})}) \|_{\infty} \le 1$.
Utilizing the behavior of $g(x)$, noting that $g(1) = e - 2 < 1$ and $\log(1+x)\le x$ for $x > 0$, we derive that
\begin{equation}  \label{6step}
\log \mathbb{E} \exp \big (t S_{\trunc_0} (\mathcal{K}_{n_{0}}^{( \bar{\leel})}) \big) \le t^2 \Var \big(S_{\trunc_0} (\mathcal{K}_{n_{0}}^{( \bar{\leel})})\big) \le t^2 \mathcal{V}_{\mathcal{K}_{n_{0}}^{(\bar{\leel})}} \, ,
\end{equation}
which directly proves (\ref{resultpropinter2}).

\noindent\textbf{Case 2:} $(2 {n_{0}})^{-\kappa_2/\kappa} < t \le c_{3} ( {n_{0}}^{ \kappa_2(\kappa -1)/\kappa} \wedge ( 2^{ \bar{\leel}} / {n_{0}})^{\kappa_2/\kappa} )$ with $c_{3} = c_{2} \wedge 2^{-\frac{\kappa_{1}}{\kappa}}$

Here we must use the decoupled sum (\ref{equ:4step}).
Notice that $\| S_{\trunc_{\tilde{\leel}+1}} (I_{\bar{\leel} , j}) \|_{\infty} \le 2 \trunc_{\tilde{\leel}+1} n_{\bar{\leel}}$.
Since $\tilde{\leel}+1 \ge (\kappa) \bar{\leel}$, the truncation level satisfies $\trunc_{\tilde{\leel}+1} \le (2n_0 2^{-\bar{\leel}})^{\kappa_2/\kappa_1}$. Combined with $n_{\bar{\leel}} \le n_0 2^{-\bar{\leel}}$, this yields $\| S_{\trunc_{\tilde{\leel}+1}} (I_{\bar{\leel} , j}) \|_{\infty} \le 2^{\kappa_2/\kappa} ({n_{0}}2^{-\bar{\leel}})^{\kappa_2/\kappa}$.
Crucially, the upper bound on $t$ guarantees that $t \le c_{3} (2^{\bar{\leel}}/ {n_{0}})^{\kappa_2/\kappa} $.
By ensuring $c_{3} \le 2^{-\kappa_2/\kappa}$, we have $t \| S_{\trunc_{\tilde{\leel}+1}} (I_{\bar{\leel} , j}) \|_{\infty} \le 1$.
Thus, using (\ref{psi}), we get
\begin{equation*}
\log \mathbb{E} \exp \big (t S_{\trunc_{\tilde{\leel}+1}} (I_{\bar{\leel} , j}) \big) \le t^2 \Var\big(S_{\trunc_{\tilde{\leel}+1}} (I_{\bar{\leel} , j})\big) \, .
\end{equation*}
Summing over all $2^{\bar{\leel}}$ blocks, the total variance is bounded by the local variance $\mathcal{V}_{\mathcal{K}_{n_{0}}^{(\bar{\leel})}}$.

To apply \eqref{equ:4step}, $t$ must formally satisfy $t \le c_{2}{n_{0}}^{\kappa_{1}(\kappa -1)/\kappa} 2^{-\bar{\leel} \kappa_2 (\frac{1}{\kappa} - \frac{1}{\kappa})}$.
Observe that for $\kappa < 1$, this condition matches $t \le c_{3} \bigl( {n_{0}}^{ \kappa_2(\kappa -1)/\kappa} \wedge ( 2^{ \bar{\leel}} / {n_{0}})^{\kappa_2/\kappa} \bigr) \le c_2 n_0^{\kappa_2(\kappa-1)/\kappa}$.
For $\kappa \ge 1$, this condition becomes $t \le c_2 (n_0 2^{-\bar{\leel}})^{\kappa_2(\kappa-1)/\kappa}$.
Since $n_0 2^{-\bar{\leel}} \ge 1$, we have $t \le c_{3} ( 2^{ \bar{\leel}} / {n_{0}})^{\kappa_2/\kappa} \le c_2 (n_0 2^{-\bar{\leel}})^{\kappa_2(\kappa-1)/\kappa}$.
Therefore, in either regime, the condition is implied by the assumption $t \le c_{3} (n_{0}^{\kappa-1}\wedge (2^{\bar{\leel}}/n_{0}))^{\frac{\kappa_{1}}{\kappa}}$ stated in the proposition, provided $c_{3}$ is chosen sufficiently small.
Thus, we obtain the intermediate bound:
\begin{equation}  \label{5step}
\log \mathbb{E} \exp \big (t S_{\trunc_{0}} (\mathcal{K}_{n_{0}}^{( \bar{\leel})}) \big) \le t^2 \mathcal{V}_{\mathcal{K}_{n_{0}}^{(\bar{\leel})}} + (2 \bar{\leel} t n_{0} + 4{n_{0}}^{\kappa}) \exp \Bigl(-\frac{1}{2}\Bigl(\frac{c_{1}{n_{0}}}{2^{\bar{\leel}}}\Bigr)^{\kappa_{1}}\Bigr) \, .
\end{equation}

To apply the standard exponential moment inequalities subsequently, the exponential decoupling error must be proportional to $t^2$.
Since we are in the regime $t > (2 {n_{0}})^{-\kappa_2/\kappa}$, we can extract a $t^2$ factor from the decoupling error by identically utilizing the lower bound inequalities $1 < t^2 (2{n_{0}})^{2\kappa_2/\kappa}$ and $t < t^2 (2{n_{0}})^{\kappa_2/\kappa}$.
Applying these universally to the error coefficients yields:
\begin{align*}
2 \bar{\leel} t n_{0} + 4{n_{0}}^{\kappa} &\le 2 \bar{\leel} n_{0} \bigl( t^2 (2{n_{0}})^{\frac{\kappa_2}{\kappa}} \bigr) + 4{n_{0}}^{\kappa} \bigl( t^2 (2{n_{0}})^{\frac{2\kappa_2}{\kappa}} \bigr) \\
&= t^2 \bigl( 2\bar{\leel} n_{0} (2{n_{0}})^{\frac{\kappa_2}{\kappa}} + 2^{2+\frac{2\kappa_2}{\kappa}}{n_{0}}^{\kappa+\frac{2\kappa_2}{\kappa}} \bigr).
\end{align*}

Substituting this back to \eqref{5step}, we completely absorb the linear and constant terms into a quadratic term:
\begin{equation*}
\log \mathbb{E} \exp \big (t S_{\trunc_{0}} (\mathcal{K}_{n_{0}}^{( \bar{\leel})}) \big) \le t^2 \mathcal{V}_{\mathcal{K}_{n_{0}}^{(\bar{\leel})}} + t^2 \bigl( 2^{1 + \frac{\kappa_2}{\kappa}} \bar{\leel} {n_{0}}^{1 + \frac{\kappa_2}{\kappa}} + 2^{2+\frac{2\kappa_2}{\kappa}}{n_{0}}^{\kappa+\frac{2\kappa_2}{\kappa}} \bigr) \exp \Bigl(-\frac{1}{2}\Bigl(\frac{c_{1}{n_{0}}}{2^{\bar{\leel}}}\Bigr)^{\kappa_{1}}\Bigr) \, .
\end{equation*}
This explicitly proves (\ref{resultpropinter2}) in the second case, preserving the structural bounds and completing the unified proof of Proposition \ref{propinter2}.

\end{proof}

Before stating the next proposition, we define several necessary constants to bound the moment generating function on $[1, {n_{0}}]$.
Let $c_{4}$ and $\mu$ be defined globally as:
\begin{equation}\label{borneA}
c_{4}=2^{\frac{\kappa_{1}}{\kappa}}3^{\frac{\kappa_{1}}{\kappa_{2}}}c_{0}^{-\frac{\kappa_{1}}{\kappa_{2}}}\,\mbox{and}\, \mu = \bigl(2(2\vee \frac{4}{c_{0}})/(1-\kappa)\bigr)^{\frac{2}{1-\kappa}}\,,
\end{equation}
and $\nu_{n_0}$ be a strict constant. Denote $\mathcal{V}_{n_0} = \mathcal{V}_{[n_0]}$ as Eq.~\eqref{def:effect_var} for short.

\begin{proposition}\label{propinter_block_unified}
Let $(X_{i})_{i\ge 1}$ be a sequence of centered real valued random variables satisfying Condition~\ref{cond:subweibull_fdm} and $0 < \kappa < 1$.
Let $n_{0}$ be an integer.
Let $\trunc=H^{-1}(\tau_{Z^{(\sigma)}}(c^{-\frac{1}{\kappa_2}}n_{0}))$ and for any $i$, set $X_{\trunc, i} = m_i \varphi_{\trunc}(Z_{i})-\mathbb{E}[m_i \varphi_{\trunc}(Z_{i})]$.
Then, if $n_{0}\ge \mu $, for any positive $t<\nu_{n_0}^{-1} \bar{m}^{-1} n_{0}^{\kappa_{1}((\kappa)-1)/\kappa}$, we get that
\begin{equation}\label{resultpropinter_unified}
\log \mathbb{E}\Bigl(\exp \Bigl(t\sum_{i=1}^{n_{0}}X_{\trunc, i}\Bigr)\Bigr)\le \frac{t^{2}\mathcal{V}_{n_{0}}'(\trunc)}{1-t\nu_{n_0} \bar{m} n_{0}^{\kappa_{1}(1-(\kappa))/\kappa}}\,,
\end{equation}
where $\mathcal{V}_{n_{0}}'(\trunc)=50\mathcal{V}_{n_{0}}+c_{5} \bar{m}^2 \exp \bigl(-c_{6}n_{0}^{\kappa_{1}(1-\kappa)}(\log n_{0})^{-\kappa}\bigr)$
and $c_{5}$, $c_{6}$ are positive constants depending only on $c$, $\kappa$ and $\kappa_{1}$.
\end{proposition}

\begin{proof}[Proof of Proposition \ref{propinter_block_unified}]
Without loss of generality, assume that $\bar{m} = 1$.
The original strategy in \cite{merlevede2011weakly} recursively constructs Cantor sets, applies Proposition~\ref{propinter2} to each constructed set, and passes the excluded samples to the next recursive depth.
However, in the heteroskedastic setting, it is unrealistic to assume that the average variance proxies are well-balanced across all these Cantor trees.
It will make the final variance bound become large.
To solve it, we will first describe the strategy about the introduction of random shifts into the recursively construction of the Cantor sets.

Let $\bar{\leel}$ be a fixed positive integer which satisfies
\begin{equation}\label{rest1l}
n_{0}2^{-\bar{\leel}}\ge (2\vee \frac{4}{c_{0}})\,.
\end{equation}
For any depth $j \ge 0$, let $n_j$ denote the deterministic size of the available sample pool.
Initially for $j=0$, $n_0$ is the original sample size and $X_{i}^{(0)}=X_{i}$ for $i=1,\dots ,n_{0}$.
We explicitly define the local structural buffer size as $b_j = \lceil c_0 n_j \rceil$.
To ensure the terminal block size satisfies the prerequisite of Proposition \ref{propinter2} unconditionally across all depths, we define the tree depth parameter $\bar{\leel}_{j}$ uniformly as:
\begin{equation}
\bar{\leel}_{j}=\inf \bigl\{h\in \mathbb{N} : (n_j - b_j) 2^{-h} \le n_{0}2^{-\bar{\leel}}\bigr\}\,. \label{restli}
\end{equation}
Notice that for any $j \ge 0$, by this infimum definition and the initial constraint \eqref{rest1l}, the value $\bar{\leel}_j - 1$ violates the inequality, which guarantees:
\begin{equation*}
(n_{j} - b_j) 2^{-\bar{\leel}_j} = \frac{(n_{j} - b_j) 2^{-(\bar{\leel}_j - 1)}}{2} > \frac{n_0 2^{-\bar{\leel}}}{2} \ge \Bigl(1 \vee \frac{2}{c_0}\Bigr) \,.
\end{equation*}
This strict algebraic bound mathematically guarantees that the elementary blocks at any depth safely satisfy the prerequisite of Proposition \ref{propinter2} on the active base of length $n_j - b_j$.
We construct the standard Cantor template $\mathcal{K}_j^{\text{std}} = \mathcal{K}_{n_j - b_j}^{(\bar{\leel}_j)}$ within $\{1, \dots, n_j - b_j\}$, leaving the tail sequence $\{n_j - b_j + 1, \dots, n_j\}$ empty.
Applying the independent random shift $U_j \sim \mathrm{Unif}(\{0, \dots, n_j - 1\})$ cyclically modulo $n_j$ yields the selected set $\mathcal{K}_j(\set{U_h}_{h\le j})$.
Because the cyclic shift is a perfect bijection, the number of selected elements is deterministic: $|\mathcal{K}_j(\set{U_h}_{h\le j})| = |\mathcal{K}_j^{\text{std}}| \ge 3(n_j - b_j)/4 \ge n_j / 2$, based on the choice of $c_0 \le 1/4$ in Proposition~\ref{propinter2}.
Unselected elements pass into the next pool, $n_{j+1} = n_j - |\mathcal{K}_j^{\text{std}}|$, which guarantees exponential decay bounded by $n_{j+1} \le n_0 2^{-(j+1)}$.
We set the original variables passing to the next level as $X_{i}^{(j+1)}=X_{h_{i}}^{(j)}$ for $i=1,\dots ,n_{j+1}$, where $\{h_{1},\dots ,h_{n_{j+1}}\}=\{1,\dots ,n_{j}\}\setminus \mathcal{K}_j(\set{U_h}_{h\le j})$.

The cyclic shift modulo $n_j$ preserves all relative gaps, with $b_j \ge d_0$ acting as the boundary separation.
Although it may split at most one Cantor block, this yields $\le 2^{\bar{\leel}_j} + 1$ chronologically ordered blocks.
Iteratively applying the decoupling Lemma \ref{lemma:mix_ineq_mgf} along this natural filtration merely alters the polynomial prefix by an absolute constant, fully preserving the exponential tail bound in Proposition \ref{propinter2}.

We now show how to bound the total variance proxy using the probabilistic method.
We first evaluate the expected variance proxy under the joint distribution of the random shift sequence $\set{U_h}$. For any fixed original index $i \in \{1, \dots, n_0\}$, its inclusion in the $j$-th selected set $\mathcal{K}_j(\set{U_h}_{h\le j})$ requires it to survive unselected through all prior pools. Conditional on reaching pool $n_j$, the random uniform shift $U_j$ gives it exactly a $|\mathcal{K}_j^{\text{std}}| / n_j$ probability of selection. Unconditionally, because at least half the available elements are deterministically selected into the Cantor set at each depth, the marginal probability of index $i$ falling into depth $j$ is identical and decays exponentially: $\mathbb{P}(1 \in \mathcal{K}_j(\set{U_h}_{h\le j})) = \cdots = \mathbb{P}(n_0 \in \mathcal{K}_j(\set{U_h}_{h\le j})) \le 2^{-j}$.

To bound the expected variance proxy, we first decouple the subset dependence inside the supremum by enlarging the inner covariance sum to the full index set $[n_0]$.
For any subset $\mathcal{K}_j \subset [n_0]$, by the exponentially decay of the functional dependence measure, we bound the local effective variance by monotonic enlargement:
\begin{equation*}
\mathcal{V}_{\mathcal{K}_j} \le \sum_{i \in \mathcal{K}_j} \Bigl(\Var(X_i) + \sum_{k\in[n_0]\setminus\{i\}} |\mathrm{Cov}(X_i, X_k)| \Bigr) := \sum_{i \in \mathcal{K}_j} v_i^2 \,.
\end{equation*}
Notice that the proxy summand $\leel_i$ is now entirely independent of the selected subset $\mathcal{K}_j$, making the upper bound additive with respect to the index set.
By definition of the total effective variance, we have $\mathcal{V}_{n_0} = \sum_{i=1}^{n_0} v_i^2$.
By applying the linearity of expectation to this additive upper bound, we rigorously obtain:
\begin{equation*}
\mathbb{E} [\mathcal{V}_{\mathcal{K}_j(\set{U_h}_{h\le j})}] \le \mathbb{E} \biggl[ \sum_{i \in \mathcal{K}_j(\set{U_h}_{h\le j})}\sigma_{\trunc,i}^2 \biggr] = \sum_{i=1}^{n_0} \mathbb{P}(i \in \mathcal{K}_j(\set{U_h}_{h\le j}))\sigma_i^2 \le 2^{-j} \mathcal{V}_{n_0} \,.
\end{equation*}
Since the square root function is strictly concave, Jensen's inequality yields:
$$\mathbb{E} \bigl[\mathcal{V}_{\mathcal{K}_j(\set{U_h}_{h\le j})}^{1/2}\bigr] \le \bigl(\mathbb{E} [\mathcal{V}_{\mathcal{K}_j(\set{U_h}_{h\le j})}]\bigr)^{1/2} \le 2^{-j/2} \mathcal{V}_{n_0}^{1/2} \,.$$
Summing these expected local standard deviations over all structural depths limits the accumulation to a convergent geometric series, entirely bypassing any depth dependency:
$$\mathbb{E} \biggl[ \sum_j \mathcal{V}_{\mathcal{K}_j(\set{U_h}_{h\le j})}^{1/2} \biggr] \le \sum_{j=0}^{\infty} 2^{-j/2} \mathcal{V}_{n_0}^{1/2} = \frac{1}{1 - 1/\sqrt{2}}\mathcal{V}_{n_0}^{1/2} \le 4 \mathcal{V}_{n_0}^{1/2} \,.$$

By the probabilistic method, there exists a deterministic sequence of optimal shifts $\set{u_0, u_1, \dots}$ such that the realized sum satisfies $\sum_j \mathcal{V}_{\mathcal{K}_j(\set{u_h}_{h \le j})}^{1/2} \le 4 \mathcal{V}_{n_0}^{1/2}$.
Let $\mathcal{K}_j = \mathcal{K}_j(\set{u_h}_{h \le j})$ be the deterministically realized subsets.
Let
\begin{equation}
 {\leel_{0}'}=\inf \{j\in \mathbb{N}\,,\,n_{j}\le n_0 2^{-\bar{\leel}}\}\,. \label{defm(n0)}
\end{equation}
Note that $ {\leel_{0}'}\ge 1$ because $n_0 > n_0 2^{-\bar{\leel}}$ and $\bar{\leel}\ge 1$.
In addition, $ {\leel_{0}'}\le \bar{\leel}$ since for all $j\ge 1$, $n_{j}\le n_0 2^{-j}$ by the recursive definition of $n_j$.
For any $j=0,\dots , {\leel_{0}'}-1$, the sequences $\{X_{i}^{(j+1)}\}$ satisfy Condition~\ref{cond:subweibull_fdm} with the same constants.
Now we set $T_{0}=\trunc=H^{-1}(\tau_{Z^{(\sigma)}}(c^{-\frac{1}{\kappa_2}}n_{0}))$, and for any integer $j=0,\dots , {\leel_{0}'}$,
\begin{equation*}
T_{j}=H^{-1}(\tau_{Z^{(\sigma)}}(c^{-\frac{1}{\kappa_2}}n_{j}))\,.
\end{equation*}
Similarly to the proof of Proposition~\ref{propinter2}, we define for all integers $j$ and truncation parameter $T$,
\begin{equation*}
X_{T, i}^{(j)}
=m^{(j)}_i \varphi_{T}\big (Z_{i}^{(j)}
\big)-\mathbb{E}[m^{(j)}_i \varphi_{T}\big (Z_{i}^{(j)}
\big)]\,,
\end{equation*}
where $Z_{i}^{(j)}
= X_{i}^{(j)} / m^{(j)}_i$ and $m^{(j)}_i$ is the scale parameter of the sample $X_{i}^{(j)}$.
Notice that by Condition~\ref{cond:subweibull_fdm}, we have that for any integer $j\ge 0$,
\begin{equation}
T_{j}\le (2n_{j})^{\frac{\kappa_{1}}{\kappa_{2}}}\,. \label{boundTl}
\end{equation}
For any $j=0,\dots , {\leel_{0}'}-1$, define $Y_{j}=\sum_{i\in \mathcal{K}_j}X_{T_{j}, i}^{(j)}\,.$
For any $j=1,\dots , {\leel_{0}'}-1$, define $W_{j}=\sum_{i=1}^{n_{j}}(X_{T_{j-1}, i}^{(j)}-X_{T_{j}, i}^{(j)})\,.$
Define $R_{{\leel_{0}'}}=\sum_{i=1}^{n_{{\leel_{0}'}}}X_{T_{{\leel_{0}'}-1}, i}^{( {\leel_{0}'})}\,.$
The following decomposition holds:
\begin{equation}
\sum_{i=1}^{n_{0}}X_{T_{0}, i}^{(0)}
=\sum_{j=0}^{ {\leel_{0}'}-1}Y_{j}+\sum_{j=1}^{ {\leel_{0}'}-1}W_{j}+R_{{\leel_{0}'}}\,. \label{P3prop3}
\end{equation}
To control the terms in the decomposition \eqref{P3prop3} by Proposition~\ref{propinter2}, we need a lower bound of the sequence $\{n_j\}_{j=0}^{ {\leel_{0}'}}$.
Notice that for any $j$ in $[0,  {\leel_{0}'})$, we have $n_{j+1} \ge \lfloor c_0 n_{j} \rfloor - 1$.
Since $c_0 n_{j} \ge 2$, we derive that
\begin{equation}\label{equ:lowerbound_Nj}
    n_{j+1} \ge \lfloor c_0 n_{j} \rfloor - 1 \ge \frac{\lfloor c_0 n_{j} \rfloor + 1}{3} \ge \frac{c_0 n_{j}}{3}.
\end{equation}
Combining \eqref{boundTl} and \eqref{equ:lowerbound_Nj}, a useful consequence is that for any $j=1, \dots,  {\leel_{0}'}$,
\begin{equation} \label{secondboundTl}
2n_{j} T_{j-1} \le c_{4} n_{j}^{\frac{\kappa_{1}}{\kappa}}
\end{equation}
where $c_{4}$ is defined by \eqref{borneA}.
Since $\bar{m} = 1$, the random variable $|R_{{\leel_{0}'}}|$ is almost surely bounded by $2n_{{\leel_{0}'}} T_{{\leel_{0}'}-1}$.
By using \eqref{secondboundTl} and \eqref{defm(n0)}, we then derive that
\begin{equation}
\norm{R_{{\leel_{0}'}}}_{\infty}\le c_{4} (n_{{\leel_{0}'}})^{\frac{\kappa_{1}}{\kappa}}\le c_{4} \bigl(n_0 2^{-\bar{\leel}}\bigr)^{\frac{\kappa_{1}}{\kappa}}\,. \label{boundRl}
\end{equation}
Hence, if $t\le c_{4}^{-1} (2^{\bar{\leel}}/n_0)^{\frac{\kappa_{1}}{\kappa}}$, by using \eqref{psi} together with the local variance bounding, we obtain
\begin{equation}
\log \mathbb{E}\bigl(\exp ({\textstyle tR_{{\leel_{0}'}}})\bigr)\le t^{2}\mathcal{V}_{R} \,, \label{P5prop3}
\end{equation}
where $\mathcal{V}_{R}$ is the variance term defined in \eqref{def:effect_var} with the index set $\mathcal{K}$ be the collection of the original sample indices of the summands in $R_{{\leel_{0}'}}$.

Now, by Proposition \ref{propinter2}, we get that for any $j \in [0,  {\leel_{0}'})$ and any $t \le c_{3} \bigl( (n_{j}-b_j)^{\kappa-1} \wedge (2^{\bar{\leel}_j} / (n_j-b_j)) \bigr)^{\kappa_2/\kappa}$,
\begin{equation*}
\log \mathbb{E} \bigl( \exp(t Y_j) \bigr) \le t^2 \mathcal{V}_{\mathcal{K}_j^{'}} + t^2 \Bigl( \bar{\leel}_j (2n_{j})^{1+\frac{\kappa_2}{\kappa}} + 4n_{j}^\kappa (2n_{j})^{\frac{2\kappa_2}{\kappa}} \Bigr) \exp \Bigl( - \frac{1}{2} \Bigl( \frac{c_1 (n_j - b_j)}{2^{\bar{\leel}_j}} \Bigr)^{\kappa_2} \Bigr) \,,
\end{equation*}
where $\mathcal{K}_j^{'}$ is the original index set in $[n_0]$ of the samples $\set{X_{i}^{(j)}}_{i \in \mathcal{K}_j}$. Note that the polynomial coefficients are safely upper-bounded using $n_j \ge n_j - b_j$.
Notice now that $\bar{\leel}_j \le \bar{\leel} \le n_0$, $n_{j} \le n_0 2^{-j}$ and $2^{-\bar{\leel}-1} n_0 \le (n_{j}-b_j) 2^{-\bar{\leel}_j} \le n_0 2^{-\bar{\leel}}$.
Taking into account these bounds, the exponential tail aligns with the global target. We then get that for any $j$ in $[0 ,  {\leel_{0}'})$ and any $t \le c_{3} \bigl( (2^j/n_0)^{1-\kappa} \wedge (2^{\bar{\leel}}/n_0) \bigr)^{\kappa_2/\kappa}$,
\begin{equation} \label{P6prop3}
\log \mathbb{E} \bigl( \exp(t Y_j) \bigr) \le t^2 \Bigl\{ \mathcal{V}_{\mathcal{K}_j^{'}}^{1/2} + \Bigl( 2^{2+\frac{\kappa_2}{\kappa}} \frac{n_0^{1+\frac{\kappa_2}{\kappa}}}{(2^j)^{\frac{\kappa}{2}+\frac{\kappa_2}{2\kappa}}} \Bigr) \exp \Bigl( - \frac{c_1^{\kappa_2}}{2^{2+\kappa_2}} \Bigl( \frac{n_0}{2^{\bar{\leel}}} \Bigr)^{\kappa_2} \Bigr) \Bigr\}^2 := t^2 \mathcal{V}_{Y,j} \,.
\end{equation}
Notice first that for any $1\le j\le  {\leel_{0}'}-1$, $W_{j}$ is a centered random variable, such that
\begin{equation*}
|W_{j}|\le \sum_{i=1}^{n_{j}}\Bigl(m_{i}^{(j)}\bigabs{(\varphi_{T_{j-1}}-\varphi_{T_{j}})(Z_{i}^{(j)})} + \mathbb{E}\big[m_{i}^{(j)}\bigabs{(\varphi_{T_{j-1}}-\varphi_{T_{j}})(Z_{i}^{(j)})}\big]\Bigr)\,.
\end{equation*}
Consequently, using \eqref{secondboundTl} and $m^{(j)}_i \le \bar{m} = 1$, we get that
\begin{equation*}
\norm{W_{j}}_{\infty}\le 2n_{j} T_{j-1}\le c_{4} n_{j}^{\frac{\kappa_{1}}{\kappa}}\,.
\end{equation*}
In addition, since $\abs{(\varphi_{T_{j-1}}-\varphi_{T_{j}})(x)}\le (T_{j-1}-T_{j})1_{\set{\abs{x}>T_{j}}}$, and the random variables $\set{Z_{i}^{(j)}}$ satisfy the sub-Weibull tail condition, by the definition of $T_{j}$, we get that
\begin{equation*}
\mathbb{E}|W_{j}|^{2}\le (2n_{j} T_{j-1})^{2}\tau_{Z^{(\sigma)}}(c^{-\frac{1}{\kappa_2}}n_{j})\le c_{4}^{2} n_{j}^{\frac{2\kappa_{1}}{\kappa}}\,.
\end{equation*}
Hence applying \eqref{psi} to the random variable $W_{j}$, we get for any positive $t$,
\begin{equation*}
\mathbb{E}\exp (tW_{j})\le 1+t^{2}g(c_{4} t n_{j}^{\frac{\kappa_{1}}{\kappa}})c_{4}^{2} n_{j}^{\frac{2\kappa_{1}}{\kappa}}\exp (-n_{j}^{\kappa_{1}})\,.
\end{equation*}
Hence, since $n_{j}\le n_0 2^{-j}$, for any positive $t$ satisfying $t\le (2c_{4})^{-1} (2^{j}/n_0)^{\kappa_{1}(1-(\kappa))/\kappa}$, we have that
\begin{equation*}
c_{4} t n_{j}^{\frac{\kappa_{1}}{\kappa}} \le n_{j}^{\kappa_{1}}/2\,.
\end{equation*}
Since $g(x)\le e^{x}$ for $x\ge 0$, we infer that for any positive $t$ with $t\le (2c_{4})^{-1} (2^{j}/n_0)^{\kappa_{1}(1-(\kappa))/\kappa}$,
\begin{equation*}
\log \mathbb{E}\exp (tW_{j})\le c_{4}^{2} t^{2}(2^{-j}n_0)^{\frac{2\kappa_{1}}{\kappa}}\exp (-n_{j}^{\kappa_{1}}/2)\,.
\end{equation*}
By taking into account that for any $1\le j\le  {\leel_{0}'}-1$, $n_{j}\ge n_{{\leel_{0}'}-1}>n_0 2^{-\bar{\leel}}$ (by definition of $ {\leel_{0}'}$), it follows that for any $j$ in $[1, {\leel_{0}'})$ and any positive $t$ satisfying $t\le (2c_{4})^{-1}(2^{j}/n_0)^{\kappa_{1}(1-(\kappa))/\kappa}$,
\begin{equation}
\log \mathbb{E}\exp (tW_{j})\le t^{2} \bigl(c_{4}(2^{-j}n_0)^{\frac{\kappa_{1}}{\kappa}}\exp (-(n_0 2^{-\bar{\leel}})^{\kappa_{1}}/4)\bigr)^{2}:=t^{2}\mathcal{V}_{W,j}\,. \label{P7prop3}
\end{equation}
With the controls of $Y, U, R$, we can derive the bound in the proposition by applying Lemma~\ref{breta} with parameters:
\begin{equation}\label{equ:C_sum}
C= c_{4}\Bigl ( \frac{n_0}{2^{\bar{\leel}}} \Bigr)^{\frac{\kappa_2}{\kappa}}+ \frac{1}{c_{3}} \sum_{j=0}^{ {\leel_{0}'}-1} \Bigl \{ \Bigl ( \frac{n_0}{2^j} \Bigr)^{1-(\kappa)} \vee \frac{n_0}{2^{\bar{\leel}}} \Bigr\}^{\frac{\kappa_2}{\kappa}}+ 2c_{4} \sum_{j=1}^{ {\leel_{0}'}-1} \Bigl ( \frac{n_0}{2^j} \Bigr)^{\frac{\kappa_2(1-(\kappa))}{\kappa}}\,,
\end{equation}
and
\begin{equation}\label{equ:var_sum}
\mathcal{V}_{n_0}'^{1/2} = \mathcal{V}_{R}^{1/2} + \sum_{j=0}^{ {\leel_{0}'}-1} \mathcal{V}_{Y,j}^{1/2} + \sum_{j=1}^{ {\leel_{0}'}-1} \mathcal{V}_{W,j}^{1/2}.
\end{equation}
We now select the maximal tree depth parameter $\bar{\leel}$.
To accommodate the distinct scaling required for the elementary block sizes in each regime, we define:
\begin{equation*}
\bar{\leel}=\inf \left\{h\in \mathbb{N} : 2^{h}\ge n_0^{\kappa } (\log n_0)^{\kappa/\kappa_{1}}
\right\}
\end{equation*}
where $c_8 > 0$ is a structural constant.
With this choice, the elementary block size satisfies $n_0 2^{-\bar{\leel}} \ge \frac{1}{2} n_0^{1-\kappa}(\log n_0)^{-\kappa/\kappa_2}$.
Furthermore, we must ensure that this selection is compatible with the initial constraint \eqref{rest1l}, which requires $n_0 2^{-\bar{\leel}} \ge (2\vee \frac{4}{c_{0}})$.
Since $\kappa < 1$, the polynomial growth of $n_0^{1-\kappa}$ dominates the logarithmic term, so the condition $n_0 2^{-\bar{\leel}} \ge (2\vee \frac{4}{c_{0}})$ holds for sufficiently large $n_0 \ge \mu$.

Substituting this selected $\bar{\leel}$ into \eqref{equ:C_sum}, we can derive a upper bound of $C$.
Since $\kappa < 1$, the exponent $1-\kappa > 0$.
The sums evaluate as convergent geometric series.
Extracting the factor $n_0^{\kappa_2(1-\kappa)/\kappa}$, we compute the aggregated coefficient:
\begin{equation}\label{boundC}
C \le \biggl( \frac{c_{4} \bigl(1+2^{(\kappa-1)\frac{\kappa_2}{\kappa}}\bigr) + c_{3}^{-1}}{1 - 2^{(\kappa-1)\frac{\kappa_2}{\kappa}}} \biggr) n_0^{\frac{\kappa_2(1-\kappa)}{\kappa}} = \nu_{n_0} \bar{m} n_0^{\frac{\kappa_2(1-\kappa)}{\kappa}},
\end{equation}
where $\nu_{n_0}$ here is a strict constant.

For the variance term, we must bound the sum of the local standard deviations $\sum_{j=0}^{\leel_0'-1} \mathcal{V}_{\mathcal{K}_j}^{1/2}$ embedded within $\sum_j \mathcal{V}_{Y,j}^{1/2}$ in \eqref{P6prop3}.
Recall that our decomposition is built upon the deterministic optimal shift sequence $\set{u_h}$ selected at the beginning of the proof.
It guarantees that:
\begin{equation*}
\sum_{j=0}^{\leel_0'-1} \mathcal{V}_{\mathcal{K}_j}^{1/2} \le 4 \mathcal{V}_{n_0}^{1/2} \,.
\end{equation*}
Summing the remaining error terms over the tree depth allows us to unify the total variance proxy as:
\begin{equation}\label{equ:var_sum2}
\mathcal{V}_{n_0}'^{1/2} \le 5\mathcal{V}_{n_0}^{1/2} + \bar{m} \biggl( \frac{2^{2+\frac{\kappa_2}{\kappa}}}{1 - 2^{-\frac{\kappa^2+\kappa_2}{2\kappa}}} + c_4 \biggr) n_0^{1+\frac{\kappa_{1}}{\kappa}}\exp \Bigl(- \frac{c_1^{\kappa_2}}{2^{2+\kappa_2}} (n_0 2^{-\bar{\leel}})^{\kappa_{1}}\Bigr)\,.
\end{equation}

Substituting $n_0 2^{-\bar{\leel}} \ge \frac{1}{2}n_0^{1-\kappa}(\log n_0)^{-\kappa /\kappa_{1}}$, we have
\begin{equation*}
\mathcal{V}_{n_0}'^{1/2} \le 5\mathcal{V}_{n_0}^{1/2} + c_5 \exp \Bigl(- \frac{2c_1^{\kappa_2}}{2^{2+2\kappa_2}} n_0^{\kappa_{1}(1-\kappa)}(\log n_0)^{-\kappa} \Bigr),
\end{equation*}
with some constant $c_5 > 0$ and $c_6$. By the basic inequality $(a+b)^2 \le 2 a^2 + 2 b^2$,  it holds that
\begin{equation}\label{sigmagene}
\mathcal{V}_{n_0}' \le 50\mathcal{V}_{n_0}+c_{5} \exp \bigl(-c_{6}n_0^{\kappa_{1}(1-\kappa)}(\log n_0)^{-\kappa}\bigr),
\end{equation}
for some constants $c_5, c_6 > 0$.
Starting from the decomposition \eqref{P3prop3} and the bounds \eqref{P5prop3}, \eqref{P6prop3} and \eqref{P7prop3}, we aggregate the contributions of the terms by using Lemma \ref{breta} given in the appendix.
Then, by taking into account the bounds \eqref{boundC} and \eqref{sigmagene}, Proposition \ref{propinter_block_unified} follows across both regimes.
\end{proof}

The following two lemmas provide the fundamental algebraic machinery for decoupling dependent sequences and subsequently aggregating their tail bounds.
Specifically, Lemma \ref{lemma:mix_ineq_mgf} quantifies the error incurred when approximating the moment generating function of a sum of dependent variables by the product of their independent marginals. It effectively translates the underlying dependence structure, measured via $L_1$ coupling distances, into an explicit exponential error term.
Complementing this decoupling step, Lemma \ref{breta} serves as a deterministic aggregation tool. It guarantees that when summing multiple random components that individually satisfy a Bernstein-type moment bound, their local variance and scale parameters accumulate linearly.
Both generic inequalities are similar to those utilized in \citet{merlevede2011weakly}.

\begin{lemma}\label{lemma:mix_ineq_mgf}
    Let $Y_{1}$, ..., $Y_{h}$ be real-valued random variables such that $|Y_i| \le \trunc_i$ almost surely for every $i \in \{1, \dots, h\}$.
    For every $i\in \{1, \dots, h\}$, let ${\mathcal{\trunc}}_{i}=\sigma (Y_{1},...,Y_{i})$ and for $i\ge 2$, let $\widetilde{Y}_{i}$ be the coupled random variable that is independent of ${\mathcal{\trunc}}_{i-1}$ and distributed as $Y_{i}$.
    Then for any real $t$,
    \begin{equation*}
    \Bigl|\mathbb{E}\exp \Big (t\sum_{i=1}^{h}Y_{i}\Big)-\prod_{i=1}^{h}\mathbb{E}\exp (tY_{i}) \Bigr| \le |t|\exp \Big(|t| \sum_{i=1}^h \trunc_i \Big)\sum_{j=2}^{h}\mathbb{E}|Y_{j}-\widetilde{Y}_{j}|.
    \end{equation*}
\end{lemma}

\begin{proof}[Proof of Lemma~\ref{lemma:mix_ineq_mgf}]
    Let $W_j = \sum_{i = 1}^j Y_i$.
    By induction, the difference can be decomposed as a telescoping sum:
    \begin{align*}
        & \mathbb{E}(e^{t W_h}) - \prod_{i=1}^h \mathbb{E}(e^{t Y_i}) \\
        = & \{\mathbb{E}(e^{t W_h}) - \mathbb{E}(e^{t W_{h - 1}}) \mathbb{E}(e^{t Y_h})\} + \{\mathbb{E}(e^{t W_{h-1}}) - \mathbb{E}(e^{t W_{h - 2}}) \mathbb{E}(e^{t Y_{h-1}}) \} \mathbb{E}(e^{t Y_h}) \\
        & + \cdots + \{\mathbb{E}(e^{t W_{2}}) - \mathbb{E}(e^{t W_{1}}) \mathbb{E}(e^{tY_2})\} \prod_{i=3}^h \mathbb{E}(e^{t Y_i}) \\
        = & \sum_{j = 2}^h \{\mathbb{E}(e^{t W_{j}}) - \mathbb{E}(e^{t W_{j - 1}}) \mathbb{E}(e^{tY_j})\} \prod_{i=j+1}^h \mathbb{E}(e^{t Y_i})
    \end{align*}
    where the product term is conventionally defined as $1$ when $j = h$.
    Notice that for any $j$, since $|Y_i| \le \trunc_i$ almost surely, the product of expectations is bounded by:
    \begin{equation*}
        \Big|\prod_{i=j+1}^h \mathbb{E}(e^{t Y_i})\Big| \le \prod_{i=j+1}^h \mathbb{E}(e^{|t| \trunc_i}) \le \exp\Bigl(|t| \sum_{i=j+1}^h \trunc_i\Bigr).
    \end{equation*}
    We also evaluate the absolute difference inside the sum. Since $\widetilde{Y}_j$ is independent of $W_{j-1}$, we have $\mathbb{E}(e^{t W_{j - 1}}) \mathbb{E}(e^{tY_j}) = \mathbb{E}(e^{t W_{j - 1}} e^{t\widetilde{Y}_j})$. Thus:
    \begin{align*}
        & |\mathbb{E}(e^{t W_{j}}) - \mathbb{E}(e^{t W_{j - 1}}) \mathbb{E}(e^{tY_j})| = |\mathbb{E}\{e^{t W_{j-1}} (e^{t Y_j} - e^{t \widetilde{Y}_{j}})\}| \\
        \le & \mathbb{E} \bigl| e^{t W_{j-1}} (t Y_j - t \widetilde{Y}_{j}) \max(e^{tY_j}, e^{t\widetilde{Y}_{j}}) \bigr|.
    \end{align*}
    Applying the almost sure bounds $|W_{j-1}| \le \sum_{i=1}^{j-1} \trunc_i$ and $|Y_j| \le \trunc_j, |\widetilde{Y}_j| \le \trunc_j$, we obtain:
    \begin{align*}
        & \mathbb{E} \bigl| e^{t W_{j-1}} (t Y_j - t \widetilde{Y}_{j}) \max(e^{tY_j}, e^{t\widetilde{Y}_{j}}) \bigr| \\
        \le & \exp\Bigl(|t| \sum_{i=1}^{j-1} \trunc_i\Bigr) |t| \mathbb{E}|Y_j - \widetilde{Y}_{j}| \exp(|t| \trunc_j) \\
        = & |t| \exp\Bigl(|t| \sum_{i=1}^{j} \trunc_i\Bigr) \mathbb{E}|Y_j - \widetilde{Y}_{j}|.
    \end{align*}
    Combining these pieces back into the telescoping sum, the exponential factors match perfectly:
    \begin{align*}
        \Bigl|\mathbb{E}\exp \Big (t\sum_{i=1}^{h}Y_{i}\Big)-\prod_{i=1}^{h}\mathbb{E}\exp (tY_{i}) \Bigr|
        &\le \sum_{j = 2}^h \biggl( |t| \exp\Bigl(|t| \sum_{i=1}^{j} \trunc_i\Bigr) \mathbb{E}|Y_j - \widetilde{Y}_{j}| \biggr) \exp\Bigl(|t| \sum_{i=j+1}^h \trunc_i\Bigr) \\
        &= \sum_{j = 2}^h |t| \exp\Bigl(|t| \sum_{i=1}^h \trunc_i\Bigr) \mathbb{E}|Y_j - \widetilde{Y}_{j}| \\
        &= |t|\exp \Bigl(|t| \sum_{i=1}^h \trunc_i\Bigr)\sum_{j=2}^{h}\mathbb{E}|Y_{j}-\widetilde{Y}_{j}|.
    \end{align*}
\end{proof}

\begin{remark}
The coupling error $\sum_{i=2}^{h}\mathbb{E}|Y_{i}-\widetilde{Y}_{i}|$ in Lemma \ref{lemma:mix_ineq_mgf} effectively measures the coupling cost. If $Y_i$ is a functional of i.i.d. innovations $\{Y_{j}^{\mathrm{ind}}\}_{j \le i}$, i.e., $Y_i = g_i(\{Y_{j}^{\mathrm{ind}}\}_{j \le i})$ for some functional $g_i$.
Then we can construct $\widetilde{Y}_i = g_i(\{Y_{i}^{\mathrm{ind}}\} \cup \{\widetilde{Y}_{j}^{\mathrm{ind}}\}_{j \le i - 1})$ where $\{\widetilde{Y}_{j}^{\mathrm{ind}}\}$ are independent copies of $\{Y_{j}^{\mathrm{ind}}\}$.
We further denote $\widetilde{Y}_{i, i - m} = g_i(\{Y_{j}^{\mathrm{ind}}\}_{i - m < j\le i} \cup \{\widetilde{Y}_{j}^{\mathrm{ind}}\}_{j \le i - m})$.
Note that $\widetilde{Y}_{i, i-1} = \widetilde{Y}_{i}$.
We have $\mathbb{E}|Y_i - \widetilde{Y}_i| \le \lim_{m \rightarrow \infty} \{\mathbb{E}|Y_i - \widetilde{Y}_{i,i - m}| + \sum_{j=1}^{m - 1} \mathbb{E}|\widetilde{Y}_{i,i-j-1} - \widetilde{Y}_{i, i-j}|\} \le \sum_{j=1}^\infty \theta_{1}(j) = \Theta_{1}(1)$.
\end{remark}

\begin{lemma}\label{lem:additive_mgf}
\label{breta} Let $Z_0 , Z_1 , \ldots $ be a sequence of real valued random variables.
Assume that there exist positive constants $\sigma_0 , \sigma_1 , \ldots$ and $c_0, c_1 , \ldots $ such that, for any positive $i$ and any $t$ in $[0, 1/c_i)$,
\begin{equation*}
\log \mathbb{E} \exp (tZ_i) \le (\sigma_i t)^2 / (1-c_it) \, .
\end{equation*}
Then, for any positive $h$ and any $t$ in $[0, 1/(c_0 + c_1 + \cdots + c_h))$,
\begin{equation*}
\log \mathbb{E} \exp (t (Z_0 + Z_1 + \cdots + Z_h)) \le (\sigma t)^2 / (1-Ct) ,
\end{equation*}
where $\sigma = \sigma_0 + \sigma_1 + \cdots + \sigma_h $ and $C= c_0 + c_1 + \cdots + c_h$.
\end{lemma}
\begin{proof}[Proof of Lemma~\ref{lem:additive_mgf}]
    The proof is identical to that provided in \citet{merlevede2011weakly}.
\end{proof}

Now we are ready to proof Theorem \ref{thm:bern_fdm}.
Without loss of generality, we can assume $\bar{m} = 1$.
We denote the partial sums by $S_j = \sum_{i=1}^j X_i$.

\begin{proof}[Proof of Theorem \ref{thm:bern_fdm}] The proof will be divided into the following three cases.

\noindent \textbf{Case 1: The large deviation regime.}
Assume that $x \ge n^{\kappa_2/\kappa}$.
We apply a global truncation strategy by setting $\trunc = x / (2n)$.
By definition, $|X_{\trunc, i}| \le 2 m_i \trunc = m_i x / n \le x / n$.
Summing over $i$, we have $\sum_{i=1}^{n}|X_{\trunc, i}| \le x$.
This implies that the event $\{\sup_{j\le n}|S_{j}| \ge 2x\}$ can only occur if the truncation residual exceeds $x$:
\begin{equation*}
\mathbb{P}\Bigl(\sup_{j\le n}|S_{j}|\ge 2x\Bigr) \le \mathbb{P}\Bigl(\sum_{i=1}^{n}|X_{i}-X_{\trunc, i}|\ge x\Bigr) \,.
\end{equation*}
Applying Markov's inequality and using the integral tail bounds:
\begin{equation*}
\mathbb{P}\Bigl(\sum_{i=1}^{n}|X_{i}-X_{\trunc, i}|\ge x\Bigr) \le \frac{1}{x}\sum_{i=1}^{n}\mathbb{E}|X_{i}-X_{\trunc, i}| \le \frac{2\sum_{i=1}^n m_i}{x}\int_{\trunc}^{\infty}H(u)du \le \frac{2n}{x}\int_{\trunc}^{\infty}H(u)du  \,.
\end{equation*}
Recall that $\log H(u) = 1 - u^{\kappa_1}$.
The function $u \mapsto \log(u^2 H(u))$ is non-increasing for $u \ge (2/\kappa_1)^{1/\kappa_1}$.
Note that if $\trunc \ge (2/\kappa_1)^{1/\kappa_1}$, we have
\begin{equation}\label{B1decST*}
    \mathbb{P}\Bigl(\sum_{i=1}^{n}|X_{i}-X_{\trunc, i}|\ge x\Bigr) \le \frac{2n}{x}\int_{\trunc}^\infty H(u)du \le \frac{2n \trunc^2 H(\trunc)}{x} \int_{\trunc}^\infty u^{-2} du = \frac{2n \trunc H(\trunc)}{x}.
\end{equation}
We now verify that $\trunc$ can be chosen sufficiently large.
Suppose, by contradiction, that $\trunc < (2/\kappa_1)^{1/\kappa_1}$, which implies $x < 2n(2/\kappa_1)^{1/\kappa_1}$.
However, we are in the regime where $x \ge n^{\kappa_2/\kappa}$.
Combining these two inequalities yields
$$n^{\kappa_2/\kappa} < 2n(2/\kappa_1)^{1/\kappa_1}.$$
Dividing both sides by $n$ and using the identity $\kappa_2/\kappa - 1 = \kappa_2/\kappa_1$, we obtain
$$n^{\kappa_2/\kappa_1} < 2(2/\kappa_1)^{1/\kappa_1}.$$
This reveals that $n$ is bounded by an absolute constant, say $N_{\max} = 2^{\kappa_1 / \kappa_2} (2 / \kappa_1)^{1/\kappa_2}$.
Consequently, $x$ is also bounded by an absolute constant $X_{\max} = 2N_{\max} (2/\kappa_1)^{1/\kappa_1}$.
For such restricted, finite configurations of $n$ and $x$, we can show that the final target bound of the theorem to be trivially greater than or equal to $1$.
Since the theorem assumes $n \ge 4$, we have $\log n \ge \log 2$.
By choosing the absolute constant $C_1 \ge X_{\max}^\kappa / \log 2$, we ensure that $$x^\kappa / C_1 \le \log 2 \le \log n.$$
This implies $n \exp(-x^\kappa / C_1 ) \ge 1$, making the probability bound hold since probabilities cannot exceed $1$.
Thus, without loss of generality, we can safely assume $\trunc \ge (2/\kappa_1)^{1/\kappa_1}$.

Under this condition, the integral is bounded by $\trunc H(\trunc)$, yielding:
\begin{equation*}
\mathbb{P}\Bigl(\sup_{j\le n}|S_{j}|\ge 2x\Bigr) \le 2n x^{-1} \trunc H(\trunc) = H(\trunc) \le \exp(1 - \trunc^{\kappa_1}) \,.
\end{equation*}
From $x \ge n^{\kappa_2/\kappa}$, we have $n \le x^{\kappa/\kappa_2}$.
Substituting this upper bound for $n$ into the definition of $\trunc$:
\begin{equation*}
\trunc = \frac{x}{2n} \ge \frac{x}{2 x^{\kappa/\kappa_2}} = \frac{1}{2} x^{1 - \kappa/\kappa_2} \,.
\end{equation*}
Using the identity $1 - \kappa/\kappa_2 = \kappa (1/\kappa - 1/\kappa_2) = \kappa/\kappa_1$, we obtain:
\begin{equation*}
\trunc \ge \frac{1}{2} x^{\kappa/\kappa_1} \implies \trunc^{\kappa_1} \ge \frac{1}{2^{\kappa_1}} x^\kappa \,.
\end{equation*}
Consequently, the tail probability is bounded by:
\begin{equation*}
\mathbb{P}\Bigl(\sup_{j\le n}|S_{j}|\ge 2x\Bigr) \le 2 \exp\Bigl(1 - \frac{x^\kappa}{2^{\kappa_1}}\Bigr) \le n \exp\Bigl(-\frac{x^\kappa}{C_1}\Bigr) \,,
\end{equation*}
as soon as $C_1$ is chosen sufficiently large.

\noindent \textbf{Case 2: Moderate $x$ regime.} Let $\zeta = \mu \vee (2/\kappa_1)^{1/\kappa_2}$ where $\mu$ is defined by (\ref{borneA}).
Assume that $(4\zeta)^{\kappa_2/\kappa} \le x \le n^{\kappa_2/\kappa}$.
Let $q = n x^{-\kappa/\kappa_2}$.
Let
\begin{equation*}
A=\Bigl\lfloor \frac{n}{2q}\Bigr\rfloor,\quad k=\Bigl\lfloor \frac{n}{2A} \Bigr\rfloor \quad \text{and} \quad \trunc=H^{-1}(\tau_{Z^{(\sigma)}}(c^{-1/\kappa_2}A)) \,.
\end{equation*}
For any set of indices $\mathcal{K}$, denote $S_{\trunc}(\mathcal{K}) = \sum_{i\in \mathcal{K}}X_{\trunc, i}$.
For integer $i \in [1, 2k]$, let $I_{i}=\{1+(i-1)A, \dots, iA\}$.
Let also $I_{2k+1}=\{1+2kA, \dots, n\}$.
Set
\begin{equation*}
S^{(1)}(j)=\sum_{i=1}^{j}S_{\trunc}(I_{2i-1})\quad \text{and}\quad S^{(2)}(j)=\sum_{i=1}^{j}S_{\trunc}(I_{2i}) \,.
\end{equation*}
We then get the following inequality:
\begin{equation}
\sup_{j\le n}|S_{j}|\le \sup_{j\le k+1}|S^{(1)}(j)|+\sup_{j\le k}|S^{(2)}(j)|+ 2A \trunc+\sum_{i=1}^{n}|X_{i}-X_{\trunc, i}| \,. \label{decST}
\end{equation}
By the choice of $\trunc$, we have that if $A \ge (2/\kappa_1)^{1/\kappa_2}$, $\trunc \ge (2 / \kappa_1)^{1/\kappa_1}$.
Using \eqref{B1decST*} together with the sub-Weibull tail condition, we get for all positive $x$ that
\begin{equation*}
\mathbb{P}\Bigl(\sum_{i=1}^{n}|X_{i}-X_{\trunc, i}|\ge x\Bigr) \le 2n x^{-1}\trunc\exp(-A^{\kappa_2})\ \text{ for }\ A\ge (2/\kappa_1)^{1/\kappa_2} \,.
\end{equation*}
By the coupling technique, we construct independent random variables $(\tilde{S}_{\trunc}(I_{2i}))_{1\le i\le k}$ with the same distribution as the random variables $S_{\trunc}(I_{2i})$ such that
\begin{equation}
\mathbb{E}|S_{\trunc}(I_{2i})-\tilde{S}_{\trunc}(I_{2i})|\le A \tau_{Z^{(\sigma)}}(A)\le A \exp\big(-cA^{\kappa_2}\big) \,. \label{coupY}
\end{equation}
The same is true for the sequence $(S_{\trunc}(I_{2i-1}))_{1\le i\le k+1}$.
Hence for any positive $x$ such that $x \ge 2A \trunc$,
\begin{equation}\label{equ:final_decomp}
    \begin{aligned}
        \mathbb{P}\Bigl(\sup_{j\le n}|S_{j}|\ge 6x\Bigr) &\le x^{-1}A (2k+2)\exp\big(-cA^{\kappa_2}\big)+2n x^{-1}\trunc\exp(-A^{\kappa_2}) \\
&\quad +\mathbb{P}\Bigl(\max_{j\le k+1}\Big|\sum_{i=1}^{j}\tilde{S}_{\trunc}(I_{2i-1})\Big|\ge x\Bigr)+\mathbb{P}\Bigl(\max_{j\le k}\Big|\sum_{i=1}^{j}\tilde{S}_{\trunc}(I_{2i})\Big|\ge x\Bigr) \,.
    \end{aligned}
\end{equation}
For any positive $t$, due to the independence and since the variables are centered, $\{\exp(t\tilde{S}_{\trunc}(I_{2i}))\}_{i}$ is a submartingale.
Hence Doob's maximal inequality entails that for any positive $t$,
\begin{equation*}
\mathbb{P}\Bigl(\max_{j\le k}\sum_{i=1}^{j}\tilde{S}_{\trunc}(I_{2i})\ge x\Bigr)\le e^{-x t}\prod_{i=1}^{k}\mathbb{E}\Bigl(\exp(tS_{\trunc}(I_{2i}))\Bigr) \,.
\end{equation*}
To bound the Laplace transform of each random variable $S_{\trunc}(I_{2i})$, we apply Proposition \ref{propinter_block_unified} to the sequence over the block $I_{2i}$.
By Proposition \ref{propinter_block_unified}, the variance proxy for the block $I_{2i}$ is bounded by $\mathcal{V}_{I_{2i}}' = 50 \mathcal{V}_{I_{2i}} + c_{5} \exp(-c_{6} A^{\kappa_2(1-\kappa)}(\log A)^{-\kappa})$.
We derive that, if $A\ge \mu$ then for any positive $t$ such that $t < \nu^{-1} A^{\kappa_2(\kappa-1)/\kappa}$,
\begin{equation*}
\sum_{i=1}^{k}\log \mathbb{E}\Bigl(\exp(tS_{\trunc}(I_{2i}))\Bigr) \le \frac{t^2 \sum_{i=1}^k \mathcal{V}_{I_{2i}}'}{1-t\nu A^{\kappa_2(1-\kappa)/\kappa}} \,.
\end{equation*}
By the definition of the average variance $\mathcal{V}_n$, the sum of local effective variances over disjoint blocks is bounded by the total effective variance, i.e., $\sum_{i=1}^k \mathcal{V}_{I_{2i}} \le \mathcal{V}_n$.
Additionally, the sum of the block sizes satisfies $\sum_{i=1}^k A = kA \le n/2 \le n$.
Thus, the aggregated variance bound becomes
\begin{equation*}
\sum_{i=1}^k \mathcal{V}_{I_{2i}}' \le 50 \mathcal{V}_n + c_{5} n \exp\bigl(-c_{6} A^{\kappa_2(1-\kappa)}(\log A)^{-\kappa}\bigr) := \mathcal{V}_n'(A) \,.
\end{equation*}
Obviously the same inequalities hold true for the sums associated to $\{-X_{i}\}_{i\in \mathbb{Z}}$.
Now some usual computations optimizing the parameter $t$ lead to
\begin{equation*}
\mathbb{P}\Bigl(\max_{j\le k}\Big|\sum_{i=1}^{j}\tilde{S}_{\trunc}(I_{2i})\Big|\ge x\Bigr) \le 2\exp\Bigl(-\frac{x^2}{4\mathcal{V}_n'(A) + 2x\nu A^{\kappa_2(1-\kappa)/\kappa}}\Bigr) \,.
\end{equation*}
Similarly, we obtain that
\begin{equation*}
\mathbb{P}\Bigl(\max_{j\le k+1}\Big|\sum_{i=1}^{j}\tilde{S}_{\trunc}(I_{2i-1})\Big|\ge x\Bigr) \le 2\exp\Bigl(-\frac{x^2}{4\mathcal{V}_n'(A) + 2x\nu A^{\kappa_2(1-\kappa)/\kappa}}\Bigr) \,.
\end{equation*}

Note that $q = n x^{-\kappa/\kappa_2}$.
It follows that $2A \le x^{\kappa/\kappa_2}$, $A \ge 4^{-1} x^{\kappa/\kappa_2} \ge \zeta$ and $A^{\kappa_2(1-\kappa)/\kappa} \ge 4^{-\kappa_2(1-\kappa)/\kappa} x^{1-\kappa}$.
Since $\trunc \le (2A)^{\kappa_2/\kappa_1}$, we obtain $2A \trunc \le (2A)^{\kappa_2/\kappa} \le x$, which ensures that Eq.\eqref{equ:final_decomp} holds. We also have that the two factors in Eq.\eqref{equ:final_decomp}, $x^{-1}A (2k+2)$ and $2n x^{-1}\trunc$ are both bounded by $n$.

Substituting the lower bound $A \ge 4^{-1}x^{\kappa/\kappa_2}$ into the exponential remainder term of $\mathcal{V}_n'(A)$, the principal variance part $50 \mathcal{V}_n$ separates from the exponentially decaying part:
\begin{equation*}
c_{5} n \exp\bigl(-c_{6} A^{\kappa_2(1-\kappa)}(\log A)^{-\kappa}\bigr) \le c_{5} n \exp\Bigl(- C_4' x^{\kappa(1-\kappa)} (\log x)^{-\kappa}\Bigr) \,.
\end{equation*}
Combining this with the sub-Gaussian form and adjusting the constants $C_1, C_2, C_3, C_4$ appropriately, we obtain the required three-term exponential inequality as stated in Theorem \ref{thm:bern_fdm}.

\noindent \textbf{Case 3: $x \le (4\zeta)^{\kappa_2/\kappa}$.}
To end the proof, we mention that if $x \le (4\zeta)^{\kappa_2/\kappa}$, then
\begin{equation*}
\mathbb{P}\Bigl(\sup_{j\le n}|S_{j}|\ge x\Bigr) \le 1 \le 3 \exp\Bigl(-\frac{x^\kappa}{(4\zeta)^{\kappa_2}}\Bigr) \le n \exp\Bigl(-\frac{x^{\kappa}}{C_1}\Bigr) \,,
\end{equation*}
where $n \ge 4$ and $C_1 \ge (4\zeta)^{\kappa_2}$.

\end{proof}

\subsection{The approximation error decomposition in Section~\ref{sec:linear}}

\begin{proof}[Proof of Proposition~\ref{prop:linear}]

For the least square loss, we have the following result,
\begin{align*}
    \xi_I &= \sum_{i \in I} [\{u_i +  x_i^\top (\f_{I}^\ast - \hf_I)\}^2 - \epsilon_i^2] - \Ebb(\norm{u_I}^2) + \sigma_{\epsilon}^2\size{I} \\
    &= \sum_{i \in I} [u_i^2 - \epsilon_i^2 + 2 u_i x_i^\top (\f_{I}^\ast - \hf_I) + \{x_i (\f_{I}^\ast -\hf_I)\}^2] - \sum_{i \in I} \norm{\f_{i}^\ast - \f_{I}^\ast}_{\Sigma}^2\\
    &= \sum_{i \in I} [2 \epsilon_i x_i^\top (\f_{i}^\ast - \f_{I}^\ast) + \{x_i^\top (\f_{i}^\ast - \f_{I}^\ast)\}^2 - \norm{\f_{i}^\ast - \f_{I}^\ast}_{\Sigma}^2] \\
    &- 2 u_I^\top X_I (\hf_I - \f_{I}^\ast) + \norm{X_I (\hf_I - \f_{I}^\ast)}^2.
\end{align*}

By Condition \ref*{cond:model_and_loss_lasso}, $\max_{i \in I}\norm{\f_{i}^\ast - \f_{I}^\ast}_{\Sigma}^2 \le C_{f} s_n$, we have $\Var[\sum_{i \in I} 2 \epsilon_i x_i^\top (\f_{i}^\ast - \f_{I}^\ast) + \{x_i^\top (\f_{i}^\ast - \f_{I}^\ast)\}^2 - \norm{\f_{i}^\ast - \f_{I}^\ast}_{\Sigma}^2] \lesssim C_f s_n \sum_{i \in I} \norm{\f_{i}^\ast - \f_{I}^\ast}_{\Sigma}^2$.
Therefore we have the conclusion that
$$\xi_I = \mathcal{O}_{p}\Bigl(\sqrt{\sum_{i \in I} \norm{\f_{i}^\ast - \f_{I}^\ast}_{\Sigma}^2 s_n \log n} \Bigr) - 2 u_I^\top X_I (\hf_I - \f_{I}^\ast) + \norm{X_I (\hf_I - \f_{I}^\ast)}^2.$$
\end{proof}

Table~\ref{tab:bias} summarizes the discussion on the order of the bias terms in Proposition~\ref{prop:linear}.

\begin{table}[H]
\setlength\tabcolsep{0pt}
\begin{threeparttable}
\caption{Terms contributing to the bias for a given homogeneous segment $I$.}
\label{tab:bias}
\centering
\begin{tabular*}{\textwidth}{c@{\extracolsep{\fill}}*{4}{c}}
\toprule
Loss & Model & $\mathsf{Cross}$ & $\mathsf{Squared}$ \\
\midrule
In-sample & $\hf_I^\mathsf{lasso}(\lambda^\ast)$ & $\mathcal{O}_p(s_n\log p)$ & $\mathcal{O}_p(s_n\log p)$ \\
\\
In-sample & $\hf_I^\mathsf{lasso}(\hlam_I^{\cv})$ & \begin{tabular}{@{}>{$}l<{$}}
\mathcal{O}_p(\|u_I^\top X_I\|_\infty \cdot \sqrt{\size{\widehat{S}_I}+s_n} \cdot \|\hf_I - \f_{I}^\ast\|_2)\\
~~~~~~=\mathcal{O}_p(\sqrt{\size{I}\log p} \cdot \sqrt{\size{\widehat{S}_I}+s_n} \cdot \sqrt{s_n(\log^2 p)/\size{I}})\\
~~~~~~=\mathcal{O}_p(\sqrt{(\size{\widehat{S}_I}+s_n)s_n}\log^{3/2}p)
\end{tabular} & $\mathcal{O}_p(s_n\log^2 p)$ \\
\\
Out-of-sample & $\hf_{J_{-m, I}}^\mathsf{lasso}(\hlam_{J_{-m, I}}^{\cv})$ & $\mathcal{O}_p(\sqrt{s_n\log^2 p})$ & $\mathcal{O}_p(s_n\log^2 p)$ \\
\bottomrule
\end{tabular*}
\end{threeparttable}
\end{table}

\subsection{Proof of Corollary~\ref{coro:recycled_cv_lasso}}

\begin{proof}[Proof of Corollary~\ref{coro:recycled_cv_lasso}]
It is sufficient to show that the requirements of Theorem~\ref{thm:cvloss} holds with $\dacc = s_n \log p$.
    Under the conditions of the corollary, it is standard to show that for $I \in \homo$ and $m \in [M]$, there is a suitable tunning parameter $\lambda$ such that $\norm{\hf_{J_{-m, I}}^{\mathrm{lasso}} - \f_{I}^\ast}_{\Sigma}^2 \lesssim s_n \log p / \size{I}$, uniformly with probability at least $1 - n^{-C}$ for some constant $C > 0$. For example, one can refer to Lemma~18 in the appendix of \citet{qian2025reliever}. It guarantees that Condition~\ref{cond:predict_exist} holds with $\dacc = s_n \log p$.
    The requirements of the loss tails (Condition~\ref{cond:model_and_loss}) and the bounded changepoints (Condition~\ref{cond:changes}(b)) are guaranteed by Condition~\ref{cond:model_and_loss_lasso}.
    Condition~\ref{cond:changes}(a) is corresponding to Condition~\ref{cond:changes_lasso}.
    And for  Condition~\ref{cond:changes}(c), we have $C_{\Delta} = 2$ for the linear models with squared loss.
\end{proof}

\subsection{Proof of Corollary~\ref{coro:cvloss_kde}}\label{app:kde_theory}

First, we will prove the following lemma which is a detailed version of Lemma~\ref{lem:kde_err}. For a function $\f$ with support set $\Zcal$, denote the $L_q$ norm by $\norm{\f}_{L_q} = (\int_{\Zcal} |\f(\z)|^q \mathrm{d}\z)^{1/q}$ for $q \in [1, \infty)$ and $\norm{\f}_{L_\infty} = \sup_{\z \in \Zcal} |\f(\z)|$.

\begin{lemma}[Kernel density estimation]\label{lem:kde_err_full}
    Under Condition~\ref{cond:model_and_loss_kernel}, there exist some constants $C_1, C_2 > 0$ such that, with probability at least $1 - \exp(- C_1 \log n)$, the following hold.
    (a) Uniformly over all segments $I$ with $\size{I} \ge M$, all $m \in [M]$, and all $h \in \Lambda$,
    \begin{equation*}
        \norm{\hf_{J_{-m, I}, h} - \f_{{J_{-m, I}}}^\ast}_{L_\infty} \le C_2 \Bigl(\sqrt{\frac{p \log n}{h^p \size{J_{-m, I}}}} + \frac{p \log n}{h^p \size{J_{-m, I}}} + h^{r}\Bigr);
    \end{equation*}
    (b) For every nearly homogeneous segment $I \in \homo$, there exists $h_I^\ast \in \Lambda$ with $h_I^\ast \asymp \size{I}^{-1/(2r+p)}$ such that for every $m\in[M]$,
    \begin{equation*}
        \max\{\norm{\hf_{J_{-m, I}, h_I^\ast} - \f_{{J_{-m, I}}}^\ast}_{L_2}, \norm{\hf_{J_{-m, I}, h_I^\ast} - \f_{{J_{-m, I}}}^\ast}_{L_\infty}\} \le C_2 \size{J_{-m, I}}^{-\frac{r}{p + 2r}} (\log n)^{{1}/{2}}.
    \end{equation*}
\end{lemma}

\begin{proof}[Proof of Lemma~\ref{lem:kde_err_full}]
The core challenge is to bound the error of an estimator trained on a specific data subset against the true density of that segment.
Let $I$ be a segment of sample size $\size{I} \ge M$.
We first consider a split of $I$ into a training set $S = J_{-m, I}$ and a validation set $J_{m, I}$ with $\lfloor\size{I} / M \rfloor \le \size{J_{m, I}} \le \lfloor\size{I} / M \rfloor + 1$.

Under Condition~\ref{cond:model_and_loss_kernel}(e), we consider a fixed bandwidth $h \in [2 n^{-\frac{1}{p} - \frac{1}{2r + p}}, C_h]$.
Let $\hf_{S, h}^{\mathrm{raw}}$ denote the standard KDE constructed using observations in $S$:
\[
\hf_{S, h}^{\mathrm{raw}}(\z) = \frac{1}{|S| h^p} \sum_{i \in S} \mathcal{K}\left(\frac{\z - \z_i}{h}\right).
\]
We study the density estimation error at $\z$: $\abs{\hf_{S, h}^{\mathrm{raw}}(\z) - \f_{S}^\ast(\z)}$.
The error decomposes into a stochastic term (variance) and a deterministic term (bias):
\begin{equation*}
    \abs{ \hf_{S, h}^{\mathrm{raw}}(\z) - \f_{S}^\ast(\z) } \le \underbrace{\left| \Ebb[\hf_{S, h}^{\mathrm{raw}}(\z)] - \f_{S}^\ast(\z) \right|}_{(\text{I})} + \underbrace{\left| \hf_{S, h}^{\mathrm{raw}}(\z) - \Ebb[\hf_{S, h}^{\mathrm{raw}}(\z)] \right|}_{(\text{II})}.
\end{equation*}

For the bias term (I), because the kernel $\Kcal$ is adaptive to the Holder class $\Hcal(r, L, \Rbb^p)$,
    we have:
        \begin{equation*}
        (\text{I}) = \abs{\Ebb[\hf_{S, h}^{\mathrm{raw}}(\z)] - \f_{S}^\ast(\z)} = \Bigabs{\int_{\Rbb^p} \Kcal(u) \bigl(\f_S^\ast(\z - hu) - \f_S^\ast(\z)\bigr) \mathrm{d}u} \lesssim h^r,
    \end{equation*}

    For the variance term (II), let $K_i(\z) = h^{-p} \Kcal((\z - \z_i)/h)$.
    Since the kernel is bounded by $C_K$, the variance is controlled by its local second moment. Applying the change of variable $u = (\z - \x)/h$:
    \begin{equation*}
        \sigma_\z^2 = \Var(K_i(\z)) \le \int_{\Rbb^p} h^{-2p} \Kcal^2\Bigl(\frac{\z - \x}{h}\Bigr) \f_S^\ast(\x) \mathrm{d}\x = h^{-p} \int_{\Rbb^p} \Kcal^2(u) \f_S^\ast(\z - hu) \mathrm{d}u \lesssim h^{-p} \f_S^\ast(\z).
    \end{equation*}
    Applying Bernstein's inequality (Lemma~\ref{lem:bernstein}) with the envelope parameter $m = C_K h^{-p}$ and the  variance proxy $\sigma_\z^2$, we have with probability at least $1 - \exp(- C_1 p \log n)$:
    \begin{equation*}
        (\text{II}) = \biggabs{ \frac{1}{\size{S}} \sum_{i \in S} (K_i(\z) - \Ebb K_i(\z)) } \lesssim \sqrt{\frac{ \f_S^\ast(\z) p \log n}{h^p \size{S}}} + \frac{p \log n}{h^p \size{S}},
    \end{equation*}
    for some constant $C_1 > 0$.

By Condition~\ref{cond:model_and_loss_kernel}(a), $\Zcal \in \Rbb^p$ is a compact set with finite diameter $D_{\Zcal}$ and Lebesgue measure $V_{\Zcal}$.
We construct an $\epsilon$-Net $\Ncal_{\epsilon}$ of $\Zcal$ with size $\size{\Ncal_{\epsilon}} \le (D_{\Zcal} \epsilon^{-1})^p$ and $\epsilon = h^{p + r + 1}$.
For every $\z \in \Zcal$, there exists $\z' \in \Ncal_{\epsilon}$ such that $\norm{\z - \z'}_2 \le \epsilon$.
By the Lipschitz continuity of the kernel function $\Kcal$ in Condition~\ref{cond:model_and_loss_kernel}(c), we have $\abs{\hf_{S,h}(\z) - \hf_{S,h}(\z')} \le h^{-(p+1)} \epsilon$.
Because $h \ge e^{\frac{1}{e}} n^{-\frac{1}{p} - \frac{1}{2r + p}}$, with probability at least $1 - (D_{\Zcal} \epsilon^{-1})^p \exp(- C_1 p \log n) \ge 1 - \exp(- C_3 \log n)$,
\begin{equation*}
    \sup_{\z \in \Zcal} (\text{II}) \le \sqrt{\frac{C_2 C_{K} p \log n}{h^p \size{S}}} + \frac{C_2 C_{K} p \log n}{h^p \size{S}} + \frac{\epsilon}{h^{p+1}},
\end{equation*}
for some constants $C_2, C_3 > 0$.

By the choice of $\epsilon$, we have $\frac{\epsilon}{h^{p + 1}} \le h^{r}$.
In summary, with probability at least $1 - \exp(- C_3 \log n)$,
\begin{equation*}
        \sup_{\z \in \Zcal} \abs{ \hf_{S, h}^{\mathrm{raw}}(\z) - \f_{S}^\ast(\z) } \lesssim \sqrt{\frac{p \log n}{h^p \size{S}}} + \frac{p \log n}{h^p \size{S}} + h^{r}.
\end{equation*}

By Condition~\ref{cond:model_and_loss_kernel}(c), the kernel function $\Kcal(\cdot)$ decreases exponentially.
Since $\f_{S}^\ast(\z) = 0$ for $\z \notin \Zcal$, it ensures that
\begin{equation}\label{equ:kde_infinity}
    \sup_{\z \in \Rbb^p} \abs{ \hf_{S, h}^{\mathrm{raw}}(\z) - \f_{S}^\ast(\z) } \lesssim \sqrt{\frac{p \log n}{h^p \size{S}}}+ \frac{p \log n}{h^p \size{S}} + h^{r}.
\end{equation}

By choosing $h = p^{\frac{1}{2 r + p}}\size{S}^{- \frac{1}{2r + p}} \le C_4 \size{S}^{- \frac{1}{2r + p}}$ for some $C_4 > 0$, it holds that
\begin{equation}\label{equ:kde_infinity_opt}
        \sup_{\z \in \Rbb^p} \abs{ \hf_{S, h}^{\mathrm{raw}}(\z) - \f_{S}^\ast(\z) } \lesssim \size{S}^{-\frac{r}{2r + p}} (\log n)^{\frac{1}{2}}.
\end{equation}

By Condition~\ref{cond:model_and_loss_kernel}(a), $\Zcal$ has finite diameter and Lebesgue measure.
Therefore, by integrating over $\Zcal$, we have
\begin{equation*}
    \int_{\z \in \Zcal} \abs{ \hf_{S, h}^{\mathrm{raw}}(\z) - \f_{S}^\ast(\z) }^2 \mathrm{d} \z \lesssim \size{S}^{-\frac{2r}{2r + p}} (\log n)^{\frac{1}{2}}.
\end{equation*}
The exponential tail behavior of $\Kcal(\cdot)$ also implies that
\begin{equation*}
    \int_{\z \in \Rbb^p \setminus \Zcal} \abs{ \hf_{S, h}^{\mathrm{raw}}(\z) - \f_{S}^\ast(\z) }^2 \mathrm{d} \z \lesssim \size{S}^{-\frac{2r}{2r + p}} (\log n)^{\frac{1}{2}}.
\end{equation*}
Jointly, we have
\begin{equation}\label{equ:kde_l2}
    \norm{\hf_{S, h}^{\mathrm{raw}} - \f_{S}^\ast}_{L_2} = \Bigl\{\int_{\z \in \Rbb^p} \abs{ \hf_{S, h}^{\mathrm{raw}}(\z) - \f_{S}^\ast(\z) }^2 \mathrm{d} \z \Bigr\}^{\frac{1}{2}} \lesssim \size{S}^{-\frac{r}{2 r + p}} (\log n)^{\frac{1}{2}}.
\end{equation}

Finally, we extend these error bounds to the clipped density estimator $\hf_{S, h} = \max\{\hf_{S, h}^{\mathrm{raw}}, t_n\}$.
Because the true density is non-negative ($\f_{S}^\ast(\z) \ge 0$), the truncation operator induces at most an additive perturbation bounded by the threshold itself:
\begin{equation*}
    \abs{\hf_{S, h}(\z) - \f_{S}^\ast(\z)} = \abs{\max\{\hf_{S, h}^{\mathrm{raw}}(\z), t_n\} - \f_{S}^\ast(\z)} \le \abs{\hf_{S, h}^{\mathrm{raw}}(\z) - \f_{S}^\ast(\z)} + t_n.
\end{equation*}
Consequently, the uniform infinity norm is bounded by $\norm{\hf_{S, h} - \f_{S}^\ast}_{L_\infty} \le \norm{\hf_{S, h}^{\mathrm{raw}} - \f_{S}^\ast}_{L_\infty} + t_n$.
For the $L_2$ norm over the bounded support $\Zcal$, the error shifts by at most $\norm{\hf_{S, h} - \f_{S}^\ast}_{L_2} \le \norm{\hf_{S, h}^{\mathrm{raw}} - \f_{S}^\ast}_{L_2} + t_n \operatorname{Vol}(\Zcal)^{1/2}$.
Because the clipping threshold decays as $t_n = n^{-\beta}$ for a sufficiently large constant $\beta$, this deterministic perturbation is of order $\mathcal{O}(n^{-\beta})$, which decays overwhelmingly fast.
Thus, the additive $t_n$ terms are safely absorbed into the absolute constants, preserving the inequalities uniformly for the clipped estimator $\hf_{S, h}$.
By taking the union bound, these bounds hold uniformly for all $I$ with $\size{I} \ge \dacc$ and $S = J_{-m, I}$ with $m \in [M]$ and all $h \in \Lambda$, with probability at least $1 - \exp(- C_5 \log n)$ for some $C_5 > 0$.
\end{proof}

We are now ready to prove Corollary~\ref{coro:cvloss_kde}.

\begin{proof}[Proof of Corollary~\ref{coro:cvloss_kde}]
First, note that we adopt a clipping operation: $$\hf_{S, h}(\z) = \max\{\hf_{S, h}^{\mathrm{raw}}(\z), t_n\}$$ with the threshold $t_n = n^{-\beta}$.
It makes the estimator $\hf_{S, h}$ not a exact density function.
We need a further scaling operation $\hf_{S,h}^{\mathrm{scaled}}(\z) = c_n^{-1} \hf_{S, h}(\z)$ with $c_n = \int_{\Zcal} \hf_{S, h}(\z) \mathrm{d}\z$ to ensure that $\hf_{S,h}^{\mathrm{scaled}}$ is a valid density function.
Since $\Zcal$ is with finite lebesgue measure, we have $c_n = 1 + O(n^{-\beta})$.
Thus, $\abs{\log \hf_{S, h}(\z) - \log \hf_{S,h}^{\mathrm{scaled}}(\z)} = \log c_n = O(n^{-\beta})$, which is negligible as $\beta > 2$ is sufficiently large.
It ensures that the NLL loss evaluated at $\hf_{S, h}$ and $\hf_{S,h}^{\mathrm{scaled}}$ are asymptotically equivalent in the sense that the changepoint detection error bound holds with the same rate in theory.
In the following, we slightly abuse the notation and directly use $\hf_{S, h}$ to denote the scaled density estimator $\hf_{S,h}^{\mathrm{scaled}}$ for simplicity.

It is sufficient to show that the requirements of Theorem~\ref{thm:cvloss}, Conditions~\ref{cond:changes}, \ref{cond:model_and_loss} and \ref{cond:predict_exist} hold with $\dacc = n^{\frac{p}{p + 2 r}} \log n$.

For the NLL loss, the change signal becomes
$$\Delta_{k} = \mathrm{KL}(\f_{(\tau^\ast_{k},\tau^\ast_{k+1}]}^\ast \| \f_{(\tau^\ast_{k-1},\tau^\ast_{k}]}^\ast) \vee \mathrm{KL}(\f_{(\tau^\ast_{k-1},\tau^\ast_{k}]}^\ast \| \f_{(\tau^\ast_{k},\tau^\ast_{k+1}]}^\ast),$$
with $\mathrm{KL}(\f || g) = \int f(\z) \log\frac{\f(\z)}{g(\z)} \mathrm{d} \z$.
This naturally satisfies Condition~\ref{cond:changes}(c) with $C_\Delta = 2$.
Meanwhile, Condition~\ref{cond:changes_kde} guarantees that the remaining requirements regarding maximum signal strength and minimum segment length in Condition~\ref{cond:changes}(a) and (b) are satisfied.

\noindent\textbf{Verification of Condition~\ref{cond:model_and_loss}(a).}
For the negative log-likelihood loss $\ell(\z; \f) = -\log \f(\z)$, the centered empirical risk is:
\begin{equation*}
    s_{i, \f} = \ell(\z_i; \f) - \ell(\z_i; \f_i^\ast) - \Ebb[\ell(\z_i; \f) - \ell(\z_i; \f_i^\ast)] = - \log \frac{\f(\z_i)}{\f_i^\ast(\z_i)} + \mathrm{KL}(\f_i^\ast \| \f).
\end{equation*}
Note that we only need to evaluate this loss on the function $\f$ that takes the scaled and clipped KDE form as $\f(\cdot) = \hf_{S, h}(\cdot)$ for some set $S$ that is independent with the sample $\z_i$.

First, we analyze the extreme deviations of $Y_i = \log(\f_i^\ast(\z_i) / \f(\z_i))$.
For the upper bound, since the estimator is bounded from below by $n^{-\beta}$ and the true density is uniformly bounded by $C_f$ (Condition~\ref{cond:model_and_loss_kernel}(a)), we deterministically have $Y_i \le \log(C_f / n^{-\beta}) \lesssim \beta \log n$.
For the lower bound, the KDE is bounded from above by some polynomial rate $C_K h^{-p} \le C_K n^{\beta'}$ for some $\beta' > 0$ by the choice of the candidate set $\Lambda$. Thus, $-Y_i \le \beta' \log n - \log \f_i^\ast(\z_i)$.
The only unbounded randomness stems from $-\log \f_i^\ast(\z_i)$. For any $x > 0$:
\begin{align*}
    \Pbb_{\z_i \sim \f_i^\ast}\big( -\log \f_i^\ast(\z_i) > x \big)
    &= \int_{\{\z:0<\f_i^\ast(\z) < e^{-x}\}} \f_i^\ast(\z) \mathrm{d}\z \\
    &\le \int_{\{\z:0< \f_i^\ast(\z) < e^{-x}\}} \frac{e^{-2x}}{\f_i^\ast(\z)} \mathrm{d}\z \le e^{-2x} \int_{\Zcal} \frac{1}{\f_i^\ast(\z)} \mathrm{d}\z \le C_{\mathsf{inv}} e^{-x}.
\end{align*}
This guarantees that $Y_i$ is inherently a sub-exponential random variable. Consequently, the centered variable $s_{i, \f}$ satisfies the sub-Weibull tail condition with $\kappa_1 = 1$. By setting the envelope parameter $m_{i, \f} = C\log n$ for a sufficiently large constant $C$, we have $\Pbb(|s_{i, \f}| / m_{i, \f} > x) \le \exp(1 - x)$. Given the optimal localization rate $\dacc = n^{p/(2r+p)} \log^2 n$, the requirement $m_{i, \f} \asymp \log n \lesssim \dacc / \log n$ holds.

We now rigorously bound the variance proxy $\sigma_{i, \f}^2 = \Var(s_{i, \f}) \le \Ebb \big[ \big( \log \frac{\f(\z_i)}{\f_i^\ast(\z_i)} \big)^2 \big]$.
Consider the non-negative function $h(x) = x - 1 - \log x \ge 0$ defined for all $x > 0$.
Substituting $x = \f(\z) / \f_i^\ast(\z)$ yields the Bregman integral:
\begin{align*}
    \int_{\Zcal} \f_i^\ast(\z) h\bigg( \frac{\f(\z)}{\f_i^\ast(\z)} \bigg) \mathrm{d}\z
    &= \int_{\Zcal} \f_i^\ast(\z) \bigg( \frac{\f(\z)}{\f_i^\ast(\z)} - 1 - \log \frac{\f(\z)}{\f_i^\ast(\z)} \bigg) \mathrm{d}\z \\
    &= \mathrm{KL}(\f_i^\ast \| \f).
\end{align*}
It is straightforward to verify that $(\log x)^2 \le 2 h(x)$ for all $x \ge 1$, and $(\log x)^2 \le (2+\abs{\log x}) h(x)$ for $x \in (0, 1)$.
Given that $\f(\z) \ge n^{-\beta}$ and the uniform upper bound of the true density $\f_i^\ast(\z) \le C_f$ (Condition~\ref{cond:model_and_loss_kernel}(a)), the density ratio is deterministically bounded from below by $\f(\z)/\f_i^\ast(\z) \ge t_n/C_f \asymp n^{-\beta}$.
Consequently, we have $\abs{\log (\f(\z)/\f_i^\ast(\z))} \lesssim \log n$, which yields :
\begin{equation*}
    \bigg(\log \frac{\f(\z)}{\f_i^\ast(\z)}\bigg)^2 \lesssim (\log n) \bigg( \frac{\f(\z)}{\f_i^\ast(\z)} - 1 - \log \frac{\f(\z)}{\f_i^\ast(\z)} \bigg).
\end{equation*}

Taking the integral with respect to $\f_i^\ast$ over the entire domain $\Zcal$, we have:
\begin{equation*}
    \sigma_{i, \f}^2 \le \int_{\Zcal} \f_i^\ast(\z) \bigg(\log \frac{\f(\z)}{\f_i^\ast(\z)}\bigg)^2 \mathrm{d}\z \lesssim (\log n) \mathrm{KL}(\f_i^\ast \| \f) .
\end{equation*}
By setting $c_n \asymp \log n$, $\beta > 2$ and $\dacc \asymp n^{p/(2r+p)} \log^2 n$, we have Condition~\ref{cond:model_and_loss}(a) holds.

\noindent\textbf{Verification of Condition~\ref{cond:model_and_loss}(b).}
For any indices $i$ and $j$, the expected excess risk evaluated across different distributions is:
\begin{equation*}
    \overline{\ell}_j(\f) - \overline{\ell}_j(\f_i^\ast) = \int_{\Zcal} f_j^\ast(\z) \log \frac{f_i^\ast(\z)}{\f(\z)} \mathrm{d} \z.
\end{equation*}
By adding and subtracting $f_i^\ast(\z)$ within the integral, we decouple it into the baseline excess risk and a distribution-shift cross-term:
\begin{equation*}
    \overline{\ell}_j(\f) - \overline{\ell}_j(\f_i^\ast) = \mathrm{KL}(f_i^\ast \| \f) + \int_{\Zcal} (f_j^\ast(\z) - f_i^\ast(\z)) \log \frac{f_i^\ast(\z)}{\f(\z)} \mathrm{d} \z.
\end{equation*}
Applying the Cauchy-Schwarz inequality to the cross-term yields:
\begin{equation*}
    \Bigabs{ \int_{\Zcal} (f_j^\ast(\z) - f_i^\ast(\z)) \log \frac{f_i^\ast(\z)}{\f(\z)} \mathrm{d} \z } \le \norm{f_j^\ast - f_i^\ast}_{L_2} \Bignorm{\log \frac{f_i^\ast}{\f}}_{L_2}.
\end{equation*}
Because the true densities are uniformly bounded from above: $\sup_{\z \in \Zcal, i \in [n]} \f_i^\ast(\z) \le C_f$ (Condition~\ref{cond:model_and_loss_kernel}(a)),
the $L_2$ distance satisfies $\norm{f_j^\ast - f_i^\ast}_{L_2}^2 \lesssim \mathrm{KL}(f_j^\ast \| f_i^\ast)$.
Furthermore, as demonstrated in the variance bounds, we have $\norm{\log(f_i^\ast/\f)}_{L_2}^2 \lesssim (\log n) \mathrm{KL}(f_i^\ast \| \f)$.

Combining the above results, we obtain:
\begin{equation*}
    \abs{ \overline{\ell}_j(\f) - \overline{\ell}_j(\f_i^\ast) } \le C_{\ell} (\log n) \abs{ \overline{\ell}_i(\f) - \overline{\ell}_i(\f_i^\ast) } + C_{\ell} \abs{\overline{\ell}_j(\f_i^\ast) - \overline{\ell}_j(\f_j^\ast)},
\end{equation*}
for some constant $C_{\ell} > 0$, thereby satisfying Condition~\ref{cond:model_and_loss}(b).

\noindent\textbf{Verification of Condition~\ref{cond:predict_exist}.}
    Let $I \in \homo$ be a nearly homogeneous segment with $\size{I} \ge d_{\mathsf{m}}$. For a given split $m \in [M]$, let $S = J_{-m, I}$ denote the training set and $J = J_{m, I}$ denote the validation set, with $\size{S} \asymp \size{J} \asymp \size{I}$.

    The expected excess risk evaluates to the Kullback-Leibler divergence:
    \begin{equation}\label{eq:risk_kl_equiv}
        \abs{\risk_{J}(\hf_{S, h_I^\ast}) - \risk_{J}(\f_{S}^\ast)} \lesssim \size{J} \int_{\Zcal} \f_{S}^\ast(\z) \log \frac{\f_{S}^\ast(\z)}{\hf_{S, h_I^\ast}(\z)} \mathrm{d}\z = \size{J} \mathrm{KL}(\f_{S}^\ast \| \hf_{S, h_I^\ast}).
    \end{equation}

    Under Condition \ref{cond:model_and_loss_kernel}, Lemma \ref{lem:kde_err_full} guarantees the existence of an optimal bandwidth $h_I^\ast \asymp \size{I}^{-1/(2r+p)}$ such that, with probability at least $1 - \exp(-C_1 \log n)$,
    \begin{equation}\label{eq:kde_linfty_bound_trunc}
        \norm{\hf_{S, h_I^\ast} - \f_{S}^\ast}_{L_\infty} \le C_2 \size{S}^{-\frac{r}{2r+p}} (\log n)^{1/2} := \delta_S,
    \end{equation}
    for some constants $C_1, C_2 > 0$.
    Here we select the truncation parameter $\beta$ sufficiently large such that $t_n = n^{-\beta} \ll \delta_S$.

    We partition the support into a dominant region $\Zcal_1 = \{\z \in \Zcal : \f_{S}^\ast(\z) > 2\delta_S\}$ and a boundary region $\Zcal_2 = \{\z \in \Zcal : \f_{S}^\ast(\z) \le 2\delta_S\}$.
    Recall the inverse integrability assumption $\int_{\Zcal} 1/\f_{i}^\ast(\z) \mathrm{d}\z \le C_{\mathsf{inv}} < \infty$. The Lebesgue measure of the boundary region is controlled:
    \begin{equation*}
        \operatorname{Vol}(\Zcal_2) \le \int_{\Zcal_2} \frac{2\delta_S}{\f_{S}^\ast(\z)} \mathrm{d}\z \le 2 C_{\mathsf{inv}} \delta_S.
    \end{equation*}

    On $\Zcal_2$, utilizing the explicit truncation $\hf_{S, h_I^\ast}(\z) \ge t_n$, we have
    \begin{equation*}
        \int_{\Zcal_2} \f_{S}^\ast(\z) \log \frac{\f_{S}^\ast(\z)}{\hf_{S, h_I^\ast}(\z)} \mathrm{d}\z \le \operatorname{Vol}(\Zcal_2) \cdot \bigg( 2\delta_S \log \frac{2\delta_S}{t_n} \bigg) \lesssim \delta_S \cdot (\delta_S \log n) = \delta_S^2 \log n.
    \end{equation*}

    On $\Zcal_1$, the uniform bound \eqref{eq:kde_linfty_bound_trunc} guarantees $\hf_{S, h_I^\ast}(\z) \ge \f_{S}^\ast(\z) - \delta_S > \f_{S}^\ast(\z)/2$, ensuring the relative error satisfies $(\hf_{S, h_I^\ast} - \f_{S}^\ast)/\f_{S}^\ast \ge -1/2$. This permits the valid application of $-\log(1+x) \le -x + x^2$:
    \begin{align*}
        \int_{\Zcal_1} \f_{S}^\ast(\z) \log \frac{\f_{S}^\ast(\z)}{\hf_{S, h_I^\ast}(\z)} \mathrm{d}\z
        &\le \int_{\Zcal_1} \big( \f_{S}^\ast(\z) - \hf_{S, h_I^\ast}(\z) \big) \mathrm{d}\z + \int_{\Zcal_1} \frac{\big( \hf_{S, h_I^\ast}(\z) - \f_{S}^\ast(\z) \big)^2}{\f_{S}^\ast(\z)} \mathrm{d}\z.
    \end{align*}
    Because the integral of the estimator over the full support cannot significantly exceed 1, the linear bias term is bounded by the mass spillover into the boundary region $\Zcal_2$, which is bounded by $\delta_S \operatorname{Vol}(\Zcal_2) \lesssim \delta_S^2$. The quadratic variance term is controlled using the inverse integrability:
    \begin{equation*}
        \int_{\Zcal_1} \frac{\big( \hf_{S, h_I^\ast}(\z) - \f_{S}^\ast(\z) \big)^2}{\f_{S}^\ast(\z)} \mathrm{d}\z \le \delta_S^2 \int_{\Zcal_1} \frac{1}{\f_{S}^\ast(\z)} \mathrm{d}\z \le C_{\mathsf{inv}} \delta_S^2.
    \end{equation*}

    Aggregating the bounds on $\Zcal_1$ and $\Zcal_2$, the overall KL divergence is safely dominated by $\mathcal{O}(\delta_S^2 \log n)$. Substituting this back into \eqref{eq:risk_kl_equiv}, with probability at least $1 - \exp(-C_1 \log n)$:
    \begin{equation*}
        \abs{\risk_{J}(\hf_{S, h_I^\ast}) - \risk_{J}(\f_{S}^\ast)} \lesssim \size{J} \cdot \delta_S^2 \log n \asymp \size{I} \cdot \size{I}^{-\frac{2r}{2r+p}} \log^2 n = \size{I}^{\frac{p}{2r+p}} \log^2 n \le n^{\frac{p}{2r+p}} \log^2 n.
    \end{equation*}
    This verifies Condition \ref{cond:predict_exist}.

In summary, all the conditions of Theorem~\ref{thm:cvloss} are satisfied with $\dacc = n^{p/(2r+p)} \log^2 n$, which guarantees the desired changepoint detection error bound.

\end{proof}

\section{Additional numerical details}

\subsection{Data generating process of the Ridgeless regression example}\label{sec:ridgeless}

We set $n = 500$ and $p = 1000$ and the changepoint $\tau^\ast = 150$.
The residuals $\set{\epsilon_i}$ are independent and identically distributed (i.i.d.) from $\Ncal(0, 1)$ and the covariates $\set{\z_i}$ are i.i.d. from $\Ncal_p(\zero, \Sigma)$ where the covariance $\Sigma$ obeys the benign setting in \citet[Theorem 2]{bartlett2020benign}.
Let $\vbf_j$ be the eigenvector corresponding to the $j$-th largest eigenvalue $\lambda_j$ of $\Sigma$.
We set $\f_{(\tau^\ast_{0},\tau^\ast_{1}]}^\ast = \sum_{j=1}^5 \vbf_j / \sqrt{5}$ and $\f_{(\tau^\ast_{1},\tau^\ast_{2}]}^\ast = \sum_{j=1}^5  (-1)^j \vbf_j / \sqrt{5}$.
The responses $y_i$ follow the linear model in Section~\ref{sec:linear}.

\subsection{Changepoint detection under high-dimensional models with heavy-tailed distributions}
\begin{figure}[H]
    \centering
    \includegraphics[width=0.85\linewidth]{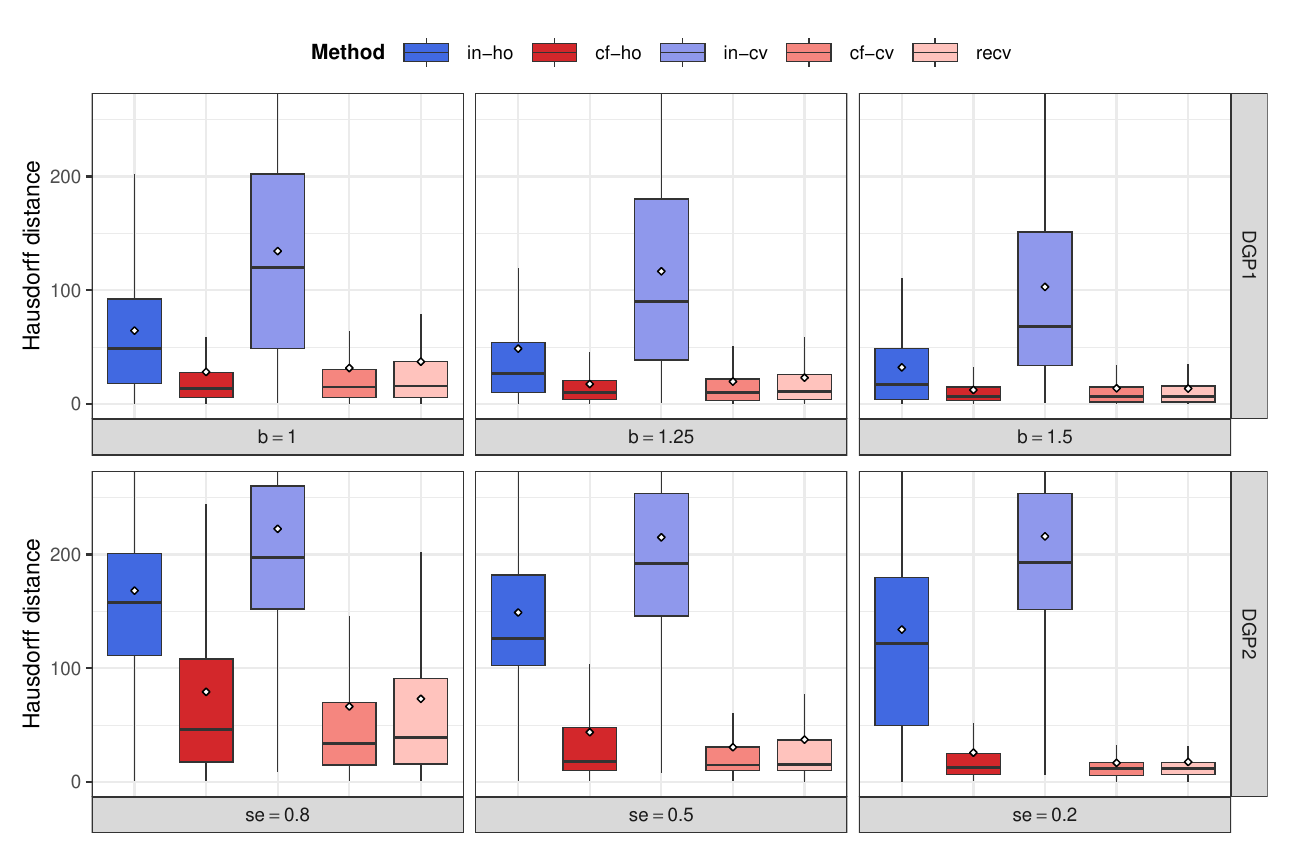}
\caption{\small Boxplot of empirical Hausdorff distances for various in-sample and cross-fitting methods in high-dimensional linear models with multiple changepoints and heavy-tailed samples.}
\label{fig:loc_err_lasso_homo_and_heter_heavy}
\end{figure}

\begin{figure}[H]
    \centering
    \includegraphics[width=0.85\linewidth]{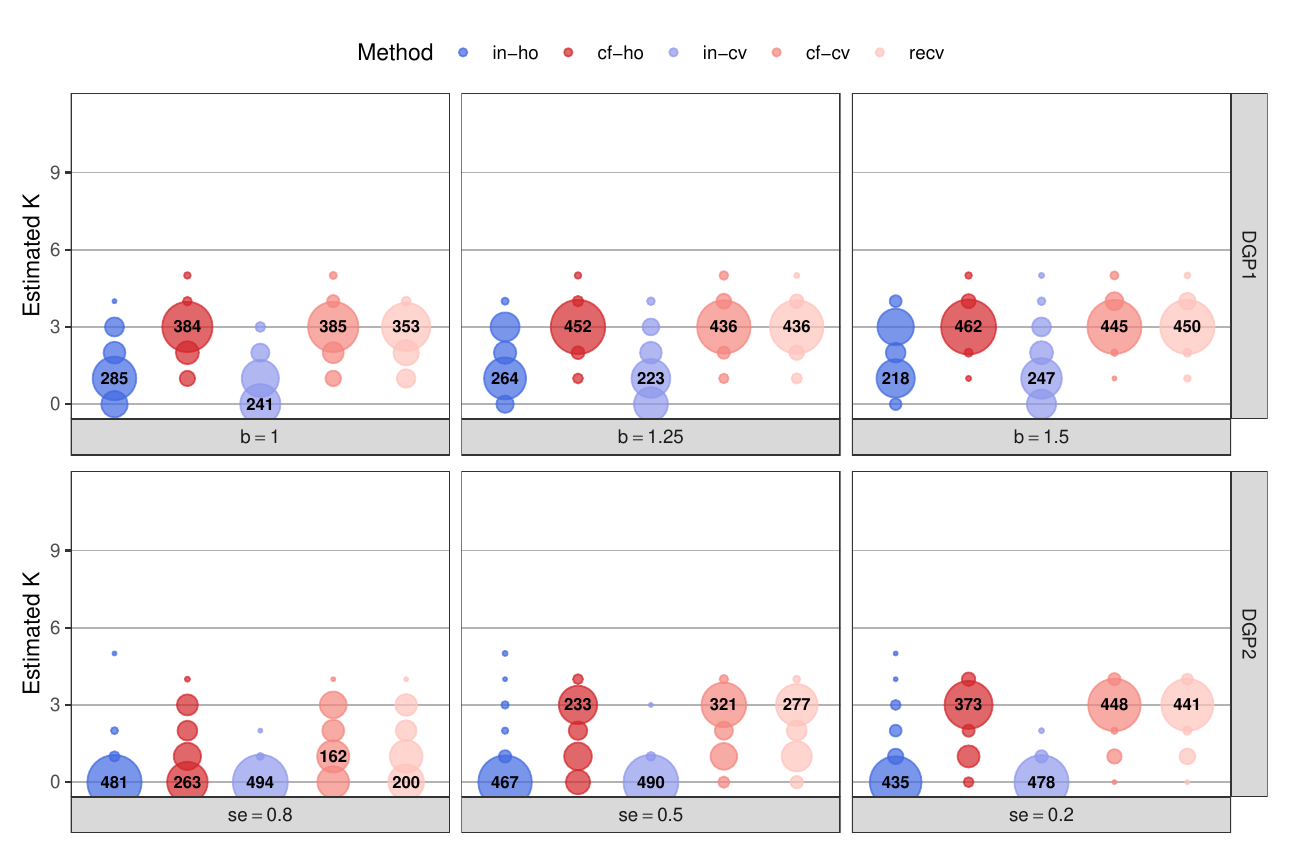}
\caption{\small Bubble plot of $\widehat{K}$ for various in-sample and cross-fitting methods in high-dimensional linear models with multiple changepoints and heavy-tailed samples.}
\label{fig:Kest_lasso_homo_and_heter_heavy}
\end{figure}

We investigate a heavy-tailed scenario under the two DGPs introduced in Section~\ref{sec:HD-common-vs-vary}.
Specifically, we replace the standard normal distributions used for generating $\{\x_i\}$ and $\{\epsilon_i\}$ with a standardized $t(4)$ distribution.
This substitution introduces heavy tails in distributions while keeping the first and second moments unchanged.
Figures~\ref{fig:loc_err_lasso_homo_and_heter_heavy} and \ref{fig:Kest_lasso_homo_and_heter_heavy} present the results under this heavy-tailed setting.
Notably, the overall performance exhibits no substantial difference from that observed in the light-tailed case.

\subsection{Estimation of the number of changepoints under high-dimensional models with temporal dependence}

\begin{figure}[H]
   \centering
   \includegraphics[width=0.85\linewidth]{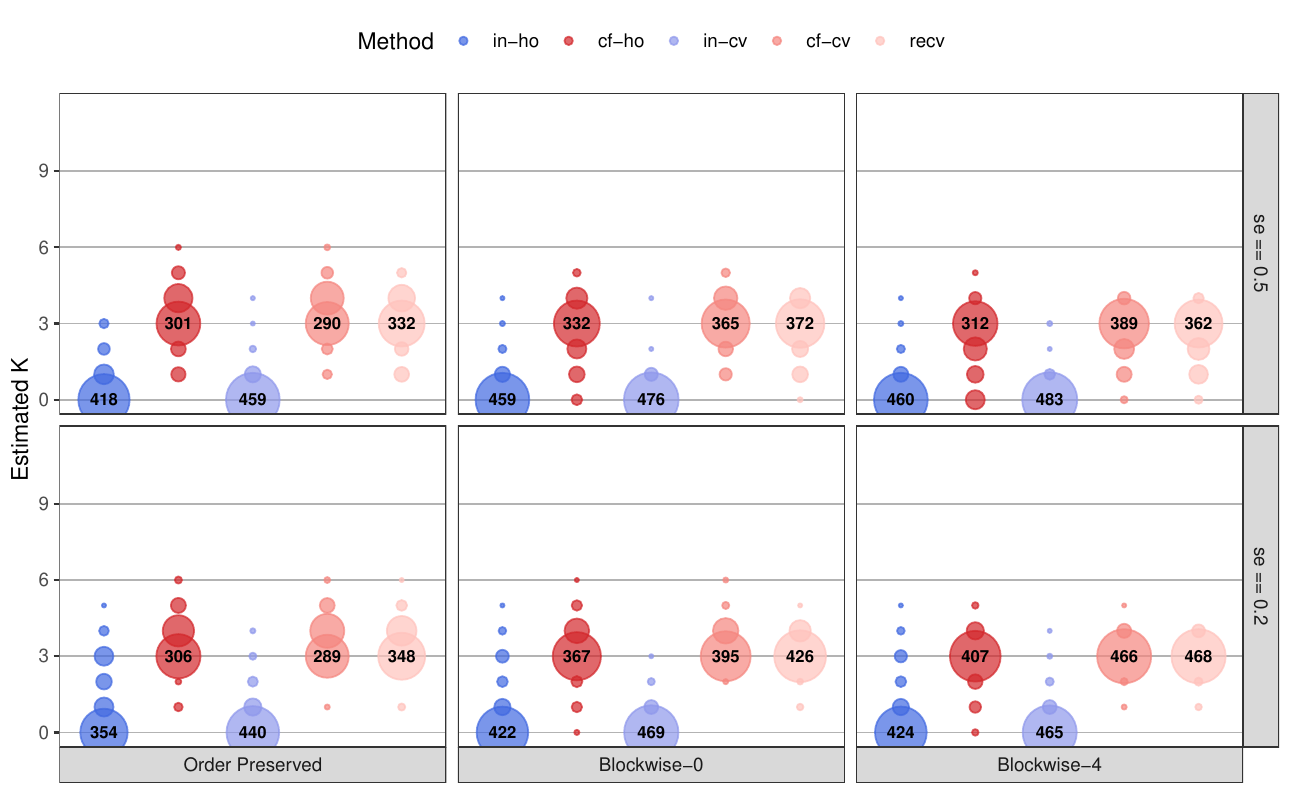}
\caption{\small Bubble plot of $\widehat{K}$ for various in-sample and cross-fitting methods in high-dimensional linear models with multiple changepoints and $AR(2)$ samples.}
\label{fig:bubble_lasso_heter_temporal}
\end{figure}

We study the fully data-driven estimates of $K^\ast$ as in Section \ref{sec:HD-common-vs-vary}. The results are consistent with those in the independent case, and the blockwise strategy also offers certain improvements. The results are summarized in Figure~\ref{fig:bubble_lasso_heter_temporal}.

\subsection{Multivariate nonparametric models}\label{sec:simul_nonpara}

We consider changepoint detection in multivariate nonparametric models under Model \((\ref{MCP})\) using the negative log-likelihood loss $\L(\z_I;\f)=-\sum_{i\in I}\log\f(\z_i)$. One direct implementation estimates the segment density $f_I^\ast$ itself, for example by the kernel density estimator in Section \(\ref{sec:kde_application}\).

Alternatively, we may work with the density ratio \(f_I^\ast(\z)/f_{(0,n]}^\ast(\z)\), since replacing \(f_I^\ast(\z)\) by this ratio changes the objective only by a term independent of the changepoint locations. This motivates a classifier-based construction. Relatedly, \cite{londschien2022changeforest} use probabilistic classifiers with binary labels for single-changepoint detection and extend the method to multiple changepoints via BS. Here we modify this idea to fit the global multiple-changepoint optimization framework.
Specifically, for each segment $I$, we we assign label 1 to observations in $I$ and label 0 to those outside $I$, and train a probabilistic classifier on the full sample $\{\z_i\}_{i=1}^n$.
Let \(\hat p_I(\z)\) denote the fitted probability that \(\z\) is assigned label 1, and define the corresponding density-ratio estimator $\hat r_I(\z):=n|I|^{-1}\hat p_I(\z)$, which targets \(f_I^\ast(\z)/f_{(0,n]}^\ast(\z)\). The corresponding in-sample empirical loss value is $-\sum_{i\in I}\log \hat r_I(\z_i)$.\textbf{}
The cross-fitting version is defined analogously.

For the probabilistic classifiers, we consider the following estimators:
\begin{itemize}
    \item \textbf{Random forest}, implemented in the R package \textsf{ranger}, with the number of trees in $\{25, 50, 100\}$ and the minimum node size in $\{5, 10, 15\}$;
    \item \textbf{Gradient boosting}, implemented in \textsf{lightgbm}, with the number of training rounds in $\{5, 15, 25\}$, the minimum gain to split in $\{0.1, 0.5\}$, and the learning rate in $\{0.05, 0.1\}$;
    \item \textbf{Multi-layer perceptron (MLP)}, with depth $3$, width $128$, and a logistic output layer, implemented in \textsf{torch}, utilizing SGD with batch size $50$ and learning rate in $\{0.01, 0.1, 1\}$.
\end{itemize}
Each hyperparameter configuration is selected using the hold-out or cross-validation strategies described in Section \(\ref{sec:simul_setup}\). We compare the methods \textbf{in-ho}, \textbf{in-cv}, \textbf{cf-ho}, and \textbf{recv}.

\begin{figure}[tb]
    \centering
    \includegraphics[width=0.85\linewidth]{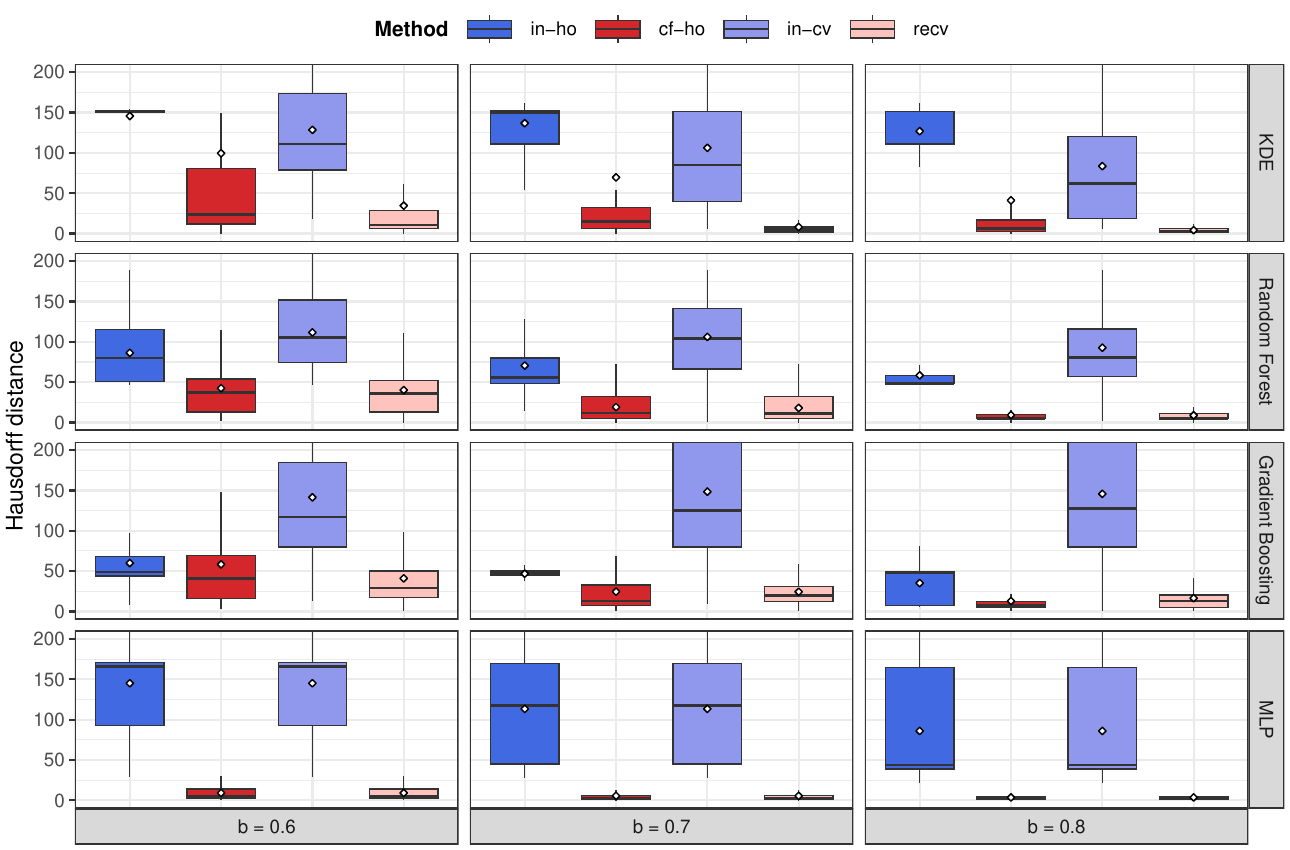}
    \caption{\small Boxplot of empirical Hausdorff distances for various in-sample and cross-fitting methods in nonparametric changepoint models.}
\label{fig:loc_err_mcp_clf}
\end{figure}

We generate independent samples $\z_i\in\mathbb{R}^p,i\in[n]$, with $(n,p)=(1000,10)$.
We set $K^\ast=3$ changepoints at locations $\{350, 500, 880\}$.
For $i \in (350, 500] \cup (880, 1000]$, $\z_i \sim \Ncal(0, \Ibf_{p})$, with $\Ibf_{p}$ being the identity matrix.
For $i \in (0, 350]$, $\z_i \sim \Ncal(0, \Sigma)$, where $\Sigma_{ij}=b^{\abs{i-j}}$, reflecting a correlation shift.
For $i \in (500, 880]$, $\z_i \sim \Ncal(b \id_{5}, \Ibf_{p})$, representing a mean shift.
The shift parameter $b$ varies over $\{0.4,0.5,0.6,0.7\}$.

\begin{figure}[tb]
   \centering
   \includegraphics[width=0.85\linewidth]{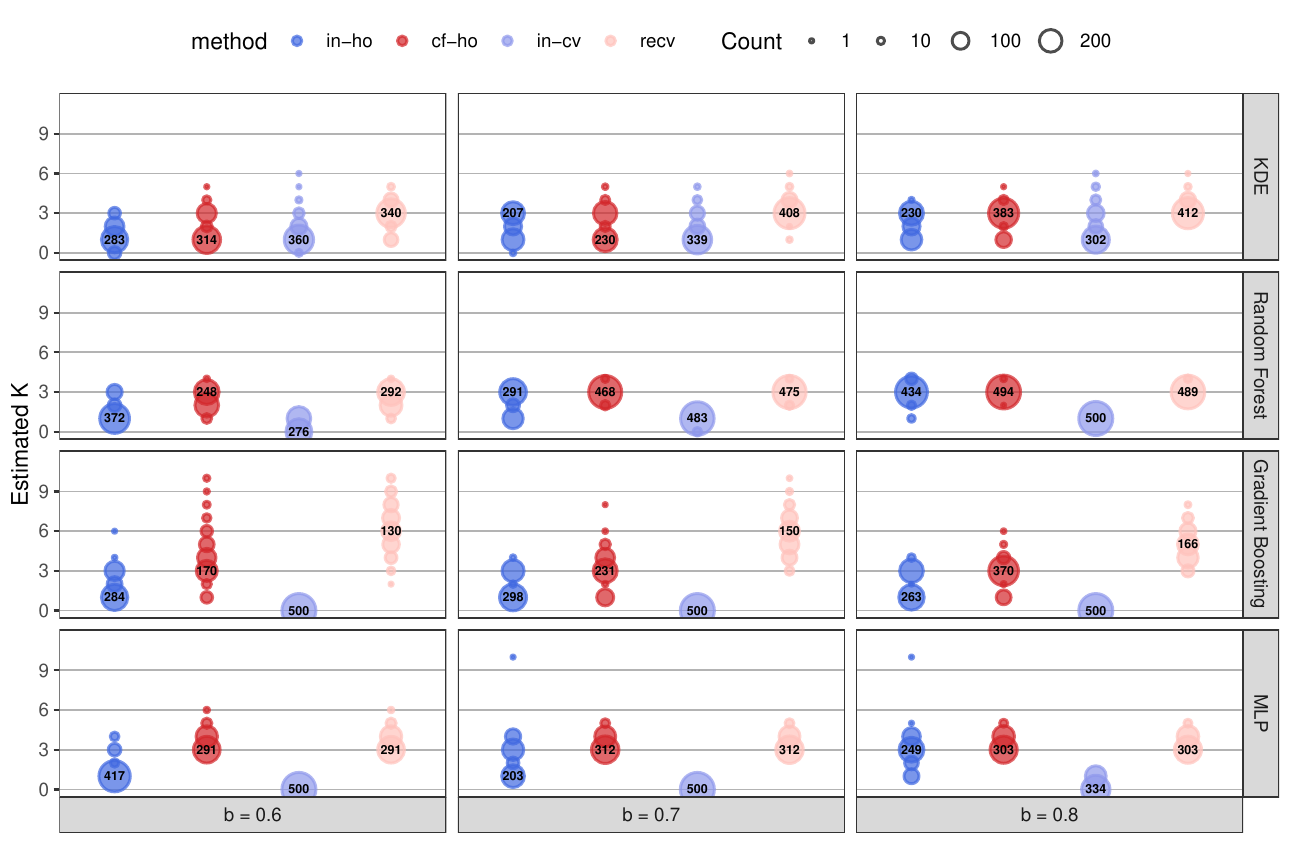}
   \caption{\small Bubble plot of $\widehat{K}$ for various in-sample and cross-fitting methods in nonparametric changepoint models.}
\label{fig:K_err_mcp_clf}
\end{figure}

As in Section \ref{sec:HD-common-vs-vary}, we assess localization accuracy with the true number of changepoints treated as known, so that the comparison focuses on changepoint localization.
Figure~\ref{fig:loc_err_mcp_clf} reports the Hausdorff distances for in-sample and cross-fitting approaches based on kernel density estimators and probabilistic classifiers.
In-sample methods often yield sub-optimal localization accuracy, reflecting the overfitting bias induced by complex fitting procedures, and their performance improves only modestly as the signal becomes stronger.
By contrast, cross-fitting methods consistently achieve better localization accuracy.
The fully data-driven results for estimating the number of changepoints are reported in Figure~\ref{fig:K_err_mcp_clf}.

\subsection{Array CGH data}

This section revisits the array comparative genomic hybridization (aCGH) dataset as an additional real-data illustration.
The dataset, accessible from the R package \texttt{ecp} \citep{MR3180567}, contains log intensity ratios of DNA copy numbers for $d=43$ bladder-tumor samples measured over $2215$ genomic loci. Our analysis is cohort-level: the goal is to identify genomic locations at which copy-number changes recur broadly across the cohort. We use this example mainly to compare in-sample and out-of-sample loss evaluations in a high-dimensional, nonparametric changepoint setting.

The data are split into two interlaced parts: $n=1108$ odd-indexed observations are used for changepoint detection, while $1107$ even-indexed observations are held out to assess the reliability of the detected changepoints.

In the detection stage, we implement both \textbf{in-cv} and \textbf{recv}, integrated with the random forest-based model fitting procedure as described in Section \ref{sec:simul_nonpara}.
Each approach optimizes the total loss with $K\in\{30, 38\}$.

Here, $K=38$ is determined using a hold-out strategy for \textbf{in-cv} by further splitting $n$ observations into two parts; one half implements \textbf{in-cv} with a sequence of candidate numbers of changepoints and generates a sequence of estimated changepoint models, while the other half calculates the prediction errors associated with each model to select the optimal number of changepoints that minimizes the prediction error.
Similarly, $K=30$ is selected by \textbf{recv}, which is consistent with the recommendation of \citet{wang+Samworth+2018+p57}.
It appears that \textbf{in-cv} tends to report more changepoints, potentially due to overfitting.

To assess the reliability of each detected changepoint $\htau_j$ for $j\in[K]$ by both \textbf{in-cv} and \textbf{recv}, we perform a sequence of local two-sample nonparametric tests comparing the test data within the intervals $((\htau_{j-1}+\htau_j)/2,\htau_j]$ and $(\htau_j,(\htau_j+\htau_{j+1})/2]$.
These tests are adjusted using Bonferroni corrections at significance level $\alpha\in\{0.10,0.05,0.01\}$, observing the independence among them.
We implement the two-sample kernel test as proposed by \cite{MR2913716}, using a Gaussian kernel with bandwidth $d$.
A rejection of this test implies that the detected changepoint is ``likely" a true positive; otherwise, it may be considered as a false positive.

Table \ref{tab:aGCH} details the number of ``false positives'' (FPs) and ``true positives'' (TPs) detected by each of the \textbf{in-cv} and \textbf{recv} approaches, with $K\in\{30,38\}$, alongside the (asymmetric) Hausdorff distance between the sets of TPs detected by both methods:
$$\Bigl\{\max_{j\in[K]} \min_{k\in[K]} |\htau_{\mathsf{incv},k} - \htau_{\mathsf{recv},j}|,\max_{k\in[K]} \min_{j\in[K]} |\htau_{\mathsf{incv},k} - \htau_{\mathsf{recv},j}|\Bigr\},$$
which measures the closest proximity of true positive detections between the two methods.
These results demonstrate that \textbf{recv} effectively identifies more reliable changepoints with fewer false positives compared to \textbf{in-cv}.
Moreover, for every TP detected by \textbf{in-cv}, \textbf{recv} tends to identify a corresponding TP nearby, as indicated by a smaller value of $\max_{k\in[K]} \min_{j\in[K]} |\htau_{\mathsf{incv},k} - \htau_{\mathsf{recv},j}|$.
Conversely, \textbf{in-cv} sometimes fails to detect a TP identified by \textbf{recv}, as reflected by a larger value of $\max_{j\in[K]} \min_{k\in[K]} |\htau_{\mathsf{incv},k} - \htau_{\mathsf{recv},j}|$.

\begin{figure}[tb]
   \centering
   \includegraphics[width=1\linewidth]{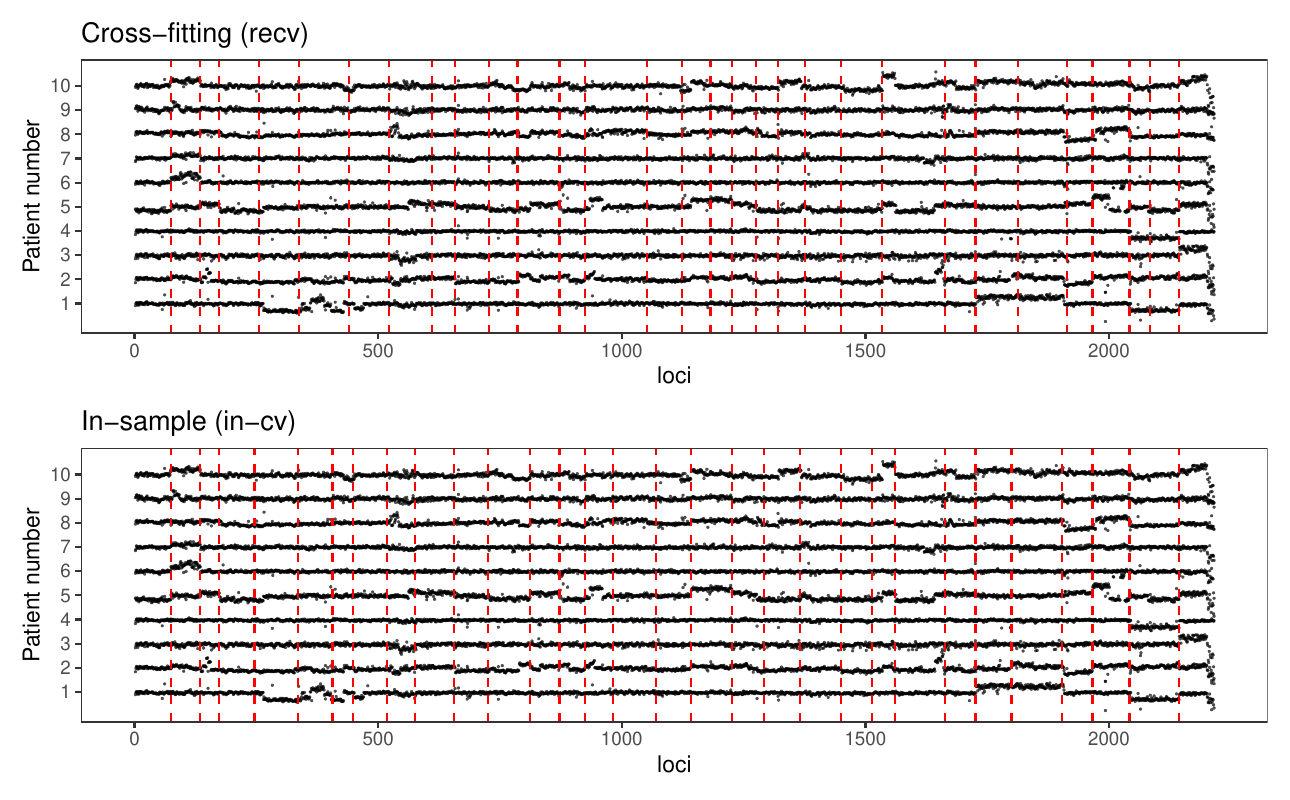}
   \caption{\small Changepoint estimates in the aCGH dataset with $K=30$.}
\label{fig:laci_acgh_est}
\end{figure}

Figure~\ref{fig:laci_acgh_est} shows the changepoint estimates for both the \textbf{in-cv} and \textbf{recv} methods using $K=30$. The changepoints detected by \textbf{recv} are broadly consistent with those reported in previous analyses of the same dataset \citep{wang+Samworth+2018+p57}.
\begin{table}[H]
\setlength\tabcolsep{0.6em}
\begin{center}
\caption{Number of FPs and TPs detected by the \textbf{in-cv} and \textbf{recv} approaches, for array CGH data.}
\begin{tabular}{lccccccc}
\toprule
&& \multicolumn{2}{c}{$\alpha = 0.1$} & \multicolumn{2}{c}{$\alpha = 0.05$} & \multicolumn{2}{c}{$\alpha = 0.01$} \\
&& \textbf{in-cv} & \textbf{recv} & \textbf{in-cv} & \textbf{recv} & \textbf{in-cv} & \textbf{recv} \\
\midrule
\multirow{3}{*}{$K=30$} & FP & 4 & 0 & 4 & 0 & 6 & 1\\
&TP & 26 & 30 & 26 & 30 & 24 & 29\\
&Hausdorff distance & \multicolumn{2}{c}{\{22, 9\}} & \multicolumn{2}{c}{\{22, 9\}} & \multicolumn{2}{c}{\{29, 9\}}\\
\cline{3-8}
\multirow{3}{*}{$K=38$}& FP & 12 & 5 & 14 & 5 & 16 & 9 \\
&TP & 26 & 33 & 24 & 33 & 22 & 29\\
&Hausdorff distance & \multicolumn{2}{c}{\{20, 6\}} & \multicolumn{2}{c}{\{20, 8\}} & \multicolumn{2}{c}{\{27, 17\}}\\
\bottomrule
\end{tabular}
\label{tab:aGCH}
\end{center}
\end{table}

These results suggest that out-of-sample loss evaluation provides more reliable cohort-level changepoint detection in this aCGH example.

\bibliographystyle{plainnat}
\bibliography{ref}

\end{document}

%% file: figures/fold_illus_reliever.tex
% --- [修正] 统一定义图例图标宏 ---
% baseline=-0.5ex 使得方块垂直居中于文字
% anchor=center 确保节点内容居中
\begin{tikzpicture}[
    % --- 全局样式 ---
    >=Latex,
    % 数据点样式
    data_point/.style={
        circle,
        draw=black!50,
        thick,
        minimum size=0.32cm,
        inner sep=0pt,
        font=\tiny\bfseries
    },
    % 轴样式
    axis_style/.style={
        ->,
        thick,
        black!80
    },
    % 文本标签
    label_text/.style={
        font=\sffamily\bfseries,
        align=right,
        anchor=east
    },
    % 区间标记样式
    proxy_marker/.style={
        thick,
        blue!80,
        dashed
    },
    % 维度标注线样式
    dim_line/.style={
        <->,
        thick
    }
]

    % --- 颜色定义 ---
    \definecolor{col1}{RGB}{102, 194, 165} % Green (Now F2 - Train)
    \definecolor{col2}{RGB}{217, 95, 2}    % Orange (Now F1 - Test, Emphasized)
    \definecolor{col3}{RGB}{141, 160, 203} % Purple (F3 - Train)

    % --- 参数设置 ---
    \def\totalwidth{14}  % 总宽度
    \def\proxystart{4}   % l
    \def\proxyend{10}    % r
    \def\sep{0.5}        % 间距

    % ==================================================
    % 1. Order-Preserved Split
    % ==================================================
    \def\yop{0}
    \node[label_text] at (-0.4, \yop + 0.5) {1. Order-Preserved};

    \draw[axis_style] (-0.2, \yop) -- (\totalwidth + 0.6, \yop);
    \node[circle, fill=black, inner sep=1.5pt] at (0, \yop) {};
    \node[circle, fill=black, inner sep=1.5pt] at (\totalwidth, \yop) {};
    % \node[circle, fill=black, inner sep=1.5pt, label=below:{$l_{end}$}] at (0, \yop) {};
    % \node[circle, fill=black, inner sep=1.5pt, label=below:{$r_{end}$}] at (\totalwidth, \yop) {};

    % 区间标注
    \draw[dim_line, black!70] (0, \yop + 1.6) -- (\totalwidth, \yop + 1.6)
        node[midway, fill=white, inner sep=2pt, font=\small\bfseries] {Original Interval $I$};
    \draw[dim_line, blue] (\proxystart, \yop + 1.1) -- (\proxyend, \yop + 1.1)
        node[midway, fill=white, inner sep=2pt, font=\small\bfseries, text=blue] {Relief Interval $R$};
    \draw[proxy_marker] (\proxystart, \yop - 0.8) -- (\proxystart, \yop + 1.1);
    \draw[proxy_marker] (\proxyend, \yop - 0.8) -- (\proxyend, \yop + 1.1);
    % \node[blue, font=\bfseries, anchor=south east] at (\proxystart, \yop - 0.5) {$l$};
    % \node[blue, font=\bfseries, anchor=south west] at (\proxyend, \yop - 0.5) {$r$};

    \foreach \i in {0, ..., 28} {
        \pgfmathsetmacro{\xpos}{\i * \sep}
        \pgfmathparse{int(mod(\i, 3) + 1)} \let\foldid\pgfmathresult
        \pgfmathsetmacro{\inproxy}{\xpos >= \proxystart - 0.1 && \xpos <= \proxyend + 0.1 ? 1 : 0}

        % [Color Swap] F1 is Test (Orange/col2), F2 is Train (Green/col1)
        \ifnum\foldid=1
            % Test (F1) - Orange (col2)
            \node[data_point, fill=col2!100, draw=col2!100] at (\xpos, \yop + 0.4) {};
        \else
            % Train (F2, F3) depends on Proxy
            \ifnum\inproxy=1
                \ifnum\foldid=2 \node[data_point, fill=col1!70, draw=col1!100] at (\xpos, \yop + 0.4) {}; \fi
                \ifnum\foldid=3 \node[data_point, fill=col3!70, draw=col3!100] at (\xpos, \yop + 0.4) {}; \fi
            \else
                \ifnum\foldid=2 \node[data_point, draw=col1!100, fill=white, dashed, thick] at (\xpos, \yop + 0.4) {}; \fi
                \ifnum\foldid=3 \node[data_point, draw=col3!100, fill=white, dashed, thick] at (\xpos, \yop + 0.4) {}; \fi
            \fi
        \fi
    }

    % % [修正] Part 1 图例 (Updated Colors)
    % \node[anchor=west, align=left, font=\scriptsize, inner sep=0pt] at (\totalwidth + 1.5, \yop + 0.5) {
    %     \legendBox{fill=col2!100, draw=col2!100} \textbf{Test (F1)}: in Full $I$\\
    %     \legendBox{fill=col1!70, draw=col1!100} \textbf{Train (F2/3)}: In Proxy $R$\\
    %     \legendBox{draw=col1!100, fill=white, dashed, thick} \textbf{Train}: Excluded \\
    %     \legendBox{draw=red!60, thick, pattern=north west lines, pattern color=red!60} \textbf{Buffer}: Removed neighbors
    % };

    % ==================================================
    % 2. Blockwise Split
    % ==================================================
    \def\yblk{-2}
    \node[label_text] at (-0.4, \yblk + 0.5) {2. Blockwise Split};

    \draw[axis_style] (-0.2, \yblk) -- (\totalwidth + 0.6, \yblk);
    \node[circle, fill=black, inner sep=1.5pt] at (0, \yblk) {};
    \node[circle, fill=black, inner sep=1.5pt] at (\totalwidth, \yblk) {};
    % \node[circle, fill=black, inner sep=1.5pt, label=below:{$l_{end}$}] at (0, \yblk) {};
    % \node[circle, fill=black, inner sep=1.5pt, label=below:{$r_{end}$}] at (\totalwidth, \yblk) {};

    \draw[proxy_marker] (\proxystart, \yblk - 0.8) -- (\proxystart, \yblk + 0.8);
    \draw[proxy_marker] (\proxyend, \yblk - 0.8) -- (\proxyend, \yblk + 0.8);
    \pgfmathsetmacro{\blkwidth}{(\proxyend - \proxystart) / 3}
    \pgfmathparse{int(floor((0 - \proxystart) / \blkwidth))} \let\startidx\pgfmathresult
    \pgfmathparse{int(floor((\totalwidth - \proxystart) / \blkwidth))} \let\endidx\pgfmathresult

    \foreach \k in {\startidx, ..., \endidx} {
        \pgfmathsetmacro{\bstart}{\proxystart + \k * \blkwidth}
        \pgfmathsetmacro{\bend}{\proxystart + (\k + 1) * \blkwidth}
        \pgfmathparse{int(mod(mod(\k, 3) + 3, 3) + 1)} \let\foldid\pgfmathresult
        \pgfmathsetmacro{\drawstart}{max(0, \bstart)}
        \pgfmathsetmacro{\drawend}{min(\totalwidth, \bend)}

        \ifdim \drawstart pt < \drawend pt
            \pgfmathsetmacro{\bcenter}{(\drawstart + \drawend) / 2}
            \pgfmathsetmacro{\inproxy}{\bcenter >= \proxystart && \bcenter <= \proxyend ? 1 : 0}

            % [Color Swap] F1 is Test (Orange)
            \ifnum\foldid=1
                \filldraw[fill=col2!50, draw=col2!100, thick] (\drawstart, \yblk + 0.1) rectangle (\drawend, \yblk + 0.7);
                \node[font=\tiny\bfseries] at (\bcenter, \yblk + 0.4) {F1};
            \else
                \ifnum\inproxy=1
                    \ifnum\foldid=2 \filldraw[fill=col1!40, draw=col1!80] (\drawstart, \yblk + 0.1) rectangle (\drawend, \yblk + 0.7); \fi
                    \ifnum\foldid=3 \filldraw[fill=col3!40, draw=col3!80] (\drawstart, \yblk + 0.1) rectangle (\drawend, \yblk + 0.7); \fi
                    \node[font=\tiny\bfseries] at (\bcenter, \yblk + 0.4) {F\foldid};
                \else
                    \ifnum\foldid=2 \filldraw[fill=white, draw=col1!50, dashed, thick] (\drawstart, \yblk + 0.1) rectangle (\drawend, \yblk + 0.7); \fi
                    \ifnum\foldid=3 \filldraw[fill=white, draw=col3!50, dashed, thick] (\drawstart, \yblk + 0.1) rectangle (\drawend, \yblk + 0.7); \fi
                    \node[font=\tiny\bfseries, text=gray] at (\bcenter, \yblk + 0.4) {F\foldid};
                \fi
            \fi
        \fi
    }

    % ==================================================
    % 3. Buffer Lag
    % ==================================================
    \def\ybuf{-4}
    \node[label_text] at (-0.4, \ybuf + 0.5) {3. Buffer Lag};

    \draw[axis_style] (-0.2, \ybuf) -- (\totalwidth + 0.6, \ybuf);
    \node[circle, fill=black, inner sep=1.5pt] at (0, \ybuf) {};
    \node[circle, fill=black, inner sep=1.5pt] at (\totalwidth, \ybuf) {};
    % \node[circle, fill=black, inner sep=1.5pt, label=below:{$l_{end}$}] at (0, \ybuf) {};
    % \node[circle, fill=black, inner sep=1.5pt, label=below:{$r_{end}$}] at (\totalwidth, \ybuf) {};

    \draw[proxy_marker] (\proxystart, \ybuf - 0.3) -- (\proxystart, \ybuf + 0.8);
    \draw[proxy_marker] (\proxyend, \ybuf - 0.3) -- (\proxyend, \ybuf + 0.8);
    \def\bufwidth{0.15}

    \foreach \k in {\startidx, ..., \endidx} {
        \pgfmathsetmacro{\bstart}{\proxystart + \k * \blkwidth}
        \pgfmathsetmacro{\bend}{\proxystart + (\k + 1) * \blkwidth}
        \pgfmathparse{int(mod(mod(\k, 3) + 3, 3) + 1)} \let\foldid\pgfmathresult
        \pgfmathsetmacro{\drawstart}{max(0, \bstart)}
        \pgfmathsetmacro{\drawend}{min(\totalwidth, \bend)}

        \ifdim \drawstart pt < \drawend pt
             \pgfmathsetmacro{\bcenter}{(\drawstart + \drawend) / 2}
             \pgfmathsetmacro{\inproxy}{\bcenter >= \proxystart && \bcenter <= \proxyend ? 1 : 0}

            % [Color Swap] F1 is Test (Orange)
            \ifnum\foldid=1
                \filldraw[fill=col2!50, draw=col2!100, thick] (\drawstart, \ybuf + 0.1) rectangle (\drawend, \ybuf + 0.7);
                \node[font=\tiny\bfseries] at (\bcenter, \ybuf + 0.4) {F1};
            \else
                \ifnum\inproxy=1
                    % Base blocks
                    \ifnum\foldid=2 \filldraw[fill=col1!40, draw=col1!80] (\drawstart, \ybuf + 0.1) rectangle (\drawend, \ybuf + 0.7); \fi
                    \ifnum\foldid=3 \filldraw[fill=col3!40, draw=col3!80] (\drawstart, \ybuf + 0.1) rectangle (\drawend, \ybuf + 0.7); \fi

                    % Buffer Logic (Neighbors of F1)
                    % F3 is LEFT neighbor of F1.
                    \ifnum\foldid=3
                        \fill[pattern=north west lines, pattern color=red!60] (\bend - \bufwidth, \ybuf + 0.1) rectangle (\bend, \ybuf + 0.7);
                        \draw[red!60, thick] (\bend - \bufwidth, \ybuf + 0.1) -- (\bend - \bufwidth, \ybuf + 0.7);
                    \fi

                    % F2 is RIGHT neighbor of F1.
                    \ifnum\foldid=2
                        \fill[pattern=north west lines, pattern color=red!60] (\bstart, \ybuf + 0.1) rectangle (\bstart + \bufwidth, \ybuf + 0.7);
                        \draw[red!60, thick] (\bstart + \bufwidth, \ybuf + 0.1) -- (\bstart + \bufwidth, \ybuf + 0.7);
                    \fi
                    \node[font=\tiny\bfseries] at (\bcenter, \ybuf + 0.4) {F\foldid};
                \else
                    \ifnum\foldid=2 \filldraw[fill=white, draw=col1!50, dashed, thick] (\drawstart, \ybuf + 0.1) rectangle (\drawend, \ybuf + 0.7); \fi
                    \ifnum\foldid=3 \filldraw[fill=white, draw=col3!50, dashed, thick] (\drawstart, \ybuf + 0.1) rectangle (\drawend, \ybuf + 0.7); \fi
                    \node[font=\tiny\bfseries, text=gray] at (\bcenter, \ybuf + 0.4) {F\foldid};
                \fi
            \fi
        \fi
    }
    % ==================================================
    % 4. 底部图例 (Global Legend) - [新增]
    % ==================================================
    % 位置计算：放在总宽度的一半处 (\totalwidth/2)，高度在最底部的图 (\ybuf) 下方
    \node[anchor=north, align=center, font=\scriptsize, inner sep=0pt] at (0.4*\totalwidth, \ybuf - 0.5) {
        % 第一行
        \legendBox{fill=col2!100, draw=col2!100} \textbf{Evaluation Folds (F1): $J_{m, I}$}
        \hspace{1em} % 控制两个图例之间的间距
        \legendBox{fill=col1!70, draw=col1!100} \legendBox{fill=col3!70, draw=col3!100} \textbf{Training Folds (F2/3): $J_{-m, R}$} \hspace{1em}
        % 第二行
        \legendBox{draw=col1!100, fill=white, dashed, thick} \textbf{Excluded Training Folds}: In $I \setminus R$ 
        \hspace{1em}
        \legendBox{draw=red!60, thick, pattern=north west lines, pattern color=red!60} \textbf{Buffer}
    };

\end{tikzpicture}

%% file: figures/cantor_set.tex
\begin{tikzpicture}[scale=0.95]
% Level 0
\draw[fill=gray!20] (0, 4) rectangle (12, 4.6);
\node at (6, 4.3) {Level 0: $I_{0,1}$ (Size ${n_{0}}$)};
% Level 1
\draw[fill=gray!20] (0, 2.5) rectangle (5.5, 3.1);
\node at (2.75, 2.8) {$I_{1,1}$};
\draw[pattern=north east lines] (5.5, 2.5) rectangle (6.5, 3.1);
\node[below] at (6, 2.5) {Gap $d_0$};
\draw[fill=gray!20] (6.5, 2.5) rectangle (12, 3.1);
\node at (9.25, 2.8) {$I_{1,2}$};

% Level 2
\draw[fill=gray!20] (0, 1) rectangle (2.5, 1.6);
\node at (1.25, 1.3) {$I_{2,1}$};
\draw[pattern=north east lines] (2.5, 1) rectangle (3.0, 1.6);
\node[below] at (2.75, 1) {$d_1$};
\draw[fill=gray!20] (3.0, 1) rectangle (5.5, 1.6);
\node at (4.25, 1.3) {$I_{2,2}$};

\draw[fill=gray!20] (6.5, 1) rectangle (9.0, 1.6);
\node at (7.75, 1.3) {$I_{2,3}$};
\draw[pattern=north east lines] (9.0, 1) rectangle (9.5, 1.6);
\node[below] at (9.25, 1) {$d_1$};
\draw[fill=gray!20] (9.5, 1) rectangle (12, 1.6);
\node at (10.75, 1.3) {$I_{2,4}$};

% Arrows
\draw[->] (2.75, 3.8) -- (2.75, 3.2);
\draw[->] (9.25, 3.8) -- (9.25, 3.2);
\draw[->] (1.25, 2.3) -- (1.25, 1.7);
\draw[->] (4.25, 2.3) -- (4.25, 1.7);
\draw[->] (7.75, 2.3) -- (7.75, 1.7);
\draw[->] (10.75, 2.3) -- (10.75, 1.7);

\end{tikzpicture}